%% file: prethesis.tex
\documentclass[12pt,oneside]{book}

\usepackage[utf8]{inputenc}
\usepackage{xcolor}
\usepackage{cite}
\usepackage{amsmath,amsbsy}
\usepackage{amssymb}
\usepackage{graphicx}
\usepackage{braket}
\usepackage{enumitem}
\usepackage[margin=2.5cm]{geometry}
\usepackage{tikz}
\usepackage{tikz-feynman}
\usepackage{contour}
\usetikzlibrary{snakes,decorations.pathmorphing}
\usepackage{comment}
\usepackage{cancel}
\usepackage[normalem]{ulem}
\usepackage{CormorantGaramond}
\usepackage{pdfpages}

\usepackage{float}
\usepackage{hyperref}
\usepackage{pifont}
\usepackage{pst-node}

\usepackage[nottoc]{tocbibind}

\newcommand{\bdm}{\begin{displaymath}}
\newcommand{\edm}{\end{displaymath}}

\newcommand{\be}{\begin{eqnarray}}
\newcommand{\ee}{\end{eqnarray}}
\newcommand{\ba}{\begin{array}}
\newcommand{\ea}{\end{array}}
\newcommand{\ds}{\displaystyle}
\newcommand{\pa}[1]{\left(#1\right)}
\newcommand{\paq}[1]{\left[#1\right]}

\newcommand{\ta}{{\tt a}}

\newcommand{\nn}{\nonumber}
\newcommand{\pp}{{\mathbf p}}
\newcommand{\Pp}{\mathbf{p}}
\newcommand{\Q}{{\mathbf q}}
\newcommand{\K}{{\bf k}}
\newcommand{\X}{{\bf x}}
\newcommand{\R}{{\bf r}}
\newcommand{\N}{{\bf\hat{n}}}
\newcommand{\XX}{{\bf\hat{x}}}
\usepackage{enumitem}

\usepackage{hyperref}

\newcommand{\GA}[1]{{\color{magenta}{#1}}}

\begin{document}

\include{includes/cover}

\include{includes/titlepage}

\includepdf{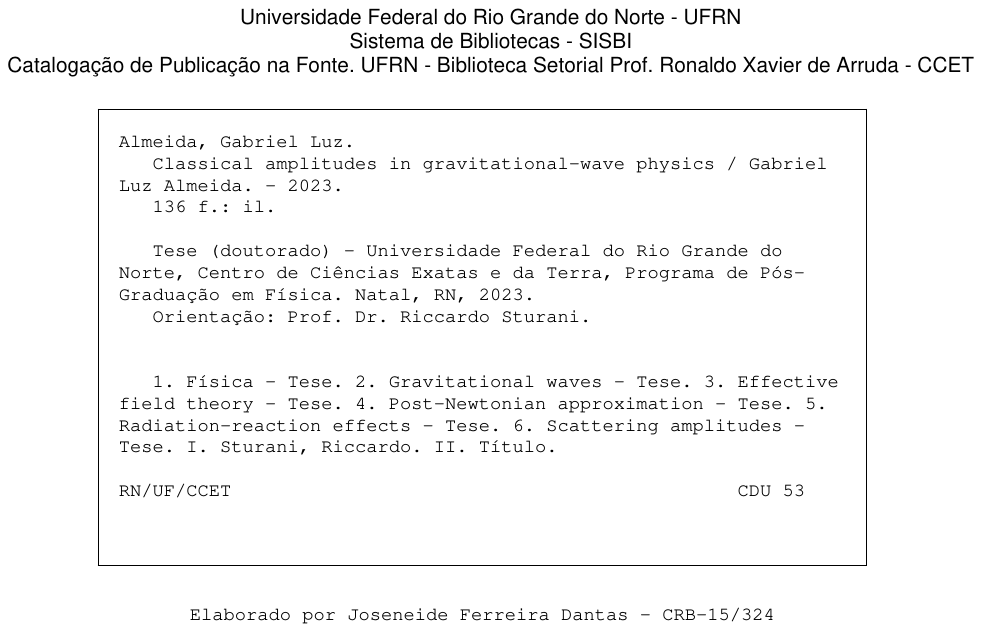}

\include{includes/boardpage}

\include{acknowledgements/acknow}

\include{abstract/abs}

\include{publicationlist/publist}

\tableofcontents

\include{chapter0/introduction}

\include{chapter1/gravitationalwaves}

\include{chapter2/eftapproach}

\include{chapter3/wardidentity}

\include{chapter4/renormalization}

\include{chapterx/perspectives}

\appendix

\include{appendices/svt}

\include{appendices/jfailedtails}

\include{appendices/eom}

\include{appendices/integrals}

\include{appendices/tailoftail}

\include{bibliography/bib}
\end{document}

%% file: includes/cover.tex

\begin{titlepage}
	\begin{center}
		
			\begin{center}
				\includegraphics[scale=0.80]{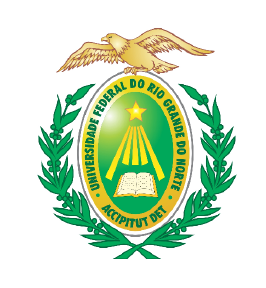}
			\end{center}
		\begin{minipage}{15.15cm}
			\begin{center}
					{ \large \ \\
                      					  UNIVERSIDADE FEDERAL DO RIO GRANDE DO NORTE	\\
                      					  \vspace{0.1cm}
							  CENTRO DE CIÊNCIAS EXATAS E DA TERRA		\\
							  \vspace{0.1cm}
							  DEPARTAMENTO DE FÍSICA TEÓRICA E EXPERIMENTAL\\
							  \vspace{0.2cm}
							  PROGRAMA DE PÓS-GRADUAÇÃO EM FÍSICA}  	\\
							\vspace{2cm}
							\textbf{DOCTORAL THESIS} 
			\end{center}
		\end{minipage}
			
		\vspace{2.5cm}
						
		{\setlength{\baselineskip}%
		{1.3\baselineskip}
		{\LARGE \textbf{Classical Amplitudes in Gravitational-Wave Physics}}\par}
			
		\vspace{3cm}
			
		{\large \textbf{Gabriel Luz Almeida}}
						
		\vspace{5cm}
		
		Natal-RN\\July 2023
	\end{center}
\end{titlepage}

%% file: includes/titlepage.tex
\newpage
\thispagestyle{empty}

\begin{center}
	{\large \textbf{Gabriel Luz Almeida}}\\
	\vspace{4cm}
	{\LARGE \textbf{Classical Amplitudes in Gravitational-Wave Physics}}\par
\end{center}
	\vspace{4cm}
\begin{flushright}
\begin{minipage}{8cm}
Thesis presented to the Graduate Program in Physics
at Universidade Federal do Rio Grande do Norte, as a partial requirement for obtaining the title of Doctor in
\mbox{Physics}.

\vspace{0.5cm}
\textbf{Supervisor}: Prof. Dr. Riccardo Sturani.

\end{minipage}
\end{flushright}
 
\vspace{7.07cm}

\begin{center}
Natal-RN\\
\vspace{-0.01cm}
July 2023
\end{center}

%% file: includes/boardpage.tex
\newpage
\thispagestyle{empty}

\begin{center}
	{\large \textbf{Gabriel Luz Almeida}}\\
	\vspace{1cm}
	{\large \textbf{Classical Amplitudes in Gravitational-Wave Physics}}\par
\end{center}
	\vspace{0.1cm}
\begin{flushright}
\begin{minipage}{8cm}
Thesis presented to the Graduate Program in Physics
at Universidade Federal do Rio Grande do Norte, as a partial requirement for obtaining the title of Doctor in
\mbox{Physics}.

\vspace{0.5cm}
\textbf{Supervisor}: Prof. Dr. Riccardo Sturani.

\end{minipage}
\end{flushright}
\vspace{0.5cm}

Accepted on: 18/07/2023.
 
\vspace{0.5cm}

\begin{center}
\textbf{EXAMINING BOARD}
\\
\vspace{1.5cm}
\rule{10cm}{0.4pt}\\
Riccardo Sturani\\
ICTP-SAIFR, UNESP
\\
\vspace{1cm}
\rule{10cm}{0.4pt}\\
Dmitry Melnikov\\
International Institute of Physics - UFRN
\\
\vspace{1cm}
\rule{10cm}{0.4pt}\\
Emanuele Orazi\\ 
Universidade Federal do Rio Grande do Norte
\\
\vspace{1cm}
\rule{10cm}{0.4pt}\\
Walter D. Goldberger\\
Yale University
\\
\vspace{1cm}
\rule{10cm}{0.4pt}\\
Laura Bernard\\
LUTH - Observatoire de Paris
\end{center}

%% file: acknowledgements/acknow.tex
\chapter*{Acknowledgements}

First and foremost, I extend my sincere gratitude to my PhD supervisor, Riccardo Sturani, for his unwavering dedication to my education over the past four years. It has been an absolute privilege to have been under your guidance, Professor. I'm indebted for all the scientific discussions we had and the lessons you taught me. I'm also thankful to you for warmly welcoming me to the ICTP-SAIFR in São Paulo, for promoting our First Brazilian Gravitational-Wave Retreat in Atibaia-SP, and finally for making my visits to the ICTP in Triste, SISSA, and INFN in Padova possible.  

I would also like to express my heartfelt appreciation to Stefano Foffa, our collaborator, for the invaluable exchanges we had during my PhD studies. Your warm welcome in Geneva and your role as a co-advisor have been truly appreciated. I will always be grateful to you for making my visit to the University of Geneva possible, and will never forget my first visit to CERN that you arranged.

During these four years, I have been lucky to have been around the most amazing people. In particular, I would like to mention the colleagues of the gravitational-wave research group, which soon became friends: Aline, Hebertt, Josiel, and Tibério from UFRN; Alan, Gabriel Macedo, Gabriel Vidal, and Isabela from IFT-UNESP/ICTP-SAIFR; and Matheus from CBPF. Also, my dear friends from UFPE: Joãozinho, Joás, Mariana, Matheus, and Michelle; The friendships I made during the months I stayed in IFT-UNESP: Abdias, Bruno, Eggon, Eric, Graciele, Isabela (Isa!), Marcelo, and many others whose contact made me feel very welcome; The IIP-UFRN friends: Daniel, Hernán, Jacinto, and Yoxara.
Thank you for all your support, conversations, lunch and ``cafézinho" we had together... Anyways, thank you for any kind of interactions we have had over these years.

I'm thankful for the kind hospitality I had both at the International Institute of Physics (IIP-UFRN) and at the IFT-UNESP/ICTP-SAIFR.
I wish to thank also the INFN in Padova, the Department of Theoretical Physics of the University of Geneva, and the ICTP in Trieste for their kind hospitality and support during the months I stayed in Europe. On that occasion, I had the great pleasure to meet and interact with Paolo Creminelli and Mehrdad Mirbabayi in ICTP; Enrico Barausse, his research group, and Guillaume Faye in SISSA; Pierpaolo Mastrolia and his research group, which include Manoj Mandal and Giacomo Brunello, in INFN Padova; And Michele Maggiore at the University of Geneva.

I'm also grateful to François Larrouturou for all the virtual meetings we had in the last few years to discuss our research projects and to the members of my PhD defense committee for the discussions: Laura Bernard, Walter Goldberger, Dmitry Melnikov, and Emanuele Orazi.

I want to thank my family for their unconditional support through this process: my mother Maria Helena, my father Givanildo, and my sister Giliane.

I also want to thank CAPES for the financial support, allowing me to continue my studies in this magnificent field that is physics. This study was financed in part by the Coordenação de Aperfeiçoamento de Pessoal de Nível Superior - Brasil (CAPES) - Finance Code 001.

%% file: abstract/abs.tex
\chapter*{Abstract}

The recent detections of gravitational waves from coalescing binaries by the LIGO and Virgo interferometers not only opened a new window of observation into the Universe but also revived interest in the relativistic two-body problem.  In this thesis, after presenting the general theory of gravitational waves, we review the effective field theory approach within the so-called Non-Relativistic General Relativity. In this framework, the different scales involved in the problem define different dynamical regions, with short scales describing the conservative interaction between the two bodies and long-distance scales accounting for gravitational-wave processes. 
At higher orders, however, radiation modes start to affect the conservative dynamics of the system through the so-called \textit{tail} effect. 

We then study emission amplitudes for the class of nonlinear processes of  \textit{tails}, 
which are processes of order $G_N^2$, and represent the effect of scattering gravitational radiation off the static background curvature, including not only mass tails but also the ``failed" tails due to the angular momentum of the source. 
As originally shown in our work, some of these amplitudes present anomalies, which are then properly understood and corrected with 
field theory tools which have straightforward counterpart in traditional methods.
We also investigate the relation between emission and self-energy diagrams and, in particular, show that a correction to the anomalous self-energy diagrams is necessary to correctly account for 
conservative terms that arise from radiative processes.
From this, we obtain the correct contribution to the conservative 5PN stemming from the electric quadrupole angular momentum failed tail, correcting previous results in the literature. Besides this, we also compute for the first time the conservative contributions from the angular momentum failed tail, for arbitrary multipole moments.

Subsequently, we explore the natural higher-order extension (of order $G_N^3$) of the mass tails, called \textit{tails of tails}, for generic electric and magnetic multipoles. As we will see, both long- and short-distance divergences are encountered, which are then properly understood and dealt with in terms of standard renormalization techniques. In this case, we are able to resum the logarithmic contributions through the integration of the renormalization group equation.

Thus, in this thesis, we have extended the post-Newtonian perturbative expansion in the particularly thorny region of processes involving the mixture of radiative and potential modes.
\\

\noindent\textbf{Keywords:} Gravitational waves. Effective field theory. Post-Newtonian approximation. Radiation-reaction effects. Scattering amplitudes.


\chapter*{Resumo}

As recentes detecções de ondas gravitacionais de coalescências de binárias pelos interferômetros LIGO e Virgo não apenas deram origem a uma nova janela de observação para o Universo, mas também reviveram o interesse pelo problema de dois corpos relativístico. Nesta tese, após apresentar a teoria geral das ondas gravitacionais, revisaremos a abordagem de teoria de campos efetiva dentro da chamada Relatividade Geral Não-Relativística. Nessa construção, as diferentes escalas envolvidas no problema definem diferentes regiões da dinâmica, com escalas curtas descrevendo a interação conservativa entre os dois corpos e escalas de longa distância descrevendo os processos de ondas gravitacionais. 
Em ordens superiores, no entanto, modos radiativos começam a afetar a dinâmica conservativa do sistema por meio do chamado efeito de \textit{cauda}.

Em seguida, estudaremos as amplitudes de emissão para a classe de processos não lineares das \textit{caudas}, que são processos de ordem $G_N^2$, e representam o efeito de espalhamento da radiação gravitacional na curvatura de fundo estática, incluindo não apenas caudas de massa, mas também as caudas ``falhas" devido ao momento angular da fonte. 
Como mostrado originalmente em nosso trabalho, algumas dessas amplitudes apresentam anomalias, que são então corretamente compreendidas e corrigidas com
ferramentas de teoria de campos que tem contra-parte direta em métodos tradicionais.
Também investigaremos a relação entre diagramas de emissão e de autoenergia e, em particular, mostraremos que correções aos diagramas anômalos de autoenergia são necessárias para contabilizar corretamente os 
termos conservativos que surgem de processos radiativos.
A partir disso, obteremos a contribuição correta para o termo conservativo a 5PN provindo da cauda falha de momento angular do quadrupolo elétrico, corrigindo resultados anteriores da literatura. Além disso, também calcularemos pela primeira vez as contribuições conservativas da cauda falha de momento angular para momentos de multipolo arbitrários.

Posteriormente, exploraremos a extensão natural de ordem superior (ordem $G_N^3$) das caudas de massa, chamadas \textit{caudas de caudas}, para multipolos elétricos e magnéticos genéricos. Conforme veremos, encontraremos divergências de longa e curta distância, que são então corretamente compreendidas e tratadas por meio de técnicas padrões de renormalização. Nesse caso, somos capazes de ressomar as contribuições logarítmicas por meio da integração da equação do grupo de renormalização.

Desta forma, nesta tese, temos extendido a expansão perturbative pós-newtoniana na região de processos particularmente difíceis envolvendo a mistura de modos radiativos e potenciais.
\\

\noindent\textbf{Keywords:} Ondas gravitacionais. Teoria de campos efetiva. Aproximação pós-newtoniana. Efeitos de reação à radiação. Amplitudes de espalhamento.

%% file: publicationlist/publist.tex
\chapter*{Publications}

During the four years of PhD, I, along with my supervisor, Riccardo Sturani, collaborator Stefano Foffa, and colleague Allan Müller, have published the following five papers \cite{Almeida2020,Almeida2021a,Almeida2021b,Almeida2022,Almeida2023}: 
\begin{enumerate}[label={[\arabic*]}]
\item G. L. Almeida, S. Foffa, R. Sturani, \textit{Classical Gravitational Self-Energy from Double Copy}, JHEP 11 (2020) \textbf{165}. arXiv:2008.06195 [gr-qc]
\item G. L. Almeida, S. Foffa, R. Sturani, \textit{Gravitational multipole renormalization}, Phys. Rev. D \textbf{104}, 084095 (2021). arXiv:2107.02634 [gr-qc]
\item G. L. Almeida, S. Foffa, R. Sturani, \textit{Tail contributions to gravitational conservative dynamics}, Phys. Rev. D \textbf{104}, 124075 (2021). arXiv:2110.14146 [gr-qc]
\item G. L. Almeida, S. Foffa, R. Sturani, \textit{Gravitational radiation contributions to the two-body scattering angle}, Phys. Rev. D \textbf{107}, 024020 (2022). arXiv:2209.11594 [gr-qc]
\item G. L. Almeida, A. Müller, S. Foffa, R. Sturani, \textit{Conservative binary dynamics from gravitational tail emission processes}, (2023). arXiv:2307.05327 [gr-qc].
\end{enumerate}
In particular, the original results presented in Chapter 4 were heavily drawn from \cite{Almeida2023}, \textit{Conservative binary dynamics from gravitational tail emission processes}, available in arXiv. Also, the work presented in Chapter 5 was based on the paper \cite{Almeida2021a}, \textit{Gravitational multipole renormalization}, published in 2021.


%% file: chapter0/introduction.tex
\chapter{Introduction}

The starting point of our discussion is the general theory of relativity. As it is more commonly known, general relativity is the theory of space, time, and gravitation formulated by Albert Einstein in 1915, originally devised in a series of four papers \cite{Einstein1915a,Einstein1915b,Einstein1915c,Einstein1915d}. In 1916, soon after the establishment of his relativistic version of a theory of gravity, Einstein realized that his field equations, when considered in the weak-field regime, had wave solutions \cite{Einstein1916}. These waves, called \texttt{gravitational waves}, are transverse perturbations traveling across spacetime at the speed of light, generated by time variations of the mass quadrupole moment of the source \cite{Einstein1918}.

Direct observation of gravitational waves, nevertheless, was only accomplished nearly a century after Einstein predicted their existence: On September 14, 2015, the two detectors of the Laser Interferometer Gravitational-Wave Observatory (LIGO) \cite{LIGO2015} simultaneously detected a transient gravitational-wave signal \cite{LIGO2016a}. 
This signal, soon after confirmed to correspond to gravitational waves emitted from the coalescence of a binary black hole (BBH) system, was dubbed GW150914\footnote{Gravitational-wave signals are named after the date of their discovery. In this case, \textit{e.g.}, the signal detected in 2015-09-14 goes by the name of GW150914.} and represented the first of many gravitational-wave events of astrophysical origin that would later be observed. The importance of GW150914 goes beyond this and also relies on the fact that it demonstrated 
the existence of 
binary stellar-mass black hole systems
and provided the first observed BBH merger. With this, a new observation window was opened, setting the beginning of the field of gravitational-wave astronomy \cite{CERNCourier}.

In the three months following the first detection, the two LIGO detectors observed two additional gravitational-wave events: GW151012 and GW151226 \cite{LIGO2016b,LIGO2016c}. Like in GW150914, these two signals were produced during the coalescence of binary systems of stellar-mass black holes. The three events GW150914, GW151012, and GW151226 were the only ones observed during the first run of observation (O1) of Advanced LIGO, which took place from the 12th of September, 2015, to the 19th of January, 2016. Then, after an upgrade and commissioning period, the second observing run (O2) started on the 30th of November, 2016. 
On August 1, 2017, Advanced Virgo \cite{Virgo2015} joined the twin LIGO detectors. By the end of O2, on August 25 of the same year, this newly-formed network of advanced interferometers had observed a total of eight events \cite{LIGO2017a,LIGO2017b,LIGO2017c,LIGO2017d,LIGO2019}. 
Among these events, seven corresponded to stellar-mass BBHs, while one, GW170817, was identified to have originated from a binary neutron star (BNS) inspiral \cite{LIGO2017d} in a joint detection of gravitational waves along with an electromagnetic counterpart \cite{GWeEM1,GWeEM2}. This latter fact makes GW170817 a unique event, important in this era of multimessenger astronomy opened by gravitational-wave observations. 
For a detailed account of the astrophysical properties of the events comprising the first and second runs of the LIGO-Virgo detector network, the reader is referred to the first Gravitational-Wave Transient Catalog (GWTC-1) in Ref. \cite{LIGO2019}.

Although GW170817 was the first BNS gravitational-wave signal to be \textit{directly} detected, it is not, however, the first time a BNS system was observed: in 1974, R. Hulse and J. Taylor discovered a  system formed by two pulsars (i.e., rapidly spinning neutron stars), called PSR B1913+16, and also known as the Hulse-Taylor pulsar \cite{HulseTaylor1975,Hulse1994}. This binary system is historically important since, later on, in 1981, Taylor, along with J. Weisberg, observed the effect of the energy loss due to the emission of gravitational radiation on the orbit of this system, indirectly proving the existence of gravitational waves \cite{TaylorWeisberg1975}.

The third and most recent observing run, O3, took place during the years of 2019 and 2020, and was split into two parts, O3a and O3b. In the first half of this observing run, O3a, with data taken from 1 April to 1 October, 2019, Advanced LIGO and Advanced Virgo were able to observe 44 new gravitational-wave binary coalescences. Such observations comprised 42 BBHs, 1 BNS and one binary system containing a black hole and a compact object whose nature could not be discerned, GW190814, being either the lightest black hole or the heaviest neutron star ever known \cite{LIGO2020}. In addition to this interesting signal, O3a counted also with 
GW190412, the first BBH observation with notably asymmetric constituent masses ($m_1=30.1 M_{\odot}, m_2 = 8.3M_{\odot}$)  \cite{LIGO2021}, although GW190814 would later turn out to be even more asymmetric, with individual masses of $ 23M_{\odot}$ and $2.6M_{\odot}$; 
GW190425, the second gravitational-wave event consistent with a BNS coalescence \cite{LIGO2021};
and GW190521, a BBH merger with a total final mass of $\sim 150M_{\odot}$, being the heaviest final stage stellar black hole to be observed by the LIGO-Virgo Collaboration \cite{LIGO2020a,LIGO2020b}. Details can be found in \cite{GWTC2,GWTC21}, in the second catalog GWTC-2, and the extended GWTC-2.1.


The second half of the third observing run, O3b, operated from 1 November 2019 to 27 March 2020 and added 35 new events to the previous observing runs. Among these signals, 3 corresponded to the first confident detections of neutron star-black hole (NSBH) coalescences, while the other 32 stemmed from BBH mergers. No BNS events were observed during O3b. For a complete analysis and details of the astrophysical properties of these detections, check the third Gravitational-Wave Transient Catalog GWTC-3 out in Ref.~\cite{GWTC3}. With this, a total of 90 gravitational-wave signals from binary system coalescences  have been detected since the beginning of the first observing run in 2015. See Table~\ref{GWTCevents} for a brief summary of the three observing runs.
\begin{table}
\centering
\begin{tabular}{c|c|c|c}
\hline
Observing run & Number of events & Start date & End date\\ 
\hline
O1 & 3 BBH & 12-Sep-2015 & 19-Jan-2016 \\
O2 & 8 = 1 BNS + 7 BBH & 30-Nov-2016 & 25-Aug-2017 \\
O3a & 44 = 1 BNS + 42 BBH + 1(?) & 1-Apr-2019 & 1-Oct-2019 \\
O3b & 35 = 32 BBH + 3 NSBH & 1-Nov-2019 & 27-Mar-2020 \\
\hline
\end{tabular}
\caption{Summary of the LIGO-Virgo  observing runs, BBH, BNS, and NSBH standing respectively for binary black hole, neutron star, and neutron star-black hole  mergers. The question mark in O3a corresponds to GW190814, which is either a BBH or NSBH merger.}
\label{GWTCevents}
\end{table}

The next observing run, O4, has recently started (on May 24) with the two LIGO detectors, which now counts, along with Virgo, with the Japanese interferometer KAGRA \cite{KAGRA} in the newly formed LIGO-Virgo-KAGRA (LVK) collaboration.
The prospects for the near future are that the new upgrades will allow us to increase considerably the number of detections as well as will allow us to see much farther, into cosmological scales. This progress calls for an immediate effort toward improving the precision of the templates currently used by the LIGO-Virgo-KAGRA collaboration. In particular, the current models of gravitational-wave templates used in these detections are especially sensitive to the phase of the gravitational wave, which has to be modeled very precisely to lead to a reliable interpretation of the data, 
especially now in face of the upcoming third-generation interferometers such as LISA \cite{LISA} and the Einstein Telescope \cite{ET}.
This phase, $\Phi(t)$, which tracks the orbital phase evolution of compact binary systems, is given in terms of the conservative energy of the system, $E$, as well as in terms of the radiated power loss due to gravitational radiation, $P$, through
\begin{equation}\label{gwphaseintro}
    \Phi(t) = 2 \int^t_{t_0} dt\, \omega(t) = - \frac{2}{G_N M} \int^{v(t)}_{v(t_0)}dv \frac{v^3}{P(v)}\frac{dE}{dv}\,.
\end{equation}
In this expression, $\omega(t)$ is the orbital phase of the binary system's motion,
$v$ the relative velocity between the two bodies,
 and $t_0$ and $t$ respectively the time where
the signal enters and exits the detector. 

In modeling the phase as above, it is assumed that the system goes through an adiabatic inspiral phase, going through a succession of quasi-circular orbits, inspiralling inward, and driven by gravitational radiation reaction. In fact, this provides a very good approximation, that optimizes the analysis of the gravitational-wave data, since it is well known that, for typical compact binary systems observed by interferometers like LIGO, by the time when the gravitational wave enters the bandwidth of the detectors, their orbits will have already been circularized long before, due to the radiation reaction on the initially elliptic orbits \cite{blanchetlivrev}.
This is achieved upon the fundamental assumption that the system's energy loss is solely due to the emission of gravitational radiation,  in accordance with the energy balance equation
\begin{equation}
\frac{dE}{dt} = - P\,.
\end{equation}

The two quantities needed to construct the gravitational-wave phase, the conserved energy and the energy flux, are appropriately modeled within the post-Newtonian approximation to general relativity. In this formalism, the Einstein field equations are solved perturbatively, iterating successive orders in powers of $v^2$, the relative velocity square, which serves as the small parameter of the expansion, and accounts for deviation of the slow-motion and weak-field regime, characteristic of linearized gravity. The most important application of this formalism is in describing the coalescence of compact binary systems, extending the Newtonian results to incorporate fast-motion and strong-field contributions as predicted by general relativity. This is essential in the construction of templates currently under use by laser-interferometric observatories to detect gravitational-wave signals emitted by inspiraling  binaries. For the most complete and up-to-date review of the post-Newtonian formalism, and in particular, applied to the problem of inspiralling compact binaries, from traditional methods, see Ref.~\cite{blanchetlivrev}.

In the post-Newtonian approximation, the $n$th correction $(v^2)^n$ beyond the Newtonian expression is said to be the $n$PN order.\footnote{Notice that, for bound systems, the virial theorem $v^2 \sim G_N M/r$ connects powers of $v^2$ to $G_N M/r$. Hence, equivalently, the $n$PN order corresponds to terms of the type $G_N^{n-j+1}v^{2j}$, with $0\leq j \leq n+1$.}
Hence, considering the motion of two spinless particles, the 1PN correction was first derived by Lorentz and Droste \cite{lorentz1917a,lorentz1917b}, and later on confirmed by Einstein, Infeld, and Hoffmann in Ref.~\cite{EIHlagrangian}, where the 1PN order for the general $N$-body system was obtained. The 2.5PN equations of motion, which include the first appearance of radiation-reaction effects, were later on derived by Damour and Deruelle \cite{damour1981,damour1983} using a post-Minkowskian approach. The 2.5PN order for the equations of motion was also obtained by Schäfer in Ref.~\cite{schafer1985} from an ADM Hamiltonian approach and within the post-Newtonian iteration of the Einstein field equations by Blanchet, Faye, and Ponsot in Ref.~\cite{blanchet1998}. The 3PN equations of motion were then obtained in Refs.~\cite{Jaranowski1999,Jaranowski1999a,Jaranowski2000,Jaranowski2001,Jaranowski2001a} using the ADM formulation of general relativity, and \cite{blanchet2000,blanchet2001,blanchet2001b,blanchet2001c} in harmonic coordinates from traditional methods, with equivalent results between the methods \cite{andrade2001,damour2001}. The equations of motion at the 4PN order was also obtained by different groups, including Ref.~\cite{damour2014}, derived within the ADM formulation, Ref.~\cite{bernard2016}, from traditional post-Newtonian methods, and, finally, Refs.~\cite{RSSF4PNa,RSSF4PNb} from effective field theory methods. See Ref.~\cite{damour2016} for a review on the complementarity of the 4PN motion from the different approaches.

The state-of-the-art of the gravitational-wave phase currently used in LVK's pipelines is the 3.5PN order and was first obtained in Ref.~\cite{blanchet2005a}, with the energy $E$ computed in \cite{blanchet2000,blanchet2001}, although with some free, undetermined parameters due to the incompleteness of the regularization technique used there, the Hadamard one, and completed in Ref.~\cite{blanchet2004} with the aid of dimensional regularization. 
More recently, however, the completion of the 4.5PN order gravitational-wave phase was accomplished by Blanchet et al in Refs.~\cite{blanchet2023,blanchet2023a}.
In this case, in addition to the energy at the 4PN order previously computed, for the computation of the emitted power loss, computed on circular orbits at the 4.5PN order in Refs.~\cite{blanchet2023,blanchet2023a}, and required to construct the orbital phase, one employs the so-called wave-generation formalism \cite{blanchet1998a}. In this formalism, which describes the exterior gravitational field to the composite system, and is based on a post-Minkowskian approximation (an expansion in powers of $G_N$), the source degrees of freedom are parametrized by a series of multipole moments, and observables at infinity are given in terms of nonlinear relations involving such quantities. The multipole moments, in their turn, can be matched in the region close to the source, to be given in terms of the orbital variables of the binary system within the post-Newtonian approximation.  

Once the phase is obtained and templates of gravitational waves are built, spanning a large region in the parameter space of the binary\footnote{These are the following 15 parameters: the masses of the constituents, $m_1$ and $m_2$; the spins of each object, ${\bf S_1}$ and ${\bf S_2}$, and hence six components; the localization of the source in the sky, $(\theta, \phi)$, and  its luminosity distance $D_L$; 
the gravitational-wave polarization angle $\psi$ and the binary inclination angle $\iota$;
the time of coalescence, $t_c$; and the phase of coalescence, $\Phi_c$.}
we can start seeking gravitational-wave signals. The usual technique employed in this case is the so-called matched filtering, which consists in correlating the detector's output to waveforms present in the template bank, and weighted over the characteristic noise of the detector. Hence, whenever a signal is detected and a correlation with a template of the bank is established, a spike emerges in the plot of this correlation, which then allows us to extract the signal within the noise. A comprehensive review on how precisely the output of the LIGO-Virgo detectors is analyzed, and gravitational-wave signals are extracted, can be found in Ref.~\cite{abbott2020}.

In particular, the process of detecting gravitational waves from merging binary systems with laser interferometers that rely on matched filtering requires accurate prediction for the gravitational waveform. In fact, a series of analyses performed in Refs.~\cite{cutler1993,cutler1993a} has shown that the 3PN, or better the 3.5PN order provides a lower-bound accuracy 
for the inspiral phase portion of the waveform
below which the errors are too large to provide a reliable interpretation of the data. At the 3.5PN order, on the other hand, we have a sufficiently accurate prediction for the waveforms, considering just the orbital phase, so that the errors are not too significant on the parameter estimation. Besides this, higher post-Newtonian orders are important to resolve and analyze signals arriving from farther distances. This justifies the need in attaining higher-order corrections to the equations of the motion and the radiation field in the post-Newtonian approximation since the main goal lies in extracting reliable predictions of the theory that can be compared against the output of the detectors.

Following the inspiral phase, which is fully described by the post-Newtonian formalism, the compact binary system subsequently undergoes two other phases as gravitational radiation is emitted:
the merger, when the two objects can no longer be resolved into two, but one single object, and then the ringdown, when the two objects have, typically, given rise to a black hole, which oscillates in quasi-normal frequencies, settling down to its unperturbed state. As the system evolves toward the merger, the post-Newtonian approximation starts to lose accuracy and should be taken over by numerical relativity \cite{pretorius2005}. The ringdown phase, on the other hand, is typically investigated within black hole perturbation theory; See Ref.~\cite{sasaki2003} for a review. Hence, the knowledge of the inspiral and ringdown phases can be put together by means of the so-called effective one-body (EOB) approach \cite{buonanno1999,buonanno2000}, a theoretical framework developed in the late 1990s by Buonanno and Damour to model waveforms without any knowledge of the merger phase; See Ref.~\cite{damour2014b} for a review. In this case, numerical relativity becomes an important tool for the calibration of the joining of inspiral-ringdown solutions.


\subsubsection{The effective field theory for gravity}

Although the origin of post-Newtonian methods dates back to the 1920s, following the early days of general relativity itself,  only in the last two decades that this approximation method gained a field-theoretical format. 
To be more precise, in a 2006 paper \cite{rothstein2006}, W. Goldberger and I. Rothstein formulated an effective field theory (EFT) to describe the dynamics of two compact objects, completely compatible with the post-Newtonian approximation. This EFT was then coined Non-Relativistic General Relativity (NRGR) as a parallel to the Non-Relativistic Quantum Chromodynamics (NRQCD) framework developed in the early 1990s to study bound states in QCD in nonrelativistic regimes, which serves as the basis for the construction. Since then, this formulation has attracted a lot of attention from until then disconnected communities: on one side, relativists interested in the classical aspects of gravity; on the other, theoretical particle physicists interested in practical applications of the most recent tools of the ever-growing field of scattering amplitudes. This synergy has  great potential, being one of the most active research fields in physics nowadays. 

To list a few of the most prominent features of this EFT implementation of the post-Newtonian approximation, we have (1) it gives separate descriptions for the conservative and dissipative dynamics of the binary system, (2) it provides a clear power counting from which the perturbative approach is implemented, in which case Feynman diagrams are used to ease the tracking of contributions, besides easily incorporating (3) nonlinearities stemming from general relativity, (4) spin effects, and even (5) modifications to general relativity. Additionally, (6) it also provides a better understanding of the UV and IR divergences present in the theory in light of the (classical) renormalization group evolution. 
Note that the treatment of the problem into separate scales is crutial to the success of NRGR.


As it soon becomes clear, the number of Feynman diagrams involved in each step increases exponentially with the perturbative order, and methods of the field of scattering amplitudes have to be considered in order to gain efficiency in computing higher-order effects. In particular, the state-of-the-art description of the conservative dynamics of binary systems lies in the 4PN order \cite{RSSF4PNa,RSSF4PNb}, i.e., fourth-order in $v^2$ beyond the Newtonian solution, in which case the binary constituents can still be described by point particles.  This is the order in which radiation-reaction terms start to contribute \cite{RSSFhereditary}. These are effects present in the interplay of different scales, between the so-called potential and radiation modes, with the corresponding imprint to the two-body potential having already been completely accounted for at the 4PN order with the leading-order \textit{tail} effect, and the pair of IR-UV divergences, present in the different regions, being mutually canceled and duly understood \cite{RSSF4PNb}.

Within this EFT framework, and using the harmonic gauge condition, the conservative dynamics of spinless compact binary systems was originally computed to the 1PN order in Ref.~\cite{rothstein2006}, and simplified with the use of the Kaluza-Klein reduction of the metric in Ref.~\cite{kol2008a}; the 2PN order was then achieved in Ref.~\cite{gilmore2008} and the 3PN in Ref.~\cite{RSSF2011}. Post-Newtonian corrections due to spin effects, which are of great relevance in any faithful description of the dynamics of compact binary systems, must, of course, be taken into account. 	
The inclusion of spin effects within NRGR was first introduced by R. Porto in 2006, in Ref.~\cite{porto2006}, where the leading order 1.5PN and, also, the 2PN corrections were reproduced. Subsequently, corrections at the 3PN order were computed in Refs.~\cite{porto2006b,porto2010,levi2010,porto2008} and the completion of the 4PN order was later on accomplished in Ref.~\cite{levi2021}, relying on several partial results contributing at this order obtained in Refs.~\cite{levi2012,levi2015,levi2016,levi2016b}.

Advances toward the complete understanding of the 5PN order represent currently the most important problem within the relativistic two-body problem. This is, in particular, the order in which finite-size effects start to appear (in the spinless case) \cite{rothstein2006} and, therefore, that will allow us to probe the strong-field regime of gravity via gravitational waves, and expected also to finally provide insights into the problem of the equation of state for neutron stars. At this order, despite the purely conservative description having already been computed for the spinless case \cite{blumlein2021}, and not introducing any further conceptual obstacle, radiation-reaction effects pose a notoriously tough and challenging problem. As a matter of fact, current results of 5PN radiation-reaction contributions to the binary system dynamics are in conflict with information stemming from self-force and \textit{Tutti-Frutti} calculations, as pointed out by Bini, Damour, and Geralico in \cite{damour2021}. 
Among the novelties at this order, there is the inclusion of higher-order terms of a particular type of radiation-reaction effect called “hereditary” \cite{blanchet1988}. These are effects entering the two-body dynamics in a nonlocal fashion, depending on the history rather than on the instantaneous state of the source at retarded time. Particular examples of such a class of phenomena are the tail and memory effects, whose contribution to the conservative sector involves a cubic multipole interaction, and hence affects the two-body dynamics inasmuch as nonlinear interactions induce gravitational radiation to back-scatter \cite{RSSFhereditary}.


Since these higher-order dissipative effects are highly non-trivial in nature, one has to be very careful on how to treat such time-asymmetric phenomena at the level of a path integral formulation.  In this case, one has to resort to the so-called \textit{in-in} formalism, also known as the \textit{closed-time-path} formalism \cite{schwinger61,keldysh65}. The in-in formalism is a field-theory formulation that appropriately incorporates causal interactions associated with nonconservative processes, like radiation-reaction, at the level of path integral and Feynman diagrams. Applications to our setup have already been done in the study of leading radiation-reaction effects \cite{galleytiglio}, as well as in the leading hereditary contribution \cite{GalleyRoss2016}, namely the tail. Nevertheless, higher-order hereditary effects entering the 5PN conservative dynamics, like the memory, are more involved and have never been fully and consistently explored. 

Another approach that has recently been pursued to study the relativistic two-body problem is a scattering amplitude implementation of the post-Minkowskian approximation (PM) \cite{zvi2019}, which has  Newton's constant as the appropriate parameter of the expansion, as a weak-field regime approximation \cite{damour2016b}. In this formulation, rather than a bound system, the focus is on the scattering of two compact objects, with the scattering angle as the most important observable to be computed. This is, in particular, a complementary and powerful gauge-invariant\footnote{It is well known in this formalism that the scattering angle provides not only a gauge-invariant quantity, but also that it can be seen as the most fundamental quantity of the problem since the Hamiltonian can be constructed from it.} way of studying radiative corrections to the conservative dynamics of binary systems, being also very important for providing independent checks of results obtained via NRGR calculations \cite{damour2021}. See also Ref.~\cite{porto2020} for an EFT adaptation of the post-Minkowskian formalism.

As part of our own attempt to solve the 5PN conservative problem, we have published two papers, \cite{Almeida2021b} and \cite{Almeida2022}. In Ref.~\cite{Almeida2021b}, 
we resolved a discrepancy that had arisen at the 5PN order between EFT and self-force results, involving the tail effect for the magnetic quadrupole moment. In this work, we also took the opportunity to compute tail contributions to the conservative dynamics of a generic self-gravitating system for every multipole order, of both electric and magnetic parities. 
In Ref.~\cite{Almeida2022}, we addressed the problem from a different direction, by including new contributions to the scattering angle in the two-body problem, at the level of the equations of motion, stemming from the square of the 2.5PN order radiation reaction, which contributes to conservative sector of the 5PN dynamics, never been included before.

In this thesis, we will present a yet novel approach to address the 5PN problem 
by consistently incorporating corrections to
the computation of self-energy diagrams, which are the diagrams responsible to describe radiation-reaction effects. As we will see, past EFT computation of such diagrams was completely blind to the fact that, for a few processes, internal subdiagrams did not correctly transfer the four-momentum due to a violation of the energy-momentum conservation 
that went unnoticed in the EFT approach. 
We will also see that this \textit{anomaly} can be better studied from the viewpoint  of emission amplitudes, which define the subdiagrams of self-energy diagrams, and from which a fixing to the violation of  energy-momentum conservation can be attained.

While the 5PN two-body dynamics still represents an open problem, we are confident that the program initiated in this thesis is essential to the complete understanding of such an important problem. In particular, the existence of anomalies in the EFT description adds interesting elements to the challenge, that not only help us fix the computation of radiation-reaction contributions but also the computation of waveforms itself, which is accomplished through radiation emission diagrams.  

\subsubsection{Outline of the thesis}

The outline of the thesis is the following: In Chapter 2, we review the basic theory of gravitational waves in the realm of the linearized theory of general relativity, building on the basis for the subsequent chapters. In Chapter 3, we review the relativistic two-body problem from an effective field theory (EFT) perspective. In this chapter, the basic construction of EFTs for the problem is based on the separation of scales characteristic of the compact system dynamics. We also perform basic computations that reproduce classical results.

Chapters 4 and 5 are devoted to the presentation of original research. In Chapter 4, we study emission amplitudes for all the tails, including, in particular, the angular momentum failed-tail. As we will see in this chapter, some of these amplitudes present anomalies, which are then properly understood and corrected, with tools adapted from traditional methods. In this chapter, we also investigate the relation between emission and self-energy diagrams, and, in particular, show that a correction to the anomalous self-energy diagrams is necessary to correctly account for radiation-reaction effects. From this, we obtain the correct contribution to the conservative 5PN stemming from the electric quadrupole angular momentum failed-tail, correcting previous results. 
Besides this, we also compute the conservative contributions from the angular momentum tail, for arbitrary multipole moments.

In Chapter 5, we study the classical renormalization of the series of multipole moments present in the far zone EFT, i.e., the effective theory that describes gravitational radiation.
This is achieved from the computation of the so-called tail-of-tail effect. In this process, both infrared and ultraviolet divergences are encountered and properly handled using the standard ideas of regularization and renormalization, which allow us to resum logarithmic contributions via the renormalization group evolution. In particular, the work presented in this chapter extends previously known results to arbitrary electric and magnetic multipole.
 
Finally, in Chapter 6, we present the conclusions and perspectives for future work.

%% file: chapter1/gravitationalwaves.tex
\chapter{The Theory of Gravitational Waves}

In this chapter, we review the most relevant topics of the theory of gravitational waves (GWs) that will be important throughout this thesis. Here we follow mainly the textbooks \cite{maggiore1,maggiore2,poissonbook} and review papers \cite{flanagan,buonanno07,buonanno02,letiec16}.

\section{Gravitational waves and the linearized theory}

The dynamics of gravitational fields in  general relativity is governed by the Einstein field equations\footnote{Here, $G_N$ is the usual Newtonian constant of gravity in three spatial dimensions.}
\begin{equation}\label{EFE}
    R_{\mu\nu} -\frac12 g_{\mu\nu}R  =  8\pi G_N T_{\mu\nu}\,.
\end{equation}
This simple tensor equation, in reality, comprises a set of ten coupled nonlinear second-order partial differential equations for the metric $g_{\mu\nu}$, which, except for a few very special geometries,
do not allow for exact solutions for most of the configurations of astrophysical interest.
Because of this, to study more realistic cases relevant to astrophysics, one needs to resort to a perturbative approach. In the so-called
linearized gravity, the spacetime metric is assumed to fluctuate weakly around the flat Minkowski background:
\begin{equation}\label{perturbation}
    g_{\mu\nu} = \eta_{\mu\nu} + h_{\mu\nu}\,, \qquad\qquad\qquad |h_{\mu\nu}| \ll 1\,.
\end{equation}
Then, thanks to the gauge freedom present in Eq.~\eqref{EFE} due to the huge symmetry group of general relativity, which consists of the infinite-dimensional set of diffeomorphisms (or coordinate transformations) $x^{\mu} \rightarrow x'^{\mu}(x)$, four of its ten components can be constrained by choosing a gauge condition. The Lorentz gauge\footnote{This condition is also commonly known as the De Donger gauge.} is the most common choice, being defined by the following expressions:
\begin{equation}\label{lorentzgauge}
    \partial^\nu \bar{h}_{\mu\nu} = 0\,, \quad\quad\qquad {\rm with}\quad\qquad  \bar{h}_{\mu\nu} \equiv h_{\mu\nu}-\frac 12\eta_{\mu\nu}h^\alpha{}_\alpha\,.
\end{equation}
In this gauge, the Einstein equations \eqref{EFE} take the simpler form of a wave equation
\begin{equation}\label{waveeq}
    \Box \bar{h}_{\mu\nu} = -16\pi G_N T_{\mu\nu}\,,
\end{equation}
with $\Box = \eta^{\mu\nu}\partial_\mu\partial_\nu= -\partial^2/\partial t^2 + \nabla^2$, the usual d'Alembertian operator in flat spacetime. 
Notice that, when applied to the wave equation above, condition \eqref{lorentzgauge} translates into the conservation of energy and momentum in the linearized theory $\partial^\nu T_{\mu\nu} = 0$. 

An enlightening approach that accounts in a more transparent way for the physical degrees of freedom of the linearized theory is the so-called SVT (Scalar-Vector-Tensor) decomposition, \cite{maggiore2,flanagan}. In this formalism, the field perturbations $h_{\mu\nu}$ are rewritten in terms of conveniently defined quantities, through four scalars $\{\phi, H, \gamma, \lambda\}$, two transverse three-vectors $\{\beta_i, \varepsilon_i\}$, and finally, a transverse-traceless rank-2 symmetric three-tensor $h_{ij}^{TT}$, that adds up to the ten original degrees of freedom\footnote{Throughout the text, we follow the convention that Greek letters refer to spacetime indices, whereas Latin indices are restricted to span just over the spatial dimensions.}. From these quantities, in turn, using clever choices of linear combinations, one can build four others, usually denoted by $\{\Theta,\Phi, \Xi_i, h_{ij}^{TT}\}$, with $\Xi_i$ being likewise a transverse three-vector, having the important property of being gauge invariant, that account for the 
gauge fixed degrees of freedom,
and, finally, that allow us to rewrite the full content of the Einstein equations in terms of the following four equations: (In Appendix \ref{svtapp} we review in detail this decomposition)
\begin{align}
    \nabla^2\Theta^e &= 4\pi G_N \rho\,, \\
    \nabla^2\Phi &= 4\pi G_N (\rho + 3P - 3\dot{S})\,,\\
    \nabla^2\Xi_i &= - 16\pi G_N S_i\,,\\
    \Box h_{ij}^{TT} &= - 16\pi G_N \sigma_{ij}.
\end{align}
The quantities on the right-hand side of these equations stem from a similar SVT decomposition of $T_{\mu\nu}$. In particular, this set of equations teaches us the important lesson that, in the linearized theory, $h_{\mu\nu}$ possesses only six physical degrees of freedom, four of them being longitudinal (in $\Theta^e$, $\Phi$, and $\Xi_i$), the remaining two (in $h_{ij}^{TT}$) being radiative degrees of freedom.  

To understand how one can access the radiative degrees of freedom of the linearized Einstein equations in the Lorentz gauge, Eq.~\eqref{waveeq}, let us first study the solutions outside the source, $T_{\mu\nu} = 0$. In this case, the resulting equation is just the equation for a freely propagating wave traveling at the speed of light:
\begin{equation}
    \Box \bar{h}_{\mu\nu} = 0 \qquad \Leftrightarrow \qquad \left(\frac{\partial^2}{\partial t^2} - \nabla^2 \right)\bar{h}_{\mu\nu} = 0 \,.
\end{equation}
It is well-known that, in this case, the Lorentz condition $\eqref{lorentzgauge}$ is not enough to fix the gauge entirely due to some residual gauge freedom. Consequently, one still has the freedom to fix four additional degrees of freedom. For instance, it is simple to show that one is free to set the components $h^{0i}$ and the trace $h^i{}_i$ to zero. In doing it so, the $\mu = 0$ component of the Lorentz gauge condition \eqref{lorentzgauge} results in $\partial^0 h_{00} = 0$, which implies that we can also set $h_{00} = 0$, as far as propagating solutions are concerned, i.e., gravitational waves. This construction defines the so-called transverse-traceless gauge, or simply TT gauge, defined for propagating vacuum solutions that are already in the Lorentz gauge by setting\footnote{For a pedagogical and detailed accounting of the steps that lead to the TT gauge, the reader is referred to sections 1.1 and 1.2 of Ref.~\cite{maggiore1}.}\footnote{We employ Einstein's notation not only to sum over repeated spacetime indices but also for spatial ones.}:
\begin{equation}\label{ttgauge}
    h^{0\mu} = 0\,, \qquad\qquad h^i{}_i =0\,, \qquad\qquad \partial^j h_{ij} = 0\,.
\end{equation}
Notice that, in this gauge, since $h^{\alpha}{}_\alpha =0$, then $\bar{h}_{\mu\nu} = h_{\mu\nu}$, and, more importantly, that we are left with just two degrees of freedom, i.e., the radiative degrees of freedom we were seeking.

In the TT gauge, the general solution for Eq.~\eqref{waveeq} outside the source, denoted by $h_{ij}^{TT}$, is a linear combination of plane waves. As an example, consider a plane wave propagating along the $z$-direction with wave vector $\mathbf{k} = \omega \bf{\hat{n}}$, being $\bf{\hat{n}}$ a unit vector in the $z$-direction. In this case, conditions \eqref{ttgauge} yield
\begin{equation}\label{httalongz}
 h_{ij}^{TT}(t,z) =  \left(
\begin{array}{ccc}
h_+ & h_{\times} & 0 \\
h_{\times} & -h_+ & 0\\ 
0 & 0 & 0
\end{array}
\right)\cos[\omega(t-z)]\,.
\end{equation}
In this expression, $h_+(t,z) = h_+ \cos[\omega(t-z)]$ and $h_\times(t,z) = h_\times \cos[\omega(t-z)]$, with constants $h_+$ and $h_\times$, are called the plus and cross polarizations, respectively, and represent the two degrees of freedom that a gravitational wave has. 

More generally, a plane wave solution $h_{\mu\nu}(t,\mathbf{x})$ of Eq.~\eqref{waveeq}, outside the source, that is propagating along a generic direction $\mathbf{\hat{n}}$, but that is not yet in the TT gauge, can be projected onto the TT gauge by applying the Lambda tensor $\Lambda_{ij,kl}$ on its spatial components:
\begin{equation}
    h_{ij}^{TT} = \Lambda_{ij,kl}(\mathbf{\hat{n}})h_{kl}\,.
\end{equation}
The Lambda tensor is a projector, in the sense that $\Lambda_{ij,mn}\Lambda_{mn,kl} = \Lambda_{ij,kl}$, with the properties of being transverse on all indices, $n^i \Lambda_{ij,kl} = 0$, $n^j \Lambda_{ij,kl} = 0$, ...; traceless with respect to either pair of indices, $(i,j)$ or $(k,l)$, $\delta^{ij}\Lambda_{ij,kl} = 0$ and $\delta^{kl}\Lambda_{ij,kl} = 0$; symmetric under the exchange of pair of indices $(i,j)\leftrightarrow (k,l)$, $\Lambda_{ij,kl} = \Lambda_{kl,ij}$; and is defined by:
\begin{equation}
 \Lambda_{ij,kl} = P_{ik} P_{jl} - \frac12 P_{ij} P_{kl}\,, \quad\quad\qquad {\rm with}\quad\qquad  P_{ij}(\mathbf{\hat{n}}) = \delta_{ij} - n_i n_j\,.
\end{equation}

\section{Interaction of GWs with test masses}

For two particles following nearby geodesics, $x^\mu(\tau)$ and $x^\mu(\tau)+\xi^\mu(\tau)$, with $\xi^\mu$ being a coordinate separation four-vector, and using the respective geodesic equations, one can derive the following expression for the evolution of $\xi^\mu$: 
\begin{equation}\label{deviationeq}
    \frac{D^2\xi^\mu}{D\tau^2} = - R^{\mu}{}_{\nu\rho\sigma}\xi^\rho \frac{d x^\nu}{d\tau}\frac{d x^\sigma}{d\tau}\,.
\end{equation}
This is the so-called geodesic deviation equation\footnote{To derive \eqref{deviationeq} one has to assume $|\xi^\mu|$ to be much smaller than the typical scale of the variation of the gravitational field.}, in which $D/D\tau$ denotes the covariant derivative along the curve $x^\mu(\tau)$, and it represents the tidal gravitational force experienced by the two nearby particles, as a result of the spacetime curvature. Together, the geodesic and geodesic deviation equations are the most fundamental results from general relativity to study the effects of the passage of a gravitational wave on the motion  of particles. Without these equations, gravitational waves would be impossible to be detected.

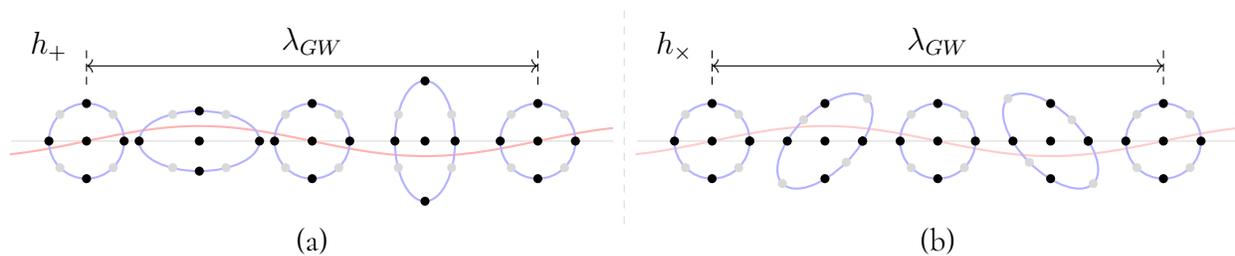
\begin{figure}[t]
\begin{center}
\begin{tikzpicture}
   \draw[red!30,thick] plot[domain=-3.2:7*pi, samples=500]  (\x/pi,{sin(\x r/3)/5}); 
    \draw[gray!30] (-1,0) -- (7,0);
    \draw[blue!30,thick] (0,0) circle (0.5);
    \draw[blue!30,thick] (1.5,0) ellipse (0.8 and 0.4);
    \draw[blue!30,thick] (3,0) circle (0.5);
    \draw[blue!30,thick] (4.5,0) ellipse (0.4 and 0.8);
    \draw[blue!30,thick] (6,0) circle (0.5); 
    \fill (0,0) circle (0.06);    
    \fill (-0.5,0) circle (0.06);
    \fill (0.5,0) circle (0.06);   
    \fill (0,-0.5) circle (0.06);
    \fill (0,0.5) circle (0.06);  
    \fill[gray!30] (0.354,0.354) circle (0.06);
    \fill[gray!30] (-0.354,0.354) circle (0.06);
    \fill[gray!30] (-0.354,-0.354) circle (0.06); 
    \fill[gray!30] (0.354,-0.354) circle (0.06);    
    \fill (1.5,0) circle (0.06);
    \fill (0.7,0) circle (0.06);
    \fill (2.3,0) circle (0.06);   
    \fill (1.5,-0.4) circle (0.06);
    \fill (1.5,0.4) circle (0.06);  
    \fill[gray!30] (1.5+0.354,0.354) circle (0.06);
    \fill[gray!30] (1.5-0.354,0.354) circle (0.06);
    \fill[gray!30] (1.5-0.354,-0.354) circle (0.06); 
    \fill[gray!30] (1.5+0.354,-0.354) circle (0.06);     
    \fill (3,0) circle (0.06);
    \fill (2.5,0) circle (0.06);
    \fill (3.5,0) circle (0.06);   
    \fill (3,-0.5) circle (0.06);
    \fill (3,0.5) circle (0.06);    
    \fill[gray!30] (3+0.354,0.354) circle (0.06);
    \fill[gray!30] (3-0.354,0.354) circle (0.06);
    \fill[gray!30] (3-0.354,-0.354) circle (0.06); 
    \fill[gray!30] (3+0.354,-0.354) circle (0.06);    
    \fill (4.5,0) circle (0.06);
    \fill (4.1,0) circle (0.06); 
    \fill (4.9,0) circle (0.06);
    \fill (4.5,-0.8) circle (0.06);
    \fill (4.5,0.8) circle (0.06);   
    \fill[gray!30] (4.5+0.354,0.354) circle (0.06);
    \fill[gray!30] (4.5-0.354,0.354) circle (0.06);
    \fill[gray!30] (4.5-0.354,-0.354) circle (0.06); 
    \fill[gray!30] (4.5+0.354,-0.354) circle (0.06);    
    \fill (6,0) circle (0.06);
    \fill (5.5,0) circle (0.06);
    \fill (6.5,0) circle (0.06);
    \fill (6,-0.5) circle (0.06);
    \fill (6,0.5) circle (0.06);  
    \fill[gray!30] (6+0.354,0.354) circle (0.06);
    \fill[gray!30] (6-0.354,0.354) circle (0.06);
    \fill[gray!30] (6-0.354,-0.354) circle (0.06); 
    \fill[gray!30] (6+0.354,-0.354) circle (0.06);    
    \draw[<->] (0,1) -- node[above] {$\lambda_{GW}$} (6,1);
    \draw[densely dashed] (0,0.75) -- (0,1.23);    
    \draw[densely dashed] (6,0.75) -- (6,1.23);
    \draw (-0.5,1.6) node[below] {$h_+$};   
    \draw (3,-1) node[below] {(a)};   
\end{tikzpicture}
\begin{tikzpicture}
   \draw[gray!30,dashed] (0,0.5) -- (0,3.4);
   \draw[white] (0,0) circle (0.03);
\end{tikzpicture}
\begin{tikzpicture}
   \draw[red!20,thick] plot[domain=-3.2:7*pi, samples=500]  (\x/pi,{sin(\x r/3)/5}); 
    \draw[gray!30] (-1,0) -- (7,0);
    \draw[blue!30,thick] (0,0) circle (0.5);
    \draw[blue!30,thick,rotate around={45:(1.5,0)}] (1.5,0) ellipse (0.8 and 0.4);
    \draw[blue!30,thick] (3,0) circle (0.5);
    \draw[blue!30,thick,rotate around={45:(4.5,0)}] (4.5,0) ellipse (0.4 and 0.8);
    \draw[blue!30,thick] (6,0) circle (0.5); 
    \fill (0,0) circle (0.06);    
    \fill (-0.5,0) circle (0.06);
    \fill (0.5,0) circle (0.06);
    \fill (0.5,0) circle (0.06);   
    \fill (0,-0.5) circle (0.06);
    \fill (0,0.5) circle (0.06);  
    \fill[rotate around={45:(0,0)},gray!30] (-0.5,0) circle (0.06);
    \fill[rotate around={45:(0,0)},gray!30] (0.5,0) circle (0.06);   
    \fill[rotate around={45:(0,0)},gray!30] (0,-0.5) circle (0.06);
    \fill[rotate around={45:(0,0)},gray!30] (0,0.5) circle (0.06);    
    \fill (1.5-0.5,0) circle (0.06);
    \fill (1.5+0.5,0) circle (0.06);
    \fill (1.5+0.5,0) circle (0.06);   
    \fill (1.5,-0.5) circle (0.06);
    \fill (1.5,0.5) circle (0.06);      
    \fill[rotate around={45:(1.5,0)}] (1.5,0) circle (0.06);
    \fill[rotate around={45:(1.5,0)},gray!30] (0.7,0) circle (0.06);
    \fill[rotate around={45:(1.5,0)},gray!30] (2.3,0) circle (0.06);   
    \fill[rotate around={45:(1.5,0)},gray!30] (1.5,-0.4) circle (0.06);
    \fill[rotate around={45:(1.5,0)},gray!30] (1.5,0.4) circle (0.06);    
    \fill (3,0) circle (0.06);
    \fill (2.5,0) circle (0.06);
    \fill (3.5,0) circle (0.06);   
    \fill (3,-0.5) circle (0.06);
    \fill (3,0.5) circle (0.06); 
    \fill[rotate around={45:(3,0)},gray!30] (2.5,0) circle (0.06);
    \fill[rotate around={45:(3,0)},gray!30] (3.5,0) circle (0.06);   
    \fill[rotate around={45:(3,0)},gray!30] (3,-0.5) circle (0.06);
    \fill[rotate around={45:(3,0)},gray!30] (3,0.5) circle (0.06);      
    \fill (4.5-0.5,0) circle (0.06);
    \fill (4.5+0.5,0) circle (0.06);
    \fill (4.5+0.5,0) circle (0.06);   
    \fill (4.5,-0.5) circle (0.06);
    \fill (4.5,0.5) circle (0.06);       
    \fill[rotate around={45:(4.5,0)}] (4.5,0) circle (0.06);
    \fill[rotate around={45:(4.5,0)},gray!30] (4.1,0) circle (0.06); 
    \fill[rotate around={45:(4.5,0)},gray!30] (4.9,0) circle (0.06);
    \fill[rotate around={45:(4.5,0)},gray!30] (4.5,-0.8) circle (0.06);
    \fill[rotate around={45:(4.5,0)},gray!30] (4.5,0.8) circle (0.06);    
    \fill (6,0) circle (0.06);
    \fill (5.5,0) circle (0.06);
    \fill (6.5,0) circle (0.06);
    \fill (6,-0.5) circle (0.06);
    \fill (6,0.5) circle (0.06);   
    \fill[rotate around={45:(6,0)},gray!30] (5.5,0) circle (0.06);
    \fill[rotate around={45:(6,0)},gray!30] (6.5,0) circle (0.06);   
    \fill[rotate around={45:(6,0)},gray!30] (6,-0.5) circle (0.06);
    \fill[rotate around={45:(6,0)},gray!30] (6,0.5) circle (0.06);      
    \draw[<->] (0,1) -- node[above] {$\lambda_{GW}$} (6,1);
    \draw[densely dashed] (0,0.75) -- (0,1.23);    
    \draw[densely dashed] (6,0.75) -- (6,1.23);
    \fill[white] (7,-0.8) circle (0.06);
    \draw (-0.5,1.6) node[below] {$h_\times$}; \draw (3,-1) node[below] {(b)};      
\end{tikzpicture}
\caption{Effect of the passage of a gravitational wave with plus (a) and cross (b) polarizations on the proper distance $s(t)$ of a set of particles displaced initially in a circular pattern. The proper distances are taken to be with respect to the center of the circle.}
\label{effectofGW}
\end{center}
\end{figure}

In the TT gauge, this equation yields the important result that, for particles initially at rest before the arrival of the gravitational wave, in which case $dx^i/d\tau = 0$ and $d\xi^i/d\tau = 0$, and hence $d^2\xi^i/d\tau^2 = 0$,  the coordinate separation $\xi^i$ remains constant. In other words, the coordinate system defined by the TT gauge is such that the coordinate separation of particles initially at rest is unaffected by the passage of gravitational waves. As a matter of fact, the geodesic equation for $x^\mu(\tau)$ alone predicts this same result for the coordinates $x^i(\tau)$  themselves. Proper distances, nevertheless, \textit{do} change. To see how they change, consider the line element for the metric associated with the gravitational wave in Eq.~\eqref{httalongz},
\begin{align}\label{lineelementttgaugez}
    ds^2 &= - dt^2 + (\delta_{ij} + h^{TT}_{ij})dx^i dx^j \nonumber\\
         &= -dt^2+ dx^2 + dy^2 + dz^2 + \left[ h_+  (dx^2 - dy^2) + 2 h_\times dx dy\right] \cos[\omega(t-z)]\,.
\end{align}
Now, assume the two particles to be on the $xy$-plane of our reference frame, with coordinate separation $\xi^i = (x_2-x_1,y_2-y_1,0)$, coordinate distance $L = \sqrt{(x_2-x_1)^2+(y_2-y_1)^2}$, and displaced along the line $(y_2-y_1)=\alpha (x_2-x_1)$. 
In this case, using the fact that $\xi^i$ is unchanged by the passage of a gravitational wave, the line element above gives for the proper distance
\begin{equation}
    s(t) = L \left\{ 1 + \frac 1{2(1+\alpha^2)} \left[ h_+ (1-\alpha^2) + 2\alpha h_\times\right] \cos(\omega t)
    \right\}\,.
\end{equation}
Notice that, for the case of particles aligned with the $x$-axis ($\alpha = 0$), just the $h_+$ contributes. Likewise, for particles lying within the $45^\circ$ line ($\alpha = 1$), just the $h_\times$ polarization affects the proper distance. Below we show, respectively, these two accounts:
\begin{equation}
    s(t)\Big|_{\alpha = 0} = L \left[1 + \frac12 h_+ \cos(\omega t)\right]\,,
    \qquad \qquad
    s(t)\Big|_{\alpha = 1} = L \left[1 + \frac12 h_\times \cos(\omega t)\right]\,.
\end{equation}
In Fig.~(\ref{effectofGW}) we show the individual effects of the $h_+$ and $h_\times$ polarizations on the physical distance $s(t)$ for a set of particles displaced initially on a circular setup, by the passage of a gravitational wave. We see that the $h_+$ polarization is responsible for stretching and shrinking the configuration in the directions along the $x$ and $y$ axes, i.e., in the shape of ``$+$", while the $h_\times$ polarization does the same, but along axes rotated by $45^\circ$, and hence in the shape of ``$\times$". This justifies the names given to these two polarizations, ``plus" and ``cross".  
In particular, particles lying along the $x$ and $y$ axes are not affected by the passage of cross-polarized gravitational waves, in the same way that particles along axes $x',y'$ rotated by $45^\circ$ with respect to $x,y$ are not affected by plus-polarized gravitational waves.

Alternatively to the frame constructed using the TT gauge, another important frame in the study of gravitational waves is the so-called detector proper frame, being the natural frame used in gravitational-wave detectors. To build such a frame, first recall the well-known fact that at any spacetime point $\mathcal{P}$, it is always possible to build a coordinate system in which $g_{\mu\nu}|_\mathcal{P} = \eta_{\mu\nu}$, $\Gamma^{\mu}_{\rho\sigma}|_\mathcal{P}=0$, being $\Gamma^{\mu}_{\rho\sigma}$ the Christoffel symbol, and hence, $d^2x^\mu/d\tau^2|_\mathcal{P} = 0$. A coordinate system with these characteristics is called a local inertial frame, since around the spacetime point $\mathcal{P}$ particles are freely falling. In other words, this frame not only provides a notion of a local inertial frame in general relativity, but also implements a realization of the equivalence principle, which states that \textit{effects of gravity are undetectable by local experiments.}\footnote{As written more precisely by James Hartle in Ref.~\cite{hartle}, the equivalence principle is the statement that ``Experiments in a sufficiently small freely falling laboratory, over a sufficiently short time, give results that are indistinguishable from those of the same experiments in  an inertial frame in empty space".}
An important example of such a frame is the so-called Riemann normal coordinates. 

Riemann normal coordinates are defined around any chosen spacetime point $\mathcal{P}$,  which serves as the origin of the coordinate system, using an orthonormal tetrad $\{e_{\mu}\}$ defined in the tangent space at $\mathcal{P}$. Then, using the important result from differential geometry that any given point $\mathcal{Q}$ sufficiently close to $\mathcal{P}$ can be reached by a geodesic leaving $\mathcal{P}$ in the direction of a unique unit vector, say $n = n^\mu e_\mu$, after running a proper distance $s$, we are able to assign a set of coordinates to the point $\mathcal{Q}$ via $x^\mu \equiv s n^\mu$. The coordinates for arbitrary points sufficiently close to $\mathcal{P}$ can be similarly defined in a one-to-one way \cite{wald,bishop}.\footnote{While proper distances are used to reach points separated from $\mathcal{P}$ by spacelike geodesics, the proper time is similarly used for the case of timelike geodesics. Null directions, on the other hand, can be filled in by continuity, as limiting cases of space- and timelike geodesics collapsing into each other \cite{hartle}.} This defines the Riemann normal coordinates and provides an example of local inertial frame centered at $\mathcal{P}$.

Starting from the Riemann normal coordinates around $\mathcal{P}$, with tetrad $\{e_\mu\}$, and timelike geodesic defined by $e_0$ and parameterized by the proper time $\tau$, 
we can define non-intersecting three-dimensional spacelike hypersurface for each point along this geodesic.
In this case, these hypersurfaces are generated by the spacelike geodesics emanating from the fiducial geodesic, following arbitrary directions spanned by ${\bf n} = n^i e_i$. Now, $\{e_\mu\}$ are orthonormal tetrad frames defined along the timelike curve, obtained from the initial orthonormal frame at $\mathcal{P}$ by Fermi-transporting \cite{frolov} the three spacelike vectors $\{e_i\}$ along the curve, and $e_0$ being its four-velocity. The Fermi transport is important here since it preserves the orthonormality condition, as well as it possesses the special property of not rotating the vectors $\{e_i\}$ as they are transported. In turn, this last property allows the condition $\Gamma^\mu_{\rho\sigma}|_\mathcal{P} = 0$, of the initial Riemann normal coordinates, to be carried all along the timelike geodesic. Thus, with each of these spacelike hypersurfaces being labeled by the proper time $\tau$ and, within these surfaces points having coordinates $x^i=s(n^1,n^2,n^3)$, we have defined a coordinate system that is a local inertial frame all along a geodesic,
by assigning spacetime points with the coordinates $(\tau,s n^1,s n^2,s n^3)$. A coordinate system built this way is called a freely falling frame.\footnote{Coordinates built this way by dragging an orthonormal tetrad, in the sense of Fermi transport, along a timelike geodesic, are also known as the Fermi normal coordinates.}

From this frame, one can derive the following metric, which represents the proper reference frame as perceived by a freely falling observer, to second order in the coordinates $x^i$\footnote{Strictly speaking, the corrections are of the order of coordinates over the typical variation scales of the metric $\lambda$, $\mathcal{O}(x^i/\lambda)$. In this case, $R_{\mu\nu\rho\sigma} = \mathcal{O}(1/\lambda^2)$.}\cite{misner}:
\begin{equation}\label{ffframe}
    ds^2 = -\left(1+R_{0i0j}x^i x^j\right) dt^2 -\left(\frac 43 R_{0jik} x^j x^k \right) dt dx^i + \left(\delta_{ij} - \frac 13 R_{ikjl} x^k x^l \right) dx^i dx^j\,.
\end{equation}
In this expression, the Riemann tensor is evaluated at the point $\mathcal{P}$, i.e., at $x^i=0$.
Notice that, the effects of curvature start to be important just at second order in $x^i$ below which the metric is flat, in accordance with the equivalence principle. A more complete metric in which the effects of acceleration, rotation, and derived inertial effects are included, also to second order in $x^i$, was obtained in 1978 by Ni and Zimmermann in 	Ref.~\cite{zimmermann}. For Earthbound detectors, this is the more appropriate metric to be used, since the Earth is rotating and everything in its surface is subjected to an acceleration ${\bf a} = {\bf -g}$ with respect to a freely falling frame. This is the frame we call proper detector frame, and the full expression can be checked out in Eq.~(188) of Ref.~\cite{maggiore1}.  Nevertheless, the effects of GWs can be isolated from these additional contributions by focusing on detections within a certain frequency window. 

Typical frequencies for interesting gravitational-wave sources lie within the frequency range $f \sim 10-10^4$Hz, being several orders higher than the frequency of Earth's own rotation ($f \sim 10^{-5}$Hz), and therefore, the effects of frame rotation can be neglected. Acceleration, on the other hand, can be compensated by suspension mechanisms. With these contributions neglected, to second order in $x^i$, the only important source of noise that can compete with gravitational waves at low frequencies are seismic and Newtonian noises, but which can reach up to  $f\lesssim 10$Hz, hence setting the lower bound frequencies for earthbound detectors \cite{badaracco}.
Thus, for frequencies higher than this lower bound, we can assume the only time-varying gravitational field that contributes to the Riemann tensor stems from gravitational waves. In this case, the metric for the proper detector frame coincides with that of a freely falling observer in Eq.~\eqref{ffframe}, which can now be taken as our proper detector frame.

In the proper detector frame, for points close to the origin of the coordinate system, in which case we can take $\Gamma^{\mu}{}_{\rho\sigma}=0$, the equation of the geodesic deviation reduces to
\begin{equation}
    \frac{d^2\xi^i}{d\tau^2} = - R^i{}_{0j0} \xi^j \left( \frac{dx^0}{d\tau}  \right)^2\,.
\end{equation}
Then, keeping just contributions up to linear order in $h$, in which case we can write $\tau = t$, this equation becomes
\begin{equation}
    \ddot{\xi}^i \equiv \frac{d^2\xi^i}{dt^2} = - R^i{}_{0j0} \xi^j\,.
\end{equation}
Moreover, using the well-known fact that, in the linearized theory, the Riemann tensor is \textit{invariant} under coordinate transformations\footnote{
For a generic coordinate transformation $x'^{\mu} = x^\mu + \xi^\mu$ in the linearized theory, the transformation law of the metric results in $h'_{\mu\nu}(x) = h_{\mu\nu}(x) - (\partial_\mu \xi_\nu + \partial_\nu \xi_\mu)$. Then, plugging this into the expression for the linearized Riemann tensor, it is easy to show that the latter is invariant, $R'_{\mu\nu\rho\sigma} = R_{\mu\nu\rho\sigma}$. See, e.g., Sec.~1.1 of \cite{maggiore1}.}, we can take the form we found for gravitational waves in the TT gauge to write:
\begin{equation}
    R^i{}_{0j0} = - \frac12 \ddot{h}^{TT}_{ij}\,.
\end{equation}
In conclusion, the effect of the passage of GWs on test masses, as measured by an experimenter in the proper detector frame, can be described by the simple, Newtonian-like equation
\begin{equation}
    \ddot{\xi}^i = \frac12 \ddot{h}^{TT}_{ij} \xi^j\,.
\end{equation}
From this equation, we see that, unlike what happens in the TT gauge, the passage of GWs changes the position of the particles. As it turns out in this case, the norm of the coordinate separation $\xi^i$ coincides with proper distances to a first approximation, thus allowing for a complete intuitive analysis on Newtonian grounds.
In particular, for the GW in Eq.~\eqref{httalongz} and separation vector given by $\xi^i(t) = (x_0+\delta x(t), y_0+\delta y(t))$, with constants $x_0,y_0$, we have:
\begin{itemize}
    \item[$\triangleright$] For the $+$ polarization 
\begin{equation}\label{plusfff}
        \delta x(t) = \dfrac{h_+}{2}  x_0 \sin(\omega t)\,, \qquad \delta y(t) = -\dfrac{h_+}{2}  y_0 \sin(\omega t)    \,.    
\end{equation}    
    \item[$\triangleright$] For the $\times$ polarization 
\begin{equation}\label{timesfff}
        \delta x(t) = \dfrac{h_\times}{2} y_0 \sin(\omega t)\,, \qquad \delta y(t) = \dfrac{h_\times}{2}  x_0 \sin(\omega t)     \,.   
\end{equation}     
\end{itemize}
Differently from what happens in the TT gauge, the plus polarization alters the $x$ and $y$ positions of particles lying in the $45^\circ$ direction, with $x_0=y_0$. Still, when computing proper distances in this case, using $\sqrt{(\xi^x)^2+(\xi^y)^2}$, and limiting ourselves to order $\mathcal{O}(h)$, we easily see that this yields a constant, consistent with the result obtained in the TT gauge.  

This description of the proper detector frame is valid as long as we keep working at order $\mathcal{O}(x^i)\sim\mathcal{O}(\xi^i)$, neglecting terms of order $\mathcal{O}(|x^i|^2) \sim \mathcal{O}(|\xi^i|^2)\ll \lambda$, with $\lambda$ the typical variation scale of the gravitational field, i.e., the wavelength of the gravitational wave, $\lambda=\lambda_{GW}$. While this is appropriate for Earthbound experiments like LIGO, in which $|\xi^i| = L \sim 4~{\rm km}$ and $\lambda_{GW} \sim 10^3~{\rm km}$ for typical GW events, and hence $(|\xi^i|/\lambda_{GW})^2\sim 10^{-6}$, it is not for space-based detectors like LISA, which has typical size $L\sim 10^6~{\rm km}$ and is designed to detect GWs with wavelengths smaller than $L$.

\section{The energy-momentum tensor for GWs}

It is well-known that gravitational radiation carries energy and momentum. In order to understand how these concepts arise in the linearized theory, we need to go beyond first-order in the perturbation $h_{\mu\nu}$, and assume a generic, nonflat background $\bar{g}_{\mu\nu}$:
\begin{equation}
    g_{\mu\nu} = \bar{g}_{\mu\nu} + h_{\mu\nu}\,, \qquad\qquad\qquad |h_{\mu\nu}| \ll 1\,.
\end{equation}
In this case, to make a clear distinction between the concepts of background and perturbation, we consider one of the following cases, in which a clear separation of scales is defined:
\begin{enumerate}
    \item \textit{Spatial separation}: $\bar{g}_{\mu\nu}$ and $h_{\mu\nu}$ scale respectively as $L_B$ and $\lambda$, with $\lambda \ll L_B$. In this case, the perturbation is regarded as a \textit{ripple} on the smooth background, with small parameter defined by $\lambda/L_B$;
    \item \textit{Time separation}: $\bar{g}_{\mu\nu}$ is defined in frequency domain up to a frequency $f < f_B$, with $h_{\mu\nu}$ being localized around a frequency $f \gg f_B$. In this case, the background is regarded as quasi-static, and the small parameter is given by $f_B/f$.
\end{enumerate}

An analysis of either of these two cases can be carried out in a similar way. Here we follow the second one since this is the natural language used in actual detections. In this case, we proceed by expanding Einstein's equation to second order in $h_{\mu\nu}$, considering the expansion in $f_B/f$.
In particular, the Ricci tensor becomes:
\begin{equation}
    R_{\mu\nu} = \bar{R}_{\mu\nu} + R^{(1)}_{\mu\nu} + R^{(2)}_{\mu\nu}\,.
\end{equation}
In this expression, $\bar{R}_{\mu\nu}$ is evaluated at the background, and hence has only low-frequency modes; $R^{(1)}_{\mu\nu}$ is linear in $h_{\mu\nu}$, and therefore contains only high-frequency modes; and finally, $R^{(2)}_{\mu\nu}$ is the Ricci tensor at second order in $h_{\mu\nu}$, and because of the mixing between positive/negative frequencies, contains both low- and high-frequency modes. From this, the low-frequency part of Einstein's can be drawn, yielding\footnote{While the low-frequency part of Einstein's equation yields energy- and momentum-related observables, from the high-frequency part, one obtains the wave equation for $h_{\mu\nu}$ in generic backgrounds.}
\begin{equation}\label{Riccibar}
    \bar{R}_{\mu\nu} = - [R^{(2)}_{\mu\nu}]^{\rm Low} + 8\pi G_N \big( T_{\mu\nu} - g_{\mu\nu} T/2\big)^{\rm Low}\,.
\end{equation}

We now introduce a time-scale $\bar{t}$ that is much larger than the period of the gravitational waves, $1/f$, and much smaller than the typical time-scale of the background, $1/f_B$, and then average the equation above over $\bar{t}$, i.e., over several periods $1/f$ of the gravitational wave: 
\begin{equation}
    \bar{R}_{\mu\nu} = - \braket{R^{(2)}_{\mu\nu}} + 8\pi G_N \braket{ T_{\mu\nu} - g_{\mu\nu} T/2 }\,.
\end{equation}
The average $\braket{\dots}$ provides a simple way of projecting Einstein's equation on the low-frequency effective theory. In modern language, this corresponds to a renormalization procedure, in which the high-frequency fluctuations (or high-energy modes) are ``integrated out", yielding an effective description of the low-energy physics at scales $\bar{t} \gg 1/f$.

In the right-hand side of Eq.~\eqref{Riccibar}, we can define an averaged energy-momentum tensor of matter $\bar{T}^{\mu\nu}$ through $\braket{ T_{\mu\nu} - g_{\mu\nu} T/2} = \bar{T}_{\mu\nu} - \bar{g}_{\mu\nu} \bar{T}/2$, corresponding to the low-frequency description, and more importantly, we define an effective energy-momentum tensor for gravitational waves $t^{\mu\nu}$ as
\begin{equation}\label{energymomentumtensor}
t_{\mu\nu} = - \frac{1}{8\pi G_N} \braket{R^{(2)}_{\mu\nu} - \frac12 \bar{g}_{\mu\nu} R^{(2)}}\,,
\qquad\qquad {\rm with}\qquad R^{(2)} = \bar{g}^{\mu\nu}R^{(2)}_{\mu\nu}\,,
\end{equation}
and hence,
\begin{equation}
        -\braket{R^{(2)}_{\mu\nu}} = 8\pi G_N \left( t_{\mu\nu} - \frac12 \bar{g}_{\mu\nu} t\right)\,.
\end{equation}
With this, the Einstein field equations take the following form:
\begin{equation}
    \bar{R}_{\mu\nu} - \frac12 \bar{g}_{\mu\nu} \bar{R} = 8\pi G_N (\bar{T}_{\mu\nu} + t_{\mu\nu})\,,
\end{equation}
which shows, in particular, how gravitational waves affect the background curvature.

Following this discussion, it is important to emphasize that the energy-momentum tensor for gravitational waves is naturally defined in terms of averages, but only in cases where a clear separation of scale exists, and hence allows us to separate background from perturbations. 
In fact, only in this effective description, a proper definition of gravitational waves is possible.

Working out explicitly the expression for $R^{(2)}_{\mu\nu}$, and considering large distances from the source, in which case we can take $\bar{g}_{\mu\nu} \rightarrow \eta_{\mu\nu}$ and $\partial_\mu$ for the covariant derivative, the energy-momentum tensor defined in Eq.~\eqref{energymomentumtensor} takes the following simple form:
\begin{equation}
    t_{\mu\nu} = \frac{1}{32\pi G_N} \braket{ \partial_\mu h_{\alpha\beta} \partial_\nu h^{\alpha\beta}}\,.
\end{equation}
To arrive at this expression, there are some assumptions that are further made: (1) we work in the Lorentz gauge, (2) we take advantage of the residual gauge choice to make $h = 0$, (3) we use the equations of motion $\Box h_{\mu\nu} = 0$, and finally, (4) we perform integration by parts inside the averages. Interestingly, $t_{\mu\nu}$ given above can be shown to be invariant under the residual gauge transformations. Consequently, it depends only on the physical modes $h_{ij}^{TT}$. In particular, the gauge invariant energy density is given by
\begin{equation}
    t^{00} = \frac 1{32\pi G_N} \braket{\dot{h}^{TT}_{ij} \dot{h}^{TT}_{ij} } = \frac 1{16\pi G_N} \braket{\dot{h}_+^2 + \dot{h}_\times^2 }\,.
\end{equation}

Defining the energy carried by gravitational waves enclosed by a sphere of volume $V$ by 
\begin{equation}
    E_V \equiv \int_V d^3x\,t^{00}\,,
\end{equation}
one can show that, at large distances from the source, the energy flux, i.e., the GW energy flowing through the spherical surface that encloses $V$, per unit time, is given by
\begin{equation}
\frac{dE}{dA dt} = t^{00}\,.    
\end{equation}
The total energy emitted by the source at a distance $D$, per unit time, is then given by
\begin{equation}\label{totalradiatedenergy}
    P \equiv   \frac{dE}{dt} = \frac {1}{32\pi G_N}D^2 \int d\Omega \, \braket{\dot{h}^{TT}_{ij} \dot{h}^{TT}_{ij} }\,,
\end{equation}
with integration performed over the solid angle $\Omega$. In GW physics, it is often useful to work in the space of positively-defined frequencies, from which we can derive the energy spectrum
\begin{equation}
    \frac{dE}{df} = \frac{\pi}{2G_N}f^2 D^2\int d\Omega\, \left( |\tilde{h}_+(f)|^2 + |\tilde{h}_\times(f)|^2 \right)\,.
\end{equation}

Similarly, we can define the momentum of the GWs inside $V$, at large distances from the source, and compute the momentum flux. In this case, one obtains:
\begin{equation}
    P_V^{i} \equiv \int d^3x\, t^{0i}\,, \qquad \frac{dP^{i}}{dA dt} = t^{0i}\,, \qquad {\rm and}\qquad 
   \frac{dP^i}{dt} = - \frac {1}{32\pi G_N} D^2 \int d\Omega \, \braket{\dot{h}^{TT}_{kl} \partial^i h^{TT}_{kl} }\,. 
\end{equation}

One can also derive the radiated angular momentum using field-theory tools. Here we present just the final result, for completeness\footnote{See, e.g., Subsec.~(2.1.3) of Ref.~\cite{maggiore1}.}:
\begin{equation}
\frac{dL^i}{dt} = \frac{1}{32\pi G_N} D^2 \int d\Omega\, \braket{-\epsilon^{ikl}\dot{h}^{TT}_{ab} x^k \partial^l h^{TT}_{ab} +  2 \epsilon^{ikl} \dot{h}^{TT}_{al} h^{TT}_{ak}}\,.
\end{equation}

\section{Generation of GWs in the linearized theory}

The starting point in our discussion on how GWs are generated in the linearized theory is by taking the wave equation and solving it formally using the notion of Green's functions. The problem of solving a differential equation by the method of Green's function is well-defined whenever we provide an equation, written as an eigenvalue equation in which the linear operator is also a differential operator, together with some chosen boundary conditions. In the case of gravitational waves, we want to solve the wave equation
\begin{equation}\label{waveeqagain}
    \Box \bar{h}_{\mu\nu} = -16\pi G_N T_{\mu\nu}\,,
\end{equation}
subject to the condition that no radiation is incoming.

The formal solution for Eq.~\eqref{waveeqagain} is given by 
\begin{equation}\label{hsolutionformal}
    \bar{h}_{\mu\nu}(x) = -16\pi G_N \int d^4x'\, G(x-x') T_{\mu\nu}(x')\,,
\end{equation}
with $G(x-x')$ being the Green's function, defined implicitly by 
\begin{equation}
   \Box G(x-x') = \delta^4(x-x') \,,
\end{equation}
and $\Box = \partial_\mu\partial^\mu$ acting just on the unprimed coordinates $x$.

The appropriate solution for $G(x-x')$ in this case is the retarded Green's function, 
\begin{equation}
    G_R(x-x') = - \frac{1}{4\pi |{\bf x-x'}|}\delta(t - t' - |{\bf x-x'}|)\,,
\end{equation}
which, when plugged into Eq.~\eqref{hsolutionformal}, gives rise to the important result\footnote{Clearly, the integral of the stress-energy tensor on the volume has the support $|\X|< d$, being $d$ the radius of a	sphere that compehends the whole system.}
\begin{equation}\label{hsolution}
    \bar{h}_{\mu\nu}(t, {\bf x}) =  4G_N \int d^3x'\,\frac{1}{|{\bf x-x'}|} T_{\mu\nu}(t-|{\bf x-x'}|,{\bf x'})\,.
\end{equation}

This result is general within the linearized theory, and valid for sources that generate a sufficiently weak gravitational field,  so that the flat background assumption can still be justified. In the more general case in which the self-gravity of the source is non-negligible, the linear theory is no longer valid, and we must resort to the complete, non-linear theory. In this case, one usually performs a weak-field expansion, by expanding the field equations in powers of $G_N$, and then proceeds by treating the problem perturbatively. In the linearized theory, on the other hand, one may expand the right-hand side of Eq.~\eqref{hsolution} in powers of $v$, the typical velocity of the source's internal dynamics, and then a natural expansion in terms of multipole moments can be constructed. Clearly, this is only valid for configurations in which the weak-field approximation is independent of the low-velocity expansion.

One of the most important sources of gravitational radiation in astrophysics are binary systems of black holes and neutron stars. In such cases, the lowest order equation for energy gives a relation between velocity (the relative velocity between the two bodies, $v$) and the Newtonian potential term linear in $G_N$ in virtue of the virial theorem, 
\begin{equation}
    v^2 \sim \frac{G_N M}{r}\,,
\end{equation}
being $r$ the orbital distance between the two bodies.
Therefore, in studying gravitationally-bound systems, the low-velocity expansion and the weak-field approximation are not independent, and a full account of their  dynamics cannot be attained within linearized gravity.


Nevertheless, it is still instructive to study the expansion of Eq.~\eqref{hsolution} in powers of $v$ since not only it continues to be valid for generic non-gravitational systems, but also it provides insightful results that closely resembles the ones encountered in the more general case where non-linearities are included. 

We then start by performing a \textit{far zone} expansion of \eqref{hsolution}, in which the source, with typical size $d$, is assumed to be far away from the detector:  $D = |{\bf x}| \gg d$. In this case we have
\begin{equation}
    |\X - \X'| = D - \X'\cdot \N + \mathcal{O}\left(\frac{d^2}{D}\right)\,,
\end{equation}
with unit vector $\N\equiv \XX$,
and then, since we are looking for solutions far outside the source, $\bar{h}_{\mu\nu}$ can be projected onto the TT gauge using $h^{TT}_{ij} = \Lambda_{ij,kl}\bar{h}_{ij}$, finally obtaining
\begin{equation}
    h^{TT}_{ij}(t, \X) = \frac{4G_N}{D} \Lambda_{ij,kl}(\N) \int d^3x'\, T_{kl}(t- D + \X'\cdot \N, \X')\,.
\end{equation}
In terms of the Fourier transform of $T_{kl}(t,\X)$, denoted by $\tilde{T}_{kl}(\omega,\K)$, this becomes
\begin{equation}\label{hTTfarFourier}
    h^{TT}_{ij}(t, \X) = \frac{4G_N}{D} \Lambda_{ij,kl}(\N) \int_{-\infty}^{\infty} \frac{d\omega}{2\pi} \, \tilde{T}_{kl}(\omega, \omega \N) e^{-i\omega (t-D)}\,.
\end{equation}
In order to perform a low-velocity expansion of Eq.~\eqref{hTTfarFourier}, we rewrite its integral as
\begin{equation}
\int \frac{d\omega}{2\pi} \, \tilde{T}_{kl}(\omega, \omega \N) e^{-i\omega (t-D)} = \int d^3x'\int \frac{d\omega}{2\pi} \tilde{T}_{kl}(\omega, \X') e^{-i \omega (t-D)} e^{-i\omega \X'\cdot \N}\,,
\end{equation}
and then expand the exponential $e^{-i\omega \X'\cdot \N}$ by noticing that, because at lowest order the frequency of gravitational waves $\omega$ scales as the typical orbital velocity of the internal motion of the source, $\omega_s$, we have $\omega \sim \omega_s \sim v/d$, and then, since $\X' \leqslant d$, we have the relation
\begin{equation}
\omega \X'\cdot \N 	\lesssim \omega_s d \ll 1\,.
\end{equation}
This expansion is easily computed and can be organized in terms of the moments of $T^{ij}$,
\begin{equation}
    S^{ij,i_1\dots i_n}(t) = \int d^3x\, T^{ij}(t,\X) x^{i_1}\cdots x^{i_n}\,,
\end{equation}
in which case, $h^{TT}_{ij}$ takes the following form:
\begin{equation}\label{mulpexpaS}
    h^{TT}_{ij}(t,\X) = \frac{4G_N}{D} \Lambda_{ij,kl}(\N) \times \left[ S^{kl} + n_m \dot{S}^{kl,m} + \frac12  n_m n_p \ddot{S}^{kl,mp} + \dots \right]_{t-r}\,.
\end{equation}

From this expression, we see that, since $d/dt \sim \mathcal{O}(\omega_s)$, and that in moving from one moment to the following one we gain an additional factor of $x^i \sim \mathcal{O}(d)$, each term in the series is one order higher in $\mathcal{O}(\omega_s d) \simeq \mathcal{O}(v)$  with respect to the previous one. Therefore, it is a series in powers of $v$. This is not, however, the multipole expansion generally used, since the moments of $T^{ij}$ can be rather exchanged by moments of $T^{00}$ and $T^{0i}$ which presents a more direct physical interpretation. To do this, we start by defining
\begin{align}
    M^{i_1\dots i_n}(t) = \int d^3x\, T^{00}(t,\X) x^{i_1}\cdots x^{i_n}\,, \label{Mij} \\
    P^{i,i_1\dots i_n}(t) = \int d^3x\, T^{0i}(t,\X) x^{i_1}\cdots x^{i_n}\,,
\end{align}
and then proceed by using the energy-momentum conservation, which in the linearized theory reads $\partial_\mu T^{\mu\nu} = 0$, to build expressions for $S^{ij,i_1\dots i_n}$ in terms of $M^{i_1\dots i_n}$ and $P^{i,i_1\dots i_n}$. In particular, using this conservation equation, it is simple to derive
\begin{equation}
    \partial_0^2 (T^{00}x^i x^j) = 2 T^{ij} + {\rm boundary\,terms}\,,
\end{equation}
from which we obtain $S^{ij} =\ddot{M}^{ij}/2$. Since it is only the traceless part of $S^{kl}$ that enters Eq.~\eqref{mulpexpaS} (because of the lambda projector $\Lambda_{ij,kl}$ in front of the expression), by defining 
\begin{equation}\label{Qij}
    Q^{ij} \equiv M^{ij} - \frac{1}{3}\delta^{ij}M_{kk}\,,
\end{equation}
the leading term in $\mathcal{O}(v)$ of the multipole expansion \eqref{mulpexpaS} becomes
\begin{equation}\label{quadformula}
   [ h^{TT}_{ij}(t,\X) ]_{\rm quad} = \frac{2G_N}{D} \Lambda_{ij,kl}(\N)  \ddot{Q}^{kl}(t-D)\,.
\end{equation}
The quantity $Q^{ij}$ is known as the mass quadrupole of the source.
From this formula, one can draw the physical polarizations $h_+$ and $h_\times$ for a GW propagating along an arbitrary direction parameterized by $\N = (\sin\theta \sin\phi, \sin\theta \cos\phi, \cos\theta)$, yielding\footnote{One way to obtain this, is by projecting $h^{TT}_{ij} = \Lambda_{ij,kl}(\N)h_{kl}$ onto the plane $(x',y')$ perpendicular to $\N = {\bf \hat{z}}'$, with unit vector given by $\N = (\sin\theta \sin\phi, \sin\theta \cos\phi, \cos\theta)$ and $({\bf \hat{x}', \hat{y}', \hat{z}'})$ being an orthonormal basis. Then, denoting by $h^{TT}_{ij} \rightarrow h'_{ab}$ such a projection, with $a,b=(x',y')$, we have that $h_+ = h'_{x'x'}$ and $h_\times = h'_{x'y'}$. }:
\begin{align}
    h_+ (t;\theta,\phi) &= \frac{G_N}{D}[ \ddot{Q}^{11}(\cos^2\phi-\sin^2\phi \cos^2\theta)
                                        +\ddot{Q}^{22}(\sin^2\phi-\cos^2\phi \cos^2\theta)
                                        -\ddot{Q}^{33} \sin^2\theta \nonumber\\
                        &\qquad \qquad \qquad   -\ddot{Q}^{12} \sin 2\phi (1+\cos^2\theta)
                                        +\ddot{Q}^{13} \sin\phi \sin 2\theta
                                        +\ddot{Q}^{23} \cos\phi \sin 2\theta ]\,, \label{hMangles}\\
    h_\times (t;\theta,\phi) &= \frac{G_N}{D}[ (\ddot{Q}^{11} - \ddot{Q}^{22}) \sin 2\phi \cos\theta 
                                             + 2 \ddot{Q}^{12} \cos 2\phi \cos \theta \nonumber\\
                        &\qquad\qquad\qquad\qquad\qquad\qquad\qquad\qquad\,\, - 2 \ddot{Q}^{13} \cos \phi \sin\theta
                                             + 2 \ddot{Q}^{23} \sin\phi \sin\theta ]\label{hmangles}\,.
\end{align}
These expressions will turn out to be particularly useful, shortly, when studying the dynamics of a two-body system.

From $[h^{TT}_{ij}]_{\rm quad}$ above, we can compute the corresponding power emission using Eq.~\eqref{totalradiatedenergy}:
\begin{equation}
    \left( \frac{dP}{d\Omega}\right)_{\rm quad} = \frac{G_N}{8\pi} \Lambda_{ij,kl}(\N) \braket{\dddot{Q}_{ij} \dddot{Q}_{kl}}\,.
\end{equation}
Then, integrating the unit vectors $n^i$ present in $\Lambda_{ij,kl}$ over the solid angle $\Omega$, we obtain the total radiated power (or total gravitational luminosity, sometimes denoted by $\mathcal{L}$):
\begin{equation}
    P_{\rm quad} = \frac{G_N}{5} \braket{\dddot{Q}_{ij} \dddot{Q}_{ij}}\,,
\end{equation}
recalling that the quadrupole moments in this expression are evaluated at the retarded time $t_r = t-D$. This is the famous quadrupole formula first derived by Einstein in 1918, \cite{Einstein1918}.

Generalization toward higher multipole moments follows similarly. In particular, from the next-to-leading term $\dot{S}^{kl,m}$ in Eq.~\eqref{mulpexpaS}, using  $\partial_\mu T^{\mu\nu} = 0$, we can derive:
\begin{equation}
    \dot{S}^{kl,m} = \frac{1}{6} \dddot{M}^{klm} + \frac{1}{3} (\ddot{P}^{k,lm} + \ddot{P}^{l,km} - 2 \ddot{P}^{m,kl})\,.
\end{equation}
The first term on the RHS of this expression generates mass octupole radiation, while the second term gives rise to the current quadrupole. As in the mass quadrupole case, trace terms will not enter $h^{TT}_{ij}$, and therefore, it is useful to define the traceless version of the above quantities. In particular, from the moment $M^{ijk}$ we define the mass octupole $\mathcal{O}^{klm}$ as
\begin{equation}
    \mathcal{O}^{ijk} = M^{ijk} - \frac{1}{5} (\delta^{ij} M^{kll} + \delta^{ik} M^{jll} + \delta^{jk} M^{ill})\,.
\end{equation}
For the current quadrupole term, we define the following quantities:
\begin{itemize}
    \item[-] Angular momentum density associated to the $(j,k)$ plane: $j^{jk} = x^j T^{0k} - x^k T^{0j}$\,.
    \item[-] Angular momentum density vector: $j^{l} = \frac 12\epsilon^l{}_{jk} j^{jk}$\,.
    \item[-] Current quadrupole: $J^{i,j} = \int d^3x\, j^i x^j$\,.
\end{itemize}
It is worth noting that this definition for the current quadrupole is traceless by default. Then, from these relations, it is easy to derive the following identity:
\begin{equation}
    P^{k,lm} + P^{l,km} - 2 P^{m,kl} = \epsilon^{mkp} J^{p,l} + \epsilon^{mlp} J^{p,k}\,. 
\end{equation}

Putting all these contributions together, including the one for the mass quadrupole, the waveform $h^{TT}_{ij}$ for the leading and next-to-leading terms takes the form:
\begin{equation}
    h^{TT}_{ij}(t,\X) = \frac{2G_N}{D} \Lambda_{ij,kl}(\N)  \left[ \ddot{Q}^{kl} + \frac{1}{3} n_m \dddot{\mathcal{O}}^{klm} + \frac{2}{3} n_m (\epsilon^{mkp} \ddot{J}^{p,l} + \epsilon^{mlp} \ddot{J}^{p,k}) + \dots \right]\,,
\end{equation}
from which the total power emission can be computed, yielding
\begin{equation}
    P = G_N\left[ \frac{1}{5}\braket{\dddot{Q}_{ij} \dddot{Q}_{ij}} + \frac{16}{45} \braket{ \dddot{\mathcal{J}}_{ij} \dddot{\mathcal{J}}_{ij} }
    +\frac{1}{189}  \left< \frac{d^4\mathcal{O}_{ijk}}{dt^4} \frac{d^4\mathcal{O}_{ijk}}{dt^4} \right> + \dots \right]\,.
\end{equation}
where we have defined $\mathcal{J}^{ij} \equiv J^{(i,j)}$\,.

The construction above can be understood from a group theory point of view by noting that the mass quadrupole arises from the purely spin-$2$ irreducible representation of the rotation group provided from $\dot{S}^{kl}$ (or, more fundamentally, from $T^{kl}$). In this case, since $S^{kl}$ is a symmetric tensor, it can be decomposed into irreducible representations of the rotation group according to
\begin{equation}
    S^{kl}: \qquad [l=2]\,\, \oplus \,\, [l=0]\,.
\end{equation}
In this case, the $l=2$ irreducible representation of the rotation group gives rise to the mass quadrupole, while the $l=0$ one is a trace term, that do not contribute to the expression for $h^{TT}_{ij}$.
Similarly, for $\dot{S}^{kl,m}$ (or, equivalente, for $T^{kl}x^m$), we have\footnote{Here we use the notation in terms of the dimension of the representation, given by $2l+1$ for a representation $l$ of the rotation group.}:
\begin{align}
    S^{kl,m}: \qquad {\bf (5\,\, \oplus \,\, 1) } \,\, \otimes \,\, {\bf 3} &= {\bf (5\,\, \otimes \,\, 3) } \,\, \oplus \,\, {\bf (1\,\, \otimes \,\, 3)} \nonumber\\
    &= {\bf (7 \,\, \oplus \,\, 5 \,\, \oplus \,\, 3) } \,\, \oplus \,\, {\bf 3}\,.
\end{align}
Then, in this case, the $l=3$ representation generates the mass octupole, while the $l=2$ produces the current quadrupole.

The explicit decomposition of higher-order terms $\dot{S}^{kl,i_1\cdots i_n}$ into irreducible representations of the rotation group leads to a mass $2^{n+2}$-pole moment, as well as to a current $2^{n+1}$-pole moment, while the remaining representations of lower $l$ contribute either with subleading terms to lower $l$ multipoles, or are trace terms that vanish in $h^{TT}_{ij}$. For instance, from the $\ddot{S}^{kl,mn}$ (or $T^{kl}x^m x^n$) term, we obtain the mass $2^{4}$-pole (hexadecapole)  (from $l=4$), the current $2^3$-pole (octupole) (from $l=3$), as well as a subleading contribution to the mass quadrupole arising from the $l=2$ irreducible representation: (the $l=1$ and $l=0$ pieces being trace terms)
\begin{align}
        S^{kl,mn}: \qquad {\bf (5\,\, \oplus \,\, 1) } \,\, \otimes \,\, {\bf 3} \,\, \otimes \,\, {\bf 3} &= {\bf 9 } \,\, \oplus \,\, {\bf (7)_2 } \,\, \oplus \,\,{\bf (5)_4} \,\, \oplus \,\, {\bf (3)_3 }\,\, \oplus \,\, {\bf (1)_2}\,.
\end{align}
In this expression, the subscripts represent the multiplicity of the corresponding irreducible representations generated in the decomposition.


As a matter of fact, general expressions for the components of the full waveform $\bar{h}_{\mu\nu}$ can be systematically constructed, being represented as multipole expansions depending on two families of symmetric-traceless (STF) tensors, the electric $I^{i_1\dots i_l}$ and magnetic multipoles $J^{i_1\dots i_l}$. Just like illustrated above, this construction is based on the tensor decomposition into irreducible representations of the rotation group, with each family being labeled by $l$ and having opposite parity, $(-1)^l$ for the electric and $(-1)^{l+1}$ for the magnetic case.\footnote{In this construction, the electric and magnetic multipole moments provide a generalization for the mass and current multipole moments introduced before to systematically account for the higher-order contributions to lower representations (such as the $l=2$ contribution from $\ddot{S}^{kl,mn}$ to the mass quadrupole moment).} This specific procedure, as well as the full expressions for $\bar{h}_{\mu\nu}$ in terms of $I^{i_1\dots i_l}$ and $J^{i_1\dots i_l}$, and for these multipole moments in terms of the energy-momentum tensor of the source can be found in Refs.~\cite{blanchet1989,damour1991}; See also Sec.~3.5.1 of \cite{maggiore1}.


\section{Inspiral of compact binary systems}

In this section, we study the emission of gravitational waves by an inspiraling binary system of compact objects moving in circular motion. As we will see, as the binary system emits gravitational radiation, its orbit gets shrunk, inducing an increase of the gravitational-wave frequency, which, in turn, back-reacts to the system dynamics, reducing even more the orbital distance between the two bodies. This succession of back-reaction eventually culminates in the coalescence of the binary system, when the two bodies plunge toward each other. 

Within the Newtonian approximation, we consider the compact objects (either neutron stars or black holes) as pointlike particles, with masses $m_1$, $m_2$, and positions $\R_1$ and $\R_2$. Then, let us define also the orbital separation $\R = \R_2 - \R_1$, with $r = |\R|$, the relative velocity ${\bf v} = {\bf v}_2 - {\bf v}_1$, with $v = |{\bf v}|$, and total mass $M = m_1+m_2$, and let us work in the center of mass frame, where $m_1 \R_1 = -m_2 \R_2$. On circular orbits, instead of the velocity, it is more convenient to talk about the orbital frequency $\omega_s$, defined by $v = \omega_s r$. With these variables, Kepler's third law takes the very simple, and elegant form 
\begin{equation}\label{keplerlaw}
    \omega_s^2 = \frac{G_N M}{r^3}\,.
\end{equation}
In particular, parameterizing the circular motion on the $(x,y)$ plane (in the center of mass frame) by 
\begin{equation}
x(t) = r \cos (\omega_s t+\pi/2)\,,\qquad y(t) = r \sin (\omega_s t+\pi/2)\,,\qquad z(t) = 0\,,
\end{equation}
we can easily compute the components of the quadrupole moment via Eqs.~\eqref{Mij} and \eqref{Qij}, with mass density $T^{00} = \rho(\X)  = m_1 \delta(\X-\R_1) + m_2 \delta(\X-\R_2)$, 
which will yield $M^{ij} = \mu r^i r^j$ in the center of mass frame,
and use  Eqs.~\eqref{hMangles}, \eqref{hmangles} to obtain the polarizations $h_+,h_\times$ of a gravitational wave emitted in an arbitrary direction $(\theta,\phi)$ with respect to the binary's frame. The results are the following:
\begin{align}
    h_+(t,\theta,\phi) &= \frac{4G_N\mu \omega_s^2 r^2}{D} \left( \frac{1+\cos^2\theta}{2} \right) \cos (2\omega_s t_{\rm ret}+2\phi)\,, \\
    h_\times(t,\theta,\phi) &= \frac{4G_N\mu \omega_s^2 r^2}{D} \cos \theta \sin (2\omega_s t_{\rm ret}+2\phi)\,.    
\end{align}
From this, notice that the frequency of the gravitational wave is twice the orbital frequency, $\omega_{\rm gw} = 2 \omega_s$, valid in the quadrupolar approximation, and
moreover, that the angle $\theta$ is just the angle between the normal to the binary's orbit and the observing line-of-sight, usually denoted by $\iota$.
Hence, defining the very important quantity called the Chirp mass by
\begin{equation}
\mathcal{M}_c = \mu^{3/5} m^{2/5} = \frac{(m_1 m_2)^{3/5}}{(m_1+m_2)^{1/5}}\,,
\end{equation}
where $\mu$ is the reduced mass given by $\mu = m_1 m_2/M$,  the above expressions for the polarizations, valid for the circular motion, become:
\begin{align}
    h_+ &= \frac{4(G_N \mathcal{M}_c)^{5/3}}{D}  (\pi f_{\rm gw})^{2/3} \left( \frac{1+\cos^2\iota}{2}\right) \cos (2\pi f_{\rm gw} t_{\rm ret} + 2\phi)\,, \\
    h_\times &= \frac{4(G_N \mathcal{M}_c)^{5/3}}{D}  (\pi f_{\rm gw})^{2/3} \cos \iota \sin (2\pi f_{\rm gw} t_{\rm ret} + 2\phi)\,,
\end{align}
after we have used Eq.~\eqref{keplerlaw} to eliminate $r$ in favor or $\omega_s$, and the definition 
$f_{\rm gw} = \omega_{\rm gw}/(2\pi)$.

The Chirp mass is an important parameter because it controls the orbital evolution of the binary system at the Newtonian level: the appearances of the masses $m_1$ and $m_2$ in gravitational-wave observables are only through this combination. For instance, by plugging the above expressions for $h_+, h_\times$ into Eq.~\eqref{totalradiatedenergy}, the emitted power takes the following form:
\begin{equation}\label{Pcircular}
P = \frac{32}{5G_N}\left( \frac12 G_N \mathcal{M}_c \omega_{\rm gw} \right)^{10/3}\,.
\end{equation}

Now, in order to investigate how the binary-system coalescence modifies the emission of gravitational waves, which then acts as a back-reaction on the binary's motion itself, we first consider the system's total energy as the source of radiated energy, which at the Newtonian level reads
\begin{align}\label{Eorbit}
E_{\rm orbit} &= E_{\rm kin} + E_{\rm pot} \nonumber\\
&= - \frac{G_N m_1 m_2}{2r}\,.
\end{align}
In this case, we see that, as the energy of the orbit decreases, i.e., becomes more negative, the separation $r$ must decrease. But then, because of $\omega^2_{\rm gw} \propto r^{-3}$, the frequency of the grativational wave increases, and so does the emitted power as a consequence, which then induces a further decrease in $r$. This process goes on for some time in this inspiraling coalescence.

The passage from the circular to a regime of quasi-circular motion, for a slowly varying orbital separation, is well justified, and provides a good approximation, as far as the condition $\dot{\omega}_s \ll \omega_s^2$ is satisfied. In this case, by taking a time derivative of Eq.~\eqref{keplerlaw}, we have, equivalently, $|\dot{r}| \ll \omega_s r$, meaning that the change in the binary's orbital distance is much smaller than its tangent velocity. As we will see later in this section, this condition is well ensured for actual astrophysical sources.

Assuming this regime of quasi-circular motion, we can compute the variation in the frequency $f_{\rm gw}$ due to the loss of energy by gravitational radiation. We do this by equating $P = -dE_{\rm orbit}/dt$, using Eq.~\eqref{Eorbit} given in terms of the gravitational-wave frequency $\omega_{\rm gw}$,
\begin{equation}
E_{\rm orbit} = -\left(\frac{G_N^2 \mathcal{M}_c^5 \omega_{\rm gw}}{32}\right)^{1/3}\,,
\end{equation}
and $P$ in Eq.~\eqref{Pcircular}. Following these steps, and writing the result in terms of $f_{\rm gw}$, we obtain
\begin{equation}\label{fdotnewton}
\dot{f}_{\rm gw} = \frac{96}{5}\pi^{8/3} (G_N \mathcal{M}_c)^{5/3} f_{\rm gw}^{11/3}\,,
\end{equation}
whose integration gives
\begin{equation}\label{fdotnewtonsol}
f_{\rm gw}(\tau) = \frac{1}{\pi} \left( \frac{5}{256 \tau}\right)^{3/8} (G_N \mathcal{M}_c)^{-5/8}\,.
\end{equation}
In this expression, $\tau \equiv t_{\rm coal} - t$ is the time to coalescence, and, in particular, it diverges as $t$ approaches the time when the coalescence happens, $t_{\rm coal}$. Nevertheless, this is only an artifact of having considered the compact objects as pointlike. For actual astrophysical sources, $f_{\rm gw}(\tau)$ should be cut off at a critical separation, after which the two bodies merge into a single object. This increase in the frequency as we move toward the coalescence is commonly known as the ``chirping" of the gravitational wave.


During the inspiral, for an interval in which the period of the gravitational radiation $T_{\rm gw} = 1/f_{\rm gw}$ is slowly varying, we can estimate the number of cycles undergone by the system, given by $d\mathcal{N}_{\rm cyc} = dt/T_{\rm gw} = f_{\rm gw}dt$, by using Eq.~\eqref{fdotnewton}: 
\begin{align}\label{numbercycles}
\mathcal{N}_{\rm cyc} &= \int^{t_{\rm max}}_{t_{\rm min}} f_{\rm gw}(t) dt = \int^{f_{\rm max}}_{f_{\rm min}} df_{\rm gw} \frac{f_{\rm gw}}{\dot{f}_{\rm gw}} \nonumber\\
&= \frac{1}{32\pi^{8/3}} (G_N \mathcal{M}_c)^{-5/3} (f^{-5/3}_{\rm min} - f^{-5/3}_{\rm max})\,.
\end{align}
Usually $f_{\rm min}$ and $f_{\rm max}$ are taken as the limits of the detector bandwidth, and hence, this quantity becomes useful to assess the sensitivity of the detector.  

Also from the expression for $f_{\rm gw}$, now using Eq.~\eqref{fdotnewtonsol} in combination with Kepler's third law \eqref{keplerlaw}, we can derive the evolution of the orbit separation as a function of $\tau$:
\begin{equation}
\frac{\dot{r}}{r} = -\frac{1}{4\tau}\qquad \rightarrow \qquad r(\tau) = r_0 \left( \frac{\tau}{\tau_0}\right)^{1/4}\,.
\end{equation} 
In this case, $r_0$ is the separation at initial time $t_0$, and similarly, $\tau_0 = t_{\rm coal} - t_0$. In particular, as we approach the time of coalescence, the curve for $r(\tau)$ becomes more and more steep, and because of this, the quasi-circular orbit approximation starts to break down. This last phase goes by the name of plunge, and, similarly to the discussion presented just below Eq.~\eqref{fdotnewtonsol}, this should be cut off at some value of $r$ for which the compact objects start to merge.

For real astrophysical sources, the cutoff value for $r$ is taken to be the \textit{inner-most stable circular orbit} (ISCO) of the Schwarzschild spacetime. This is the smallest value for $r$ such that stable circular orbits can still exist. This value is derived in standard general relativity books (See, for instance, Sec.~6.3 of \cite{wald} or Sec.~5.4 of \cite{carroll}), and has the following value:\footnote{To be more precise, this formula is valid just in the extreme-mass ratio limit, in which, say, $m_1 \gg m_2$, and hence gets corrected in powers of the symmetric mass $\nu = m_1 m_2/M^2$ as we move toward equal mass	 systems.}
\begin{equation}
r_{\rm ISCO} = 6 G_N M\,.
\end{equation} 


\subsection{Estimate for real astrophysical sources}

Using as reference the masses $m_1 = m_2 = 1.4M_\odot$, which is the typical mass for neutron stars and gives $\mathcal{M}_c = 1.21M_\odot$, we can use Eqs.~\eqref{fdotnewtonsol} and \eqref{numbercycles} along with Kepler's law to estimate astrophysical values for gravitational-wave quantities in binary coalescences.

For typical gravitational-wave frequencies, we have
\begin{equation}
f_{\rm gw} (\tau) \simeq 134{\rm Hz} \left( \frac{1.21 M_\odot}{\mathcal{M}_c}\right)^{5/8} \left( \frac{1\,{\rm s}}{\tau}\right)^{3/8}\,.
\end{equation}
Equivalently, inverting for $\tau$, we have
\begin{equation}
\tau \simeq 2.18{\rm s} \left( \frac{1.21M_\odot}{\mathcal{M}_c}\right)^{5/3} \left( \frac{100\,{\rm Hz}}{f_{\rm gw}}\right)^{8/3}\,.
\end{equation}
From this, we see that at $f_{\rm gw} = 10$ Hz, which is the typical lowest frequency in the bandwidth for ground-based interferometers, we have access to the last $\sim 17$ min of the binary coalescence, while for $f_{\rm gw} = 100$Hz and $f_{\rm gw} = 1$ kHz, we have $\tau \simeq 2.18$ s and $\tau \simeq 1$ ms, respectively. For these same frequencies, 
the orbital separation takes the following values: $r=700\,{\rm km}, 156\,{\rm km}$, and $34\,{\rm km}$, respectively.  Note that, only compact objects such as neutron stars and black holes can achieve such small orbital distances.
 
Similarly, for the number of cycles, we have 
\begin{equation}
\mathcal{N}_{\rm cyc} \simeq 1.6\times 10^4 \left( \frac{10\,{\rm Hz}}{f_{\rm min}} \right) \left( \frac{1.2M_\odot}{\mathcal{M}_c}\right)^{5/3}\,.
\end{equation}
So, for $f_{\rm min}=10$ Hz and $\mathcal{M}_c = 1.21M_\odot$, the detector typically observes thousands of cycles. 

Notice that, using Eq.~\eqref{fdotnewton}, along with Kepler's third law, the condition for quasi-circular motion $\dot{\omega}_s \ll \omega_s^2$ gets translated into $G_N \mathcal{M}_c \pi f_{\rm gw} \ll 0.5$, and hence
\begin{equation}
f_{\rm gw} \ll 13\,{\rm kHz} \left(\frac{1.2 M_\odot}{\mathcal{M}_c}\right)\,.
\end{equation}
On the other hand, combining $r_{\rm ISCO} = 6 G_N M$, our estimated value for when the plunge phase starts, with $\omega_s^2 = G_N M/r^3$, we get
\begin{equation}
(f_{s})_{\rm ISCO} \simeq 2.2\,{\rm kHz} \left( \frac{M_\odot}{M}\right)\,.
\end{equation}
We conclude from this that the plunge always happens earlier than the threshold value of the quasi-circular limit can be reached. 
Therefore, as far as we keep ourselves in the inspiral phase, we can safely consider the quasi-circular orbit approximation.
As some examples, we have $(f_{s})_{\rm ISCO} \sim 800\,{\rm Hz}$ for a NS-NS coalescence of masses $m_1 = m_2 = 1.4M_\odot$, and $(f_{s})_{\rm ISCO} \sim 200\,{\rm Hz}$ for the case of two black holes with masses $m_1 = m_2 = 10M_\odot$.

\subsection{The waveform in the quasi-circular orbit approximation}

Similarly to the circular case, for an inspiraling binary system 
in a quasi-circular orbit with time-varying frequency $\omega_s(t)$, we parameterize the motion in Cartesian coordinates by  
\begin{equation}
x(t) = r(t) \cos (\Phi(t)/2)\,, \qquad
y(t) = r(t) \sin (\Phi(t)/2)\,, \qquad
z(t) = 0\,, 
\end{equation}
with $\Phi$ being the gravitational-wave phase defined by 
\begin{align}
\Phi(t) &= 2 \int^{t}_{t_0} dt' \omega_s(t')\nonumber\\
&=\int^t_{t_0} \omega_{\rm gw}(t')\,.\label{accumulatesGWphase}
\end{align}

In order to take into account the effects of the inspiral coalescence into the waveform, we assume the quasi-circular limit $\dot{\omega}_s \ll \omega_s^2$. This limit amounts to considering negligible appearances of $\dot{r}$ when deriving the expressions for $h_+$ and $h_\times$. In this case, the only difference in comparison to the circular case is that in the argument of the trigonometric functions we have $2\omega_s t \rightarrow \Phi(t)$, while $\omega_s \rightarrow \omega_s(t)$ anywhere else. We thus obtain
\begin{align}
    h_+(t) &= \frac{4(G_N \mathcal{M}_c)^{5/3}}{D}  (\pi f_{\rm gw}(t_{\rm ret}))^{2/3} \left( \frac{1+\cos^2\iota}{2}\right) \cos [ \Phi(t_{\rm ret}) ]\,, \label{hMdirect} \\
    h_\times(t) &= \frac{4(G_N \mathcal{M}_c)^{5/3}}{D}  (\pi f_{\rm gw}(t_{\rm ret}))^{2/3} \cos \iota \sin [ \Phi(t_{\rm ret}) ]\,. \label{hXdirect}
\end{align}

However, for data analysis purposes, to correlate the signals that enter the detector to theoretical predictions, one must have instead the Fourier transform of the gravitational-wave amplitude. This is achieved using the stationary phase method, necessary in this case to circumvent the fact that $h_{+}(t)$ and $h_{\times}(t)$ diverge in direct space at $t = t_c$.\footnote{To see this, one should plug $f_{\rm gw}(\tau)$, in Eq.~\eqref{fdotnewtonsol}, into the polarizations $h_{+,\times}(t)$ in Eqs.~\eqref{hMdirect}-\eqref{hXdirect}.}
The stationary phase method is widely known and can be found in, e.g., Refs.~\cite{maggiore1,buonanno07}. With this, one obtains:  
\begin{align}
\tilde{h}_{+}(f) &= A e^{i \Psi_{+}(f)} \frac{(G_N \mathcal{M}_c)^{5/6}}{D} f^{-7/6} \left( \frac{1+\cos^2\iota}{2}\right) \,, \label{hMFourier}\\
\tilde{h}_{\times}(f) &= A e^{i \Psi_{\times}(f)} \frac{(G_N \mathcal{M}_c)^{5/6}}{D} f^{-7/6} \sin\iota \,, \label{hXFourier}
\end{align}
where
\begin{equation}
A = \frac{1}{\pi^{2/3}} \left( \frac{5}{24}\right)^{1/2}\,,
\end{equation}
and phases given by $\Psi_\times = \Psi_+ + (\pi/2)$ and
\begin{equation}\label{psiplus}
\Psi_+(f) = 2\pi f(t_c + r) - \Phi_0 - \frac{\pi}{4} + \frac{3}{4} (8\pi f G_N \mathcal{M}_c)^{-5/3}\,.
\end{equation}
In this expression, $\Phi_0$ is the gravitational-wave phase at the coalescence.

The waveforms given in Eqs.~\eqref{hMdirect}-\eqref{hXdirect} in direct space and Eqs.~\eqref{hMFourier}-\eqref{hXFourier} in Fourier space describe the gravitational radiation emitted during the coalescence of a binary system of neutron stars or black holes undergoing a slow adiabatic inspiral, going through a succession of quasi-circular orbits. Note that these are valid only during the inspiral phase evolution of the system before the two compact objects start to plunge toward each other.


\subsection{Beyond the Newtonian approximation}

As argued before, the linearized theory of gravity is valid for arbitrarily large velocities, but only within the weakly-field regime, in which the particles' motion lies in the flat spacetime. Nevertheless, astrophysical systems of interest in gravitational-wave physics, like binary systems of black holes and neutron stars, are held together by gravitational forces, which implicates a deviation from the flat spacetime as we go to higher and higher velocities, in accordance with the virial theorem, that relates $v^2$ to $G_N M/r$.
Hence, corrections in velocity are not independent of corrections in the spacetime geometry. In this case, a consistent framework that takes into account both kinds of contribution, that holds beyond the linearized theory, is necessary. The post-Newtonian (PN) formalism does precisely this job. Having this in mind, the present subsection is devoted to presenting general aspects and important results of this formalism. In Chapter~\ref{ChaptEFT} we introduce a way of performing actual computation of post-Newtonian corrections, within an effective field theory approach.

The post-Newtonian approximation provides general-relativistic corrections to the Newtonian trajectory and to gravitational-wave observables in powers of $v$, that encapsulate the deviations from the slow motion and flat spacetime, valid during the orbital phase of inspiraling compact binaries. In this perturbative approximation, it is convenient to introduce the dimensionless gauge-invariant parameter $x$, defined in terms of the orbital frequency by 
\begin{equation}
x \equiv (G_N M \omega_s)^{2/3}\,.
\end{equation}
This parameter is of order $x \sim O(v^2)$, with the power $x^n$ defining the $n$PN order. In particular, one can compute PN-corrected expressions in terms of $x$ for all the important gravitational-wave quantities. For instance, up to the first PN correction beyond the Newtonian trajectory, the energy of the system is given by
\begin{equation}\label{E1PN}
E = - \frac12 \mu x  \left[ 1 + \left( -\frac{3}{4} - \frac{1}{12}\nu \right) x + O(x^2) + O (x^3) + \dots \right]\,,
\end{equation}
while the radiated power reads
\begin{equation}\label{P1PN}
P_{\rm gw} = \frac{32}{5G_N} \nu^2 x^5 \left[ 1 + \left( -\frac{1247}{336} - \frac{35}{12}\nu\right) x + O (x^{3/2}) + O (x^2) + O (x^{5/2}) + \dots \right]\,.
\end{equation}
In these expressions, $\nu$ is the symmetric mass ratio defined by $\nu \equiv \mu/M = m_1 m_2/M^2$. Note that,
while in the expression for the total energy of the system we only see integer powers of $x$, and hence, even scalings in $v^2$, as expected by parity, we see half-integer contributions in the radiated power. The latter odd-behaving contributions stem from radiation reaction. In particular, in Eq.~\eqref{P1PN}, the $O(x^{3/2})$ term stems from the tail effect.

Equations \eqref{E1PN} and \eqref{P1PN} can be used, combined into the energy balance equation
\begin{equation}\label{Ebalance}
\frac{dE}{dt} = - P_{\rm gw}\,,
\end{equation}
to derive an expression for the accumulated orbital frequency $\phi$, defined by $\int dt\, \omega_s$, and which gives $\Phi = 2 \phi$ in the quadrupole approximation (See Eq.~\eqref{accumulatesGWphase}). By doing it so, one obtains\footnote{Note that this procedure is equivalent to taking the so-called TaylorT4 approximant \cite{taylorF4}, which consists in Taylor-expanding the ratio $(dE/dv)/P$ appearing in Eq.~\eqref{gwphaseintro} in the relevant PN order.}
\begin{equation}
\phi = - \frac{x^{-5/2}}{32\nu} \left[ 1 + \left( \frac{3715}{1008} + \frac{55}{12}\nu \right)x + O(x^{3/2}) + O(x^2) + \dots \right]\,.
\end{equation}
For the complete 4PN energy and 4.5PN energy flux expressions, as well as full derivation of the 4.5PN order orbital phase $\phi$, cf. Refs.~\cite{blanchet2023,blanchet2023a}; See also Ref.~\cite{blanchet2005a} for the 3.5PN derivation.
Notice that, in the first term on the right-hand side of this equation, which corresponds to the Newtonian level, the masses of the two constituents appear only through the combination $x^{-5/2}\nu^{-1}\propto \mathcal{M}_c^{-5/3}$. Only by taking into account at least the next term, corresponding to the 1PN correction, is that this degeneracy is broken, and one is able to resolve separately the values of the two masses with observation.


In general, when considering the post-Newtonian approximation, the first ingredient that one should consider in the theoretical analysis of the binary's dynamics is its equation of motion. This provides a generalization of the Kepler's third law, from which the orbital frequency can be derived, as well as the energy of the system, that is conserved in the absence of radiation-reaction forces. The second ingredient in this analysis refers to the gravitational waveform generated by the binary. Our interest in such a quantity lies in the fact that we can compute the radiated energy from it. In particular, one has to take into account relativistic corrections to the source multipole moments, that will enter the expression for the energy flux. But not only this, one has also to include contributions stemming from nonlinear interactions of the gravitational field \cite{blanchet2005a}. Once this is done, the energy balance equation \eqref{Ebalance} is used to derive the orbital phase, which is, in turn, directly related to the gravitational-wave phase.
In particular, the state-of-the-art for the gravitational-wave phase for inspiraling binaries (neglecting spin effects) is in the 4.5PN order. In this case, Eq.~\eqref{psiplus} gets corrected into the following expression \cite{buonanno07}, with coefficients obtained in Refs.~\cite{blanchet2023,blanchet2023a}:
\begin{equation}
\Psi_+(f) = 2\pi f (t_c + r) -  \Phi_0 - \frac{\pi}{4} + \frac{3}{128 \nu x^{5/2}} \sum_{k=0}^9 \psi_{(k/2) \rm PN} x^{k/2}\,,
\end{equation}
with
\begin{align}
\psi_{0\rm PN} &= 1\,, \\
\psi_{0.5\rm PN} &= 0\,, \\
\psi_{1\rm PN} &= \left( \frac{3715}{756} + \frac{55}{9}\nu\right)\,, \\
\psi_{1.5\rm PN} &= -16\pi \,, \\
\psi_{2\rm PN} &= \left( \frac{15293365}{508032} + \frac{27145}{504}\nu + \frac{3085}{72}\nu^2\right) \,,\\
\psi_{2.5\rm PN} &= \pi\left(\frac{38645}{756} - \frac{65}{9}\nu\right) \left(1+\frac{3}{2}\log x\right) \,, \\
\psi_{3\rm PN} &= \left( \frac{11583231236531}{4694215680} - \frac{640\pi^2}{3} - \frac{6848 \gamma_E}{21} \right) \nonumber\\
&+ \nu \left( - \frac{15737765635}{3048192} + \frac{2255\pi^2}{12}\right) \nonumber\\
&+ \frac{76055}{1728} \nu^2 - \frac{127825}{1296}\nu^3 - \frac{3424}{21}\log(16x)\,,\\
\psi_{3.5\rm PN} &= \pi \left( \frac{77096675}{254016} + \frac{378515}{1512}\nu - \frac{74045}{756} \nu^2 \right)\,,\\
\psi_{4\rm PN} &= \frac12 \bigg[\frac{90490}{189} \pi^2+\frac{36812}{63} \gamma_E+\frac{1011020}{1323} \log 2+\frac{78975}{196} \log 3+\frac{9203}{63} \log x \nonumber\\
& +\left(-\frac{109295}{224} \pi^2+\frac{3911888}{1323} \gamma_E+\frac{9964112}{1323} \log 2-\frac{78975}{49} \log 3+\frac{977972}{1323} \log x\right) \nu \nonumber\\
& +\left(-\frac{7510073635}{3048192}+\frac{11275}{144} \pi^2\right) \nu^2-\frac{1292395}{12096} \nu^3+\frac{5975}{96} \nu^4 \nonumber\\
& - \frac{2550713843998885153}{276808510218240} + \frac{680712846248317}{42247941120} \nu \bigg] \log x \,,\\
\psi_{4.5\rm PN} &= \pi \left[\frac{105344279473163}{18776862720}-\frac{640}{3} \pi^2-\frac{13696}{21} \gamma_E-\frac{6848}{21} \ln (16 x)\right. \nonumber\\
&+\left(-\frac{1492917260735}{134120448}+\frac{2255}{6} \pi^2\right) \nu+\frac{45293335}{127008} \nu^2+\frac{10323755}{199584} \nu^3\bigg]  \,.
\end{align}

To detect gravitational-wave signals emitted from astrophysical sources, laser interferometer observatories such as LIGO, Virgo, and KAGRA, 
which employ
matched-filtering techniques, usually make use of templates built from the simple extension of the quadrupole formula given in Eqs.~\eqref{hMdirect} and \eqref{hXdirect}. This phase, being twice the binaries' orbital frequency in this approximation, is then corrected to higher post-Newtonian orders. Nevertheless, while this optimizes the detection of gravitational-wave signals, and roughly estimates the binaries' properties, a further refined analysis is necessary for a more accurate parameter estimation. This is done by PN-correcting the amplitudes of the gravitational-wave components $h_+$, $h_\times$ themselves, the state-of-the-art lying at the 3.5PN order in this sector \cite{henry2023}; See also \cite{blanchet2008}. Templates modelled from the quadrupole formula (the ``dominant harmonic"), and neglecting higher-order multipole moments, is usually called the restricted PN approximation.

The knowledge of spin effects in the compact binary inspiral is of crucial importance for astrophysics, as it is one of the two unique features black holes have, along with the mass, being also relevant for neutron stars in order to learn on their internal structure. In addition, the presence of spins crucially affects the dynamics of the binary, in particular leading to orbital plane precession if they are not aligned with the orbital angular momentum and thereby to strong modulations in the observed signal frequency and phase. The complete dynamics and waveform for spinning and non-precessing binaries in quasi-circular motion is currently known at the 3.5PN order \cite{quentin2022}.

%% file: chapter2/eftapproach.tex
\chapter{The Effective Field Theory Approach}\label{ChaptEFT}

Effective field theories (EFTs) can be constructed whenever there is a clear separation of scales involved in the physical problem one is interested in studying. This happens because of the well-known fact that the observables of a long-distance/low-frequency theory decouple from the short-distance/high-frequency physics. This property, which goes under the name of \textit{decoupling}, is the backbone behind the construction of EFTs. This is why, for instance, in Newtonian physics, we do not need to have any prior knowledge of the (quantum) internal structure of the bodies to be able to describe their dynamical evolution. In this way, an effective description can be very powerful to describe the physics at different scales within the same problem. As we will see in some detail, the effective field theory approach applied to describing the dynamics of gravitationally bound binary systems will be very advantageous in such a setting, as the gravitational radiation and conservative dynamics can be studied separately\footnote{At higher orders, there is a mix of scales involving radiation-reaction processes (long-distance physics), that will affect the conservative dynamics of the system (short-distance physics). See Sec. \ref{secfarzone}}. Here, we follow the Non-Relativistic General Relativity construction developed  by W. Goldberger and I. Rothstein in the seminal paper \cite{rothstein2006}, which provides an EFT framework in which the inspiral phase of coalescing binary systems can be studied. 
For reviews on the EFT application to such systems, the reader is referred to Refs.~\cite{porto2016,levi2018,sturani2021,sturani2014,goldberger2022a,goldberger2022b,goldberger2007,rothstein2014,brunello2022}. 
In this chapter, for simplicity, we deal with the nonspinning case.

\section{Hierarchy of scales}\label{subsec:hierarchy}

To justify the use of an effective field theory description in the case of inspiralling compact binaries, we need first to study the typical scales involved in this classical two-body problem. We start from the smallest of the scales, which corresponds to the internal structure of the constituents of the binary system, characterized by their Schwarzschild radius $r_s=2G_Nm$ in the case of black holes, and $\mathcal{R}\gtrsim \mathcal{O}(r_s)$ for neutron stars. Then, going up in the scale of distances, we arrive at the orbital scale, which has a typical size given by the separation distance $r$ between the two constituents. The next relevant scale is given by $\lambda$,  which describes the scale of long-wavelength physics, i.e., the physics of gravitational-wave emission. 

In the case of gravitationally bound binary systems, the virial theorem $v^2\sim G_N m/r$, relating the relative velocity $v$ between the two compact objects to its orbital separation $r$, is used to establish a hierarchy of scales by noting that  $r_s = 2G_N m\sim r v^2$ and, since $\lambda\sim r/v$, as can be easily checked using basic kinematics arguments\footnote{Recall from the previous chapter that, for the circular motion, we had $\omega_{\rm gw} = 2\omega_s$. Hence, using $v = r \omega_s$ and $\omega_{\rm gw} = 2\pi f = 2\pi/\lambda$, we arrive at $\lambda = \pi r/v$, and then $\lambda \sim r/v$.}, we have $\mathcal{R} \sim r_s \sim rv^2 \sim \lambda v^3$. Thus, for nonrelativistic systems, in which  $v\ll 1$, we have a well-defined separation of scales:
\begin{equation}
r_s \,\,  \lesssim \mathcal{R}  \ll \,\, r \,\, \ll \,\, \lambda\,\,.
\end{equation}  
In this case, the parameter $v$ not only controls the hierarchy, but also serves as the small parameter of the perturbative expansion that will follow the construction of the effective description for the binary system. Such a perturbative treatment of a gravity theory, as expanded in powers of $v^2$, is the so-called post-Newtonian (PN) approximation, each order $(v^2)^n$ being labeled by $n$PN, 0PN being the leading (Newtonian) order, 1PN the next-to-leading order, and so on for each of the corrections beyond the Newtonian level.

Then, in principle, we can use the hierarchy of scales defined above to construct a tower of effective field theories, one for each of the relevant scales: an effective field theory for the physics at length scale $r_s$, designed to describe the internal properties of the compact objects; another one valid at the scale $r$, that describes the orbital dynamics of bound systems, assuming the size of the objects as negligible; and, finally, an effective field theory for the physics at the scale $\lambda$, which describes the gravitational field and gravitational waves generated by the composite system, considered as a point-particle at this scale. Hence, the physical effects that emerge at different length scales can be investigated separately.

\section{Building the EFT for the binary system}\label{buildingEFTs}

The most common approach employed to build EFTs, the \textit{bottom-up approach}, consists in writing down a general Lagrangian containing all possible types of \textit{local} terms (called \textit{operators} in this context) consistent with the symmetries of the problem. In general, this will comprise an infinite number of terms that should be truncated at some relevant order in the PN expansion. Here we follow the framework originally proposed in \cite{rothstein2006} for the construction of the EFT describing the dynamics of nonrelativistic extended objects coupled to gravity. The starting point is that of a theory of relativistic point particles coupled to gravity
\begin{equation}\label{EHPPaction}
S=S_{\rm EH}+S_{\rm pp}\,,
\end{equation}
where $S_{\rm EH}$ is the Einstein-Hilbert Lagrangian
\begin{equation}\label{EHaction}
S_{\rm EH}=\frac{1}{16\pi G_N}\int d^4x \sqrt{-g}\,R[g]
\end{equation}
and $S_{\rm pp}$ the effective action for the point-like objects. For the latter, we construct the action by adding to the minimal coupling all the operators consistent with the symmetries of \textit{diffeomorphism} and \textit{reparameterization} invariance. We have, in this case,
\begin{equation}\label{ppaction}
S_{\rm pp}=-\sum_a m_a \int d\tau_a+\sum_a c_R^{(a)}\int d\tau_a\, R(x_a)+\sum_a c_V^{(a)}\int d\tau_a\, R_{\mu\nu}(x_a)u_a^{\mu}u_a^{\nu}+\dots\,,
\end{equation}
where $u_a^\mu = dx_a^\mu/d\tau$ is the four-velocity of particle $a$, and $R$ and $R_{\mu\nu}$ respectively the Ricci scalar and Ricci tensor.
The idea is that by adding all these infinitely many operators allowed by the symmetries of the problem, finite size effects would be systematically taken into account. In this case, the coefficients $c_R^{(a)}$, $c_V^{(a)}$, $\dots$, carry information about the ultra-violet (UV) limit of the theory, which in this case is the physics of the internal structure of the compact objects, valid at the scale $r_s$. In the EFT setup, have we had the complete theory, the effective action \eqref{ppaction} would then be obtained by \textit{integrating out} the modes corresponding to the scale $r_s$ in this full theory, yielding the effective theory at orbital scale $r$ described by Eqs. \eqref{EHPPaction}-\eqref{ppaction}. This is simply the implementation of the Wilsonian paradigm of decoupling of short distance scales, through the renormalization group evolution, with coefficients $c^{(a)}_R, c^{(a)}_V, \dots$, being the so-called Wilson coefficients of the effective action, that encode the internal structure of the objects in the UV limit. In this case, the effective action is organized in powers of the small parameter $(r_s/r)$.

As it turns out, all the operators built up from the Ricci tensor $R_{\mu\nu}$ and Ricci scalar $R$ can be removed by field redefinitions \cite{rothstein2006,porto2016}, and therefore will not play any role in our discussion. Physically speaking, this can be understood as a consequence of vacuum solutions having a vanishing Ricci tensor. Therefore, finite size effects are rather incorporated through terms constructed out of the Riemann tensor (or, rather, the Weyl tensor). The first two of these operators entering Eq. \eqref{ppaction} are quadratic in the curvature tensor and read
\begin{equation}\label{finite}
S_{\text{finite}} =\frac{1}{2}\int d\tau\, (c_E\, E_{\mu\nu}E^{\mu\nu}+c_B\, B_{\mu\nu}B^{\mu\nu})\,,
\end{equation} 
where $E_{\mu\nu}$ and $B_{\mu\nu}$ are respectively the electric and magnetic components of the Weyl tensor\footnote{As discussed above, terms proportional to $R_{\mu\nu}$ or $R$ in the effective action Eq. \eqref{ppaction} are not physically relevant, and thus it is equivalent to use either $R_{\mu\nu\rho\sigma}$ or $C_{\mu\nu\rho\sigma}$.}:
\begin{equation}\label{EB}
E_{\mu\nu} \equiv C_{\mu\alpha\nu\beta}u^\alpha u^\beta\,,		\qquad	  B_{\mu\nu} \equiv\frac{1}{2}\epsilon_{\mu\alpha\beta\sigma} C^{\alpha\beta}{}_{\nu\rho}u^\sigma u^\rho\,.
\end{equation}
We recall below the expression for the Weyl tensor in terms of the Riemann tensor, Ricci tensor and Ricci scalar, recalling also that it is just the tracefree version of $R_{\mu\nu\rho\sigma}$:
\begin{equation}
C_{\mu\nu\rho\sigma} = R_{\mu\nu\rho\sigma} - \frac{2}{D-2}(g_{\rho[\mu}R_{\nu]\sigma} - g_{\sigma[\mu}R_{\nu]\rho}) + \frac{2}{(D-1)(D-2)}g_{\rho[\mu}g_{\nu]\sigma}R\,. 
\end{equation}

The coefficients $c_E$ and $c_B$ appearing in $S_{\rm finite}$ provide a gauge invariant definition for the $\ell = 2$ static tidal ``Love numbers" which characterize the gravitational
response to tidal deformations \cite{goldberger2022b}. Thus, in order to consistently 
account for tidal effects in the binary system, higher-order operators should be subsequently included. 
Interestingly, it has been shown that these numbers are vanishing for nonspinning black holes and, hence, finite size effects are pushed to a yet higher order in the small parameter $(r_s/r)$ when the two compact objects are taken to be Schwarzschild black holes \cite{damour2009}.

\section{The path integral approach in the classical problem}\label{pathintegral}

Once we have built our effective theory for the constituents of a binary system, which, as we have seen, can also incorporate finite size effects, we are in position to compute the relevant physical quantities of interest in the description of the gravitational waveforms: (1) the binding potential that holds the two astrophysical objects together; and (2) the radiated power loss of GWs emitted by the sources. To this end, we make use of the path integral formulation. One could immediately object to the use of such a tool, usually used in the realm of a quantum theory, to describe a classical system. As it turns out, the presence of the constant $\hbar$ drops out in all classical computations for a class of diagrams as we will see below.

In the classical setting the path integral is dominated by the saddle-point\footnote{Here we normalize $Z[0]=1$ since this constant factor enters the worldline effective action as an additive term, and therefore plays no role in our effective description.}:
\begin{equation}
Z[J]=e^{iW[J]} = \int \mathcal{D}\phi\, e^{i S[\phi,J]}\quad\xrightarrow{\text{Classical limit}}\quad e^{iS[\phi_J,J]}\,,
\end{equation}
where $S[\phi,J]$ is the field theory for a generic field $\phi$ and source $J$, and  $\phi_J$ is a classical solution, i.e., a field configuration that minimizes the action $S[\phi,J]$. From this, we see that $W[J]$ becomes the most important quantity in this discussion, as opposed to $Z[J]$, since in the classical limit $W[J]\rightarrow S[\phi_J,J]$, $W[J]$ becomes the 
action that describes the classical
dynamics of the source $J$. Then, when applied to our particular case,  defined by Eqs. (\ref{EHPPaction})-(\ref{ppaction}), we arrive at what we call a \textit{worldline effective theory}. Before proceeding  any further, there are important points worthwhile mentioning regarding the perturbative approach based on path integrals and about the use of Feynman diagrams in classical settings:
\begin{enumerate}[label=(\roman*)] 
\item While $Z[J]$ is given by the sum of all Feynman diagrams, which in particular include also disconnected diagrams, $W[J]$, on the other hand, (and, therefore, $S[\phi_J]$ in the classical limit) contains only connected diagrams \cite{zee};
\item In this worldline approach, diagrams are represented by solid lines, one for each source $J$, and interactions (mediated by the field $\phi$) are represented by lines attached to the source worldlines (See, for instance, the diagrams in Fig.~\ref{correction1PN} below);  
\item In classical theories, i.e., theories in which $S[\phi_J,J]\sim L \gg \hbar$, where $L$ is the angular momentum of the system, loop diagrams are not present since they are $\hbar$-suppressed with respect to the tree-level diagrams\footnote{To have an idea, systems detected by LIGO have a typical angular momentum of $L\sim 10^{77}\hbar$.}, the latter being the only ones to enter the classical description; (here, by ``loop", we mean closed loops made exclusively of graviton modes; See the discussion in Sec.~\ref{secmassive}.) 
\item Once we have obtained an expression for $W[J]\equiv\text{Re}W[J]+i \text{ Im}W[J]$, its real part will give information on the conservative dynamics of the classical system, while from its imaginary part, it will be possible to extract the radiated power via the optical theorem, as we will see in Sec.~\ref{opticaltheoremsec}.\footnote{As we also discuss in Sec.~\ref{opticaltheoremsec}, this is only valid when employing Feynman's boundary condition, that is not, however, appropriate for higher-order processes where the complete dissipative nature of radiation has to be properly dealt with.}
\end{enumerate}

When applied to our particular classical system, this last point plays the role of utmost importance, for it already isolates the two physical quantities used in the construction of waveform templates: (1) the circular orbit energy, which follows from the conservative dynamics of the binary system, and (2) the radiated power loss, encoded in the imaginary part of $W[J]$. Moreover, as we will shortly see, the near zone physics, i.e., the effective theory description at scale $r$, is devoid of imaginary part, leaving the complete characterization of energy loss due to gravitational wave emission to the far zone description, namely the effective field description of gravitational waves at scale $\lambda\sim r/v$.
In other words, the complete description of gravitational waves, as well as any sort of dissipative effects, is contained within the far zone description. In contrast, the conservative dynamics is almost completely described by the near zone effective theory, just missing both local and nonlocal (logarithmic) corrections stemming 
from processes of radiative nature (such as the so-called tail and memory effects) that have to be drawn from the far zone description \cite{RSSFhereditary}.

\section{The method of regions}

The study of different momentum regimes of a physical problem that presents an important separation of scales is well suited within the application of the so-called \textit{method of regions} \cite{beneke1997,smirnov2002}. 
As we will shortly see, the method of regions allows us to build different effective field theories, one for each of the different regions of momentum our theory presents, by allowing us to perform asymptotic expansions of loop integrals. 

Consider a theory that presents a clear separation of scales, characterized by the low-energy scale $m$ and high-energy scale $M$, and a cutoff scale $\Lambda$ such that $m \ll \Lambda \ll M$. 
Then, according to the method of regions, a loop integral $I(k)$ of the complete  theory,  say
\begin{equation}\label{loopintegral}
I(k) = \int_0^{\infty} dk\, \frac{f(k)}{(k^2 + m^2) (k^2 + M^2)}\,, 
\end{equation}
will have the same result either by direct evaluation of the loop integral in the full theory or by splitting the integration domain and expanding the propagators in powers of the small parameter of that particular region. 

In the above example, the low-energy region $[0,\Lambda]$ has $k\sim m \ll M$, and therefore, the $(k^2+M^2)$ propagator can be expanded in powers of $k/M$, whereas the high-energy interval $[\Lambda,\infty]$ is characterized by $k\sim M$, and hence, the propagator for $(k^2+m^2)$ can be expanded in powers of $m/k$. Thus, the above statement tells us that the loop integral in Eq.~\eqref{loopintegral} can be split into $I(k) = I_1(k) + I_2(k)$, with
\begin{equation}
I_1(k) = \int_0^{\Lambda} dk\, \frac{f(k)}{(k^2 + m^2) (k^2 + M^2)}  \qquad \text{and} \qquad    
I_2(k) = \int_{\Lambda}^{\infty} dk\, \frac{f(k)}{(k^2 + m^2) (k^2 + M^2)}\,,
\end{equation}
and the propagators can be expanded asymptotically in their small parameter, yielding
\begin{equation}\label{I1expanded}
I_1(k) = \int_0^{\Lambda} dk\, \frac{f(k)}{(k^2 + m^2) M^2 (1+(k^2/M^2))} 
       = \sum_{n=0}^\infty  \int_0^{\Lambda} dk\, \frac{f(k)}{(k^2 + m^2) M^2} \left(-\frac{k^2}{M^2}\right)^n  
\end{equation}
and
\begin{equation}\label{I2expanded}
I_2(k) = \int_{\Lambda}^{\infty} dk\, \frac{f(k)}{k^2(1 +(m^2/k^2)) (k^2 + M^2)}
       = \sum_{n=0}^\infty  \int_{\Lambda}^{\infty} dk\, \frac{f(k)}{k^2 (k^2 + M^2)}  \left(-\frac{m^2}{k^2}\right)^n\,.     
\end{equation}

If we take, for instance, $f(k) = k$, the leading-order terms for $I_1$ and $I_2$ in the small parameters $m/\Lambda$, for the low-energy region, and $\Lambda/M$, for the high-energy one, are given by
\begin{align}
I_1(k) &= -\frac{1}{M^2} \log \left( \frac{m}{\Lambda}\right) - \frac{\Lambda^2}{2M^4} + \mathcal{O}\left( \frac{\Lambda^4}{M^6}, \frac{m^2}{M^4} \log\left( \frac{\Lambda}{m}\right)\right)\,,\\	
I_2(k) &= -\frac{1}{M^2} \log \left( \frac{\Lambda}{M}\right) + \frac{\Lambda^2}{2M^4} + \mathcal{O}\left( \frac{\Lambda^4}{M^6}, \frac{m^2}{M^4} \log\left( \frac{M}{\Lambda}\right)\right)\,.
\end{align}
On the other hand, the result for the original integral $I(k)$ in the full domain $[0,\infty]$ with $f(k) = k$ is easy to derive and results in
\begin{equation}\label{Iwithk}
I(k) = \frac{1}{M^2-m^2} \log \left( \frac{M}{m} \right)
     = \frac{1}{M^2} \log \left( \frac{M}{m} \right)  \sum_{n=0}^{\infty} \left( \frac{m^2}{M^2}\right)^n\,.
\end{equation}
By summing the expressions for $I_1(k)$ and $I_2(k)$ above, we immediately notice that: (1) the leading-order term precisely matches the term of the same order in $I(k)$, and (2) the momentum scale $\Lambda$ cancels out after summing the contributions from the two regions. As a matter of fact, these properties are not restricted to the leading-order contributions but hold true generally, and can be checked order by order, see Ref.~\cite{brunello2022}. 

Notice that the naive attempt to integrate the RHS of $I_2(k)$ in Eq.~\eqref{I2expanded} over the complete domain  $[0,\infty]$, would lead to IR divergences due to the non-analytic behavior (in $m/M$) we see in the exact result for $I(k)$, given in Eq.~\eqref{Iwithk}. Likewise, we would encounter UV divergences in $I_1(k)$. Nevertheless, if instead of using the cuttoff scale $\Lambda$, we wanted to regularize the integrals $I_1(k)$ and $I_2(k)$ via dimensional regularization, and integrated $I_1$ and $I_2$, separately, over the whole domain $[0,\infty]$, we would find, remarkably, that the IR divergences in $I_2(k)$ would precisely get canceled out from the UV ones found in $I_1(k)$. Moreover, the expression obtained after summing the two contributions would lead precisely to the result presented in Eq.~\eqref{Iwithk}, without introducing any double counts. 

The example above illustrates the rather general conclusions underlying the beauty of the method of regions: for a theory that presents a clear separation of scales, which then allows us to define EFTs for these different scales, not only we can expand asymptotically the propagators inside loop integrals in each of these regions, but also we can perform integration in the full momentum domain employing  dimensional regularization. In this case, when summing the results for the low-energy and high-energy regions, the unphysical scale $\mu$ defined in this scheme cancels out, with final result being precisely the one we would obtain had the loop integral been evaluated in its full range without the splitting into regions. 

For the particular case of our interest, namely in describing the dynamics of bound binary systems, we follow a perturbative approach by first expanding the full metric around the Minkowski spacetime 
\begin{equation}
g_{\mu\nu}=\eta_{\mu\nu}+\frac{h_{\mu\nu}}{\Lambda}\,,
\end{equation}
where we introduce $\Lambda = (32\pi G_N)^{-1}$.
Then, Eqs. (\ref{EHPPaction})-(\ref{ppaction}) are used to build up information about both the conservative dynamics of the binary system, describing the physics at the orbital scale $r$, and dissipative effects, namely the physics of gravitational waves, that holds at the long-distance scale $\lambda$. Nevertheless, the decoupling between these two scales is not transparent in the momentum domain of the full theory and may become cumbersome when the nonrelativistic limit $v\ll 1$ is taken. To circumvent such a difficulty, we make use of the method of regions presented above, which can be implemented by splitting the metric perturbation $h_{\mu\nu}$ into two mode \textit{regions}: $H_{\mu\nu}$, the short-distance modes, or \textit{potential modes} - defining the \textit{near zone}; and $\bar{h}_{\mu\nu}$, the long-distance modes, or \textit{radiation modes} - defining the \textit{far zone}. This splitting of modes into two non-overlapping regions, justified by the decoupling of scales discussed in Sec.~\ref{subsec:hierarchy} for nonrelativistic sources, reads
\begin{equation}
h_{\mu\nu} = \underbrace{\,\,\,H_{\mu\nu}\,\,\,}_\text{\parbox{1.5cm}{\centering potential\\ \vspace{-0.2cm}modes}}+\underbrace{\,\,\,\bar{h}_{\mu\nu}\,\,\,}_\text{\parbox{1.5cm}{\centering radiation\\ \vspace{-0.2cm}modes}}\,.
\end{equation}    
In this case, the potential modes $H_{\mu\nu}$ are off-shell modes with momenta scaling as $(k^0,|\K|)_{\rm pot }\sim (v/r,1/r)$, that are responsible for describing the conservative dynamics of the binary system at the orbital scale, i.e., in the near zone. In other words, these modes are the mediators of the binding interaction between the constituents of the binary system. The radiation modes $\bar{h}_{\mu\nu}$, on the other hand, are on-shell modes describing gravitational waves in the far zone and have momenta scaling as $(k^0,|\K|)_{\rm rad}\sim (v/r,v/r)$.

Therefore, while in the near zone each of the individual constituents of the binary system can be resolved (since they are treated as point particles placed at a typical distance $\sim r$ apart from each other), in the far zone the binary system as a whole can be treated as a single point endowed with structure (since, at this scale, $r\ll\lambda$). As we will see in Sec.~\ref{secfarzone}, the far zone description of the binary system can be expressed as a multipolar expansion, each multipole moment being given in terms of orbital scale quantities such as $r$ and $v$.

\section{The near zone}\label{nearzonesec}

As discussed in the previous section, the near zone is the scale where the potential modes $H_{\mu\nu}$ are exchanged between the constituents of the binary system and accounts for the conservative sector of the theory. At this scale, the long-distance modes $\bar{h}_{\mu\nu}$ are not important to describe the orbital dynamics and therefore can be set to zero, at least for the moment. We will come back later on to this point when discussing the far zone.

Then, from the path integral tools discussed above, we have that the effective action describing the orbital dynamics, $S_{\rm eff}\equiv W[x_a]$, is given by
\begin{equation}
e^{i S_{\rm eff}[x_a]}\equiv e^{i W[x_a]} = \int\mathcal{D}H_{\mu\nu}\, \exp\{
i S_{\rm EH+GF}[H_{\mu\nu}]
+i S_{\rm pp}[x_a(t),H_{\mu\nu}]
\}\,,
\end{equation} 
where, here, the worldline coordinates $x_a(t)$, for $a=1,2$, parameterize the trajectory of each of the binary system constituents and play the role of the source term $J$. 
In this case, by integrating out the graviton degrees of freedom corresponding to the potential modes in the saddle point approximation, we are left with a classical effective action for the conservative dynamics of the two interacting particles, with degrees of freedom given by the worldline coordinates  $x_1(t)$ and $x_2(t)$. 
Besides this, $S_{\rm EH+GF}[H_{\mu\nu}]$ is the Einstein-Hilbert action \eqref{EHaction} with the addition of a gauge-fixing term, chosen here to be the harmonic gauge:
\begin{equation}
S_{\rm GF}=-\frac{1}{32\pi G_N} \int d^4x\,\sqrt{-g}\, \Gamma_\mu[g] \Gamma^\mu[g]\,,
\end{equation}
where $\Gamma^\mu \equiv \Gamma^\mu{}_{\alpha\beta}g^{\alpha\beta}$ is the contracted Christoffel symbol, and, recall, $g_{\mu\nu}=\eta_{\mu\nu}+H_{\mu\nu}/\Lambda$.
The introduction of a gauge-fixing term, which is equivalent to selecting a particular coordinate system in our spacetime, is fundamental for perturbative computations in gauge and gravity theories, since, otherwise, no Green's function (and hence, propagators) would be possible to the defined. The introduction of such a non-covariant term at the level of a path integral can be carried out using the Faddev-Popov procedure. In this case, however, the presence of ghosts will not influence our computations since they contribute only at the quantum level.

\begin{figure}
\centering
\begin{tikzpicture}
\begin{feynman}
\vertex (a);
\vertex[small,dot,right=1.5cm of a] (b) {};
\vertex[right=1.5cm of b] (c);
\vertex[below=1.8cm of a] (d);
\vertex[small,dot,right=1.5cm of d] (e) {};
\vertex[right=1.5cm of e] (f);
\vertex[above=0.6cm of b] (g);	
\vertex[right=1.2cm of g] (h);
\vertex[opacity=0,dot,below=2cm of a] (k){};
\vertex[below=0.5cm of e] (z1){(a)};
\vertex[right=1.25cm of c] (a2);
\vertex[small,dot,right=1.5cm of a2] (b2) {};
\vertex[right=1.5cm of b2] (c2);
\vertex[below=1.8cm of a2] (d2);
\vertex[opacity=0,dot,right=1.5cm of d2] (e2) {};
\vertex[right=1.4cm of e2] (f2);
\vertex[above=0.6cm of b2] (g2);	
\vertex[right=1.2cm of g2] (h2);
\vertex[small,dot,right=0.6cm of d2] (x1){};
\vertex[small,dot,left=0.6cm of f2] (x2){};
\vertex[below=0.5cm of e2] (z2){(b)};
\vertex[right=1.25cm of c2] (a3);
\vertex[small,dot,right=1.5cm of a3] (b3) {};
\vertex[right=1.5cm of b3] (c3);
\vertex[below=1.8cm of a3] (d3);
\vertex[opacity=0,dot,right=1.5cm of d3] (e3) {};
\vertex[right=1.5cm of e3] (f3);
\vertex[small,dot,below=0.9cm of b3] (g3) {};
\vertex[right=1.5cm of g3] (h3);
\vertex[small,dot,right=0.6cm of d3] (y1){};
\vertex[small,dot,left=0.6cm of f3] (y2){};
\vertex[below=0.5cm of e3] (z3){(c)};
\diagram*{
(a) --[thick,plain] (c),
(d) --[thick,plain] (f),
(b) -- [thick,dashed] (e),
(a2) --[thick,plain] (c2),
(d2) --[thick,plain] (f2),
(b2) -- [thick,dashed] (x1),
(b2) -- [thick,dashed] (x2),
(a3) --[thick,plain] (c3),
(d3) --[thick,plain] (f3),
(b3) -- [thick,dashed] (g3),
(g3) -- [thick,dashed] (y1),
(g3) -- [thick,dashed] (y2),
};
\end{feynman}
\end{tikzpicture}
\caption{Diagrams contributing to the 1PN correction to the Newtonian potential. Diagram (a) alone also gives rise to the leading-order term, i.e., the Newtonian potential. In these diagrams, the two horizontal solid lines represent the worldlines of the two particles, and the dashed lines the potential graviton interaction between them.}
\label{correction1PN}
\end{figure}
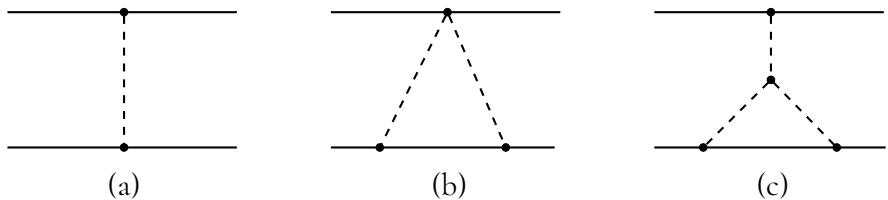

Computations are then performed perturbatively in powers of $v^2$ and also in terms of $G_N$, whose power counting with $v$ is achieved by means of the virial theorem $v^2 \sim G_N M/r$. In particular, the expansion for the Einstein-Hilbert (plus gauge fixing) action, necessary for this perturbative computation, is given schematically by
\begin{equation}
S_{\rm EH+GF} \sim \int d^4x \left( h\partial^2 h + \frac{1}{\Lambda}h^2\partial^2 h + \frac{1}{\Lambda^2}h^3\partial^2 h + \cdots \right)\,.
\end{equation}
Then, diagrams such as the ones in Fig.~\ref{correction1PN} can be evaluated, where the two horizontal solid lines represent the non-propagating worldline degrees of freedom of the sources, $x_1(t)$ and $x_2(t)$, and dashed lines represent the exchange of potential modes between the two objects. In this case, we must include all diagrams with fully connected graviton modes whose external legs are all connected to vertices in either of the two solid lines. This is equivalent to the ``one-particle-irreducible" diagrams one encounter in quantum field theory textbooks. Moreover, only the diagrams which do not present internal graviton loops contribute to classical processes. 

Thus, evaluation of diagrams following the procedure just described gives rise to an effective action $S_{\rm eff}$ for the conservative dynamics of the binary system, given by the real part of $W[x_a]$, which can be split into
\begin{equation}
\text{Re}(W[x_a])=\int dt\,(K[x_a]-V[x_a])\,,
\end{equation}
where $K[x_a]$ and $V[x_a]$ are, respectively, the kinetic and potential energies of the system. Moreover, as it was pointed out before, and will shortly become clear, in the near zone we have no trace of dissipative phenomena, and hence $S_{\rm eff} = \text{Re}(W[x_a])\equiv W[x_a]$.

To see this, first let us consider the simplest of the cases, in which we have static sources: $v_a(t)=0$. In this case, working perturbatively in the gravitational modes $H_{\mu\nu}$, and assuming just leading-order contributions in the scaling of $v$, we obtain:
\begin{equation}
W[x_a]=\int dt \left( \frac{G_N m_1 m_2}{r}\right)\quad \longrightarrow \quad V[x_a] = -\frac{G_N m_1 m_2}{r}\,,
\end{equation}
the latter expression being easily recognized as the Newtonian potential. This expression was obtained from diagram (a) in Fig.~\ref{correction1PN}, considering the leading contributions in the point-particle action \eqref{ppaction} and in the bulk action $S_{EH+GF}$, in the static limit.

In the full expression for $W[x_a]$ corresponding to the first of the diagrams in Fig.~\ref{correction1PN}, before integrating in the momenta, we have an expression with the following behavior: 
\begin{equation}\label{Wbehavior}
W[x_a]\sim m_1 m_2\int dt dt' \int \frac{dk^0 d\K}{(2\pi)^4}\frac{e^{-i k^0(t-t')+i\K\cdot(\X_1(t)-\X_2(t'))}}{k_0^2-\K^2+i \epsilon}\,,
\end{equation}
where the denominator $k_0^2-\K^2+i \epsilon$ clearly comes from the graviton propagator with Feynman prescription. There are important points we can draw from this expression. First of all, since these are off-shell modes scaling as $(k^0,\K)_{\rm pot}\sim (v/r,1/r)$, we see that the Feynman's $i\epsilon$ (or any other prescription for that matter) plays no role in this sector, rendering vanishing the imaginary part of this expression.\footnote{Since in our case  we can neglect the $i\epsilon$ in the Feynman Green's function $G_{F}$, which entails to pick the real part of $G_F$, the identity ${\rm Re}[G_{F}(t,\X)] = \frac 12 [(G_R(t,\X) + G_A(t,\X))]$ shows us that we are necessarily computing time-symmetric quantities (and hence, conservative), since the retarded and advanced Green's functions are related by $G_A(t,\X) = G_R(-t,\X)$.} 
Therefore, this proves the statement that potential modes in the near zone do not contribute to dissipative effects.  Secondly, note that when the sources are considered static, the exponential $e^{i k^0 t}$ in Eq. \eqref{Wbehavior} becomes a delta function $\delta(k^0)$, which kills the $k^0$ in the denominator and immediately gives rise to the instantaneous Newtonian potential. This, in particular, 
shows how departure from instantaneity are handled when nonrelativistic sources are considered, by noticing that since $k^0/|\K|\sim v$, and since we are assuming $v\ll 1$, we can expand the graviton propagator as
\begin{equation}\label{k0overk}
\frac{1}{k_0^2-\K^2} = -\frac{1}{\K^2\left(1-{k_0^2}/{\K^2}\right)} = -\frac{1}{\K^2}\left(1+\frac{k_0^2}{\K^2}+\frac{k_0^4}{\K^4}+\cdots\right)\,.
\end{equation}
In this case, the first term on the RHS gives rise to the instantaneous potential, while each additional power of $k^0/|\K|$ parameterizes departures from instantaneity.
Indeed, when integrated in $k^0$, assuming \eqref{k0overk}, the exponential $e^{-ik^0 (t-t')}$ gives $\delta(t-t')$, resulting in instantaneous interaction corrected by trajectory derivative: retardation effect are reconstructed via Taylor expansion of the trajectory.


The above expansion in powers of $(k^0/|\K|)^2$ is precisely the asymptotic expansion of the propagators presented before, particular of the method of regions. Note, however, that this expansion in $(k^0/|\K|)^2$ is actually an expansion in powers of $v^2$, and this is the heart of the so-called Non-Relativistic General Relativity (NRGR) built in \cite{rothstein2006}, which follows in complete analogy to the so-called Non-Relativistic QCD (NRQCD) to study bound states in QCD. This way of treating gravitationally bound nonrelativistic systems provides a natural way of incorporating the post-Newtonian (PN) approximation to general relativity, within an EFT setup. But not only, it also defines a clear power counting scheme for the perturbative expansion, given in terms of the small parameter of the theory: $v^2$. In particular, moving on to the computation of the next-to-leading order contribution to $W[x_a]$, i.e., the 1PN correction, whose contributing diagrams are displayed in Fig.~\ref{correction1PN}, one obtains
\begin{equation}
L_{1PN} = \frac{m_a v_a^4}{8} 
+\frac{G_N m_1 m_2}{2r}\left[
3(v_1^2+v_2^2)-7(v_1\cdot v_2)-\frac{(v_1\cdot r)(v_1\cdot r)}{r^2}
\right]
-\frac{G_N^2 m_1 m_2(m_1+m_2)}{2r^2}\,.
\end{equation} 
This result, which should be added to Newtonian Lagrangian, has been long known in the literature by the name of Einstein-Infeld-Hoffmann Lagrangian, and was derived in 1938 \cite{EIHlagrangian}.

Going further in perturbation theory, we can in principle compute every order, one by one, attaining better and better precision in the description of the orbital dynamics of the bound binary system. Nevertheless, as one might expect, the number of Feynman diagrams grows exponentially with every additional step. In particular, the state-of-the-art computation of the conservative dynamics of the binary system lies in the 5PN order \cite{blumlein2021b}, although still missing radiative-reaction corrections coming from far-zone computations \cite{blumlein2021}. The 4PN order, on the other hand, has already been completed within EFT methods \cite{RSSF4PNa,RSSF4PNb}. As it turns out, finite size effects start to enter at the 5PN order, as guaranteed by the \textit{effacement principle} \cite{rothstein2006}, which prevents finite-size effects to enter at a lower order in the binary system dynamics, hence being the main motivation for going forward to the completion of this order. Up to now, no consensus has been reached on how to consistently include radiation-reaction effects at this order, and therefore, the 5PN order presents the current challenge in high precision gravity.

\section{Classical and quantum terms from massive mode propagation}\label{secmassive}

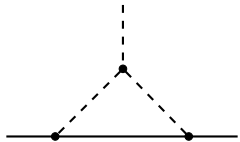
\begin{figure}
\centering
\begin{tikzpicture}
\begin{feynman}
\vertex (a);
\vertex[right=1.25cm of a] (a3);
\vertex[opacity=0,small,dot,right=1.5cm of a3] (b3) {};
\vertex[right=1.5cm of b3] (c3);
\vertex[below=1.8cm of a3] (d3);
\vertex[opacity=0,dot,right=1.5cm of d3] (e3) {};
\vertex[right=1.5cm of e3] (f3);
\vertex[small,dot,below=0.9cm of b3] (g3) {};
\vertex[right=1.5cm of g3] (h3);
\vertex[small,dot,right=0.6cm of d3] (y1){};
\vertex[small,dot,left=0.6cm of f3] (y2){};
\diagram*{
(d3) --[thick,plain] (f3),
(b3) -- [thick,dashed] (g3),
(g3) -- [thick,dashed] (y1),
(g3) -- [thick,dashed] (y2),
};
\end{feynman}
\end{tikzpicture}
\caption{One-loop diagram containing a massive particle, represented by the solid line, and two massless propagators, represented by dashed lines. This is a subamplitude of the diagram (c) in Fig.~\ref{correction1PN}.}
\label{massiveloop}
\end{figure}

Before moving on to the discussion of the long-distance/low-frequency scale description of the relativistic two-body problem, in this section we discuss the problem of separating classical from quantum contributions that arise in loop integrals containing the propagator of a massive particle. To this end, in this discussion, we treat our compact objects as propagating degrees of freedom, rather than the nondynamical sources that we have introduced above, characteristic of NRGR.

Consider the standard elastic scattering of two massive particles in quantum field theory. In the center of mass frame, the incoming and outcoming four-wave numbers\footnote{Recall that, for a given four-wave number $p^\mu$, we have a corresponding four-momentum given by $P^\mu = \hbar p^\mu$. Note that the use of this quantity, instead of four-momentum, is desired to avoid appearances of $\hbar$ in the exponential of the Fourier transform, as well as in the integration measure.} of the two particles can be parametrized respectively by $p_1 = (\omega_1, {\bf p})$, $p_2 = (\omega_2, -{\bf p})$ and $p'_1 = (\omega_1, {\bf p'})$, $p'_2 = (\omega_2, -{\bf p'})$. In this case, the transfer four-wave number in the scattering process is given by
\begin{equation}
    q = p_1 - p'_1 = (0, {\bf p} - {\bf p'}) = (0, {\bf q})\,.
\end{equation}

Now, consider in particular processes in this scattering involving a loop made of the massive propagator of particle 1, closed by massless modes that are being exchanged between the two particles; See Fig.~\ref{massiveloop}. Focusing just on this part of the scatting process, we obtain the following amplitude\footnote{The factor of $m_1^3/\hbar^2$ in front of the integral comes from two mass insertions of the diagram, hence giving $(m_1^2/\hbar^2)^2$, times a factor of $\hbar^2/m_1$ stemming from the normalization of the relativistic and nonrelativistic amplitudes, related by $\braket{\Pp| T_{\mu\nu}|\Pp'}^{(R)} = (\hbar^2/m_1) \braket{\Pp | T_{\mu\nu}|\Pp'}^{(NR)}$ \cite{sturani2021}, where $\braket{\Pp | T_{\mu\nu}|\Pp'}$ is the matrix element between states $\ket{\Pp}$ and $\ket{\Pp'}$ mediated by the energy momentum tensor $T_{\mu\nu}$ for a particle of mass m.}
\begin{equation}\label{clasloop}
\mathcal{A} \sim  \frac{m_1^3}{\hbar^2}  \int\frac{d^4 k}{(2\pi)^4} 
    \frac{1}{(-k^2+i \epsilon) [-(k+q)^2+i \epsilon]
    [-(p_1+k)^2-m_1^2 / \hbar^2+i \epsilon]}\,, 
\end{equation}
where $k = (\omega,\K)$ is the loop wave number that are being integrated over. Notice that we have kept the Planck constant $\hbar$, so we can investigate what is quantum and what is classical about this amplitude.
Notice that, for the diagram in Fig.~\ref{massiveloop}, this is the only factors of $\hbar$ that appears in this process, since, in a generic Feynman diagram, each vertex will contribute with a factor of $1/\hbar$ (coming from the expansion of $e^{i S/\hbar}$), while each propagator will contribute with a factor of $\hbar$ (since propagators are obtained from the inversion of the kinetic term in $e^{i S_0/\hbar}$, being $S_0$ the action quadratic in the fields). Hence, for diagram \ref{massiveloop}, having three propagators and three vertices, we get $(1/\hbar)^3 \times \hbar^3 \sim 1$.

Now, considering the nonrelativistic limit $\hbar \omega_1 \sim m_1$ and $m_1 \gg  \hbar|{\bf p}|$, and focusing on the large mass region $m_1 \gg \hbar |\K|$, the last of the three propagators in Eq.~(\ref{clasloop}) can be expanded, giving $(p_1+k)^2 + m_1^2 / \hbar^2 = k^2 + 2k\cdot p_1 \simeq - 2m_1 \omega/\hbar (1-\hbar |{\bf q}|/m_1)$ \cite{bjerrum2018}. Thus, Eq.~(\ref{clasloop}) takes the form
\begin{equation}
\mathcal{A} \sim   \frac{m_1^2}{2 \hbar}  \int\frac{d^4 k}{(2\pi)^4} 
    \frac{1}{(-k^2+i \epsilon) [-(k+q)^2+i \epsilon]
    (\omega +i \epsilon)} \left[ 1 + \mathcal{O}\left( \frac{\hbar |{\bf q}|}{m_1}\right) \right]\,.
\end{equation}
Now, we integrate this expression in $\omega$ using the residue theorem, by closing the integration contour in the upper half-plane to pick the contribution from the poles $\omega=-|\K|-i \epsilon$ and $\omega=-|\K+\mathbf{q}|-i \epsilon$ to get
\begin{equation}
\mathcal{A} \sim   -i\frac{m_1^2}{4 \hbar}  \int\frac{d^3 \K}{(2\pi)^3} 
    \frac{1}{\K^2 (\K+{\bf q})^2} 
    \left[ 1 + \mathcal{O}\left( \frac{\hbar |{\bf q}|}{m_1}\right) \right]\,.
\end{equation}
Interestingly, this result is proportional to the loop integral obtained in the near zone.

This simple example shows us that the loop integral in Eq.~(\ref{clasloop}) has both classical and quantum contributions. And, moreover, that the quantum piece can be isolated, and further neglected, by taking the large mass limit $\hbar{\bf q}/m_1 \rightarrow 0$ before the loop integral in $\K$ is performed. More importantly, this example shows us how the treatment of nonpropagating sources in NRGR is recovered from more fundamental scattering amplitude computations, by taking the large mass approximation, which, in turn, contain only classical contributions. 
In other words, this justify why in NRGR the quantum contributions are automatically neglected since the beginning. Note that, in this case, for a given NRGR diagram, we have the following quantum scaling:
\begin{equation}
\hbar^{I-V} = \hbar^{L-1}\,,
\end{equation}
where $I$ is the number of internal lines, $V$ the number of vertices, and $L$ the number of (internal) loops, again, considering that the solid lines for the massive sources are not propagating. Note that any quantity of internal loops in a diagram within NRGR results in a completely quantum contribution, since the classical scaling\footnote{Recall, the classical effective action is obtained by computing $W[J]/\hbar$, and hence, only contributions scaling as $1/\hbar$ are classical.} $1/\hbar$ is surpassed.

\section{The far zone}\label{secfarzone}

As already pointed out in  Sec.~\ref{buildingEFTs}, in order to obtain an EFT for the long-distance/low-frequency scale  of a given ``full'' theory in which there exists a clear separation of scales, we \textit{integrate out} the short-distance modes. The resulting effective theory is then given as a series of field \textit{operators}, each being coupled to coefficients that encode the short-distance physics. Applying these concepts to our description of the dynamics of binary systems corresponds to integrating out the potential modes to get access to the effective description at the long-wavelength scale. In this description, 
\begin{equation}\label{Efffar}
e^{i S_{\rm eff}[x_a]} = \int\mathcal{D}H_{\mu\nu}\, \exp\left\{
i S_{\rm EH+GF}[g_{\mu\nu}]
+i S_{\rm pp}[x_a(t),g_{\mu\nu}]
\right\}\,,
\end{equation} 
\begin{equation}
S_{\rm GF}=-\frac{1}{32\pi G_N} \int d^4x\,\sqrt{-g}\, \Gamma_\mu[g] \Gamma^\mu[g]\,,
\end{equation}
where now $g_{\mu\nu}=\eta_{\mu\nu}+(H_{\mu\nu}+\bar{h}_{\mu\nu})/\Lambda$.
Then, by computing the path integral \eqref{Efffar}, in addition to the results already derived in the near zone, one obtains terms with couplings to all powers of the external radiation field $\bar{h}_{\mu\nu}$. Nevertheless, since the binary systems detected by LIGO/Virgo emit gravitational waves with typical amplitudes of $\bar{h}\sim 10^{-22}$, terms beyond linear coupling can be safely discarded. The result has the general form given by
\begin{equation}\label{Tmunu}
S_{\rm eff} = \frac 12 \int d^4x\, T^{\mu\nu}(x)h_{\mu\nu}(x)\,,
\end{equation}
where $T^{\mu\nu}$ is the pseudo stress-energy tensor including contributions not only from the (matter) point-particles but also from the binding gravitational field itself, the latter following from the nonlinearities present in the Einstein-Hilbert action.\footnote{From now on we drop out the bar above $\bar{h}$, since we will no longer have ambiguities between $h_{\mu\nu}=H_{\mu\nu}+\bar{h}_{\mu\nu}$ and $\bar{h}_{\mu\nu}$.} 

\begin{figure}[t] \label{fromneartofar}
\begin{equation*}
\underbrace{\,\,\,\,    
\begin{tikzpicture}
\begin{feynman}
\vertex (a);
\vertex[small,dot,right=1.5cm of a] (b) {};
\vertex[right=1.5cm of b] (c);
\vertex[below=1.8cm of a] (d);
\vertex[right=1.5cm of d] (e);
\vertex[right=1.5cm of e] (f);
\vertex[above=0.6cm of b] (g);	
\vertex[right=1.2cm of g] (h);
\vertex[opacity=0,dot,below=2cm of a] (k){};
\vertex[right=0.5cm of c] (a2);
\vertex[small,dot,right=1.5cm of a2] (b2) {};
\vertex[right=1.5cm of b2] (c2);
\vertex[below=1.8cm of a2] (d2);
\vertex[small,dot,right=1.5cm of d2] (e2) {};
\vertex[right=1.5cm of e2] (f2);
\vertex[above=0.6cm of b2] (g2);	
\vertex[right=1.2cm of g2] (h2);
\vertex[right=0.5cm of c2] (a3);
\vertex[small,dot,right=1.5cm of a3] (b3) {};
\vertex[right=1.5cm of b3] (c3);
\vertex[below=1.8cm of a3] (d3);
\vertex[small,dot,right=1.5cm of d3] (e3) {};
\vertex[right=1.5cm of e3] (f3);
\vertex[small,dot,below=0.9cm of b3] (g3) {};
\vertex[right=1.5cm of g3] (h3);
\diagram*{
(a) --[thick,plain] (c),
(d) --[thick,plain] (f),
(b) -- [thick,photon] (h),
(a2) --[thick,plain] (c2),
(d2) --[thick,plain] (f2),
(b2) -- [thick,dashed] (e2),
(b2) -- [thick,photon] (h2),
(a3) --[thick,plain] (c3),
(d3) --[thick,plain] (f3),
(b3) -- [thick,dashed] (g3),
(e3) -- [thick,dashed] (g3), 
(g3) -- [thick,photon] (h3),
};
\end{feynman}
\end{tikzpicture}
\,\,\,\,}_{\text{Near zone diagrammatic representation}}
\hspace{0.1cm}\text{\parbox{1cm}{\,\,\,\,\,$\Longrightarrow$\\ \\ \\ \\ \vspace{0.15cm}}}
\underbrace{\,\,\,\,
\begin{tikzpicture}
\begin{feynman}
\vertex (a) at (1,0);
\vertex[right=1.5cm of a] (b);
\vertex[right=1.5cm of b] (c);
\vertex[below=0.08cm of a] (d);
\vertex[right=3cm of d] (e);
\vertex[dot,below=-0.027cm of b] (f){};
\vertex[above=1.5cm of b] (h){};
\vertex[opacity=0,dot,below=0.2cm of a] (k){};
\diagram*{
(a) --[thick,plain] (c),
(d) --[thick,plain] (e),
(b) --[thick,double,photon] (h),
};
\end{feynman}
\end{tikzpicture}\,\,\,\,}_{\text{\parbox{4cm}{\centering Far zone diagrammatic \\ \vspace{-0.2cm}representation}}}
\end{equation*}
\vspace{-0.8cm}
\caption{Diagrammatic representation of how the far zone diagrams are obtained within the full theory from multipolar expansions of diagrams defined in the near zone.}
\end{figure}
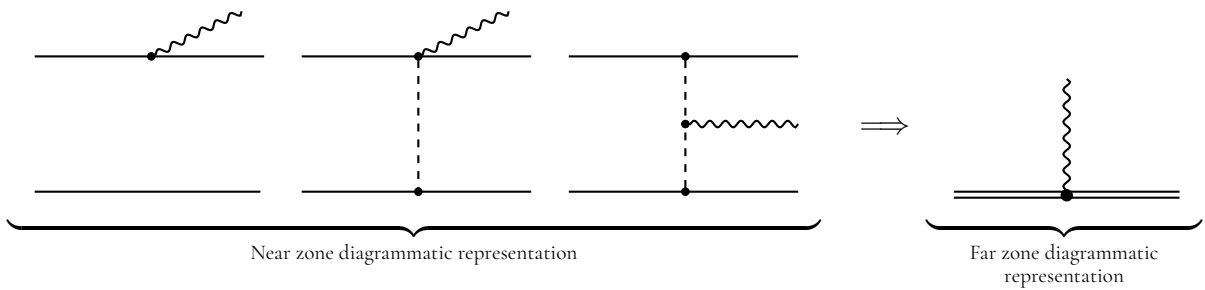

In addition to this, since now we are assuming the effective descrition at scale $\lambda\gg r$, our formalism is optimized by expanding $h_{\mu\nu}$ around the center of mass of the binary system. This expansion is represented by the diagrams in Fig.~\ref{fromneartofar}, as we depart from a theory where the two bodies can be resolved - the near zone, to a description where the binary system can be treated as a single point endowed with some structure - the far zone. This is a consequence of the method of regions again, where now we are making an expansion in $r/\lambda$. 

This way of constructing an effective theory by integrating out the short-distance modes of the full theory goes under the name of \textit{top-down} approach, as opposed to the \textit{bottom-up} already presented. Then, although we have all the ingredients to construct the long-distance effective description starting from the ``full" theory given by Eq. \eqref{Efffar}, the analysis is rather simplified by considering the effective description obtained from a bottom-up manner since the beginning. As a matter of fact, this has already been considered in reference \cite{goldberger2010}, where the authors construct such an effective theory to describe the long-wavelength gravitational radiation from compact objects. In this description, the binary system takes the form of a point-particle, with internal dynamics described by a series of multipole moments localized on the particle worldline, with effective action given by:
\begin{align}\label{EFfarzone}
S_{\rm mult} &= - M \int d\tau - \frac12 \int dx^\mu L^i \epsilon_{ijk} \omega_\mu^{jk} \nonumber\\
&- \int d\tau \sum_{\ell =2}^{\infty} \frac 1{\ell !}I^{L} \nabla_{L-2} E_{k_{\ell-1}k_\ell} + \int d\tau \sum_{\ell =2}^{\infty} \frac{2\ell}{(\ell+1)!} J^{L} \nabla_{L-2} B_{k_{\ell-1}k_\ell}\,.
\end{align}

In the above expression, $M$ and $L^i$ stand respectively for the conserved energy and angular momentum of the compact system, $E_{ij}$ and $B_{ij}$ the electric and magnetic components of the Weyl tensor previously presented in Eq.~\eqref{EB}, and the index $L$ represents a multi-index notation for $L = i_1 i_2 \dots i_\ell$. The tensors $I^{L}$ and $J^{L}$ are respectively the electric and magnetic parity multipole moments, these latter being $SO(3)$ symmetric and traceless tensors, with $\ell$ representing the multipolarity of the moments, and, finally, $\omega_\mu^{jk}$ the components of the spin connection.

Now, let us assume in this effective action that the source is localized at a fixed spatial position $\X_{\rm cm}$ so that our interest lies in describing gravitational waves emitted by the source at rest, at a distance $D = |\X_{\rm cm}|$ far away from the detector/observer. In this case, we have
\begin{align}\label{farzoneactionstatic}
S_{\rm mult} = \int dt \left[ \frac 12 E h_{00} - \frac12 J^{b|a} h_{0b,a} -\sum_{r\geq 0}  \left( c_r^{(I)} I^{ijR} \partial_R R_{0i0j} - \frac 12 c_r^{(J)} J^{b|iRa} \partial_R R_{0iab}\right)
\right]\,,
\end{align}
with coefficients $c_r^{(I)}$ and $c_r^{(J)}$ given by
\begin{equation}\label{cIcJ}
    c_r^{(I)} = \frac{1}{(r+2)!}, \qquad c_r^{(J)} = \frac{2(r+2)}{(r+3)!}\,. 
\end{equation}
In this expression, $r=0$ corresponds to the quadrupole, $r=1$ to the octupole, and so on, and we have also introduced a new notation for the magnetic moments, $J^{b|iRa}$, first introduced in Ref.~\cite{henry2021} as a generalization of $J^{ijR}$ to arbitrary dimensions. This object is antisymmetric in the indices $a,b$, symmetric in $i,R$, and completely tracefree,  
and in particular, in $d=3$, we have $J^{a|iRb}=\epsilon_{jab} J^{ijR}$. Likewise, for the angular momentum, we have $J^{a|b}= L^{i} \epsilon_{iab}$.

The way in which the effective descriptions obtained from the bottom-up and top-down approaches connect to each other is by performing a multipolar expansion on expression \eqref{Tmunu} and identifying the multipole moments. In this case, we obtain expressions for each of the multipoles in terms of kinetic variables of the binary system ($x_a, v_a$, and so on). This is usually referred to as a \textit{matching} procedure, in which, once we know the UV completion of a long-distance effective theory, the coefficients of the field operators can then be obtained in terms of the corresponding short-distance variables. In our case, we actually happen to know the UV physics (the near zone) of our effective long-wavelength theory (the far zone). Nevertheless, a complete analysis of the gravitational-wave physics can still be carried out without the knowledge of the internal structure - this is the realization of the decoupling theorem. In this case, observables will be given universally in terms of the important degrees of freedom in our theory, namely the multipole moments.

The power counting of this theory follows from the scalings 
\begin{equation}
x^\mu\sim\lambda\,, \quad \partial_\mu\sim\lambda^{-1}\,, \quad I^{ijR}\sim m r^{r+2}\,,\quad J^{ijR}\sim m v r^{r+2}\,,
\end{equation} 
the last two being given in terms of internal scale quantities, $m$, $r$, and $v$. From this, we can see that each successive term in this multipolar expansion is suppressed by one power of $r/\lambda\sim v$ with respect to the preceding one, for either electric or magnetic types:
\begin{align}
\left\{
\begin{array}{cc}
&I^{ijR}\partial_{R}E_{ij}
\quad\,\,\,\,\sim m \left(\dfrac{r}{\lambda}\right)^{r+2} \sim m v^{r+2}\,,\\
&I^{ij(R+1)}\partial_{R+1}E_{ij}
\sim m \left(\dfrac{r}{\lambda}\right)^{r+3} \sim m v^{r+3}\,,
\end{array}
\right.
\end{align}
and likewise for the magnetic case.
From this well-defined power counting, we are able to work on a perturbative approach, where again $v$ plays the important role of the small parameter of the theory just like in the near zone. Notice, however, that at the level of the far zone, perturbation theory is naturally carried out in powers of $G_N$, but which can be ultimately connected to the parameter $v$ with the help of the virial theorem.

In  Sec.~\ref{pathintegral} it was argued that the imaginary part of the effective action gives information about the energy loss, and in Sec.~\ref{nearzonesec} we saw that this information is fully comprehended within the far zone. In this case, the effective action is obtained through the path integral
\begin{equation}
e^{i S_{\rm eff}[I(t)]} = \int\mathcal{D}h_{\mu\nu}\, \exp\{
i S_{\rm EH+GF}[h_{\mu\nu}]
+i S_{\rm mult}[h_{\mu\nu},I(t)]
\}\,,
\end{equation} 
with diagrammatic representation for the leading-order contribution given by
\newline
\parbox{2cm}{$\qquad\qquad i S_{\rm eff}$\,\,=\vspace{1.5cm}}
\hspace{0.7cm}
\begin{tikzpicture}
\begin{feynman}
\vertex (a);
\vertex[right=0.5cm of a] (b);
\vertex[right=3cm of a] (c2);
\vertex[left=0.5cm of c2] (c);
\vertex[below=0.08cm of a] (d);
\vertex[right=3cm of d] (e);
\vertex[dot,below=-0.027cm of b] (f1){};
\vertex[dot,below=-0.027cm of c] (f2){};
\vertex[below=0.03cm of b] (g1){$I^{ij}$};
\vertex[below=0.03cm of c] (g2){$I^{ij}$};
\vertex[above=1.5cm of b] (h){};
\vertex[opacity=0,dot,below=0.2cm of a] (k){};
\diagram*{
(a) --[thick,plain] (c2),
(d) --[thick,plain] (e),
(b) --[thick,half left,photon] (c),
};
\end{feynman}
\end{tikzpicture}
\parbox{0.5cm}{\,\,+\vspace{1.5cm}}
\begin{tikzpicture}
\begin{feynman}
\vertex (a);
\vertex[right=0.5cm of a] (b);
\vertex[right=3cm of a] (c2);
\vertex[left=0.5cm of c2] (c);
\vertex[below=0.08cm of a] (d);
\vertex[right=3cm of d] (e);
\vertex[dot,below=-0.027cm of b] (f1){};
\vertex[dot,below=-0.027cm of c] (f2){};
\vertex[below=0.03cm of b] (g1){$I^{ijk}$};
\vertex[below=0.03cm of c] (g2){$I^{ijk}$};
\vertex[above=1.5cm of b] (h){};
\vertex[opacity=0,dot,below=0.2cm of a] (k){};
\diagram*{
(a) --[thick,plain] (c2),
(d) --[thick,plain] (e),
(b) --[thick,half left,photon] (c),
};
\end{feynman}
\end{tikzpicture}
\parbox{0.5cm}{\,\,+\vspace{1.5cm}}
\begin{tikzpicture}
\begin{feynman}
\vertex (a);
\vertex[right=0.5cm of a] (b);
\vertex[right=3cm of a] (c2);
\vertex[left=0.5cm of c2] (c);
\vertex[below=0.08cm of a] (d);
\vertex[right=3cm of d] (e);
\vertex[dot,below=-0.027cm of b] (f1){};
\vertex[dot,below=-0.027cm of c] (f2){};
\vertex[below=0.03cm of b] (g1){$J^{ij}$};
\vertex[below=0.03cm of c] (g2){$J^{ij}$};
\vertex[above=1.5cm of b] (h){};
\vertex[opacity=0,dot,below=0.2cm of a] (k){};
\diagram*{
(a) --[thick,plain] (c2),
(d) --[thick,plain] (e),
(b) --[thick,half left,photon] (c),
};
\end{feynman}
\end{tikzpicture}
\parbox{2cm}{\,\,+ \,\,$\cdots$\,\,. \vspace{1.5cm}}

Besides these leading-order \textit{self-energy} diagrams, as we go to higher order in the Newtonian constant $G_N$, other types of diagrams start to contribute. In particular, the next-to-leading order class of diagrams, which entails the so-called \textit{tail} and \textit{memory} processes, include diagrams that involve three-graviton vertex insertions\footnote{As we will see in the following two chapters, the terminologies ``tail" and ``memory" have physical grounds, and are related to how these nonlinear interactions particularly affect the wave propagation. For now, it suffices to have in mind that diagrams such as these allow a deeper mapping of fundamental processes into phenomenological results.}, e.g.:
\newline
\parbox{4.6cm}{\qquad\qquad\qquad\qquad\qquad\qquad\qquad\qquad\qquad\qquad\qquad\qquad}
\begin{tikzpicture}
\begin{feynman}
\vertex (a);
\vertex[right=1.5cm of d] (p0);
\vertex[dot,above=-0.03cm of p0] (p) {};
\vertex[dot,above=0.9cm of p] (q) {};
\vertex[right=0.5cm of a] (b);
\vertex[right=3cm of a] (c2);
\vertex[left=0.5cm of c2] (c);
\vertex[below=0.08cm of a] (d);
\vertex[right=3cm of d] (e);
\vertex[dot,below=-0.027cm of b] (f1){};
\vertex[dot,below=-0.027cm of c] (f2){};
\vertex[below=0.03cm of b] (g1){$I^{ij}$};
\vertex[below=0.03cm of c] (g2){$I^{ij}$};
\vertex[below=0.3cm of p] (g3){$E$};
\vertex[above=1.5cm of b] (h){};
\vertex[opacity=0,dot,below=0.2cm of a] (k){};
\diagram*{
(a) --[thick,plain] (c2),
(d) --[thick,plain] (e),
(b) --[thick,half left,photon] (c),
(p) --[thick,dashed] (q), 
};
\end{feynman}
\end{tikzpicture}
\parbox{0.5cm}{\,\,,\vspace{1cm}}
\begin{tikzpicture}
\begin{feynman}
\vertex (a);
\vertex[right=1.5cm of d] (p0);
\vertex[dot,above=-0.03cm of p0] (p) {};
\vertex[dot,above=0.9cm of p] (q) {};
\vertex[right=0.5cm of a] (b);
\vertex[right=3cm of a] (c2);
\vertex[left=0.5cm of c2] (c);
\vertex[below=0.08cm of a] (d);
\vertex[right=3cm of d] (e);
\vertex[dot,below=-0.027cm of b] (f1){};
\vertex[dot,below=-0.027cm of c] (f2){};
\vertex[below=0.03cm of b] (g1){$I^{ij}$};
\vertex[below=0.03cm of c] (g2){$I^{kl}$};
\vertex[below=0.3cm of p] (g3){$I^{mn}$};
\vertex[above=1.5cm of b] (h){};
\vertex[opacity=0,dot,below=0.2cm of a] (k){};
\diagram*{
(a) --[thick,plain] (c2),
(d) --[thick,plain] (e),
(b) --[thick,half left,photon] (c),
(p) --[thick,photon] (q), 
};
\end{feynman}
\end{tikzpicture}
\parbox{0.5cm}{\,\,.\vspace{1cm}}
\newline
The first of these diagrams is an example of a tail process, which is characterized by the insertion of an energy vertex in the worldline, whereas the second one is an example of memory, where the new insertion is another multipole moment. 
These nonlinear effects turn out to be of great relevance in our formalism, as they contribute directly to the near-zone conservative dynamics with the inclusion of local as well as long-range logarithmic terms to the effective action. 
Physically speaking, these kinds of processes can be understood as the system emitting gravitational waves that are backscattered to the system as they interact with either the Newtonian field generated by the source or by the gravitational waves themselves and constitute examples of the so-called radiation-reaction effects.

\subsection{The optical theorem and the radiated power loss}\label{opticaltheoremsec}

As we will see below, the total power loss can be obtained from self-energy diagrams in the $i\epsilon$-Feynman's prescription. Notice that, formally speaking, the path integral formulation is consistent only under the assumption of the Feynman boundary condition, which guarantees convergence of the integral. Nonetheless, although this boundary condition is not appropriate to study dissipative phenomena, as is the case in gravitational-wave physics since the Feynman boundary condition is time-symmetric and, hence, does not allow for outgoing radiation, it becomes important if we focus on certain time-averaged observables.

In particular, within this prescription, the imaginary part of the effective action $S_{\rm eff} = W[x_a]$ encodes the power loss through the optical theorem. 
It should be emphasized here that we must use necessarily the Feynman propagator, as will be shortly justified.
To investigate this, we start by first identifying ${\rm Im}\,S_{\rm eff}$ with the differential rate of radiation, or ``decay width", $\Gamma$, through\footnote{It is worthwhile mentioning that, while $\Gamma$ is an intrinsically quantum object, the emitted radiation has a clear classical interpretation, since here we have a macroscopically large number of emitted gravitons. Note, In particular, that, being $\Gamma(\omega)$ the spectrum of number of emitted particles, the energy spectrum of emitted particles is thus given by $\hbar \omega \Gamma(\omega)$.} \cite{rothstein2006}
\begin{equation}
\frac{1}{T}\text{Im}\, S_{\rm eff}\rightarrow \frac{1}{2}\int\frac{d^2\Gamma}{d\omega d\Omega} d\omega d\Omega\,.
\end{equation}
In this expression, $d\Omega$ is the differential solid angle, $\omega = k^0$ is the energy of the emitted graviton and $T$ represents the time interval for which the emitted gravitons are being measured, the latter being useful to the computation of time-averaged observables by defining the average	  
\begin{equation}
\braket{f(t)} = \lim_{T\rightarrow \infty}\frac{1}{T} \int_0^T f(t)\,.
\end{equation}

The differential rate of radiation yields the total radiated energy by integrating it over the full integration domain, weighted by the single-graviton energy $\omega$:
\begin{equation}
P \equiv \int dP = \int \omega d\Gamma = \int \omega \frac{d^2\Gamma}{d\omega d\Omega} d\omega d\Omega\,.
\end{equation} 

Now, we investigate how one can derive an expression for $d\Gamma$ and how imposing the Feynman boundary condition becomes a crucial stone in this construction. Consider the self-energy diagram constructed from the source coupling given by
\begin{equation}
S_{\rm source} = \frac 12 \int dt\, T^{\mu\nu} h_{\mu\nu}\,, \qquad \text{or, equivalently,} \qquad i \mathcal{A} = \frac{i}{2} T^{\mu\nu} h^*_{\mu\nu}\,. 
\end{equation}
Computation of this diagram yields\\
\parbox{2cm}{$\,\, i S_{\rm eff}$\,\,=\vspace{1.5cm}}
\hspace{-1.75cm}
\hspace{0.7cm}
\begin{tikzpicture}
\begin{feynman}
\vertex (a);
\vertex[right=0.5cm of a] (b);
\vertex[right=3cm of a] (c2);
\vertex[left=0.5cm of c2] (c);
\vertex[below=0.08cm of a] (d);
\vertex[right=3cm of d] (e);
\vertex[dot,below=-0.027cm of b] (f1){};
\vertex[dot,below=-0.027cm of c] (f2){};
\vertex[below=0.03cm of b] (g1){$T^{\mu\nu}$};
\vertex[below=0.03cm of c] (g2){$T^{\rho\sigma}$};
\vertex[above=1.5cm of b] (h){};
\vertex[opacity=0,dot,below=0.2cm of a] (k){};
\diagram*{
(a) --[thick,plain] (c2),
(d) --[thick,plain] (e),
(b) --[thick,half left,photon] (c),
};
\end{feynman}
\end{tikzpicture}
\parbox{10cm}{$=-\dfrac 12 {\displaystyle\int} d^4x {\displaystyle\int} d^4x'\, \dfrac{1}{4} T^{\mu\nu}(t,\X) T^{\rho\sigma}(t',\X') 
{\displaystyle\int} \dfrac{d^4k}{(2\pi)^4}\dfrac{-i e^{i k\cdot(x-x')}}{k^2+i\epsilon}P_{\mu\nu,\rho\sigma}\,,
$ \vspace{1.5cm}}

\vspace{-1cm}
\noindent where $P_{\mu\nu,\rho\sigma} = (1/2)(\eta_{\mu\rho} \eta_{\nu\sigma} + \eta_{\mu\sigma} \eta_{\nu\rho} - \eta_{\mu\nu}\eta_{\rho\sigma})$,
which then becomes
\begin{align}
S_{\rm eff} &= \frac 12 \int \frac{d^4k}{(2\pi)^4} \frac{1}{( k^2+i \epsilon )}\times \frac 14 P_{\mu\nu,\rho\sigma}  T^{\mu\nu *}(\omega,\K) T^{\rho\sigma}(\omega,\K) \nonumber\\
&= \frac 12 \int \frac{d^4k}{(2\pi)^4} \frac{1}{( k^2+i \epsilon )} |\mathcal{A}(\omega,\K)|^2\,. 
\end{align}
In the last line we have made use of the sum over graviton polarizations, $\sum h^*_{\mu\nu}h_{\rho\sigma}=P_{\mu\nu,\rho\sigma}$.  
Now, we make use of the so-called Sochocki-Plemejl identity
\begin{equation}
{\rm Im}\left( \frac{1}{k^2 + i\epsilon} \right) = -\pi \delta(k^2)\,,
\end{equation}
which allow us to compute the imaginary part of $S_{\rm eff}$ in an easy way:
\begin{align}
{\rm Im}\,S_{\rm eff} &= \frac 12 \int \frac{d^4k}{(2\pi)^4} \left[-\pi \delta(k^2)\right] |\mathcal{A}(\omega,\K)|^2 \nonumber\\
&= -\frac 12 \int \frac{d^4k}{(2\pi)^4} \frac{\pi}{2|\K|} \left[ \delta(\omega-|\K|) + \delta(\omega+|\K|) \right] |\mathcal{A}(\omega,\K)|^2  \nonumber\\
&= -\frac 12 \int \frac{d^3\K}{(2\pi)^3 2|\K|} |\mathcal{A}(|\K|,\K)|^2\,.
\end{align}
From this, we immediately draw that, for on-shell gravitons, in which $\omega^2 = |\K|^2$, we have:
\begin{equation}\label{gravitonrate}
d\Gamma(\K)=\dfrac{1}{T}\dfrac{d^3\K}{(2\pi)^32\omega}|\mathcal{A(\omega,\K)}|^2\,.
\end{equation}
This derivation, representing the optical theorem, is depicted schematically in Fig.~\ref{opticaltheorem}.

\begin{figure}[t]\label{opticaltheorem}
\parbox{1cm}{\,\,2 Im\vspace{1.5cm}}
\begin{tikzpicture}
\begin{feynman}
\vertex (a);
\vertex[right=0.5cm of a] (b);
\vertex[right=3cm of a] (c2);
\vertex[left=0.5cm of c2] (c);
\vertex[below=0.08cm of a] (d);
\vertex[right=3cm of d] (e);
\vertex[dot,below=-0.027cm of b] (f1){};
\vertex[dot,below=-0.027cm of c] (f2){};
\vertex[above=1.5cm of b] (h){};
\vertex[opacity=0,dot,below=0.2cm of a] (k){};
\diagram*{
(a) --[thick,plain] (c2),
(d) --[thick,plain] (e),
(b) --[thick,half left,photon] (c),
};
\end{feynman}
\end{tikzpicture}
\parbox{0.5cm}{\,\,=\vspace{1.5cm}}
\begin{tikzpicture}
\begin{feynman}
\vertex (a) at (1,0);
\vertex[right=0.5cm of a] (b);
\vertex[right=1.5cm of a] (c);
\vertex[below=0.08cm of a] (d);
\vertex[right=1.5cm of d] (e);
\vertex[dot,below=-0.027cm of b] (f){};
\vertex[above=1.5cm of b] (h){};
\vertex[opacity=0,dot,below=0.2cm of a] (k){};
\vertex[left=0.cm of e] (h1);
\vertex[above=1cm of h1] (h1);
\diagram*{
(a) --[thick,plain] (c),
(d) --[thick,plain] (e),
(b) --[thick,double,photon] (h1),
};
\end{feynman}
\end{tikzpicture}
\parbox{1cm}{\centering\,$\times$\vspace{1.5cm}}
\begin{tikzpicture}
\begin{feynman}
\vertex (a) at (1,0);
\vertex[right=1cm of a] (b);
\vertex[right=1.5cm of a] (c);
\vertex[below=0.08cm of a] (d);
\vertex[right=1.5cm of d] (e);
\vertex[dot,below=-0.027cm of b] (f){};
\vertex[above=1.5cm of b] (h){};
\vertex[opacity=0,dot,below=0.2cm of a] (k){};
\vertex[right=0.cm of d] (l1);
\vertex[above=1cm of l1] (h1);
\diagram*{
(a) --[thick,plain] (c),
(d) --[thick,plain] (e),
(b) --[thick,double,photon] (h1),
};
\end{feynman}
\end{tikzpicture}
\parbox{7cm}{\quad $\Longleftrightarrow$ \quad $d\Gamma(\K)=\dfrac{1}{T}\dfrac{d^3\K}{(2\pi)^32\omega}|\mathcal{A(\K)}|^2$\,.\vspace{1.5cm}}
\vspace{-0.8cm}
\caption{Schematic representation of the optical theorem: the imaginary part of self-energy diagrams computed with the Feynman boundary condition, and has the interpretation of graviton emission rate, can be recovered by squaring on-shell emission amplitudes.}
\end{figure}

In Eq.~\eqref{gravitonrate}, the emission amplitude $\mathcal{A(\K)}$ is general, and, in particular, may also include higher-order contributions in $G_N$ stemming from the nonlinearities present in the Einstein-Hilbert action. For the leading-order amplitudes, which are linear in $G_N$, we are able to derive the well-known expression for the radiated power loss $P$:
\begin{equation}
P = \frac{G_N}{5}\bigg<\left(\frac{d^3}{dt^3}I^{ij}(t)\right)^2\bigg>
+\frac{16G_N}{45}\bigg<\left(\frac{d^3}{dt^3}J^{ij}(t)\right)^2\bigg>
+\frac{G_N}{189}\bigg<\left(\frac{d^4}{dt^4}I^{ijk}(t)\right)^2\bigg>
+\cdots\,.
\end{equation}
This expression will be derived systematically below for all the multipole moments, both electric and magnetic.

\subsubsection{A few comments on the choices of boundary conditions}

As we have seen above, the Feynman boundary condition is indispensable when one is interested in deriving averaged physical observables from self-energy diagrams. Nevertheless, the physics behind the coalescence of compact binary systems is necessarily a radiation problem, and therefore, in general, the Feynman boundary condition for the propagators is not applicable, since it leads to a non-causal evolution of the system. Hence, while the prescription for the $i\epsilon$ is irrelevant in near-zone calculations, because of the asymptotic expansion
\begin{equation}
\frac{1}{\K^2 - k_0^2} = \frac{1}{\K^2} \sum_{n \geq 0} \left(\frac{k_0^2}{\K^2}\right)^n\,,
\end{equation}
which is valid in the limit $k_0^2 \ll \K^2$ and, hence, the poles in the propagator are never hit, the choice of boundary conditions becomes crucial when radiation is considered. 
Therefore, the appropriate causal boundary conditions must be taken into account in the far zone. 
To this end, one should resort to so-called in-in formalism, which is a framework that consistently implements causal boundary conditions by selecting automatically the correct Green's function, at the level of the path integral.

Nevertheless, for the processes of interest to us in this thesis, namely the tails, tails of tails and angular-momentum failed tails, the prescription given by the in-in formalism corresponds, equivalently, to naively replacing the Feynman propagator $G_F$ by the retarded one $G_R$ in the diagrams derived from the path integral formulation. This happens because, in all these processes, the orientation of the propagator for radiative gravitons, corresponding to retarded propagators, is trivial, in both self-energy and emission diagrams. For more intricate diagrams, on the other hand, all the machinery of the in-in formalism becomes necessary.

Below, we justify the important result that, for diagrams containing exactly two gravitational radiative modes, the propagator choice, for either Feynman's, retarded or advanced boundary conditions, is irrelevant for the accounting of conservative contributions from farzone computations. This result is particularly important since this is the case of the self-energy diagrams for all the tails, including also the angular momentum failed tail, that will be discussed in Chapter \ref{wardidentity}.

Let us start by displaying the relevant types of propagators, Feynman's, retarded and advanced, given respectively by
\begin{align}
 G_{F}(t,\X) &= \int \frac{d^3\K}{(2\pi)^3} \frac{dk_0}{2\pi} \frac{e^{-i k_0 t+i \K\cdot\X}}{\K^2-k_0^2+i\epsilon}\,, \\
 G_{R}(t,\X) &= \int \frac{d^3\K}{(2\pi)^3} \frac{dk_0}{2\pi} \frac{e^{-i k_0 t+i \K\cdot\X}}{\K^2-(k_0 + i\epsilon)^2}\,,\\
 G_{A}(t,\X) &= \int \frac{d^3\K}{(2\pi)^3} \frac{dk_0}{2\pi} \frac{e^{-i k_0 t+i \K\cdot\X}}{\K^2-(k_0 - i\epsilon)^2}\,.
\end{align}
In particular, these Green's functions can be combined into the important relation \cite{RSSFnearfar}
\begin{equation}\label{GFGRGA}
G_F(t,\X) = \frac 12 (G_R(t,\X) + G_A(t,\X)) - \frac{i}{2} (\Delta_+(t,\X) + \Delta_-(t,\X))\,,
\end{equation}
where $\Delta_{\pm}(t,\X)$ are the Wightman functions defined by
\begin{equation}
\Delta_{\pm}(t,\X) = \int \frac{d^3\K}{(2\pi)^3}\frac{dk_0}{2\pi} \theta(\pm k_0) \delta(k_0^2-\K^2) e^{-i k_0 t+i \K\cdot\X}\,.
\end{equation}
From Eq.~\eqref{GFGRGA}, one can derive the following important relation\cite{RSSFnearfar,galleytiglio}:
\begin{equation}
G_F^2 = \frac 12 (G_A^2 + G_R^2) - \Delta_+ \Delta_- - \frac{i}{2} (G_R + G_A) (\Delta_+ + \Delta_-)\,.
\end{equation}
Finally, from this relation we have the following behavior for its Fourier transform counterpart:
\begin{equation}
\tilde{G}_F^2(\omega) - \frac 12 (\tilde{G}_A^2(\omega) + \tilde{G}_R^2(\omega)) \sim  - \tilde{\Delta}_+(\omega) \tilde{\Delta}_-(\omega) - \frac{i}{2} (\tilde{G}_R(\omega) + \tilde{G}_A(\omega)) (\tilde{\Delta}_+(\omega) + \tilde{\Delta}_-(\omega))\,.
\end{equation}
The real part of this difference vanishes since $\tilde{\Delta}_+(\omega)$ and $\tilde{\Delta}_-(\omega)$ have no common support, and since both products $(\tilde{G}_R(\omega) + \tilde{G}_A(\omega))$ and $(\tilde{\Delta}_+(\omega) + \tilde{\Delta}_-(\omega))$ are real, as can be easily checked. Because of this, and because we have $\tilde{G}_A^2(\omega) = \tilde{G}_R^2(\omega)$, as can also be easily checked, we see that, for dissipative processes with Feynman diagram involving exactly two radiative graviton propagators, the result for the conservative contributions is the same computed with either $\tilde{G}_F^2(\omega)$, $\tilde{G}_R^2(\omega)$ or $\tilde{G}_A^2(\omega)$.

\subsection{The multipolar expansion and the matching}

As mentioned before, in order to derive expressions for the multipole moments in terms of the orbital variables present in the near-zone dynamics, we need to perform a matching procedure between the near and far regions. This is done by computing diagrams that couple linearly to one external radiative mode $h_{\mu\nu}$, after the potential modes have been integrated out, and then performing a multipolar expansion on the resulting effective action. In this case, the coupling to the radiative modes is given in a generic form by Eq.~\eqref{Tmunu}, by taking into account the energy-momentum tensor of matter and curvature.  

Then, the multipolar expansion consists in Taylor expanding the effective action \eqref{Tmunu} around the center of mass of the source, chosen for simplicity  as the origin $\X_{\rm cm} = 0$: (with $N\equiv i_1 \dots i_n$)
\begin{align}\label{Tmunuexpanded}
S_{\rm eff} =\frac 12 \int dt \sum_{n=0}^{\infty} \int d^3x T^{\mu\nu}(t,\X) x^N \partial_N h_{\mu\nu}(t,0)\,. 
\end{align}

From the conservation law $\partial_\mu T^{\mu\nu} = 0$ and nearzone scaling $\partial_\mu \sim (v/r, 1/r)$, we have that $\partial_\mu T^{0\mu} = 0$ gives $T^{0i} \sim \mathcal{O}(v T^{00})$, and, similarly, from the $\nu = i$ component, we obtain $T^{ij} \sim \mathcal{O}(v T^{0j})$. Thus, we have the following power law in the relative velocity $v$:
\begin{equation}
T^{ij}  \quad \sim \quad v T^{0i} \quad \sim \quad v^2 T^{00}\,.
\end{equation}
We can use this equivalence relation to collect terms in Eq.~\eqref{Tmunuexpanded} in powers of $v$, recalling that the long-wavelength scaling for the derivatives in the external graviton field goes as $\partial_\rho h_{\mu\nu} \sim h_{\mu\nu}/\lambda \sim (v/r) h_{\mu\nu}$, and hence, $x^N \partial_N h_{\mu\nu} \sim v^n h_{\mu\nu}$. Thus, up to $v^2$, we have:
\begin{align}
S_{\rm eff} &= \frac 12 \int dt \int d^3x\bigg\{ T^{00}(t,\X) h_{00}(t,0) \bigg|_{v^0} + \Big[ 2 T^{0i}(t,\X) h_{0i}(t,0) + T^{00}(t,\X) x^i \partial_i h_{00}(t,0)\Big]_{v^1} \nonumber\\
            &+ \Big[ \tfrac 12 T^{00}(t,\X) x^i x^j \partial_i\partial_j h_{00}(t,0) + 2 T^{0i}(t,\X) x^j \partial_j h_{0i}(t,0) + T^{ij}(t,\X) h_{ij}(t,0) \Big]_{v^2} \bigg\}\,.  
\end{align}
Then, since the external graviton field $h_{\mu\nu}(t,0)$ does not depend on the position $\X$, the integrals in the three-volume define moments of the energy-momentum tensor $T^{\mu\nu}$:
\begin{align}\label{Seffmultv2}
S_{\rm eff} &= \frac 12 \int dt \bigg[\left( \int d^3x\, T^{00} \right) h_{00} + 2\left( \int d^3x\, T^{0i} \right) h_{0i} + \left( \int d^3x\, T^{00} x^i\right) \partial_i h_{00} \nonumber\\
            &+ \left( \frac 12 \int d^3x\, T^{00} x^i x^j \right) \partial_i\partial_j h_{00} + 2 \left( \int d^3x\, T^{0i} x^j \right) \partial_j h_{0i} + \left( \int d^3x\, T^{ij}\right) h_{ij}  \bigg]\,.  
\end{align}
In particular,  from the $v^0$ term, we identify the total energy of the system, given by
\begin{equation}
E = \int d^3x\, T^{00}(t,\X)\,, 
\end{equation}
while from the $v^1$ terms we obtain the total momentum and center of mass of the system
\begin{align}
P^i = \int d^3x\, T^{0i}(t,\X) \qquad \text{and} \qquad 
X^i = \frac1{E}\int d^3x\,T^{00}(t,\X) x^i \,,
\end{align}
which are both zero in the center of mass frame: $X^i = P^i = 0$.

As for the $v^2$ contributions, we treat them by first defining the quadrupole moment
\begin{equation}\label{Qijmatching}
Q^{ij} = \int d^3x\, T^{00} x^i x^j\,,  
\end{equation}
which gives directly the first term in the second line of Eq.~\eqref{Seffmultv2}. At this stage, it is also convenient to introduce the total angular momentum
\begin{equation}
L^k = \frac 12 \epsilon_{ijk} \int d^3x\, (x^i T^{0j} - x^j T^{0i})\,.
\end{equation}
Hence, for the second term in the second line of Eq.~\eqref{Seffmultv2}, we can write
\begin{equation}
2\int d^3x\, T^{0i} x^j  = \int d^3x\, ( T^{0i} x^j + T^{0j} x^i) + \int d^3x\, ( T^{0i} x^j - T^{0j} x^i)\,,
\end{equation}
where we identify the appearance of the angular momentum in the last term.
Then, using the conservation law for the energy-momentum tensor, we obtain
\begin{equation}
\dot{Q}^{ij} = \int d^3x\, \frac{\partial}{\partial t}(T^{00}x^i x^j) = - \int d^3x\, \partial_k T^{0k} x^i x^j = \int d^3x\, (T^{0i} x^j + T^{0j} x^i)\,,
\end{equation}
which then gives
\begin{equation}
2\int d^3x\, T^{0i} x^j  = \dot{Q}^{ij} - \epsilon_{ijk} L^k\,.
\end{equation}
Finally, for the third piece of the second line, using twice the conservation law of the energy-momentum tensor, we obtain
\begin{align}
\ddot{Q}^{ij} &= \int d^3x\, \frac{\partial^2}{\partial t^2} (T^{00}x^i x^j) = \int d^3x\, \frac{\partial}{\partial t}\left( x^i T^{0j} + x^j T^{0i} \right) \nonumber\\
              &= - \int d^3x\, (x^i \partial_k T^{jk} + x^j \partial_k T^{ik}) = 2 \int d^3x\, T^{ij}\,.
\end{align}

Collecting all the above results and performing integration by parts in $t$ to move all time derivatives to the external graviton field, we finally obtain:
\begin{equation}
S_{\rm eff} = \frac 12 \int dt \bigg[E h_{00}-\epsilon_{ijk} L^k \partial_j h_{0i} + \frac 12 Q^{ij} \left( \ddot{h}_{ij} - \partial_i \dot{h}_{0j} - \partial_j \dot{h}_{0i} + \partial_i\partial_j h_{00} \right) \bigg]\,.  
\end{equation}
Notice that, at the linear level $\mathcal{O}(h)$,
\begin{equation}
E_{ij} = R_{0i0j} = -\frac 12 \left( \ddot{h}_{ij} - \partial_i \dot{h}_{0j} - \partial_j \dot{h}_{0i} + \partial_i\partial_j h_{00} \right)\,.
\end{equation}
With this, we have reproduced the far zone effective action \eqref{EFfarzone} at order $v^2$ and at the linear level $\mathcal{O}(h)$, while identifying the quadrupole moment $Q^{ij}$ as given in terms of near-zone information (contained in the energy-momentum tensor of the system) via Eq.~\eqref{Qijmatching}. In particular, $Q^{ij}$ can be replaced with the trace-free tensor $I^{ij}$ since its trace will  couple to the Ricci tensor, which vanishes in vacuum. 

The steps followed above illustrate a general procedure for matching the far-zone degrees of freedom, i.e., the multipole moments, with the near-zone degrees of freedom given by orbital variables via the energy-momentum tensor of the system. In fact, this procedure has already been carried out systematically in Ref.~\cite{ross2012}, in which the author was able to give explicit expressions for all the multipole moments, of either parity, in terms of the energy-momentum tensor of the system, that are valid to all orders in the velocity. This construction follows a group-theory point of view, in which, to every order in the velocity, the moments of $T^{\mu\nu}$ are reduced into irreducible representations of the rotation group, $SO(3)$, with the $\ell = 2$ mode giving rise to electric quadrupole moment, the $\ell = 3$ generating the electric octupole and magnetic quadrupole moments, and so on.


\section{Linearized theory of gravity from amplitudes}

At the linear level, gravitational waves are generated by the simple on-shell (i.e., for $\K^2 = \omega^2$) emission amplitudes from the whole series of source multipole moments, in the far zone. As we will see shortly, this corresponds to the linearized theory of gravity presented in the previous chapter. These lowest-order processes are represented diagrammatically by
\begin{equation}
i{\cal A}(k) = \sum_{r=0}^\infty
\,\, \left(
      \begin{tikzpicture}[baseline={([yshift=-.5ex]current bounding box.center)}]
      \draw [black, thick] (0,0) -- (2,0);
      \draw [black, thick] (0,-0.07) -- (2,-0.07);
      \draw[decorate, decoration={snake},segment length=6.5pt,segment amplitude=1.2pt, line width=1pt,shorten >= 0.8pt] (1,0) -- (1,1);
      \filldraw[black] (1,-0.035) circle (2pt) node[anchor=north] {$I^{ijR}$};
    \end{tikzpicture}
\,\,
    +
\,\,
      \begin{tikzpicture}[baseline={([yshift=-.5ex]current bounding box.center)}]
      \draw [black, thick] (0,0) -- (2,0);
      \draw [black, thick] (0,-0.07) -- (2,-0.07);
      \draw[decorate, decoration={snake},segment length=6.5pt,segment amplitude=1.2pt, line width=1pt,shorten >= 0.8pt] (1,0) -- (1,1);
      \filldraw[black] (1,-0.035) circle (2pt) node[anchor=north] {$J^{i|jRl}$};
    \end{tikzpicture}
\right)\,.    
\end{equation}
In particular, to each external graviton leg there is a polarization tensor $\epsilon_{ij}^{*}(\K,h)$ attached to it, of polarization state $h =\pm 2$, that satisfies the transverse and trace-free conditions $k^i \epsilon_{ij}(\K,h) = 0, \delta^{ij} \epsilon_{ij}(\K,h) = 0$. Hence, diagrammatic rules derived from Eq.~\eqref{farzoneactionstatic} yields
\begin{equation}\label{LOemissamp}
i \mathcal{A}_h(\K)=
\sum_r\frac{(-i)^{r+1}}{2\Lambda}\epsilon_{ij}^{*}(\K,h) k_R\left[
c_r^{(I)} \omega^2 I^{ijR}(\omega)
-
c_r^{(J)} \omega k_l  J^{i|jRl} (\omega)
\right]\,.
\end{equation}

More generally, for a generic gravitational process involving the emission of gravitational waves, possibly containing nonlinear interactions, we have an amplitude  
\begin{equation}
i{\cal A}(k) =
\,\,
      \begin{tikzpicture}[baseline={([yshift=-.5ex]current bounding box.center)}]
      \begin{feynman}
      \draw [black, thick] (0,0) -- (2,0);
      \draw [black, thick] (0,-0.07) -- (2,-0.07);
      \draw[decorate, decoration={snake},segment length=6.5pt,segment amplitude=1.2pt, line width=1pt,shorten >= 0.8pt] (1,0) -- (1,1);
      \filldraw [white, thick] (1,0) circle (10pt);
      \vertex [blob,thick] (m) at (1,0) {\contour{black}{}};
      \end{feynman}
    \end{tikzpicture}\,.    
\end{equation}
From this, we can directly compute gravitational-wave observables, like the energy power loss, as well as the gravitational waveform itself. In particular, the on-shell energy-momentum tensor of the gravitational wave can be drawn from this amplitude, by means of the relation
\begin{equation}\label{Tijonshell}
i \mathcal{A}_h(\K) = \frac{i}{2\Lambda} T^{ij}(\K) \epsilon^*_{ij}(\K,h)\,. 
\end{equation}

The waveform for this generic process can be obtained by plugging the corresponding energy-momentum tensor into the following expression for the TT part of $h_{\mu\nu}$:
\begin{equation}\label{hTTfromTij}
h_{ij}^{TT}(t,\X) = \frac{4G_N}{D} \Lambda_{ij,kl}({\bf \hat{n}}) \int_{-\infty}^{\infty} \frac{d\omega}{2\pi} T^{kl}(\omega,\omega {\bf \hat{n}}) e^{-i\omega t_{\rm ret}}\,,
\end{equation}
where, as usual, $t_c = t -D$ is the retarded time, $D= |\X|$ is the distance from the detector to the source and ${\bf \hat{n}} = \X/D$.
Also from the amplitude $i \mathcal{A}_h (\K)$ we can compute the graviton emission rate for the polarization $h$, recalling here the general expression
\begin{equation}
d\Gamma_h (\K) = \frac{1}{T} \frac{d^3\K}{(2\pi)^3 2 \omega} | \mathcal{A}_h (\K)|^2\,,
\end{equation} 
from which, the total rate of radiated linear four-momentum can be obtained via
\begin{equation}
\frac{dP^\mu}{dt}\bigg|_{h=\pm 2} = \int k^\mu d\Gamma_h(\K)\,, 
\end{equation}
where $k^\mu = (\omega,\K)$, with $\K^2 = \omega^2$, is the four-momentum of the emitted graviton.
In particular, for the radiated energy, i.e., $P^0 = E$, we have
\begin{equation}\label{EfluxA}
\frac{dE}{dt}\bigg|_{h=\pm 2} = \int \omega d\Gamma_h(\K) = \frac{1}{2T}\int \frac{d^3\K}{(2\pi)^3} | \mathcal{A}_h (\K)|^2\,.
\end{equation}

In general, any observables related to the emission of gravitational radiation, like the emitted energy in the expression above, are obtained from the square of the amplitude $| \mathcal{A}_h (\K)|^2$ via $d\Gamma_h(\K)$. Moreover, because these quantities are always integrated over the phase volume $d^3\K$, they are necessarily averaged quantities. In particular, as we are usually interested in computing the total averaged quantity, we have to make use of the standard expression for the sum over graviton polarizations, which is given by
\begin{align}\label{sumoverpols}
\sum_h \epsilon_{ij}(\K,h)\epsilon^*_{kl}(\K,h) = 
\frac 12 & \left[
\delta_{ik}\delta_{jl}+\delta_{il}\delta_{jk}-\frac{2}{d-1}\delta_{ij}\delta_{kl}
+\frac{2}{(d-1)\omega^2}(\delta_{ij}k_k k_l+\delta_{kl}k_i k_j)\right.
\nonumber\\
-&
\left.\frac{1}{\omega^2}(\delta_{ik}k_j k_l+\delta_{il}k_j k_k+\delta_{jk}k_i k_l+\delta_{jl}k_i k_k)
+\frac{4}{c_d \omega^4}k_i k_j k_k k_l
\right]\,.
\end{align}
This expression generalizes Eq.~(11) of Ref.~\cite{goldberger2010} to arbitrary spatial dimension $d$. As it will be explained in Chapter
\ref{wardidentity}, this $d$-dependence is very important when one proceeds to compute self-energy contributions starting from emission amplitudes that contain divergences in $d=3$, like it happens in the tail processes.

For completeness, we present below the expression for the total loss of the system's angular momentum in terms of the emission amplitude $\mathcal{A}_h$:
\begin{equation}
\frac{d{\bf J}}{dt} = \sum_h \int h\, {\bf \hat{n}}\, d\Gamma_h(\K)\,,
\end{equation}
where, here, ${\bf \hat{n}} = \K/\omega$.

\subsection{Waveform at the lowest order}

Now, consider the leading-order emission amplitude given in Eq.~\eqref{LOemissamp}. The energy-momentum tensor for the gravitational wave derived from it, drawn with the help of Eq.~\eqref{Tijonshell}, reads
\begin{equation}
T^{ij}(\K) = - \sum_r (-i)^r k_R \left[ c_r^{(I)} \omega^2 I^{ijR}(\omega) - c_r^{(J)} \omega k_l J^{(i|j)Rl}(\omega) \right]\,.
\end{equation}
Notice the symmetrization in the indices $i,j$ in $J^{i|jRl}$ in the equation above, necessary so that we have a consistent symmetric tensor $T^{ij}$. With this expression in hand, we can then plug it into Eq.~\eqref{hTTfromTij} and build an expression of $h^{TT}_{ij}$ for all the source multipole moments, corresponding precisely to the complete waveform in the linearized theory, which is a theory valid at order $\mathcal{O}(G_N)$.  In this case, we obtain, after performing the inverse Fourier transform,
\begin{equation}
h^{TT}_{ij}(t,\X) = \frac{4G_N}{D} \Lambda_{ij,kl} \sum_r n_R \left[ c_r^{(I)} \frac{d^{r+2}}{dt^{r+2}}I^{klR}(t) - c_r^{(J)} n_m \frac{d^{r+2}}{dt^{r+2}}J^{(k|l)Rm}(t) \right]\bigg|_{t\rightarrow t_r}\,.
\end{equation}
In this equation, ${\bf n} = \X/D$.
Then, in particular, using the fact that, in $d=3$, we have the identification $J^{l|jRi} = \epsilon_{ikl} J^{jkR}$, and plugging the expressions for $c_r^{(I)}$ and $c_r^{(J)}$ given in Eq.~\eqref{cIcJ}, using also the standard notation $\ell = r + 2$, where $\ell$ represents the multipolarity of the contribution, with $\ell = 2$ representing the quadruple, $\ell = 3$ for the octupole, and so on, the above result  becomes (recalling the notation for the collective indices $L-2 = i_1 i_2 \dots i_{\ell-2}$)
\begin{equation}\label{hijTTLOell}
h^{TT}_{ij}(t,\X) = \frac{4 G_N}{D} \Lambda_{ij,kl} \sum_{\ell = 2}^{\infty} n_{L-2} \left[ \frac{1}{\ell !} \frac{d^{\ell}}{dt^{\ell}}I^{klL-2}(t) + \frac{2\ell}{(\ell + 1)!} n_n \epsilon^{mn(k}  \frac{d^{\ell}}{dt^{\ell}}J^{l)mL-2}(t) \right]\,.
\end{equation}
This equation can be directly compared with Eq.~(4.8) of Ref.~\cite{thorne1980} or to Eq.~(66) of Ref.~\cite{blanchetlivrev}, in whichever cases a complete agreement is immediatally found.

\subsection{Total power radiated}

The expression for the total energy radiated as gravitational waves, or shortly, the energy flux, $P$, is given by summing Eq.~\eqref{EfluxA} over the polarizations, 
\begin{equation}
P \equiv \sum_h \frac{dE}{dt}\bigg|_{h}  = \frac{1}{2T}\int \frac{d^3\K}{(2\pi)^3} \sum_h | \mathcal{A}_h (\K)|^2\,.
\end{equation}
In particular, for the sum over polarizations of the amplitude squared, we have:
\begin{multline}
\sum_h |\mathcal{A}_h(\K)|^2 = \sum_{r,r'} \frac{1}{4\Lambda^2} k_R k_{R'}
\Big[c_r^{(I)} \omega^2 I^{ijR}(\omega)-c_r^{(J)} \omega k_m  J^{i|jRm} (\omega)\Big] \\ 
\times \Big[c_{r'}^{(I)} \omega^2 I^{klR'*}(\omega)-c_{r'}^{(J)} \omega k_n  J^{k|lR'n*} (\omega)
\Big] 
\sum_h \epsilon_{ij}^{*}(\K,h) \epsilon_{kl}(\K,h)\,.
\end{multline}
Note that, for terms involving multipole moments of different parity, after integration in $\K$ the only case that is not trivially vanishing is when we have $r=r'+1$ in
\begin{equation}
\int_\K k_R k_{R'} k_n I^{ijR}(\omega) J^{k|lR'n*}(\omega) \sum_h \epsilon_{ij}^{*}(\K,h) \epsilon_{kl}(\K,h)\,.  
\end{equation}
(Or, similarly, $r'=r+1$ for the other crossed term)
Otherwise, there will be an excess of Kronecker deltas being contracted to either $I^{ijR}$ or $J^{k|lR'n*}$, which vanish immediately. Hence, for the case $r = r'+1$, the above integral has the following types of terms:
\begin{equation}\label{IJRRnvanishing}
I^{ijR}(\omega) J^{i|jR'n*}(\omega) \delta_{R R'n}  \qquad \text{and} \qquad
I^{ijR}(\omega) J^{i|lR'n*}(\omega) \delta_{jln R R'}  \,,  
\end{equation}
where $\delta_{R R'n}$ and $\delta_{jln R R'}$ are generalized Kronecker deltas, defined by a sum over all products of simple Kronecker deltas for all possible index permutations, e.g.,
\begin{equation}
\delta_{ijkl} = \delta_{ij} \delta_{kl} + \delta_{ik} \delta_{jl} + \delta_{il} \delta_{jk}\,.
\end{equation} 
As it turns out, both contributions in Eq.~\eqref{IJRRnvanishing} vanish since they are fully contracted: $I^{ijR}(\omega) J^{i|jR*}(\omega)$, involving, in particular, contractions of symmetric and antisymmetric indices. Because of this, there is no mix in $\sum_h |\mathcal{A}_h|^2$ among multipole moments of different parities. As for multipoles of the same type, there are contributions only when $r = r'$, following a similar analysis in the symmetries of the indices. Thus, we have
\begin{align}
P &= \frac{1}{2T}\int \frac{d^3\K}{(2\pi)^3} \sum_h | \mathcal{A}_h (\K)|^2 \nonumber\\
  &= \sum_{r} \frac{[c_r^{(I)}]^2}{4\Lambda^2} 
\frac{1}{2T}\int\frac{d^3\K}{(2\pi)^3} \omega^4 I^{ijR}(\omega) I^{klR'*}(\omega) 
k_R k_{R'}\sum_h \epsilon_{ij}^{*}(\K,h) \epsilon_{kl}(\K,h) \nonumber\\
  &+
\sum_{r} \frac{[c_r^{(J)}]^2}{4\Lambda^2} 
\frac{1}{2T}\int\frac{d^3\K}{(2\pi)^3}\omega^2 J^{i|jRm}(\omega) J^{k|lR'n*} (\omega)
k_R k_{R'} k_m k_n \sum_h \epsilon_{ij}^{*}(\K,h) \epsilon_{kl}(\K,h)\,.
\end{align}
We deal with the integrals in $\K$ by using the following important relation, that can be derived in $d$ dimensions using the generalization of the solid angle to arbitrary dimensions, and reads
\begin{equation}\label{kintegrals}
\int \frac{d^d\K}{(2\pi)^d} f(\omega)
k_{i_1}\cdots k_{i_{2r}} = \frac{a_{2r}}{(2\pi)^d} \frac{2\pi^{d/2}}{\Gamma(d/2)}
\int_0^{\infty} d\omega\, \omega^{2r+d-1}\delta_{i_1\cdots i_{2r}} f(\omega)
\end{equation}
for a generic function $f$ of $\omega$, with coefficient $a_{2r}$ given by 
\begin{equation}\label{kintegrals2r}
a_{2r}=\frac 1{d(d+2)(d+4)\cdots [d+(2r-2)]}=\frac{(d-2)!!}{[d+2(r-1)]!!}\,.
\end{equation}
Then, we plug into the above equation for $P$ the sum over polarizations given in Eq.~\eqref{sumoverpols}, perform the integration in $\K$, and expand the generalized Kronecker deltas resulting from this integration using the symmetries of the multipole moments. For instance, when contracted to $I^{ijR}(\omega) I^{klR'*}(\omega)$, we can make the following replacement for the generalized delta $\delta_{R R'}$, since the $I^{ijR}$ are traceless and symmetric in the indices $R = i_1 \dots i_r$,
\begin{equation}
\delta_{R R'} = r! \delta_{i_1 i'_1}\delta_{i_2 i'_2}\cdots \delta_{i_r i'_r}+\text{trace terms}\quad \rightarrow \quad r! \delta_{i_1 i'_1}\delta_{i_2 i'_2}\cdots \delta_{i_r i'_r}\,.
\end{equation}

Following the procedure described above, we eventually end up with 
\begin{multline}
P
= \frac{2\pi^{d/2}}{\Gamma(d/2)} \sum_r \frac{r!}{(2\pi)^d}
\bigg\{ \frac{[c_r^{(I)}]^2}{8\Lambda^2} 	
\frac{(d-2)(d+r)(d+r+1)a_{2r}}{(d-1)(d+2r)(d+2r+2)} \times \frac{1}{T} \int_0^{\infty} d\omega \, \omega^{2r+d+3} I^{ijR} I_{ijR}^{*} \\
+
\frac{[c_r^{(J)}]^2}{8\Lambda^2} \frac{(r+3)\left[a_{2r+2} - (r+1)a_{2r+4} \right]}{2(d-1)} 
\times \frac{1}{T}\int_{0}^{\infty}d\omega\, \omega^{2r+d+3} J^{i|jRl} J_{i|jRl}^{*} \bigg\}\,.
\end{multline}
Then, setting $d=3$, replacing $\Lambda^2 = (32\pi G_N)^{-1}$, and using the following important relation, valid in the limit as $T \rightarrow \infty$,
\begin{equation}
\frac{1}{\pi T}\int_{0}^{\infty} d\omega\, \omega^{2 s} | A(\omega)|^2 = \frac{1}{T} \int_{0}^{T} dt\, \left( \frac{d^s}{dt^s}A(t) \right)^2 \quad \rightarrow \quad
\bigg< \left(\frac{d^s}{dt^s}A(t)\right)^2 \bigg>\,,
\end{equation}
we obtain
\begin{multline}
P = G_N \sum_r \bigg[  
\frac{(r+3)(r+4)}{(r+1)(r+2)(r+2)!(2r+5)!!}
\times
\bigg< \left(\frac{d^{r+3}}{dt^{r+3}}I^{ijR}(t)\right)^2 \bigg> \nonumber\\
+ 
\frac{4(r+4)}{(r+1)(r+3)(r+1)!(2r+5)!!}
\times
\bigg<\left(\frac{d^{r+3}}{dt^{r+3}}J^{ijR}(t)\right)^2\bigg> \bigg]\,,
\end{multline}
or, in terms of $\ell$, 
\begin{multline}\label{powerLOell}
P = G_N \sum_{\ell = 2}^\infty \bigg[  
\frac{(\ell+1)(\ell+2)}{(\ell-1)\ell \ell!(2\ell+1)!!}
\times
\bigg< \left(\frac{d^{\ell+1}}{dt^{\ell+1}}I^{L}(t)\right)^2 \bigg> \\
+ 
\frac{4\ell(\ell+2)}{(\ell-1)(\ell+1)! (2\ell+1)!!}
\times
\bigg<\left(\frac{d^{\ell+1}}{dt^{\ell+1}}J^{L}(t)\right)^2\bigg> \bigg]\,,
\end{multline}
which can be directly compared to, e.g., Eq.~(3.210) of Ref.~\cite{maggiore1}.

Interestingly, although the formulas above have been derived for the leading amplitude, they are also valid, in the following sense, when general-relativistic nonlinearities are assumed. When graviton interactions are included, i.e., by accounting for higher-order processes in $G_N$, we can always write our on-shell amplitude in the form of Eq.~\eqref{LOemissamp}, in which case we will have new moments, $I^{ijR}_{\rm rad}$ and $J^{i|jRl}_{\rm rad}$, called the \textit{radiative multipole moments}, in the place of the source moments, having precisely the same properties as the old ones. Hence, in particular, the formulas derived above for the waveform, Eq.~\eqref{hijTTLOell}, and energy flux, Eq.~\eqref{powerLOell}, present the same functional form, but with $I^{ijR}$ and $J^{i|jRl}$ replaced by $I^{ijR}_{\rm rad}$ and $J^{i|jRl}_{\rm rad}$.

%% file: chapter3/wardidentity.tex
\chapter{Gravitational Scattering Amplitudes and the Ward Identity}\label{wardidentity}

\section{Introduction}

The problem of how to include radiation-reaction contributions to the description of the conservative sector of the relativistic two-body problem, completing near-zone computations, is currently one of the most important problems 
in the context of the analytic gravitational two-body modeling.
The post-Newtonian order in which the first of such contributions enter is the 4PN order \cite{RSSF4PNb}, in which the only contribution stems from the self-energy diagram for the simple (mass) tail of the electric quadrupole moment computed in the far zone. The tail process \cite{blanchet1988}, which physically represents the emission of gravitational radiation that is then back-scattered into the system, and hence modifies the dynamics of the system, has been known for a long time within an EFT perspective \cite{RSSF2013}, and presents matching results with respect to traditional computations \cite{blanchet1993}.

For the conservative dynamics at the 5PN order, on the other hand, several are the processes that will contribute, stemming from far zone self-energy diagrams \cite{RSSFhereditary}. In addition to the simple tail for the 1PN-corrected electric quadrupole moment, we also have the simple tails for the electric octupole and magnetic quadrupole, the angular momentum failed tail, and finally, the memory effect. While the tail and memory effects are hereditary, i.e., their emission depends nonlocally on the binaries' whole history (by means of an integral on the multipoles evaluated from $t=-\infty$ to $t = t_{\rm ret}$), the angular momentum failed tail depends only locally on the multipoles, even though the topology of the diagrams are the same. 

All the contributions that enter the 5PN order have been previously computed in Ref.~\cite{RSSFhereditary}, and confirmed later on in Refs.~\cite{blumlein2021,brunello2022}, but whose complete understanding is still not satisfactory, as they are in conflict with self-force considerations \cite{damour2021}. Indeed, by taking into account the effective action for such radiation-reaction processes, quantities such as the scattering angle, proper of the gravitational scattering problem, can be computed. Such quantity, in particular, should be devoid of $\nu^2$ terms (being $\nu$ the symmetric mass ratio $\nu = m_1 m_2/(m_1+m_2)^2$) in the conservative dynamics, as expected from simple analysis of its dependence on $\nu$ \cite{sturani2021}.




Thus, the inconsistency involved in the nonvanishing of $\nu^2$ terms in the scattering angle motivated us to investigate deeper questions regarding the own construction of self-energy diagrams, that could be affecting radiation-reaction computations. 
The presentation of this program is carried out
by studying the self-consistency in the computation of emission processes, by investigating the compatibility of the Ward identity, gauge condition, and energy-momentum tensor conservation. As we will see, while for the leading-order processes and simple tails, these ideas follow trivially, the same does not happen in the case of the angular momentum failed tail. Indeed, for the latter, we will encounter Ward-violating amplitudes that will naturally lead to an incomplete description of the corresponding radiation-reaction effects.

Along the chapter, we build on the violation of the Ward identity and learn that this comprises a ``consistent anomaly." We also discuss how the self-energy results are affected in the cases where anomalies are present by presenting a consistent way of getting rid of them, after which a complete account for the radiation reaction is obtained. Besides this, in the construction, the proper connection between emission and self-energy amplitudes is established, while limiting ourselves to the investigation of  the leading order processes, simple tails, and angular momentum failed tails. We will learn that just the quadrupole cases of the latter process present anomalies.


Finally, the research developed in this chapter is original and provides an essential step toward the full completion of the 5PN conservative dynamics for binary systems.\footnote{The paper with the results presented in this chapter is currently under preparation.}

\section{Setup}

The starting point in our discussion is the worldline effective action for the far zone, introduced in the previous chapter. This effective action describes the coupling of a compact system to the external gravitational field generated by it and is valid at length scales $\lambda \gg r$, with $r$ denoting the characteristic size of the system. As we have seen, this effective action can be expressed in terms of an infinite series of multipole moments that carry information about the internal physics of the system and is given by
\begin{equation}\label{sourcelag}
    S_{\rm source} = \int dt\,\left[ \frac12 E h_{00} -\frac12 J^{b|a} h_{0b,a} - \sum_{r\ge 0}\left( c_r^{(I)} I^{ijR} \partial_R R_{0i0j} + \frac{c_r^{(J)}}{2} J^{b|iRa} \partial_R R_{0iab} \right) \right]\,,
\end{equation}
with
\begin{equation}
c_r^{(I)} = \frac{1}{(r+2)!}\,, \qquad c_r^{(J)} = \frac{2(r+2)}{(r+3)!}\,.
\end{equation}
Recalling that, in this expression,  $R$ represents the collective index $R=i_1\dots i_r$, 
$E$ and $J^{b|a}$ are the conserved energy and angular momentum, and $I^{ijR}$ and $J^{b|iRa}$ the set of electric-type and magnetic-type multipole moments of the source, with $r=0$ standing for the quadrupole, $r=1$ for the octupole, and so on.

The dynamics of the gravitational field is dictated by the bulk action, given by the Einstein-Hilbert action plus a gauge-fixing term, chosen to be the harmonic gauge as before:
\begin{align}\label{bulkgravity}
    S_{\rm bulk} &= 2\Lambda^2 \int d^{d+1}x \sqrt{-g} \left[ R(g) - \frac12 \Gamma_\mu \Gamma^\mu \right]\,.
\end{align}
Notice that we have kept this action in arbitrary spacetime dimension $n=d+1$ since dimensional regularization will be employed throughout the chapter in the particular cases of tail processes, in which divergences are present in $d=3$. Recall also that $\Lambda = (32\pi G_N)^{-1/2}$\,.	

In our calculations, we find it useful to use the Kaluza-Klein decomposition of the metric \cite{kol2008a,kol2008b},
\begin{equation}\label{KKdecompward}
g_{\mu\nu}=e^{2\phi/\Lambda}\left(
  \begin{array}{cc}
  -1 & \dfrac{A_i}\Lambda\\
  \dfrac{A_j}\Lambda & \quad e^{-c_d\phi/\Lambda}\left(\delta_{ij}+\dfrac{\sigma_{ij}}\Lambda\right)-
  \dfrac{A_iA_j}{\Lambda^2}
  \end{array}
  \right)\,,
\end{equation}
in which the $n(n+1)/2$ degrees of freedom of the metric $g_{\mu\nu}$ (10 in $n=4$) gets replaced by a scalar field $\phi$, a spatial vector $A_i$, and a spatial symmetric tensor $\sigma_{ij}$, summing up the correct number of degrees of freedom. We also have that $c_d\equiv 2(d-1)/(d-2)$, and we adopt the mostly-plus signature of the metric, as can be realized from the expression above.

The most interesting property of this decomposition is that the propagators for each of these fields are independent of each other.
Derivation is simple and follows from the quadratic term in Eq.~\eqref{bulkgravity}, resulting in
\begin{equation}
\left.
\begin{array}{rcl}
\mathcal{D}[\phi,\phi]&=&-\dfrac 1{2c_d}\\
\mathcal{D}[A_i,A_j]&=& \dfrac{\delta_{ij}}2\\
\mathcal{D}[\sigma_{ij},\sigma_{kl}]&=& -\dfrac 12\left(\delta_{ik}\delta_{jl}+\delta_{il}\delta_{jk}-\dfrac 2{d-2}\delta_{ij}\delta_{kl}\right)
\end{array}
\right\}\times \frac i{\K^2-\omega^2}\,.
\end{equation}


\section{The gravitational field and the gauge condition}\label{secgravfield}

The classical gravitational field at a spacetime position $x$ is given by the one-point function of the field $h_{\mu\nu}$, denoted here as $\braket{h_{\mu\nu}(x)}$ and given by the path integral\footnote{The correct way of treating this problem is by employing the in-in formalism, since time-asymmetry is in the heart of radiative processes. However, for the processes considered in this chapter, as discussed at the end of Sec.~\ref{opticaltheoremsec}, it is equivalent to just replace the Feynman propagator by the retarded one.} 
\begin{equation}\label{onepointdef}
    \braket{h_{\mu\nu}(x)} = \int\mathcal{D}h\, e^{iS[h]}h_{\mu\nu}(x)\,.
\end{equation}
Rather than $h_{\mu\nu}$, the object we want ultimately to compute and which will play the most relevant role in our discussion is $\bar{h}_{\mu\nu}$, the trace-reversed version of $h_{\mu\nu}$, which is defined by
\begin{equation}\label{projetctorTT}
\bar{h}_{\mu\nu} = P_{\mu\nu}{}^{\alpha\beta} h_{\alpha\beta}\,, \qquad \qquad \text{with} \qquad\qquad P_{\mu\nu}{}^{\alpha\beta} = \frac12 \left(\delta_\mu^{\alpha} \delta_\nu^{\beta} + \delta_\mu^{\beta} \delta_\nu^{\alpha} - \eta_{\mu\nu}\eta^{\alpha\beta}\right)\,.
\end{equation}

In particular, before jumping into nonlinear effects, the classical solution $\braket{\bar{h}_{\mu\nu}(x)}$ can be computed separately for each of the couplings of the action \eqref{sourcelag}, namely $E$, $J^{b|a}$ $I^{ijR}$ and $J^{b|iRa}$. This is precisely what we do below. Moreover, for each of the solutions generated by these terms, we check their consistency with respect to the Lorentz condition, $\partial^\mu \bar{h}_{\mu\nu}=0$, which is the leading-order equation for the harmonic gauge, defined by the equation $\Gamma^\mu =0$. 

For higher-order processes, the perturbative expansion of the harmonic gauge condition $\Gamma^\mu =0$ will give rise to an inhomogeneous generalization of the Lorentz gauge condition in the form of $\partial^\mu \bar{h}^{(n)}_{\mu\nu}=\lambda[h^{(n-1)},h^{(n-2)},\dots, h^{(1)}]$, where $h^{(n)}$ represents a process of order $(G_N)^n$. Nevertheless, as it will be shown in Sec.~\ref{onshelness}, the RHS of this equation represents longitudinal modes that do not go on-shell and, hence, will not play any role in the emission of gravitational radiation and self-energy computations for all the tail processes (including the failed tail of the angular momentum), which are of order $G_N$. We can, therefore, neglect such terms.
The complete justification for this will also be presented in Sec.~\ref{onshelness}.


\subsection{Leading-order energy coupling}

From the monopole coupling, for the leading-order contribution, i.e., at order $G$, we have:
\begin{equation}
    \braket{h_{\mu\nu}(x)} = \int dt'\,\frac{i E}{2} \braket{h_{00}(t') h_{\mu\nu}(x)},
\end{equation}
from which we obtain
\begin{equation}
    \braket{h_{00}(x)} = \frac{2G_N E}{|\X|}\,, \quad \braket{h_{0i}(x)} = 0\,, \quad \braket{h_{ij}(x)} = \frac{2G_N E}{|\X|}\delta_{ij}\,, \quad \text{and}\quad \braket{h(x)} = \frac{4G_N E}{|\X|} \,. 	
\end{equation}
The last of the terms above is just the trace $h=\eta^{\mu\nu}h_{\mu\nu}$\,.
Notice that this solution is just the well-known Newtonian result $h_{00} = -2\Phi$ and $h_{ij} = -2\Phi\delta_{ij}$\,, with $\Phi = -G_N E/|\X|$ the Newtonian potential.
In this calculation, the following integral was needed:
\begin{equation}
\int_\K\frac{e^{-i\K\cdot\X}}{\K^2} = \frac{1}{4\pi |\X|}\,,
\end{equation}
where we use the notation $\int_\K \equiv \int d^3\K/(2\pi)^3$.

From the components above, the trace-reversed components can be evaluated, yielding:
\begin{equation}
    \braket{\bar{h}_{00}(x)} = \frac{4G_N E}{|\X|}\,,\qquad\qquad \braket{\bar{h}_{0i}(x)} = 0\,,\qquad\text{and}\qquad \braket{\bar{h}_{ij}(x)} = 0\,. 	
\end{equation}
This solution is easily seem to be consistent with the Lorentz gauge condition, given by
\begin{equation}
\partial_\mu \bar{h}^{\mu\nu} = 0\,,
\end{equation}
since $\braket{\bar{h}_{\mu\nu}}$ is time independent (assuming $dE/dt = 0$) and $\braket{\bar{h}_{i0}}=\braket{\bar{h}_{ij}}=0$:  
\begin{align}
    \partial_\mu \braket{\bar{h}^{\mu 0}} &= \partial_0 \braket{\bar{h}_{00}} - \partial_i \braket{\bar{h}_{i0}} = 0\,,\\
    \partial_\mu \braket{\bar{h}^{\mu j}} &= -\partial_0 \braket{\bar{h}_{0j}} + \partial_i \braket{\bar{h}_{ij}} = 0\,.    
\end{align}


\subsection{Leading-order angular momentum coupling}

Similarly to the previous case, for the second static coupling, of $J^{b|a}$ to $h_{0b,a}$, we have:
\begin{equation}
\braket{h_{\mu\nu}(x)} = -\frac{i}{2} J^{b|a} \int dt'\,\braket{h_{0b,a}(t')h_{\mu\nu}(x)}\,,
\end{equation}
from which we derive
\begin{equation}\label{Lcoupling}
\braket{h_{00}(x)} = \braket{h_{ij}(x)} = 0\,, \qquad \text{and} \quad \braket{h_{0i}(x)} = 2 G_N J^{i|a} \frac{x^a}{|\X|^3}\,.
\end{equation}

In this case, since $h = 0$, we have $\bar{h}_{\mu\nu} = h_{\mu\nu}$. Hence, the Lorentz gauge condition can be verified directly from the components in Eq.~\eqref{Lcoupling}, where it gives (assuming $dJ^{i|a}/dt = 0$)
\begin{align}
    \partial_\mu \braket{\bar{h}^{\mu 0}} &= \partial_0 \braket{\bar{h}_{00}} - \partial_i \braket{\bar{h}_{i0}} = -2 G_N J^{i|a} \frac{1}{|\X|^3}\left(
    \delta^{ia}-3\frac{x^i x^a}{|\X|^2}\right) = 0\,,\\
    \partial_\mu \braket{\bar{h}^{\mu j}} &= -\partial_0 \braket{\bar{h}_{0j}} + \partial_i \braket{\bar{h}_{ij}} = 0\,,   
\end{align}
which shows that this result also respects the gauge condition. 

For the calculation above, we used the following integral in $\K$:
\begin{equation}
\int_\K \frac{k^a}{\K^2}e^{-i\K\cdot\X} = -\frac{i}{4\pi}\frac{x^a}{|\X|^3}\,.
\end{equation}


\subsection{Higher electric and magnetic multipole leading-order couplings}

\subsubsection{Electric multipole moments}

Now, for the coupling of the time-varying multipole moments of electric parity, we have
\begin{equation}
\braket{\bar{h}_{\mu\nu}(x)} = -i c_r^{(I)} \int dt'\,I^{ijR}(t')\braket{\partial_R R_{0i0j}(t')P_{\mu\nu}{}^{\alpha\beta}h_{\alpha\beta}(x)}\,.
\end{equation}
From this, we obtain\footnote{Throughout the chapter, we use $\omega \equiv k^0$.}:
\begin{equation}
\braket{\bar{h}_{00}(x)} = \mathcal{F}^{(I)}_0 [k_i k_j]\,, \qquad
\braket{\bar{h}_{0k}(x)} = \mathcal{F}^{(I)}_0 [-\omega k_j \delta_{ik}]\,, \qquad
\braket{\bar{h}_{kl}(x)} = \mathcal{F}^{(I)}_0 [\omega^2\delta_{ik}\delta_{jl}]\,, \label{hijIij}
\end{equation}
where we have defined
\begin{equation}
\mathcal{F}^{(I)}_0[f] = 
- 16\pi G_N (-i)^r c_r^{(I)} \int\frac{d\omega}{2\pi} I^{ijR}(\omega)\int_\K\frac{e^{-i\omega t+i\K\cdot\X}}{\K^2-\omega^2} k_R \times f(\omega,\K)\,,
\end{equation}
for arbitrary functions $f =f(\omega,\K,\Q)$.

Then, we immeditaly have:
\begin{align}
    \partial_\mu \braket{\bar{h}^{\mu 0}} = \mathcal{F}^{(I)}_0 [i\omega k_i k_j-i\omega k_i k_j] = 0 \quad \text{and} \quad
    \partial_\mu \braket{\bar{h}^{\mu l}} = \mathcal{F}^{(I)}_0 [i\omega^2 k_i\delta_{jl}-i\omega^2 k_i \delta_{jl}]=0\,.    
\end{align}

\subsubsection{Magnetic multipole moments}

As for the magnetic-type multipole moments, we have similarly:
\begin{equation}
\braket{\bar{h}_{\mu\nu}(x)} = -i \frac{c_r^{(J)}}{2} \int dt'\,J^{b|iRa}(t')\braket{\partial_R R_{0iab}(t')P_{\mu\nu}{}^{\alpha\beta}h_{\alpha\beta}(x)}\,,
\end{equation}
from which we obtain
\begin{align}
\braket{\bar{h}_{00}(x)} &= 0\,, \\
\braket{\bar{h}_{0k}(x)} &= -8\pi G_N (-i)^r c_r^{(J)} \int \frac{d\omega}{2\pi} J^{b|iRa}(\omega) \int_\K \frac{e^{-i \omega t +i \K \cdot \Q}}{\K^2-\omega^2}\times k_R k_i k_a \delta_{bk}\,, \\
\braket{\bar{h}_{kl}(x)} &= -8\pi G_N (-i)^r c_r^{(J)} \int \frac{d\omega}{2\pi} J^{b|iRa}(\omega) \int_\K \frac{e^{-i \omega t +i \K \cdot \Q}}{\K^2-\omega^2}\times k_R \omega k_b 
(\delta_{ik}\delta_{al} + \delta_{il}\delta_{ak})\,.
\end{align}
And, similarly to the electric case, the Ward identity can be easily checked, holding thanks to the anti-symmetry of $J^{b|iRa}$ in the indices $a,b$.

The results of this section show the consistency of leading-order effects with the Lorentz gauge condition, which, as we have seen in the previous chapter, is completely equivalent to the linearized theory of gravity. The verification of this condition for higher-order effects will turn out to be relevant for the consistency of the solutions within Einstein's general relativity. 

Before studying the nonlinear effects corresponding to all the tail processes, in the following sections, we define the important quantity of the gravitational scattering amplitude $\mathcal{A}_{\mu\nu}$, and derive general properties, important for the continuity of the chapter.


\section{The gravitational scattering amplitude}

\subsection{The gravitational field from scattering amplitudes}\label{gravscatamplsec}


In the general case when interactions are considered, the field $h_{\mu\nu}$ will have the generic form
\begin{equation}\label{hmunu}
\braket{h_{\mu\nu}(x)} = \int_\K \frac{d\omega}{2\pi} \frac{e^{-i\omega t+i\K\cdot\X}}{\K^2-(\omega+i\varepsilon)^2}\times i\mathcal{A}_{\mu\nu}(\omega,\K)\,,
\end{equation}
which can be used to define the quantity $i \mathcal{A}_{\mu\nu}$, that we shall call the \textit{gravitational scattering amplitude} of a given process.
In this expression, the gravitational field is described causally with the use of the retarded Green's funtion.
In particular, writing this equation in terms of the trace-reversed quantities $\bar{h}_{\mu\nu}$ and $\bar{\mathcal{A}}_{\mu\nu}$, and recalling the defition for the retarded Green's function 
\begin{equation}
G_{R}(\omega,\K) = \frac{1}{\K^2-(\omega+i\varepsilon)^2}\,,
\end{equation}
we have the following expression in direct space, derived using the convolution theorem, which simply corresponds to the general solution of a wave equation with retarded boundary condition and source term $T_{\mu\nu}(x)$: 
\begin{equation}\label{hbarasTmunu}
\braket{\bar{h}_{\mu\nu}(x)} = -16G_N\pi\int d^{d+1}x' G_{R}(t-t',\X-\X') T_{\mu\nu}(x')\,.
\end{equation}
In this construction, we identify the energy-momentum tensor of the gravitational field $T_{\mu\nu}(x)$ with the inverse Fourier transform of $i \bar{\mathcal{A}}_{\mu\nu}(\omega,\K)$, 
\begin{equation}
\frac{1}{2\Lambda^2} T_{\mu\nu}(x) \qquad
\Leftrightarrow \qquad
-i\bar{\mathcal{A}}_{\mu\nu}(\omega,\K)\,.
\end{equation}

As it will become clear along the chapter, the gravitational scattering amplitude $i\mathcal{A}_{\mu\nu}$ plays a fundamental role in all the analysis of higher-order processes. Investigation of this quantity for a few processes will show us that some of them suffer from classical anomalies.

\subsection{The far field approximation}

In this subsection, we derive in great detail an important relation that shows how one can compute the gravitational waveform for a given process in the regime of long distances from the source, using the corresponding gravitational scattering amplitudes on-shell. 

We start by plugging in the expression for the retarded Green's function in direct space into Eq.~\eqref{hbarasTmunu} and performing the integration in $t$:
\begin{align}
\braket{\bar{h}_{\mu\nu}(x)} &= -16G_N\pi\int d^{d+1}x' G_{R}(t-t',\X-\X') T_{\mu\nu}(x')\nonumber\\
&= -16G_N\pi\int d^{d+1}x' \frac{-1}{4\pi|\X-\X'|} \delta((t-t')-|\X-\X'|) T_{\mu\nu}(x')\nonumber\\
&= 4G_N\int d^{d}\X' \frac{T_{\mu\nu}(t-|\X-\X'|,\X')}{|\X-\X'|}\,.
\end{align}
We then take the far field approximation, defined by taking the limit $|\X|\gg d$, with $d \gtrsim |\X'|$ being the typical size of the source. In this case we have that $|\X-\X'| = r- \X'\cdot{\bf n} + \mathcal{O}(d^2/r)$, where $r\equiv |\X|$ and ${\bf n}\equiv \X/|\X|$, and hence, the equation above becomes
\begin{align}
\braket{\bar{h}_{\mu\nu}(x)} &=\frac{4G_N}{r}\int d^d\X'\,T_{\mu\nu}(t-r+\X'\cdot{\bf n},\X') + \mathcal{O}\left(\frac{1}{r^2}\right)\nonumber\\
&=\frac{4G_N}{r}\int d^d\X'\int\frac{d\omega}{2\pi}\,T_{\mu\nu}(\omega,\X') e^{-i\omega(t-r+\X'\cdot{\bf n})} + \mathcal{O}\left(\frac{1}{r^2}\right) \nonumber\\
&=\frac{4G_N}{r}\int\frac{d\omega}{2\pi}e^{-i\omega(t-r)} \left[ \int d^d\X'\,T_{\mu\nu}(\omega,\X') e^{-i\omega(\X'\cdot{\bf n})} \right] + \mathcal{O}\left(\frac{1}{r^2}\right)\,.
\end{align}
We immediatelly notice that the term enclosed by square brackets is just the Fourier transform of $T_{\mu\nu}$, namely the gravitational scattering amplitude $i\bar{\mathcal{A}}_{\mu\nu}$, evaluated on-shell, at $(\omega,\omega{\bf n})$. Thus, the distant field expression for $\braket{h_{\mu\nu}(x)}$ is given by the relation
\begin{equation}
    \braket{\bar{h}_{\mu\nu}(x)} = -\frac{1}{4\pi r}\int^{+\infty}_{-\infty}\frac{d\omega}{2\pi}\,i\bar{\mathcal{A}}_{\mu\nu}(\omega,\omega{\bf n}) e^{-i\omega t_{\rm ret}}\,.
\end{equation}

\subsection{Relating emission and gravitational scattering amplitudes}


In the previous chapter, we have introduced the emission amplitude $i\mathcal{A}$, which is obtained directly by the Feynman rules generated from the effective action \eqref{sourcelag} and Einstein-Hilbert action, and by attaching the polarizations $\sigma_{ij}, A_i, \phi$ to the external legs of the diagram. Here, we relate this amplitude, which will be called by the name of ``emission amplitude," to the gravitational scattering amplitude defined in Sec.~\ref{gravscatamplsec}.

Consider a generic gravitational-wave process, with off-shell amplitude containing not only the radiative degrees of freedom given by $\sigma_{ij}$ but also the longitudinal ones  $A_i$ and $\phi$: 
\begin{equation}\label{genericemissA}
  i\mathcal{A} = \alpha^{cd}_\sigma \sigma^{*}_{cd} + \alpha^c_A A^{*}_c + \alpha_\phi \phi^{*}\,.
\end{equation}
In this expression, all the quantities have the same argument, $(\omega,\K)$: the emission amplitude $i\mathcal{A}$, coefficients $\alpha$'s, and polarizations $\sigma_{ij},A_i, \phi$.

The way in which the two types of amplitudes are related to each other is by means of
\begin{equation}\label{emissvsscattering}
i\mathcal{A}_{\mu\nu}(\omega,\K) = \braket{h_{\mu\nu}\times i\mathcal{A}}^\sim\,,
\end{equation}
where we use the notation $\braket{\cdots}^\sim$ to refer to the coefficients of the graviton propagator with the term $1/(\K^2-\omega^2)$ removed, namely 
\begin{equation}\label{brakettilde}
\left\{
\begin{array}{rcl}
\braket{\phi \phi^*}^\sim &=& -\dfrac{i}{2c_d}\,, \\
\braket{A_a A_b^*}^\sim &=& \dfrac{i}{2} \delta_{ab}\,,\\ 
\braket{\sigma_{ab} \sigma_{cd}^*}^\sim &=& -\dfrac{i}{2}\left( {\delta}_{ac}{\delta}_{bd}+{\delta}_{ad}{\delta}_{bc}-\dfrac{2}{d-2}{\delta}_{ab}{\delta}_{cd}\right)\,,\\
\end{array}
\right.
\end{equation}
and using $h_{00}=-2\phi/\Lambda$, $h_{0i}=A_i/\Lambda$, and $h_{ij}=[\sigma_{ij}+(2-c_d)\phi \delta_{ij}]/\Lambda$.

We can derive explicit expressions for the components of $i\mathcal{A}_{\mu\nu}$ in terms of the $\alpha's$ of a given process, using Eqs.~\eqref{genericemissA}, \eqref{emissvsscattering}, and \eqref{brakettilde}. By doing this, we  derive
\begin{align}
\mathcal{A}_{00} &= \frac{1}{c_d \Lambda}\alpha_\phi\,, \\
\mathcal{A}_{0c} &= \frac{1}{2\Lambda} \alpha^c_A\,, \\
\mathcal{A}_{kl} &= -\frac{1}{2\Lambda} \left( \delta_{kc} \delta_{ld} + \delta_{kd} \delta_{lc} - \frac{2}{d-2} \delta_{kl} \delta_{cd} \right) \alpha^{cd}_\sigma + \frac{1}{2(d-1)\Lambda} \alpha_\phi \delta_{kl}\,.
\end{align}
From this, we can build the trace $\mathcal{A}^\mu{}_\mu$ in order to compute the trave-reversed version of $i\mathcal{A}_{\mu\nu}$,
\begin{equation}
\mathcal{A}^\mu{}_\mu = \frac{2}{(d-2)\Lambda}\delta_{cd}\alpha^{cd}_\sigma + \frac{1}{(d-1)\Lambda} \alpha_\phi\,.
\end{equation}

The trace-reversed gravitational scattering amplitude $\bar{\mathcal{A}}_{\mu\nu}$ is built out of the components of $\mathcal{A}_{\mu\nu}$ by means of the projector $P_{\mu\nu}{}^{\alpha\beta}$ defined in Eq.~\eqref{projetctorTT}, through $\bar{\mathcal{A}}_{\mu\nu} = P_{\mu\nu}{}^{\alpha\beta} \mathcal{A}_{\mu\nu}$,
and leads to the following simple expressions given in terms of the $\alpha$ coefficients:	
\begin{align}
\bar{\mathcal{A}}_{00} &= \frac{1}{2\Lambda} \alpha_\phi +\frac{1}{(d-2)\Lambda} \delta_{cd} \alpha^{cd}_\sigma\,, \label{AbarwrtalphaS}\\
\bar{\mathcal{A}}_{0c} &= \frac{1}{2\Lambda} \alpha^c_A\,, \label{AbarwrtalphaA}\\
\bar{\mathcal{A}}_{kl} &= -\frac{1}{2\Lambda} (\delta_{kc} \delta_{ld} + \delta_{kd} \delta_{lc}) \alpha^{cd}_\sigma \,. \label{AbarwrtalphaP} 
\end{align}
Note that we can rewrite the emission amplitude $i\mathcal{A}$ in terms of $i\bar{\mathcal{A}}_{\mu\nu}$ by inverting these relations to give the $\alpha$'s in terms of the components of $\bar{\mathcal{A}}_{\mu\nu}$, in which case we obtain:
\begin{align}
  i\mathcal{A} &= \Lambda \left[-\bar{\mathcal{A}}_{cd} \sigma^{*}_{cd} + 2    \bar{\mathcal{A}}_{0c}  A^{*}_c + 2  \left( \bar{\mathcal{A}}_{00} + \frac{1}{d-2}\delta^{kl}\bar{\mathcal{A}}_{kl}  \right) \phi^{*} \right]\nonumber\\
               &=-\Lambda^2 \left( \bar{\mathcal{A}}_{cd} h_{cd}^{*}  - 2 \bar{\mathcal{A}}_{0c} h_{0c}^{*} + \bar{\mathcal{A}}_{00}  h_{00}^{*} \right)\,,
\end{align}
and, hence, 
\begin{equation}
  i\mathcal{A} =-\Lambda^2 \bar{\mathcal{A}}^{\mu\nu} h_{\mu\nu}^{*}\,.
\end{equation}

In the notation introduced above, and in terms of the emission amplitude $i\mathcal{A}$, we have
\begin{equation}
\braket{h_{\mu\nu}(x)} = \int_\K \frac{d\omega}{2\pi} \frac{e^{-i\omega t+i\K\cdot\X}}{\K^2-(\omega+i\varepsilon)^2}\times \braket{\tilde{h}_{\mu\nu}\times i\mathcal{A}}^\sim\,,
\end{equation}
for the general gravitational field, and
\begin{equation}
\braket{h_{\mu\nu}(t,\X)} = - \frac{1}{4\pi |\X|} \int \frac{d\omega}{2\pi} e^{-i\omega t_{ret}} \braket{\tilde{h}_{\mu\nu}\times i\mathcal{A}}^\sim\,,
\end{equation}
for the gravitational field in the far field approximation.

\subsection{The Ward identity}

It follows directly from the trace-reversed version of \eqref{hmunu} that
\begin{equation}
\partial_\mu\braket{\bar{h}^{\mu\nu}(x)} = -\int_\K \frac{d\omega}{2\pi} \frac{e^{i k\cdot x}}{\K^2-(\omega+i\varepsilon)^2}\times k_\mu\mathcal{\bar{A}}^{\mu\nu}(\omega,\K)\,.
\end{equation}
Hence, we immediately see that, if the condition $k^\mu \bar{\mathcal{A}}_{\mu\nu} = 0$ is satisfied, we have
\begin{equation}
k_\mu\mathcal{\bar{A}}^{\mu\nu}(\omega,\K) = 0 \qquad \Longrightarrow \qquad \partial_\mu\braket{\bar{h}^{\mu\nu}(x)} = 0 \qquad \text{and} \qquad  
\partial_\mu T^{\mu\nu}(x) = 0   \,.
\end{equation}

The condition $k^\mu \bar{\mathcal{A}}_{\mu\nu} = 0$ is called the Ward identity and will play the most fundamental role in the discussion throughout the chapter. As we can see in the relation above, the Ward identity is intimately connected to the Lorentz gauge condition, as well as to the conservation of the energy-momentum tensor. In particular, for any gravitational-wave process, this identity should always hold on-shell, and hence, give $\partial_\mu\braket{\bar{h}^{\mu\nu}(x)} = 0$ and $\partial_\mu T^{\mu\nu}(x) = 0$ in this case.

An expression for $k^\mu \bar{\mathcal{A}}_{\mu \nu}$ given in terms of the $\alpha$'s previously defined can be easily obtained:
\begin{align}
(\nu = 0)& \qquad \qquad k^\mu \bar{\mathcal{A}}_{\mu 0} = \frac{\omega}{\Lambda} \left[ \frac12 \alpha_\phi + \frac{1}{(d-2)}\delta_{cd}\alpha^{cd}_\sigma \right] + \frac{1}{2\Lambda} k_c \alpha^c_A\,,\label{nu0}\\
(\nu = l)& \qquad \qquad k^\mu \bar{\mathcal{A}}_{\mu l} = -\frac{1}{2\Lambda} (k_c \delta_{ld} + k_d \delta_{lc}) \alpha_\sigma^{cd} + \frac{\omega}{2\Lambda} \alpha^l_A \label{num}\,.
\end{align}


Notice that, we can easily obtain the gravitational scattering amplitudes $i\bar{\mathcal{A}}_{\mu\nu}$ for the leading-order processes of Sec.~\ref{secgravfield}, and directly check whether the Ward identity holds in this case. It is easy to see that the validity of the Lorentz condition in this case, explicitly evaluated in the same section, implies necessarily  $k^\mu \bar{\mathcal{A}}_{\mu\nu}=0$.


\section{The simple mass tail for arbitrary multipoles}



\begin{figure}
\centering
\begin{equation}
\braket{h_{\mu\nu}(x)}_{{\rm tail}} = \,\,
      \begin{tikzpicture}[baseline={([yshift=-.5ex]current bounding box.center)},scale=1]
      \draw [black, thick] (0,0) -- (3,0);
      \draw [black, thick] (0,-0.07) -- (3,-0.07);
      \draw[decorate, decoration={snake},segment length=6.5pt,segment amplitude=1.2pt, line width=1pt] (0.6,0) -- (1.5,1) node [pos=0.6] {};
      \draw[dashed, thick,line width=1pt] (2.37,0.03) -- (1.5,1) node [pos=0.56] {};
      \draw[decorate, decoration={snake},segment length=6.5pt,segment amplitude=1.2pt, line width=1pt,shorten >= 0.8pt] (1.5,1.01) -- (1.5,1.84) node [pos=0.5] {};
      \filldraw[black] (0.6,-0.035) circle (2pt) node[anchor=north] {$I^{ijR},J^{b|iRa}$};
      \filldraw[black] (2.4,-0.035) circle (2pt) node[anchor=north] {$E$};
      \filldraw[black] (1.5,1) circle (2pt) node[anchor=north] {};
      \filldraw[black] (1.5,1.89) circle (2pt) node[anchor=west] {};
      \node at (1.45,1.89) [anchor=east] {$x$};
      \node at (1.55,1.4) [anchor=west] {$\mu\nu$};
    \end{tikzpicture}\nonumber
\end{equation}
\vspace{-0.7cm}
\caption{Diagram representing the gravitational-wave process called the simple tail. This process presents two worldline vertices, one being either an electric- or magnetic-type multipole moment, while the other is a coupling to the conserved energy of the system. This diagram also presents a three-graviton vertex, following from the Einstein-Hilbert action.}
\label{taildiagram}
\end{figure}
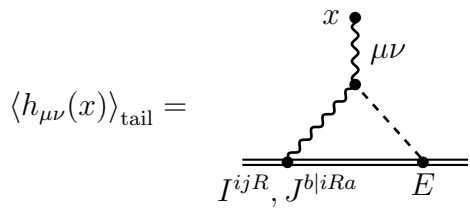

In this section, we study the emission properties of the so-called tail effect, a gravitational-wave process that involves the interaction of a multipole moment with the conserved energy of the system. This process is also usually called simple tail or mass tail. In this case, we compute the emission amplitude in its full generality for both electric and quadrupole cases, from which we derive the gravitational scattering amplitude and study the Ward identity. As expected, no amplitude presents anomalies, and thus, the process is physically consistent.
The emission process by simple tails is represented in Fig.~\ref{taildiagram}.

The general expression for arbitrary multipole moments of the electric type reads
\begin{equation}\label{tailelecbeta}
i\mathcal{A}^{(I)}_{E-\rm tail} = \frac{(-i)^{r+1} c_r^{(I)}E}{\Lambda^3} I^{ijR}(\omega) \int_\Q \frac{(k+q)_{R}}{[(\K+\Q)^2-\omega^2]\Q^2}\times \left( \beta_{\sigma}^{ab} \sigma^{*}_{ab} + \beta_{A}^a A^{*}_a + \beta_{\phi} \phi^{*}\right)\,,
\end{equation}
where the coefficients $\beta_\sigma^{ab}$, $\beta_A^a$, and $\beta_\phi$ are given by expressions \eqref{betastail}, \eqref{betaatail}, and \eqref{betaptail} of Appendix \ref{appward}.
\footnote{The coefficients $\beta$ should, of course, carry the indices $ij$, [$\beta^{ab}_\sigma \rightarrow (\beta^{ab}_\sigma)_{ij}$, $\beta^{a}_A \rightarrow (\beta^{a}_A)_{ij}$ and $\beta_\phi \rightarrow (\beta_\phi)_{ij}$], but we have kept them hidden here to make the notation lighter. }

Comparing Eqs.~\eqref{genericemissA} and \eqref{tailelecbeta}, we can easily see how the definitions of the $\alpha$'s, introduced before, and $\beta$'s above are related:
\begin{equation}
\alpha = \frac{(-i)^{r+1} c_r^{(I)}E}{\Lambda^2} I^{ijR}(\omega) \int_\Q \frac{(k+q)_{R}}{[(\K+\Q)^2-\omega^2]\Q^2}\times  \beta\,,
\end{equation}
from which we can directly use Eqs.~\eqref{AbarwrtalphaS}, \eqref{AbarwrtalphaA}, \eqref{AbarwrtalphaP} to derive the expressions for the scattering amplitudes for the tail processes of electric parity. 
The explict expressions for the components $\bar{\mathcal{A}}_{00}$, $\bar{\mathcal{A}}_{0k}$, and $\bar{\mathcal{A}}_{kl}$ are shown in Appendix \ref{appward}, in Eqs.~\eqref{A00taileapp}, \eqref{A0ktaileapp}, and \eqref{Akltaileapp}.
Then, with these components in hands, we can test whether the condition $k^\mu\bar{\mathcal{A}}_{\mu\nu} = 0$ is satisfied. As we showed before, this condition is equivalent to the Lorentz condition, $\partial^\mu\braket{\bar{h}_{\mu\nu}}=0$, and to the conservation of the energy-momentum tensor, $\partial^\mu T_{\mu\nu}(x)=0$.
In particular, neglecting terms $(\omega^2 - \K^2)$, which vanish on-shell and will not play any role in our discussion, as it will be properly discussed in Sec.~\ref{onshelness}, it is easy to show that we indeed obtain $k^\mu \bar{\mathcal{A}}_{\mu 0} = 0$ and $k^\mu \bar{\mathcal{A}}_{\mu l}=0$. See \ref{appwardtestingtaile} for the details.

The analysis for the magnetric case follows similarly and is reported in \ref{appwardtestingtailm}. In particular, like in the electric case, it satisfies $k^\mu \bar{\mathcal{A}}_{\mu \nu}=0$ for arbitrary magnetic multipole moments.


\section{The angular momentum failed tail for arbitrary multipoles}



Now, let us consider the case of the angular momentum failed tail, that we shall call, for simplicity, the $J$-failed tail. 
The diagram for this process presents the same topology as for the tails in Fig.~\ref{taildiagram}, but differs by having an angular momentum coupling in place of the energy.
As we have alread antecipated in the introduction of the chapter, we will find a Ward-violating scattering amplitude in the cases where the interaction involves a quadrupole for either of the types, electric or magnetic. We start our analysis by considering the electric case.

\subsection{$J$-failed tail for the electric multipole moments}

Like in the case of the simple tail, for arbitrary multipole moments of the electric type, we have
\begin{equation}\label{ALarbitrary}
i\mathcal{A}^{(I)}_{J-\rm tail} = (-i)^r\frac{c_r^{(I)}}{2\Lambda^3}J^{b|a} I^{ijR}(\omega) \int_\Q \frac{q_a(k+q)_R}{[(\K+\Q)^2-\omega^2]\Q^2}\times
\left( \beta_{\sigma}^{cd} \sigma^{*}_{cd} + \beta_{A}^c A^{*}_c + \beta_{\phi} \phi^{*}\right)\,,
\end{equation}
with coefficients $\beta_{\sigma}^{cd}, \beta_{A}^{c}$, and $\beta_{\phi}$ given by Eqs.~\eqref{betasigma}, \eqref{betaA}, and \eqref{betaphi}.
Hence, from the above expression, we can immediately identify the corresponding $\alpha_{\sigma}^{cd}$, $\alpha_{A}^{c}$, and $\alpha_{\phi}$ as in Eq.~\eqref{genericemissA}, and, hence, use the expressions for the scattering amplitude components $i \bar{\mathcal{A}}_{\mu\nu}$ given in terms of the $\alpha$'s,  Eqs.~\eqref{AbarwrtalphaS}, \eqref{AbarwrtalphaA}, \eqref{AbarwrtalphaP}. Derivation is simple, with result presented in \ref{jtaileapp}.

Now, we analyze this result in the same way we did in the cases above for the mass tail.
In particular, for the $\nu = 0$ component, we are left with just a term with the following behavior: 
\begin{align}
k^\mu \bar{\mathcal{A}}_{\mu 0} \sim J^{i|a} I^{ijR}(\omega) \int_\Q \frac{(q-k)_a q_j q_R}{(\Q^2-\omega^2)}. 
\end{align}
The integral in $\Q$ on the right-hand side of this expression has two terms, proportional to $\delta_{ajR}$ and $\delta_{jR}$, that lead to zero when contracted to $J^{i|a} I^{ijR}$, irrespective of the value for $r$. For $\nu=l$, the complete remaining expression, after simplification, is given by
\begin{equation}
	k^\mu \mathcal{\bar{A}}_{\mu l} = (-i)^{r+1}\frac{c_r^{(I)}}{2\Lambda^4} \left( \frac{i\omega}{4} \right)  
\bigg[ k_a \omega^2 J^{i|a} I^{iRl}(\omega) \int_\Q \frac{q_R}{(\Q^2-\omega^2)}  \bigg]\,.
\end{equation}
Hence, we notice that, since the integral in $\Q$ is proportional to $\delta_R$, this result vanishes on account of the tracelessness of $I^{iRl}$, unless $r=0$, in which case we have:
\begin{align}
	k^\mu \mathcal{\bar{A}}_{\mu l}\big|_{r=0} &= \frac{1}{16\Lambda^4}   
k_a \omega^3 J^{i|a} I^{il}(\omega) \int_\Q \frac{1}{(\Q^2-\omega^2)} \nonumber\\
	&= 16\pi i G_N^2   k_a \omega^4 J^{i|a} I^{il}(\omega) \,,
\end{align}
where for the integral in $\Q$ we used \ref{I0omega}.

In summary, we have shown that, while the Ward identity is satisfied for all electric multipole moments with $r \ge 1$, is is not for the quadrupole. For the latter, we have
\begin{align}
	k^\mu \mathcal{\bar{A}}_{\mu 0} \big|_{r=0} &= 0\,, \\
	k^\mu \mathcal{\bar{A}}_{\mu l} \big|_{r=0} &= 16\pi i G_N^2   k_a \omega^4 J^{i|a} I^{il}(\omega) \,.
\end{align}

\subsubsection{A direct consequence of the Ward-violating amplitude}

Waveform solutions presenting a violation of the Ward identity are not proper solutions of general relativity since it also represents a violation of the Lorentz gauge condition. Hence, one cannot simply apply the TT projector $\Lambda_{ij,kl}$ to this solution since only solutions consistent with the Lorentz gauge condition can be put in this gauge. In Sec.~\eqref{seceinsteineq}, we will discuss the consistency behind the anomalous solution, as well as we will also provide a fix for it, and, hence, bring the solution to a valid solution of general relativity.

\subsection{$J$-failed tail for the magnetic multipole moments}

For the magnetic case, the angular momentum failed tail can be analyzed in the same way we did above for the electric case. Below we present the solutions one obtains in this case:
\begin{align}\label{offemJL}
i\mathcal{A}^{(J)}_{J-\rm tail} = (-i)^r \frac{c_r^{(J)}}{8\Lambda^3} J^{l|k} J^{b|iRa}(\omega) \int_\Q \frac{(k+q)_R (k+q)_b }{[(\K+\Q)^2 - \omega^2]\Q^2}  \times q_k 
\left( \beta_{\sigma}^{cd} \sigma^{*}_{cd}  + \beta_A^c A^{*}_c  + \beta_\phi \phi^{*} \right)\,,
\end{align}
with $\beta$'s given by \eqref{jtailmsigmaapp}, \eqref{jtailmaapp}, and \eqref{jtailmphiapp}, and components of $\bar{\mathcal{A}}_{\mu\nu}$ by \ref{A00jtailmapp}, \ref{A0kjtailmapp}, and \ref{Akljtailmapp}.
Hence, analyzing the Ward identity $k^\mu\bar{\mathcal{A}}_{\mu\nu} =0$ (See Sec.~\ref{jtailmapp}), we end up with the same qualitative behavior we found in the electric case, namely that no anomalies are present for $r\ge 1$, while the Ward identity is violated in the quadrupole case, in which case we have the following result:
\begin{align}
k^\mu\bar{\mathcal{A}}_{\mu 0} &=0\,, \\
k^\mu \bar{\mathcal{A}}_{\mu d} &= -\frac{128}{45}i G_N^2 \pi J^{l|k} J^{b|ia}(\omega) (\delta_{al}\delta_{id} + \delta_{ad} \delta_{il}) \delta_{bk} \omega^5\,.
\end{align}

In the following section, we discuss the consistency of these anomalous solutions within our framework. We also work out the fix to it by first going back to the Einstein field equations and checking that these are not indeed complete solutions to the problem. Nevertheless, the full solution can be obtained from it by adding a fixing term.


\section{The Einstein’s equation in perturbation theory}\label{seceinsteineq}


The main goal of this section is to show that, while the anomalous solutions obtained for the angular momentum failed tail are solutions of the equations of motion derived from the Einstein-Hilbert action plus gauge fixing, it is not a complete solution to our problem. We start by deriving such equations of motion with the aid of the Mathematica package \textit{xPert} \cite{xpert}, which is a package developed to perform perturbative calculations, and by expanding the metric as 
\begin{equation}
g_{\mu\nu} = \eta_{\mu\nu} + h_{\mu\nu} + \frac 12 h^{(2)}_{\mu\nu}\,,
\end{equation}
to accommodate the Kaluza-Klein decomposition. By doing so, we obtain the Einstein-Hilbert action expanded to third order in the fields \eqref{SEHGexpandedxxx} (See Appendix \ref{eomapp} for details), represented schematically by 
\begin{equation}\label{SEHGexpanded}
S_{\rm EH+GF} \sim \Lambda^2 \int d^4 x\, (h\partial^2 h + h^2\partial^2 h)\,.
\end{equation}

Besides performing perturbative expansions of general relativity quantities, the package xPert can also be used to perform variations of an action. Then, we use it again to obtain the equations of motion for the action in Eq.~\eqref{SEHGexpanded}, in which occasion we obtain Eq.~\eqref{boxheqN}, which, in turn, can be written in the following compact form, using a notation by Blanchet \cite{blanchetlivrev}:
\begin{equation}\label{heqNhh}
\Box \bar{h}_{\mu\nu} = N_{\mu\nu}[h,h]\,.
\end{equation}

In terms of perturbative expansions, this equation should be understood in the following sense: the $h_{\mu\nu}$ on the left-hand side of this equation represents processes of order $G_N^2$, while the two $h$ on the right-hand side are each of order $G_N$. Hence, in order to test whether a given process at order $G_N^2$ is solution to this equation, one should plug it in the left-hand side and compare it to the right-hand side computed from the lower order processes (at order $G_N$).

\subsubsection{Checking the $J$-failed tail solution}

For the electric $J$-failed tail, the $h_{\mu\nu}$ at order $G_N^2$ that enters the LHS of Eq.~\eqref{heqNhh} is the one constructed out of components given in Eqs.~\eqref{LtailAbar00}, \eqref{LtailAbar0k}, and \eqref{LtailAbarkl}. As for the lower-order processes (i.e., of order $G_N$) that enter the right-hand side of Eq.~\eqref{heqNhh}, we consider the $J$ coupling given in Eq.~\eqref{Lcoupling} and the electric quadrupole's in Eqs.~\eqref{hijIij}, with $r = 0$.

Checking this is easy but tedious. In Appendix \ref{eomapp} we present the results, which turn out to have matching expressions, either by computing $\Box \bar{h}_{\mu\nu}$ using the electric $J$-failed tail solution or $N_{\mu\nu}$ from the lower-order solutions. Thus, we have explicitly checked that, for our solution of the electric $J$-failed tail, we indeed have $\Box \bar{h}_{\mu\nu} = N_{\mu\nu}[h,h]$. The same also holds true for the magnetic-type solution.

As we will see in the following section, although we have obtained an expression of $\bar{h}_{\mu\nu}$ for the $J$-failed tail that provides a solution for the Einstein's equation with gauge fixing term, it is not a solution fully consistent with general relativity. This happens because not only this solution has to satisfy the Einstein's equation but also the gauge condition equation.

\subsection{Equations of motion for the full problem}


The problem of solving perturbatively the Einstein field equations can be posed in the following format: it can be written separately as the two equations 
\begin{equation}\label{eineqsss}
\Box \bar{h}_{\mu\nu} = \Lambda_{\mu\nu} \qquad\qquad \text{and} \qquad\qquad \partial^\mu \bar{h}_{\mu\nu} = 0\,. 
\end{equation}
Here, $\Lambda_{\mu\nu}$ represents the energy-momentum tensor of the source plus curvature, including the terms coming from the gauge fixing term, that should be truncated to the desired order in field perturbation. At this level, these equations are considered independent, and the problem is only fully solved once both equations are simultaneously satisfied.

In practice, what our field theory approach is effectively solving is the first of these equations, subject to the retarded boundary condition. Notice, however, that, in general, the solution will not satisfy the second equation automatically, as it turned out to happen in the quadrupolar $J$-failed tail emission of both electric and magnetic parities.

Nevertheless, once we have obtained a particular solution for the first of these equations, $\Box\bar{h}_{\mu\nu} = \Lambda_{\mu\nu}$, we can still look for a solution of the homogeneous wave solution, $\Box\bar{h}_{\mu\nu} = 0$, that cancels the divergence in $\partial^\mu\bar{h}_{\mu\nu}$\,. Below, we systematize these statements by closely following Blanchet's Ref.~\cite{blanchetlivrev}.

\subsection{General Framework}

The complete solution to our problem consists in obtaining a particular solution of the Einstein equation $\Box\bar{h}_{\mu\nu} = \Lambda_{\mu\nu}$, denoted here by $\bar{h}^{p}_{\mu\nu}$, in which case
\begin{equation}
\Box\bar{h}^{p}_{\mu\nu} = \Lambda_{\mu\nu}\,.
\end{equation}
This is done through our ordinary way of computing amplitudes. Nevertheless, in general, this particular solution will not satisfy the second equation of \eqref{eineqsss}, as 
\begin{equation}
\partial^\mu \bar{h}^{p}_{\mu\nu} = h_{\nu} \neq 0\,,
\end{equation} 
for some function $h_{\nu}$.
In this case, the complete solution to our problem must include a homogeneous solution, $\bar{h}^{h}_{\mu\nu}$, satisfying
\begin{equation}
\Box\bar{h}^{h}_{\mu\nu} = 0 \qquad\qquad \text{and} \qquad\qquad \partial^{\mu}\bar{h}^{h}_{\mu\nu} = -h_{\nu}\,. 
\end{equation} 
A solution for such $\bar{h}^h_{\mu\nu}$ exists whenever $\Box h_\nu =0$, as it will be shown in the next subsection, occasion where we will provide an explicit fix for this problem.
Then, the full solution reads
\begin{equation}
\bar{h}_{\mu\nu} = \bar{h}^{p}_{\mu\nu} + \bar{h}^{h}_{\mu\nu}\,. 
\end{equation}
With this, the equation $\Box\bar{h}_{\mu\nu} = \Box\bar{h}^{p}_{\mu\nu} + \Box\bar{h}^{h}_{\mu\nu} = \Lambda_{\mu\nu}$ is preserved, while the gauge condition is recovered:
\begin{equation}
\partial^{\mu} \bar{h}_{\mu\nu} = \partial^{\mu} \bar{h}^{p}_{\mu\nu} + \partial^{\mu} \bar{h}^{h}_{\mu\nu} = h_\nu - h_\nu = 0\,. 
\end{equation}

\subsection{The homogeneous solution}\label{buildinga}

It follows from the divergence of the source term $\Lambda_{\mu\nu}$, which can be checked explicitly, that the expression $h_\nu = \partial^\mu\bar{h}^{p}_{\mu\nu}$, together with the wave equation $\Box\bar{h}_{\mu\nu} = \Lambda_{\mu\nu}$, yields
\begin{equation}\label{divergenceward}
\Box\left( \partial^\mu\bar{h}^p_{\mu\nu} \right) = 0\,.
\end{equation} 
(Alternatively, for the cases at hand, one can check directly that this equation holds) 
In other words, what Eq.~\eqref{divergenceward} is telling us is simply that the divergence $h_\nu = \partial^\mu\bar{h}^{p}_{\mu\nu}$ is solution for the source-free wave equation (with flat d'Alembertian). On the other hand, the most general solution for this equation can be parametrized by four symmetric-tracefree (STF) $SO(3)$ tensors, being functions of the retarded time $u=t-r$. Following the notation of Ref.~\cite{blanchetlivrev}, we call these tensors $N_L(u)$, $P_L(u)$, $Q_L(u)$, and $R_L(u)$. Then, the general solution reads (See Eqs.~(47a) and (47b) of Ref.~\cite{blanchetlivrev})
\begin{align}
h^0 &= \sum_{l=0}^{\infty} \partial_L \left[ r^{-1} N_L(u) \right]\,, \label{h0general} \\
h^i &= \sum_{l=0}^{\infty} \partial_{iL} \left[ r^{-1} P_L(u) \right] + \sum_{l=1}^{\infty} 
\left\{
\partial_{L-1}\left[ r^{-1} Q_{iL-1}(u) \right] + \epsilon_{iab} \partial_{aL-1} \left[ r^{-1} R_{bL_1}(u)
\right]
\right\}\,.\label{higeneral}
\end{align}

The statement that this is the most general solution to the homogenous wave equation follows from the fact that, for any function $F_L (t-r)$, we have automatically
\begin{equation}
\Box\left[ r^{-1} F_L(t-r) \right] = 0\,,
\end{equation}  
and that the set of tensors $F_L$, with all possible ranks $l$, provides a complete set of representations of the rotation group.
Hence, from the divergence $\partial^\mu\bar{h}_{\mu\nu}$ of our particular solutions, we can use Eqs.~\eqref{h0general} and \eqref{higeneral} to identify the explicit form of the tensors $N_L(u)$, $P_L(u)$, $Q_L(u)$, and $R_L(u)$ of our problem. The only thing left to do after this procedure has been carried out is to invert the equation $\partial^\mu h^h_{\mu\nu}= -h_\nu$ so that the homogenous solution can be added to the particular one, providing the complete solution to our problem, hence fully consistent with general relativity. 
Nevertheless, a solution for $h^h_{\mu\nu}$ is not unique, with each solution differing from one another by a coordinate transformation \cite{Larrouturou2021}. The simplest general solution for the homogeneous contribution reads\footnote{This choice was made to be such that the 00 component could be written just in terms of some monopolar and dipolar terms, and that the spatial trace was monopolar: $h^{h}_{ii} = -3r^{-1}P$.} \cite{blanchetlivrev}
\begin{align}
h^h_{00} &= -r^{-1} N^{(-1)} + \partial_a \left[ r^{-1} \left( - N_a^{(-1)} + Q_a^{(-2)} - 3 P_a\right) \right]\,. \label{generichh00}\\
h^h_{0i} &= -r^{-1} \left(-Q_i^{(-1)}+3P_i^{(1)}\right) +\epsilon_{iab} \partial_a \left[ r^{-1}R_b^{(-1)} \right]
+ \sum_{l=2}^\infty \partial_{L-1}\left[ r^{-1} N_{iL-1} \right]\,, \label{generichh0i}\\
h^h_{ij} &= -\delta_{ij} r^{-1} P + \sum_{l=2}^\infty \bigg\{ 2\delta_{ij}\partial_{L-1}\left[ r^{-1} P_{L-1}\right]
-6 \partial_{L-2(i}\left[ r^{-1} P_{j)L-2}\right] \nonumber\\
&\qquad\qquad+\partial_{L-2} \left[ r^{-1}(N_{ijL-2}^{(1)} + 3 P_{ijL-2}^{(2)} - Q_{ijL-2})\right] -2\partial_{aL-2}\left[ r^{-1} \epsilon_{ab(i} R_{j)bL-2}
\right]
\bigg\} \label{generichhij}\,.
\end{align}
In these expressions, the negative exponents are anti-derivatives: $N_L^{(-1)}(u) = \int^{u}_{-\infty} dv N_L(v)$.

Inspired by this construction, we can bring all this to a scattering amplitude level and work in momentum space. In this case, we rewrite the above expressions, moving from the direct-space quantities $h^\nu$ and $h^h_{\mu\nu}$ to the Fourier transformed $\bar{\mathcal{A}}_\nu$ and $\bar{\bf a}_{\mu\nu}$, respectively, where we define $\bar{\mathcal{A}}_\nu \equiv i k^\mu \bar{\mathcal{A}}_{\mu\nu}$ and $i k^\mu\bar{\bf a}_{\mu\nu} \equiv -\bar{\mathcal{A}}_{\nu}$\,. In this case, we have
\begin{align}
\bar{\mathcal{A}}^0 &= \sum_{l=0}^{\infty} i^l k_L N_L(\omega)\,, \label{Amu0} \\
\bar{\mathcal{A}}^i &= \sum_{l=0}^{\infty} i^{l+1} k_i k_L P_L(\omega) + \sum_{l=1}^{\infty} 
\left[
i^{l-1} k_{L-1} Q_{iL-1}(\omega) + \epsilon_{iab} i^{l} k_a k_{L-1} R_{bL-1}(\omega)
\right]\,, \label{Amul}
\end{align}
and, for $\bar{\bf a}_{\mu\nu}$, we have
\begin{align}
\bar{\bf a}_{00} &= -\frac{i}{\omega} N(\omega) + i k_a \left[ - \frac{i}{\omega}  N_a(\omega) -\frac{1}{\omega^2} Q_a(\omega) - 3 P_a(\omega) \right]\,.\label{a00}\\
\bar{\bf a}_{0i} &= \frac{i}{\omega}Q_i(\omega) + 3 i \omega P_i(\omega) - \epsilon_{iab} \frac{k_a}{\omega} R_b(\omega) 
+ \sum_{l=2}^\infty i^{l-1} k_{L-1} N_{iL-1}(\omega)\,,\label{a0i}\\
\bar{\bf a}_{ij} &= -\delta_{ij} P(\omega) + \sum_{l=2}^\infty i^{l-1} \bigg\{ 2 \delta_{ij} k_{L-1} P_{L-1}(\omega) 
-6 k_{L-2} k_{(i} P_{j)L-2}(\omega) \nonumber\\
&-i k_{L-2} \left[ -i \omega N_{ijL-2}(\omega) - 3\omega^2 P_{ijL-2}(\omega) - Q_{ijL-2}(\omega)\right] -2 k_{aL-2}  \epsilon_{ab(i} R_{j)bL-2}(\omega)
\bigg\}\,. \label{aij}
\end{align}
In particular, from these components, we checked explicitly that we indeed obtain $i k^\mu \bar{\bf a}_{\mu\nu} = - \bar{\mathcal{A}}_{\nu}$, for the expressions of $\bar{\mathcal{A}}_{\nu}$ given in Eqs.~\eqref{Amu0} and \eqref{Amul}.

Finally, in this framework, the complete gravitational scattering amplitude that is consistent with general relativity and which is the full solution we were looking for is the sum
\begin{equation}
i \bar{\mathcal{M}}_{\mu\nu} = i \bar{\mathcal{A}}_{\mu\nu} + i \bar{\bf a}_{\mu\nu}\,. 
\end{equation}

\subsection{The homogeneous solution for the electric $J$-failed tail}\label{sechomogeneouselectricJ}

For the electric $J$-failed tail, with solution for $\bar{\mathcal{A}}_{\mu\nu}$ in Eqs.~\eqref{LtailAbar00}, \eqref{LtailAbar0k}, \eqref{LtailAbarkl}, $\bar{\mathcal{A}}^\nu \equiv i k_\mu \bar{\mathcal{A}}^{\mu\nu}$ yields
\begin{align}
\bar{\mathcal{A}}^0 &= 0\,,\\
\bar{\mathcal{A}}^l &= -16\pi G_N^2 k_a \omega^4 J^{i|a} I^{il}(\omega)\,.
\end{align}
In particular, because of the spatial momenta $k_a$, the box of the direct-space version of this expression vanishes since the spatial derivative stemming from $k_a$ will be acting just on time-dependent quantities. Then, comparing this result against Eqs.~\eqref{Amu0} and \eqref{Amul}, we immediately conclude that, in this case, $N_L = 0$ and $P_L = 0$. Moreover, we also see that, for the multipoles $Q_L$ and $R_L$, the only ones that are nonvanishing are $Q_{ab}$ and $R_{b}$. Hence, using the symmetries of the structure present in $\bar{\mathcal{A}}^l$, we finally obtain
\begin{align}
Q_{al} &= 16\pi i G_N^2 \omega^4 J^{i|(a}I^{l)i}\,, \\
R_b &= -8\pi i G_N^2 \omega^4 \epsilon_{bcd} J^{i|c}I^{id}\,.
\end{align}

From these results, plugging them into Eqs.~\eqref{a00}, \eqref{a0i}, and \eqref{aij}, we get
\begin{align}
\bar{\bf a}_{00} &= 0\, \label{correctedia00}\,\\
\bar{\bf a}_{0i} &= -8\pi i G_N^2 \omega^3 J^{b|k} (k_j \delta_{ib} - k_b \delta_{ij})I^{jk}\,, \label{correctedia0k}\\
\bar{\bf a}_{ij} &= -16\pi i G_N^2 \omega^4 J^{m|(i}I^{j)m}\,. \label{correctediakl}
\end{align}

The above components, summed to the components of $\bar{\mathcal{A}}_{\mu\nu}$ given in Eqs.~\eqref{LtailAbar00}, \eqref{LtailAbar0k}, and \eqref{LtailAbarkl}, provide the complete solution we were after for the electric quadrupole $J$-failed tail process.

\subsubsection{On-shell integration}

By integrating on-shell  the components of $\bar{\mathcal{A}}_{\mu\nu}$ in Eqs.~\eqref{LtailAbar00}, \eqref{LtailAbar0k}, and \eqref{LtailAbarkl}, we obtain
\begin{align}
i \mathcal{\bar{A}}_{00} &= \frac{1}{4\Lambda^4} J^{b|a}I^{ij} \times \bigg\{ \frac{\omega^2}{48\pi} \delta_{aj}k_i k_b \bigg\}\,, \nonumber\\
i \mathcal{\bar{A}}_{0k} &= \frac{1}{4\Lambda^4} J^{b|a}I^{ij} \times \bigg\{ -\frac{\omega}{384\pi} 
\left[ 5k_i k_j k_a \delta_{bk} +2 \omega^2 \delta_{ai} (6k_j \delta_{bk}-2\delta_{jk}k_b) \right] \bigg\}\,,\nonumber\\
i \mathcal{\bar{A}}_{kl} &= \frac{1}{4\Lambda^4} J^{b|a}I^{ij} \times \bigg\{ -\frac{1}{384\pi}
\left[ 2k_i k_j k_a k_{(k}\delta_{l)b} - 2\delta_{jb} k_a k_i k_k k_l + 16\omega^2 \delta_{ib} k_a k_{(k}\delta_{l)j} \right.\nonumber\\
&\qquad\qquad\qquad\qquad\qquad\,\,\, \left.
-12 \omega^2 k_i k_a \delta_{b(k}\delta_{l)j} +24\omega^4 \delta_{ib} \delta_{a(k}\delta_{l)j} -6\omega^2 \delta_{jb} \delta_{kl} k_i k_a\right]
\bigg\}\,.
\end{align}

Now, for the complete solution $i\bar{\mathcal{M}}_{\mu\nu}$ computed on-shell, we sum the original amplitude, $i\bar{\mathcal{A}}_{\mu\nu}$, given above, with the correcting one, $i{\bf a}_{\mu\nu}$, given in Eqs.~\eqref{correctedia00}, \eqref{correctedia0k}, \eqref{correctediakl}.
This sum corresponds to the complete solution:
\begin{equation}
i \bar{\mathcal{M}}_{\mu\nu} = i \bar{\mathcal{A}}_{\mu\nu} + i \bar{{\bf a}}_{\mu\nu}\,.
\end{equation}
From this, we can directly compute the waveform in the far field approximation using
\begin{equation}
    \braket{\bar{h}_{\mu\nu}(x)} = -\frac{1}{4\pi r}\int^{+\infty}_{-\infty}\frac{d\omega}{2\pi}\,i\bar{\mathcal{M}}_{\mu\nu}(\omega,\omega{\bf n}) e^{-i\omega t_{\rm ret}}\,.
\end{equation}
We then obtain
\begin{align}
\bar{h}_{00}(x) &= \dfrac{G_N^2}{r} \times \dfrac{4}{3} n_b n_c J^{a|b} I_{ac}^{(4)} \,, \nonumber\\
\bar{h}_{0i}(x) &= \dfrac{G_N^2}{r} \times \left( J^{b|k}  \dfrac{4}{3}n_b \delta_{ij} + J^{i|a} \dfrac 56 n_k n_j n_a \right) I_{jk}^{(4)}\,,\nonumber\\
\bar{h}_{ij}(x) &= -\dfrac{G_N^2}{r}  \bigg[ \dfrac{1}{3} n_l n_a n_i n_j J_{k|a} I_{kl}^{(4)} - \dfrac 83 n_b n_{(i} I_{j)k}^{(4)} J_{k|b}
- 2 n_k n_a J_{a|(i} I_{j)k}^{(4)} +   J_{l|a} n_k n_a\delta_{ij} I_{kl}^{(4)} \nonumber\\
&\qquad\qquad+\dfrac{1}{3} n_k n_l n_a J_{a|(i} n_{j)} I_{kl}^{(4)}  \bigg]\,,
\end{align}
where the quadrupole moments appearing in these expressions are evaluated at the retarded time $t_{\rm ret}$.
As it turns out, this solution is precisely the one obtained by Blanchet in  Ref.~\cite{blanchet1998b} (See Eqs.~(B.3a), (B.3b), and (B.3c)).

\section{Self-energy diagrams from emission amplitudes}

By performing cuts in the self-energy diagrams, we can see how this type of diagram is related to the emission amplitudes present in the remaining subdiagrams.
For instance, for the simplest self-energy diagram, involving the electric-type quadrupole-quadrupole interaction, the usual Feynman rules yield
\begin{align}
i S_{\rm eff} &= - \frac{1}{16\Lambda^2} \int \frac{d\omega}{2\pi} \omega^4 I_{ij}(\omega)I_{kl}^{*}(\omega) \int_{\K} \frac{1}{\K^2-\omega^2} \left( -\frac{i}{2} \right) \nonumber\\
&\qquad \times \frac 12\left[ \delta_{ik} \delta_{jl} + \delta_{il} \delta_{jk} - \frac{2}{(d-1)}\delta_{ij}\delta_{kl} + \frac{2}{(d-1)\omega^2} (k_i k_j \delta_{kl} + k_k k_l \delta_{ij})\right. \nonumber\\
&\qquad\left.- \frac{1}{\omega^2} (k_i k_k \delta_{jl} + k_i k_l \delta_{jk} + k_j k_k \delta_{il} +k_j k_l \delta_{ik}) + \frac{4}{c_d \omega^4} k_i k_j k_k k_l \right]\,.
\end{align}
From this expression, we immediately notice that the content of the last two lines is simply the $d$-dimensional sum of the physical polarizations $\epsilon_{ij}(\K,h)$ over $h = +,\times$, Eq.~\eqref{sumoverpols}, computed on the mass-shell $|\K|^2 = \omega^2$.
With this, $i S_{\rm eff}$ can be written more compactly as 
\begin{equation}
    i S_{\rm eff} = - \frac{1}{16\Lambda^2} \int \frac{d\omega}{2\pi} \omega^4 I_{ij}(\omega)I_{kl}^{*}(\omega) \int_{\K} \frac{1}{\K^2-\omega^2} \left( -\frac{i}{2} \right) \left[ \sum_h \epsilon_{ij}(\K,h) \epsilon_{kl}^{*}(\K,h) \right]\,,
\end{equation}
or, more generally, as
\begin{equation}\label{selfenergydiag}
    i S_{\rm eff} =   \frac 12 \int_\K \frac{d\omega}{2\pi} \mathcal{A}^{TT}_{ij}(\omega,\K) \mathcal{D}[\sigma_{ij},\sigma_{kl}] \mathcal{A}^{TT}_{kl}(-\omega,-\K)\,, 
\end{equation}
where $\mathcal{A}^{TT}_{ij}(\omega,\K) = \Lambda_{ij,kl}(\N) \mathcal{A}_{kl}(\omega,\K)$, with $i\mathcal{A}_{kl}(\omega,\K)$ being the $\sigma_{kl}$ component of the standard emission amplitude $i\mathcal{A}(\omega,\K)$ and $\Lambda_{ij,kl}(\N)$ the standard TT projector. In particular, this expression shows us how to reconstruct self-energy diagrams from the emission amplitudes and that the knowledge of just the transverse-traceless part of it is sufficient information to reconstruct the full expression for the self-energy diagram. As we will see later, this can be extended more generally to the cases of all the tails, with equality holding as a consequence of the Ward identity. 

The generality of Eq.~\eqref{selfenergydiag} for processes beyond the leading-order self-energy diagrams, extended to all the tails, including the angular momentum failed tail, will be proven at the end of this section. For now, we take this result for granted in the following computations, first for the simple tails, and subsequently to the angular momentum failed tail, in which case we will see that the inclusion of the correcting Ward piece is fundamental in this construction.

\subsection{Simple tail self-energy diagrams from emission amplitudes}

For the reconstruction of the self-energy diagrams for generic tails, we start from the on-shell emission amplitudes for both the tails and leading-order emission for arbitrary moments. For the latter, it is enough to read the amplitude, in Fourier space, from the multipole moment couplings of the effective action \eqref{sourcelag}, which gives
\begin{align}
 i \mathcal{A}_0(\omega,\K) &= [ i \mathcal{A}_0(\omega,\K)]^{ij} \epsilon_{ij}^{*}(\K,h)  \nonumber\\
 	&= \frac{(-i)^{r+1}c_r^{(I)}}{4\Lambda}\omega^2 I^{ijR}(\omega) k_{R}\epsilon_{ij}^{*}(\K,h)\,.
\end{align}
For the simple tail, the emission amplitude for generic multipole moments was first derived within EFT methods in Ref.~\cite{Almeida2021a} for both electric and magnetic types and will be presented in the following chapter. In particular, the expression for the electric case reads
\begin{align}
 i \mathcal{A}_{\rm tail}(\omega,\K) &= [ i \mathcal{A}_{\rm tail}(\omega,\K)]^{ij} \epsilon_{ij}^{*}(\K,h)  \nonumber\\
 	&= \left[ (-i)^r \omega^3 \left( \frac{G_N E c_r^{(I)}}{4\Lambda} \right) I^{ijR}(\omega) k_{R}\times \left( \frac{2}{\epsilon} - 2\kappa_{r+2}+\log x\right) \right] \epsilon_{ij}^{*}(\K,h) \,,
\end{align}
with
\begin{equation}
\kappa_{l} = \frac{2l^2+5l+4}{l(l+1)(l+2)} + \sum_{i=1}^{l-2} \frac{1}{i}\,.
\end{equation}
This result was computed within dimensional regularization and presents an IR divergence in  $d=3$, as we will see in full detail in the next chapter. 
Then, from Eq.~\eqref{selfenergydiag}, the self-energy amplitude for generic moments can be computed using the expressions above for $\mathcal{A}_0$ and $\mathcal{A}_{\rm tail}$, in which case it is given by
\begin{equation}\label{tails}
    i S^{E}_{\rm eff} = \frac 12 \int \frac{d\omega}{2\pi} \int_\K \frac{-i}{\K^2-\omega^2} 
    \left[ i \mathcal{A}_{\rm tail}(\K)\right]^{ij}\times\left[ i \mathcal{A}_0(-\K) \right]^{kl}
    \left( \sum_h \epsilon_{ij}^{*}(\K,h)\epsilon_{kl}(\K,h)\right)\,,
\end{equation}
where $x\equiv - e^{\gamma}\omega^2/(\pi \mu^2)$, with $\gamma$ being the Euler-Mascheroni constant, $\gamma = 0.5772156649$, and $\mu$ the scale introduced due to the use of dimensional regularization.
Thus, we notice that, in order to fully evaluate this expression, we must compute an integral in $\K$ for the sum of polarizations, whose general expression in arbitrary dimensions $d$ was given in Eq.~\eqref{sumoverpols}, containing an arbitrary number of momenta $k_R k_{R'}$ in their numerator. 
An expression for this can be derived generally and reads 
\begin{align}\label{sumpolgeneric}
    \int_\K \frac{k_{R} k_{R'}}{\K^2-\omega^2}  \left( \sum_h \epsilon_{ij}^{*}(\K,h)\epsilon_{kl}(\K,h)\right) &= (d+r)(d+r+1)\frac{(d-2)}{(d-1)}\frac{r!(d-2)!!}{(d+2r+2)!!} \nonumber\\
    &\qquad\qquad\times\omega^{2r} \delta_{ik}\delta_{jl} \delta_{i_1 i'_{1}}\dots \delta_{i_r i'_r} I_0(\omega) + \text{trace terms}\,,
\end{align}
where $I_0(\omega)$ is the integral given in Eq.~\eqref{I0omega}, and by ``trace terms" we mean terms with Kronecker deltas that, when contracted to $I^{ijR}(\omega)$, vanish.
Note that, in this expression, we have to keep the dependence in the spatial dimension $d$ so that the expansion in $\epsilon = d - 3$ can be carried out up to linear order, important here since,  on account of the $(2/\epsilon)$ dependence in $i\mathcal{A}_{\rm tail}$, we will have terms of order $\mathcal{O}(1)$ when multiplying $\epsilon \times (2/\epsilon)$.

Then, plugging the above expressions for $i \mathcal{A}_{\rm tail}$ and $i \mathcal{A}_0$ into Eq.~\eqref{tails}, as well as using the result in Eq.~\eqref{sumpolgeneric} expanded to $\mathcal{O}(\epsilon^1)$, we obtain to order $\mathcal{O}(\epsilon^0)$:
\begin{equation}
S^{E}_{\rm eff} = - G_N^2 E \frac{2^{r+2}(r+3)(r+4)}{(r+1)(r+2)(2r+5)!} \times \int \frac{d\omega}{2\pi} (\omega^2)^{r+3}I^{ij R}(\omega) I^{ij R}(-\omega) \left( \frac{1}{\epsilon} - \gamma_r^{(e)} + \log x\right)\,,
\end{equation}
in which 
\begin{equation}
\gamma_{r}^{(e)} \equiv \frac 12 (H_{r+\frac 52} - H_{\frac 12} + 2H_r +1) + \frac{2}{(r+2)(r+3)}\,,
\end{equation}
with $H_n$ being the harmonic number. This result, valid for arbitrary electric tails of electric type, was obtain for the first time by us in Ref.~\cite{Almeida2021b}, using standard EFT methods, in which we had to perform a two-loop integration with an arbitrary number of momenta in the numerator of the expression.

For the mass tail of the quadrupole moment, in particular, the scheme presented above allow us to understand the connection between the coefficients $11/12$ of the emission and $41/30$ present in the self energy: expanded to linear order in $\epsilon$, Eq.~\eqref{sumpolgeneric} gives a terms proportional to $[1-(9/20)\epsilon]$ that, when multiplied by the factor $(1/\epsilon-11/12)$ present in the emission amplitude, gives $(1/\epsilon - 41/30)$.

Similarly, for the self-energy diagrams for the tails involving generic magnetic multipoles, we start with the on-shell emission amplitude (also computed in the following chapter)
\begin{equation}
i \mathcal{A}_{\rm tail} = \frac{(-i)^{r+1}c_r^{(J)}}{4\Lambda^3} E J^{b|iRa}(\omega) \times 
\bigg[ \frac{i}{16\pi} \omega^2 k_R k_b \left( \frac{2}{\epsilon} - 2 \pi_{r+2} + \log x \right) \epsilon^*_{ia} \bigg]\,,
\end{equation}
where
\begin{equation}
\pi_l = \frac{l-1}{l(l+1)} + \sum_{i=1}^{l-1}\frac{1}{i}\,,
\end{equation}
and follow precisely the same steps as in the electric case. By doing this, we eventually get
\begin{multline}
S^{E}_{\rm eff} = - G_N^2 E \frac{2^{r+2}(r+2)^2(r+4)(r!)^2}{(2r+1)(2r+3)(2r+5)(2r)![(r+3)!]^2} \nonumber\\
\times \int \frac{d\omega}{2\pi} 
J^{b|iRj}(\omega) J^{b'|kR'l}(-\omega) \omega^{2r+6} [\delta_{bb'}\delta_{ik} + (r+1) \delta_{ib'}\delta_{kb}] \left(
\frac{1}{\epsilon} - \gamma^{(m)}_r + \log x \right)\,,
\end{multline}
with
\begin{equation}
\gamma^{(m)}_r = \frac{2}{r+3} + \frac{1}{2r+5} - \frac{1}{r+2} - \frac{1}{r+4} + H_{r+1} + \frac12 H_{r+\frac 32} + \log 2\,.
\end{equation}
And, like in the electric case, this matches the expression we first obtained in Ref.~\cite{Almeida2021b}.

\subsection{$J$-failed tail self-energy diagrams from emission amplitudes}

The same reasoning as followed for the simple tails could be, in principle, used to compute self-energy diagrams for the angular momentum failed tail by either directly performing standard EFT computations, in which we would need to perform integrals corresponding to two-loop integrals or by gluing two on-shell emission amplitudes and performing the simple one-loop integrals in $\K$. Of course, these two directions should yield equivalent results. However, this will only be the case for those processes whose gravitational scattering amplitude satisfies the Ward identity, which is the case not only for all the simple tails but also for all the $J$-failed tails with $r\ge 1$. 

This correspondence should break down in the quadrupole cases since the Ward identity is violated, and therefore, the physical polarizations cannot be obtained by simply applying the TT projector $\Lambda_{ij,kl}$ on the original emission amplitude. Likewise, the result for the self-energy diagram for such processes, computed from standard EFT methods, should not be taken as correct since conservation of the energy-momentum tensor is not consistently carried over along the diagram and, thus, will generally yield a physically inconsistent result.

Let us start by analyzing the electric quadrupole angular momentum failed tail. The computation following standard EFT methods was first performed in Ref.~\cite{RSSFhereditary}, where the authors obtained the following result, 
\begin{equation}\label{ltail815}
i S^J_{\rm eff} = -\frac{8}{15}G_N^2 J^{i|k} \int \frac{d\omega}{2\pi}\, I^{ij}(\omega)I^{jk}(-\omega) \omega^7 \,.  
\end{equation}
Notice that this scalar result is transparent to inconsistencies that one would observe in studying the properties of gravitational scattering amplitudes.

On the other hand, performing the computation by gluing the on-shell emission amplitudes for the leading-order quadrupole emission to the electric $J$-failed tail, in which we use
\begin{equation}\label{Jtailquadrupoleschem}
i S^J_{\rm eff} = \frac 12 \sum_h \int_\K\frac{d\omega}{2\pi} [i\mathcal{M}_{J-\rm tail}^{r=0} (\K) ]^{TT,ij} \mathcal{D}[\sigma_{ij},\sigma_{kl}] [i \mathcal{A}_0^{r=0} (-\K)]^{TT,kl}\,, 
\end{equation} 
we obtain after integrating in $\K$\footnote{Interestingly, this same result has also been recently obtained within traditional methods by Quentin Henry and François Larrouturou in \cite{francs2023}.}
\begin{equation}\label{self30}
i S^J_{\rm eff} = -\frac{1}{30} G_N^2 J^{i|k} \int \frac{d\omega}{2\pi} I^{ij}(\omega)  I^{jk}(-\omega)  \omega^7\,.
\end{equation}
As expected, this result differs from the one presented in Eq.~\eqref{ltail815}. In particular, in Eq.~\eqref{Jtailquadrupoleschem}, $i\mathcal{M}_{J-\rm tail}^{r=0}$ is the emission amplitude obtained from the Ward-corrected scattering amplitude derived in Sec.~\ref{sechomogeneouselectricJ}, recalling that, in our conventions, the TT part of an emission amplitude $i\mathcal{A}$ is simply given by $i\mathcal{A} = -\Lambda\bar{\mathcal{A}}_{ij} \sigma_{ij}^*$. In this case, we obtain
\begin{align}
[i\mathcal{M}_{J-\rm tail}^{r=0} (\K) ]^{TT} &= [i\mathcal{M}_{J-\rm tail}^{r=0} (\K) ]^{TT,ij} \epsilon_{ij}^* \nonumber\\
&= \left[ \frac{i G_N}{4\Lambda} J^{i|a} I^{jl} \omega^2 k_l k_a \right] \epsilon_{ij}^*\,.
\end{align}

\subsubsection{A simple consistency test}

Computing the complete expression for the emission amplitude corresponding to the Ward-fixing amplitude $i{\bf a}_{\mu\nu}$, we obtain
\begin{equation}
{\bf a}(\omega,\K) = \alpha_\phi \phi^{*}+ \alpha_A^i A_i^{*} + \alpha_\sigma^{kl}\sigma_{kl}^{*}\,,
\end{equation}
with
\begin{align}
\alpha_\phi &= 0 \,, \\
\alpha^i_A &= -16\pi \Lambda G_N^2 \omega^3 J^{b|k} (k_j \delta_{ib} - k_b \delta_{ij}) I^{jk}\,,\\
\alpha^{kl}_{\sigma} &= 16\pi \Lambda G_N^2 \omega^4 J^{m|(l}I^{k)m}\,. 
\end{align}
Putting these expressions together, we are left with
\begin{equation}
i{\bf a}(\omega,\K) = 16\pi i \Lambda G_N^2 \omega^3
\left[
- J^{b|k} (k_j \delta_{ib} - k_b \delta_{ij}) I^{jk}(\omega) A_i^*(\omega,\K)
+ \omega J^{m|(l} I^{k)m} \sigma^*_{kl}
\right] \,.
\end{equation}
Then, notice that we can glue this amplitude to the complete leading-order emission amplitude for the electric quadrupole, which is given by
\begin{equation}
i\mathcal{A}_0(\omega,\K) = \frac{(-i)^{r+1}c_r^{(I)}}{\Lambda} I^{ijR}(\omega) k_R \left[ \frac12 \omega^2 \sigma_{ij}^*(\omega,\K) + \omega k_j A_i^*(\omega,\K) - k_i k_j \phi^*(\omega,\K) \right]\,,
\end{equation}
to account for the complete contribution of the correcting term to the original result of Eq.~\eqref{ltail815}. Be aware that, here, we cannot consider the TT part of the emission amplitude $i{\bf a}$, since it, alone, is not consistent with the Ward identity (just the sum of orignal plus Ward fixing amplitude is!). Hence, it is simply inconsistent to take the TT part. Then, we use
\begin{equation}
i S^J_{\rm eff}\Big|_{\bf a} = \frac 12 \int_\K \frac{d\omega}{2\pi} \frac 1{\K^2-\omega^2}\braket{[i{\bf a}(\omega,\K)]\times[i\mathcal{A}_0(-\omega,-\K)]}^{\sim}\,,
\end{equation}
in which case, we arrive at
\begin{equation}
i S^J_{\rm eff}\Big|_{\bf a} = \frac {G_N^2}{2} J^{i|k} \int \frac{d\omega}{2\pi}  I^{ij}(\omega) I^{jk}(-\omega) \omega^7 \,.
\end{equation}
Now, summing this term, which, again, is the contribution solely from the Ward-fixing term, to the original amplitude obtained from EFT methods, we find the interesting result
\begin{equation}
i S^J_{\rm eff} = i S^J_{\rm eff}\Big|_\text{stand} + i S^J_{\rm eff}\Big|_{\bf a}
= - \frac {G_N^2}{30} J^{i|k} \int \frac{d\omega}{2\pi}  I^{ij}(\omega) I^{jk}(-\omega) \omega^7\,,
\end{equation}
which is the same result obtained in Eq.~\eqref{self30} from the on-shell emission amplitude for the angular momentum failed tail after the anomaly had been corrected.

\subsubsection{$J$-failed tail for arbitrary multipole moments}

As we have seen above in this section since the gravitational scattering amplitudes for the angular momentum failed tail is anomaly-free for multipole moments with $r\ge 1$, we can directly compute the self-energy diagrams without any problem. Below, we do this in the two available ways and see the agreement between the methods. In particular, for the gluing of on-shell emission amplitudes, we need the generic expression of $i \mathcal{A}$ for the $J$-failed tail, valid for $r\ge 1$. Computation of this amplitude yields
\begin{multline}
i\mathcal{A}_{J-\rm tail}(\K) = (-i)^{r+1} \frac{c_r^{(I)}}{2\Lambda^3} J^{b|a} I^{ijR}(\omega)\omega^2\epsilon^*_{id}(\K) \nonumber\\
\times \left[\frac{r(r^2+5r+10)\delta_{i_1 b}\delta_{jd} k_a k_{i_2}\cdots k_{i_r}\omega^2 -2(r^2+4r+6)\delta_{bd} k_a k_j k_R}{16\pi(r+1)(r+2)(r+3)(r+4)}
\right]\,.
\end{multline}

With this amplitude, gluing it to leading-order multipole moments of the same multipolarity, we obtain
\begin{equation}\label{eLtailFromEmiss}
i S^J_{\rm eff} = \frac{2^r [c_r^{(I)}]^2 G_N^2 (12+50r+35r^2+10r^3+r^4) (r!)^2}{(r+1)(r+2)(r+3)(1+2r)(3+2r)(5+2r)(2r)!}
J^{b|a} \int\frac{d\omega}{2\pi}\, \omega^{7+2r} I^{aiR}(\omega) I^{biR}(-\omega)\,.
\end{equation}

The self energy computed from standard EFT methods gives the following expression for the same process, given in terms of two-loop integrals:
\begin{align}\label{SEarbitrarye}
i S^J_{\rm eff} &= \frac{[c_r^{(I)}]^2}{4\Lambda^4} J^{b|a}\int \frac{d\omega}{2\pi}\, I^{ijR}(\omega)I^{klR'}(-\omega)
\int_{\K\Q}\frac{(k+q)_R k_{R'}q^a}{(\K^2-\omega^2)[(\K+\Q)^2-\omega^2]\Q^2} \times\bigg\{ \nonumber\\
&-\frac{\omega^5}{4} \delta_{jl} (k_b \delta_{ik} + 2q_i \delta_{kb})   +\frac{\omega^3}{2} \left[ -k_k q_i k_b \delta_{jl} 
 + \delta_{jb} k_l \K\cdot\Q \delta_{ik} \right] \nonumber\\
&+\frac{\omega^3}{2} \left[ (k+q)_j k_l (q_k \delta_{ib} + k_b \delta_{ik})  - k_k k_l q_i \delta_{bj} 
+ (2k_i k_l q_j + k_l q_i q_j - k_i(k+q)_j q_l) \delta_{kb}\right]  \nonumber\\
&+\frac{\omega}{2}(k+q)_j k_l \big[ k_b k_k q_i + (k+q)_i \K\cdot\Q \delta_{kb}  \big]  -\frac{\omega}{2c_d} (k+q)_i (k+q)_j k_k k_l k_b  \bigg\}\,.
\end{align}
Performing the two-loop integrals of this expressions, we are able to show that this gives precisely  the result of Eq.~\eqref{eLtailFromEmiss} for arbitrary multipole moments of the electric type.

Incidently, when setting $r=0$ in Eq.~\eqref{eLtailFromEmiss}, which is formally valid for $r\ge 1$, it gives
\begin{equation}
i S^J_{\rm eff} = - \frac{G_N^2}{30}
J^{i|k} \int\frac{d\omega}{2\pi}  I^{ij}(\omega) I^{jk}(-\omega) \omega^{7}\,,
\end{equation}
which is precisely the value obtained for the Ward-corrected self energy.

\subsection{On the on-shell property of the tail processes}\label{onshelness}

\subsubsection{Why just on-shell emission amplitudes contribute to $i S_{\rm eff}$ in the tails}

Consider an emission amplitude that satisfies the Ward identity, given by
\begin{equation}
i\mathcal{A} = \alpha_{\sigma}^{cd} \sigma_{cd}^* + \alpha_A^c A_c^* + \alpha_\phi \phi^*\,.
\end{equation}
As a consequence of the Ward-identity expressions in Eqs.~\eqref{nu0} and \eqref{num} in terms of the $\alpha$'s, we obtain:
\begin{equation}
\omega \left( \alpha_\phi + \frac{2}{d-2}\delta_{cd}\alpha_\sigma^{cd} \right) + k_c \alpha_A^c = 0\,,\qquad \omega \alpha_A^a = k_c \alpha_\sigma^{ca} + k_c \alpha_\sigma^{ac}\,.
\end{equation}
This can be solved for $\alpha_\phi$ and $\alpha_A^c$ in terms of $\alpha_\sigma^{cd}$:
\begin{align}
\alpha_\phi &= -\frac{2}{d-2}\delta_{cd}\alpha_\sigma^{cd} - \frac{2}{\omega^2} k_c k_d \alpha_\sigma^{cd}\,, \\
\alpha_A^a  &= \frac{1}{\omega} (k_c \alpha_\sigma^{ac} + k_c \alpha_\sigma^{ca})\,.
\end{align}
Then, plugging this back to the generic expression for $i \mathcal{A}$ in Eq.~\ref{genericemissA}, we obtain:
\begin{align}
i \mathcal{A} &= \frac{2}{\omega^2} \alpha_\sigma^{ij} \left[ \frac 12 \omega^2 \sigma_{ij}^* + \frac12 \omega (k_i A_j^* + k_j A_i^*) - k_i k_j \phi^* - \frac{\omega^2}{d-1} \delta_{ij} \phi^* 
\right] \nonumber\\
&= \frac{2}{\omega^2} \alpha_\sigma^{ij} R_{0i0j}^*\,.
\end{align}

Now, considering another process that also satifies the Ward identity, with coefficients $\beta$, 
\begin{equation}
i\mathcal{A}_0 = \beta_{\sigma}^{cd} \sigma_{cd}^* + \beta_A^c A_c^* + \beta_\phi \phi^*\,,
\end{equation}
we have
\begin{align}
i S_{\rm eff} &= \frac12 \int_\K \frac{d\omega}{2\pi} \frac{1}{\K^2-\omega^2} \braket{[i\mathcal{A}(\omega,\K)]\times [i\mathcal{A}_0(-\omega,-\K)]}^\sim \nonumber\\
              &= \frac12 \int_\K \frac{d\omega}{2\pi} \frac{1}{\K^2-\omega^2} \times \frac{4}{\omega^4} \alpha_\sigma^{ij}\beta^{kl*}_\sigma  \braket{R^*_{0i0j}\times R_{0k0l}}^\sim\,,
\end{align}
and then, it can be checked explicitly that the following relations are equivalent:
\begin{equation}
\frac{1}{\K^2-\omega^2}\braket{R^*_{0i0j}\times R_{0k0l}}^\sim = \frac{-i}{\K^2-\omega^2} \frac{\omega^4}{4} \sum_{h} \epsilon^*_{ij} \epsilon_{kl} = \frac{1}{4} \omega^4 \Lambda_{ij,ab} \mathcal{D}[\sigma_{ab}, \sigma_{cd}] \Lambda_{kl,cd}\,.  
\end{equation}
Hence, we have the following equivalent ways of writing $i S_{\rm eff}$, that we have been using above:
\begin{align}
i S_{\rm eff} &= \frac12 \int_\K \frac{d\omega}{2\pi} \frac{1}{\K^2-\omega^2} \braket{[i\mathcal{A}(\omega,\K)]\times [i\mathcal{A}_0(-\omega,-\K)]}^\sim \\
              &= \frac12 \int_\K \frac{d\omega}{2\pi} \frac{-i}{\K^2-\omega^2} \times \alpha_\sigma^{ij}\beta^{kl*}_\sigma  \left( \sum_h \epsilon^*_{ij} \epsilon_{kl} \right) \\
              &= \frac12 \int_\K \frac{d\omega}{2\pi} (\alpha_\sigma^{ab})^{TT} \mathcal{D}[\sigma_{ab},\sigma_{cd}] (\beta^{cd*}_\sigma)^{TT}  \,.          
\end{align}
Finally, from the last of these equations, we can derive
\begin{align}
i S_{\rm eff} &= -\frac{i}{2} \int_\K \frac{d\omega}{2\pi} \frac{1}{\K^2-\omega^2}(\alpha_\sigma^{ab})^{TT}(\omega,\K) (\beta^{ab*}_\sigma)^{TT}(-\omega,-\K)  \nonumber\\
              &= -\frac{i}{32\pi^2} \int d\Omega \int \frac{d\omega}{2\pi}(i \omega) (\alpha_\sigma^{ab})^{TT}(\omega,\omega \N) (\beta^{ab*}_\sigma)^{TT}(-\omega,\omega \N) \,,
\end{align}
where, in the passage from the first to the second line, we have used the following result 
\be
\int_{\K}\frac{k_{i_1}\dots k_{i_{2l}}}{\K^2 -(\omega\pm i\tt{a})^2}=\pa{\pm i\frac{\omega}{4\pi}}\delta_{i_1\dots i_{2l}}\frac{\omega^{2l}}{(2l+1)!!}=\pa{\pm i\frac{\omega}{16\pi^2}}\omega^{2l}\int{\rm d}\Omega n_{i_1}\dots n_{i_{2l}}\,.
\ee
This result shows that the computation of self-energy diagrams can be performed on-shell, with just the TT components of the emission amplitude. However, the passage from one line to the other in the above expression is not general. This happens because when we perform a tensor reduction to write Feynman integrals containing momenta in its numerator in terms of scalar integrals, the structure $(\K^2-\omega^2)/\K^2$ usually appears, which gives a nonvanishing contribution in general. For the case in which one of the amplitudes represents a tail process (either for the mass or angular momentum cases), with the other being a leading-order amplitude, we have the following type of loop-integral coming from the tail:
\begin{equation}
\int_\Q \frac{1}{[(\K+\Q)^2-\omega^2]\Q^2}\,.
\end{equation}  
Hence, when performing tensor reduction on the integrals that appear in the tails, we will have appearance of a factor of $(\K^2-\omega^2)$ in the form of
\begin{equation}
\frac{(\K^2-\omega^2)}{\K^2}\int_\Q \frac{1}{[(\K+\Q)^2-\omega^2]\Q^2}\,.
\end{equation}
This will enter the expression of $i S_{\rm eff}$ as
\begin{align}
i S_{\rm eff} &\sim \int_\K \frac{1}{\K^2-\omega^2} \times \frac{(\K^2-\omega^2)}{\K^2}    \int_\Q \frac{ k_{i_1}\cdots k_{i_{2l}} }{[(\K+\Q)^2-\omega^2]\Q^2} \nonumber\\
&\sim \int_{\K\Q} \frac{k_{i_1}\cdots k_{i_{2l}}}{[(\K+\Q)^2-\omega^2]\K^2\Q^2} = c_0 \delta_{i_1\dots i_{2l}}\,.
\end{align}
The proportionality factor $c_0$ on the right-hand side of this equation can be obtained by contracting, e.g., the indices $i_1 i_2$. By doing this, we have
\begin{align}
(d+2l-2) c_0 \delta_{i_3\dots i_{2l}} &= \int_{\K\Q} \frac{\K^2 k_{i_3}\cdots k_{i_{2l}}}{[(\K+\Q)^2-\omega^2]\K^2\Q^2} \nonumber\\
&= \int_{\K} \frac{1}{(\K^2-\omega^2)} \int_\Q \frac{(k-q)_{i_3}\cdots (k-q)_{i_{2l}}}{\Q^2} = 0\,.
\end{align}
The integral on the right-hand side vanishes since we are left with scaleless integrals, which are zero in dimensional regularization. Notice that, have we had a massive propagator $(\Q^2-\omega'^2)$ instead of $\Q^2$, this integral would not be vanishing, and thus, factors of $(\K^2-\omega^2)$, which are otherwise zero on-shell, would contribute to $i S_{\rm eff}$. This is the case of the memory effect. And this is why we can use just the on-shell part of the emission amplitudes in the case of the tails, both simple and angular momentum ones, to reconstruct the self-energy diagrams. 

In what follows, we justify the use of the Lorentz gauge condition $\partial^\mu \bar{h}_{\mu\nu}=0$ instead of the more general harmonic condition $g^{\mu\nu}\Gamma^\alpha_{\mu\nu} = 0$ in obtaining the Ward-correcting amplitudes. As we will see, this happens because the longitudinal terms that make these two equations inequivalent do not go on-shell and, therefore, do not contribute to $i S_{\rm eff}$, as we have seen above, for the case of leading-order and tail processes.

\subsubsection{Why the Lorentz and harmonic gauge conditions are equivalent in the tails}

The gauge we have been using to compute amplitudes is the harmonic gauge, defined by
\begin{equation}\label{harmgauge}
\Gamma^\alpha_{\mu\nu} g^{\mu\nu} = 0\,.
\end{equation}
When expanding in first order in the fields, we obtain the equivalent to the Lorentz gauge
\begin{equation}
\partial^\mu \bar{h}_{\mu\nu} = \partial^\mu \left( h_{\mu\nu} - \frac12 \eta_{\mu\nu} h \right) = 0\,.
\end{equation}
This condition holds only for the leading-order processes. For higher orders, like the tails, we would rather have to solve Eq.~\eqref{harmgauge} iteratively in $G_N$, like it was done for Einstein's equation itself.
In particular, the general structure of the problem is given by
\begin{align}
\partial^\mu \bar{h}^{(1)}_{\mu\nu} &= 0\,,\nonumber\\
\partial^\mu \bar{h}^{(2)}_{\mu\nu} &= \lambda^{(2)}_{\mu\nu}(h^{(1)},h^{(1)})\,,\nonumber\\
\partial^\mu \bar{h}^{(3)}_{\mu\nu} &= \lambda^{(3)}_{\mu\nu}(h^{(1)},h^{(1)},h^{(1)}) + \gamma^{(3)}_{\mu\nu}(h^{(2)},h^{(1)})\,,\nonumber\\
\cdots\,,
\end{align}
where $\bar{h}^{(n)}_{\mu\nu}$ represents processes of order $(G_N)^n$. In particular, we have at order $G_N^2$:
\begin{equation}\label{gaugecondcorrect}
\partial^\mu \bar{h}^{(2)}_{\mu\alpha} = h^{(1)\mu\nu} \left( h^{(1)}_{\alpha\mu,\nu} -\frac12 h^{(1)}_{\mu\nu,\alpha}\right) - \frac12 \left( H^{(2)}_{\alpha\mu}{}^{,\mu} -\frac12 H^{(2)}_{,\alpha}\right)\,.
\end{equation}
On the right-hand side of this equation, we should plug in processes of order $G_N$. The second term on the RHS includes $(G_N-{\rm processes})^2$ which appears because we have been using the Kaluza-Klein parameterization. In this case, we have
\begin{align}
H^{(2)}_{\mu\nu} 
&= \left(
\begin{array}{cc}
-(h^{(1)}_{00})^2 & -2 h^{(1)}_{00} h^{(1)}_{0i} \\
-2 h^{(1)}_{00} h^{(1)}_{0i} & h^{(1)}_{00}[(c_d-2)h^{(1)}_{ij}-\frac{1}{4}(c_d-2)^2\delta_{ij}h^{(1)}_{00}]-2 h^{(1)}_{0i} h^{(1)}_{0j}
\end{array}
\right)\,.
\end{align}

Now, consider two processes of order $G_N$,  given generically by
\begin{equation}
h^{(1)}_{\mu\nu} \sim \int\frac{d\omega}{2\pi} A^{R}(\omega) \int_\K \frac{e^{-i\omega t+ i \K\cdot\X}}{\K^2-\omega^2} K_L\,,
\quad \text{and} \quad
h^{(1)}_{\mu\nu} \sim \int\frac{d\omega}{2\pi} B^{R'}(\omega') \int_\Q \frac{e^{-i\omega' t+ i \Q\cdot\X}}{\Q^2-\omega'^2} Q_{L'}\,,
\end{equation} 
where $A^{R}(\omega)$ and $B^{R'}(\omega')$ represent generic order-$G_N$ processes, including the ones related to conserved multipoles, by making, e.g., $A^{R}(\omega) \rightarrow E \delta(\omega)$; $K_L$ represents any combination of the momenta $k$'s, and likewise for $Q_{L'}$. By plugging this into the right-hand side of Eq.~\eqref{gaugecondcorrect}, we will always have the following behavior:
\begin{align}
\partial^\mu \bar{h}_{\mu\nu} &\sim \int\frac{d\omega}{2\pi} A^{R}(\omega) \int_\K \frac{e^{-i\omega t+ i \K\cdot\X}}{\K^2-\omega^2} K_L
\times
\int\frac{d\omega'}{2\pi} B^{R'}(-\omega') \int_\Q \frac{e^{i\omega' t- i \Q\cdot\X}}{\Q^2-\omega^2} Q_{L'} \nonumber\\
&= \int_\K\frac{d\omega}{2\pi}  \frac{e^{-i \omega t+i \K\cdot\X} }{\K^2-\omega^2}
 \times \left[  k^\mu \bar{\mathcal{A}}_{\mu\nu}
 \right]\,, 
\end{align}
with
\begin{equation}
k^\mu \bar{\mathcal{A}}_{\mu\nu} = (\omega^2-\K^2)
 \int\frac{d\omega'}{2\pi} A^{R}(\omega+\omega')  B^{R'}(-\omega')\int_{\Q} \frac{(KQ)_{LL'}}{[(\K + \Q)^2-(\omega+\omega')^2](\Q^2-\omega'^2)}\,,
\end{equation}
which always vanishes on-shell, and hence, will not apply any role in the construction of self-energy diagrams. This is the reason behind the use of the Ward identity, which is consistent with the Lorentz condition $\partial^\mu\bar{h}_{\mu\nu}=0$, even for the tails, which are processes of order $G_N^2$.


\section{Conclusions}

In this chapter, we have studied the emission amplitudes and what we called the gravitational scattering amplitudes for the leading-order processes, simple tails, and angular momentum failed tails for arbitrary multipole moments. As we have seen, the cases of electric and magnetic quadrupole $J$-failed tails presented anomalies, but whose fixing at the level of the amplitudes could be implemented by the introduction of counter-terms, within a consistent framework, which could be used to bring the original solutions to valid solutions of general relativity. 

The construction presented above is essential if one is interested in correctly accounting for radiation-reaction effects, especially for the 5PN order, for which the new electric $J$-failed tail contribution is indispensable. Moreover, with the intention of computing valid far zone self-energy diagrams, which, again, represent radiation-reaction effects, we developed a way of connecting the emission amplitudes to the - different in nature - self-energy diagrams.

The same analysis must also be carried out for the memory effect in order to fully account for radiation reaction at 5PN and finally solve the problem. In this case, however, the memory presents a much more intricate structure, where a more careful analysis beyond the on-shell level becomes necessary:
As it turns out, in the case of the memory, the emission amplitude is not analytic in $\K^2$, so when glued with a term $1/(\K^2-\omega^2)$ one cannot simply substitute $\K^2=\omega^2$.
In face of this, the importance of the present chapter lies  in (i) identifying the presence of anomalies, that were otherwise providing an incomplete description for certain radiation-reaction effects and had never been realized before, (ii) and also paving the way toward the possible solution for the 5PN problem, as we present all the necessary ingredients for a consistent computation of higher-order self-energy diagrams.

%% file: chapter4/renormalization.tex
\chapter{Gravitational Multipole Renormalization}\label{renormachapter}

In this chapter, we study the effect of scattering gravitational radiation off the  static background curvature, up to second order in Newton constant, known in literature as tail and tail-of-tail processes, for generic electric and magnetic multipoles. 
This reports on original research published in 2021 \cite{Almeida2021a}.
Starting from the multipole expansion of composite compact objects, and as expected due to the known electric quadrupole case, both long- and short-distance (UV) divergences are encountered. The former disappear from properly defined observables, the latter are renormalized and their associated logarithms give rise to a classical renormalization group flow. UV divergences alert for incompleteness of the multipolar description of the composite source, and are expected not to be present in a UV-complete theory, as explicitly derived in literature for the case of conservative dynamics. Logarithmic terms from the tail of tail of generic magnetic multipoles are computed in this work for the first time.

\section{Introduction}
\label{sec:intro}

Building on the multipole expansion, we study a specific class of post-Minkowskian (PM) corrections up to second order in the Newton's constant $G_N$ beyond the Newtonian level. At $O(G_N)$ beyond leading order emission, one encounters leading non-linear \emph{hereditary} effects, i.e., terms depending on the history of the source rather than on an instantaneous state at retarded time. Historically, these  have been divided into \emph{memory} and \emph{tail} effects \cite{blanchet1992}, the former arising from scattering of radiation onto radiation \cite{christ1991}, the latter from scattering of radiation onto the static background curvature sourced by the total mass $E$ of the system \cite{blanchet1988}. The denominations are related to the nature of the phenomenological effects they have on the waveform: the tail part of the waveform arrives later than the ``wavefront'', being delayed by the scattering, and then smoothly fades off with time; the memory part is a persistent zero-frequency effect which is still present well after the wavefront has passed.

While hereditary in the waveform, on circular orbits, radiation-radiation scattering leads to a vanishing effect in the emitted flux \cite{blanchetlivrev}, and to an instantaneous (i.e., non-hereditary) contribution to the conservative energy \cite{RSSFhereditary}\footnote{Actually, this is the knowledge based on the results of Ref.~\cite{RSSFhereditary}. Nevertheless, in face of the framework presented in the previous chapter, this might not be the case, being currently under investigation.}; tail effects on the other hand give a hereditary contribution to the waveform \cite{blanchet1988} and to the conservative energy \cite{RSSF2013} (later confirmed in \cite{GalleyRoss2016}), while giving an instantaneous contribution to the flux emission from circular orbits \cite{goldberger2010}. The scattering of radiation off the angular-momentum dependent static background curvature leads to instantaneous terms both in the waveform \cite{Faye2012} and in the conservative energy shift \cite{RSSFhereditary}, and no contribution to the flux.

In particular only the (mass) tail-corrected emission process involves a large-distance, or infra-red (IR),
divergence, as thoroughly explained in \cite{goldberger2010}, which however disappears from suitably defined observables. In the waveform, the IR tail divergences are relatively imaginary with respect to the leading order, and they exponentiate to a pure phase, so disappearing from the flux. While in principle still showing up in the waveform, analogous to the well-known infinite phase shift induced by the Coulomb potential in scattering amplitudes \cite{Weinberg1965}, one has to consider that actual detections do not measure the instantaneous absolute value of the phase, but phase differences between different times, and the infinity cancels out of any observable quantity \cite{Porto2012}. Note however that finite contributions of the tail effect for different multipoles are different, and their non-zero difference \emph{is} physical, while the IR divergent part is common to all multipoles \cite{Faye2015} and cancels out in the difference.

Note that observability of the finite shift in the waveform phase generated by tail effect has already been investigated long ago in \cite{blanchet1994,blanchet1995}, and unfortunately the possibility of it being measured is scarce, as such an effect appears as $G_NE\omega\sim v^3$ correction to the leading order phase which goes as $v^{-5}$, hence a 4th order post-Newtonian (PN) effect \cite{blanchet1995}, where $v^2\sim G_NE/r$ is the expansion parameter of the PN approximation. Current knowledge of PN-expanded waveforms stops at 4PN order, see \cite{blanchetlivrev} for a review, and \cite{LIGOScientific:2020ros} for the most recent tests on real data. Note that finite contributions of the tail affects the waveform phase at the same order as a shift $\Delta t$ in the arrival time of the signal, which enters the phase with a term $\sim 2\pi f\Delta t\sim v^3 (\Delta t/G_NE)$.

In the present chapter, we focus on the analysis of (mass) tail-of-tail effects at waveform level or equivalently, in the language of field theory, in one-point amplitudes. IR and UV divergences are both present, the former are consistent with the exponentiation to a phase of the simple tail IR divergences, the latter have associated logarithmic terms that give rise to renormalization group equations, which can be integrated to compute all orders leading logarithmic corrections, as already done for the electric quadrupole case \cite{BlanchetRSSF2019}.

In particular we generalize the computation of logarithmic terms in tail-of-tail processes, already known in the electric case from the results obtained in \cite{anderson1982} for the mass quadrupole and in \cite{blanchet1988} for all the electric multipoles, to magnetic multipole at all orders. While sharing the same topology, diagrams of increasing multipole order become more intricate because of the presence of an increasing number of momenta. In PN scaling, moving from a multipole to the following one adds a power of $v$ to the coupling, hence tail diagrams involving the electric (magnetic) $2^{n}$-multipole affects one-point amplitudes starting at $1/2+n/2$ ($1+n/2$) PN order. Multipoles corrected by gravitational self-interactions are also called in literature \emph{radiative} multipoles \cite{blanchetlivrev}, to differentiate from \emph{source} multipoles which instead designate the source terms in the fundamental multipolar expansion.

Note however that when multipoles of composite objects like binary systems are expressed in terms of individual binary constituents, they can naturally be expanded in $v^2$, i.e., in a PN series, whose terms are determined by a \emph{matching} procedure, as discussed in Chapter 3, which for the mass quadrupole has been completed in an effective field theory framework up to second PN order \cite{Maia2020}, and to fourth PN order in the multipolar-post-Minkowskian approach \cite{Marchand2020}.

By analogy with the conservative dynamics case treated in detail in \cite{RSSF4PNb}, it is expected that the UV divergence in the tail-of-tail process will be cancelled by analogous divergences in the expression of the PN-corrected source multipoles, to leave a finite, consistent result. After all, the multipole expansion is bound to fail at short enough distance, i.e., when the actual internal structure of the composite system becomes important. In fact, Larrouturou et. al. has explicitly shown this cancellation for the electric quadrupole and electric octupole, as well as for the magnetic quadrupole moment in Refs.\cite{blanchet4PNrada,blanchet4PNradb}.

The chapter is structured as follows: in Sec.~\ref{sec:method} we give an overview of the method treating in detail the known case of the tail process, building on which we obtain new results for the tail-of tail in Sec.~\ref{sec:res}. Sec.~\ref{sec:sdisc} concludes with a discussion of the results.

\section{Method}
\label{sec:method}

\subsection{Basic setup}

Here, we use the basic setup introduced in the previous chapter, that we briefly replicate for the sake of clarity and proceed along the lines of \cite{goldberger2010}, which applies to the radiative gravitational sector the EFT approach developed in \cite{rothstein2006}, i.e., NRGR.
In particular, as we have seen,
at large distances from the source, its interaction with gravity can be encoded
in terms of multipoles as in the following effective Lagrangian, whose form is
uniquely dictated by the symmetries and scaling of the theory
\begin{align}
\label{eq:smult}
S_{\rm mult} &=\int {\rm d}t\left[\frac 12 Eh_{00}-\frac 12\epsilon^{ijk}L_ih_{0j,k} 
     -\sum_{r\geq 0}\left(c^{(I)}_r I^{ijR}\partial_{R}{\cal E}_{ij}-c^{(J)}_r J^{ijR}\partial_{R}{\cal B}_{ij}\right)
\right]\,,
\end{align}
with \cite{ross2012}
\be
c_r^{(I)}=\ds\frac 1{\pa{r+2}!}\,,\quad c_r^{(J)}=\ds\frac {2\pa{r+2}}{\pa{r+3}!}\,,
\ee
where, like before, $E,L_i$ are respectively energy and angular momentum, $I^{ijR}$ ($J^{ijR}$) are generic electric (magnetic) source $2^n$-poles for $n\geq 2$, $n=r+2$, ${\cal E}_{ij}$ and ${\cal B}_{ij}$ denoting respectively the electric and magnetic part of the Riemann tensor and $R = i_1\cdots i_r$ the multi-index notation.

In case one is interested in applications to compact binary systems, the source multipoles appearing in (\ref{eq:smult}) can be explicitly related to individual constituents' parameters by means of a matching procedure, as done up to 2PN for the mass quadrupole $I^{ij}$ within the EFT approach in \cite{Maia2020}, and to higher orders within the multipolar Minkowskian formalism; See Refs.~\cite{Faye2015,Marchand2020,blanchet2005,blanchet2008,henry2021}.

In the present chapter, we are mainly interested in the  universal properties (i.e., not depending on the short-scale features of the source) of the gravitational waveform, so our focus will not be on the matching procedure, but rather on the study of emission amplitudes, as expressed in terms of the generic multipoles $I^{ijR}$ and $J^{ijR}$, with particular emphasis on the divergences appearing in dimensional regularization, and on the associated logarithmic terms. We are also not studying here conservative effects associated to emission and re-absorption of radiative modes, for which we refer to Refs.~\cite{RSSFhereditary,Almeida2020,Almeida2021b}.

We work in the harmonic gauge, as in \cite{blanchetlivrev,bernard2016}, which is equivalent to using the following form for the pure (bulk) gravity action 
\be\label{az_bulk}
S_{\rm bulk}= 2 \Lambda^2 \int {\rm d}^{d+1}x\sqrt{-g}\left[ R(g)-\frac{1}{2}\Gamma_\mu\Gamma^\mu\right]\,,
\ee
where $R(g)$ is the Ricci scalar, $\Gamma^\mu \equiv g^{\rho\sigma}\Gamma^\mu_{\rho\sigma}$, being $\Gamma^\mu_{\rho\sigma}$ the standard Christoffel
coefficients,
and $\Lambda^{-2}\equiv 32 \pi G_N \mu^{3-d}$. Note that for
the number of purely spatial dimensions $d\neq 3$ an
inverse length $\mu$ appears, as it is necessary to relate $\Lambda$,
which has dimensions $(\mathrm{mass}/\mathrm{length}^{d-2})^{1/2}$, to the ordinary 3+1 dimensional
Newton's constant $G_N$.

In the work presented in this chapter we also find it useful to decompose the metric via a Kaluza-Klein parameterization presented in the previous chapter, in Eq.~\eqref{KKdecompward}. 
In particular, in this decomposition one can write at linear order
\be
\ba{rcl}
\ds\Lambda{\mathcal E}_{ij}&\simeq&\ds-\frac 12\pa{\ddot\sigma_{ij}-\dot A_{i,j}-\dot A_{j,i}}
+\phi_{,ij}+\frac{\delta_{ij}}{d-2}\ddot\phi+O(h^2)\\
\ds\Lambda{\mathcal B}_{ij}&\simeq&\ds\frac 14\epsilon_{ikl}\paq{\dot\sigma_{jk,l}-\dot\sigma_{jl,k}+A_{l,jk}-A_{k,jl}+\frac 2{d-2}\pa{\dot\phi_{,k}\delta_{jl}-
\dot\phi_{,l}\delta_{jl}}}+O(h^2)\,,
\ea
\ee
where $h$ denotes the generic metric perturbation around Minkowski spacetime and overdots represent time derivatives.

\begin{figure}
\centering
\begin{tikzpicture}[baseline={([yshift=-.5ex]current bounding box.center)},scale=1.7]
      \draw [black, thick] (0,0) -- (2,0);  
      \draw [black, thick] (0,-0.07) -- (2,-0.07); 
      \draw[decorate, decoration={snake},segment length=6.5pt,segment amplitude=1.2pt, line width=1pt,shorten >= 0.8pt] (1,0) -- (1,1) node [pos=0.53, label=right:{$(\omega,\K)$}] {}; \
      \filldraw[black] (1,-0.035) circle (2pt) node[anchor=north] {$I\,,J$};
\end{tikzpicture}
\caption{Feynman diagram representing the leading order emission amplitude.} 
\label{fig:emLO}
\end{figure}
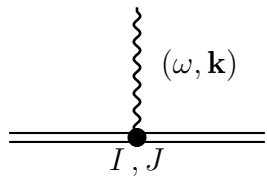

The radiative, transverse-traceless part of the metric
perturbation corresponds to the transverse-traceless part of $\sigma_{ij}$
(also denoted $\sigma_{ij}$ for simplicity) and the leading order amplitude
for emission of gravitational mode with on-shell 4-momentum $(\omega,\K)$,
with $\omega^2=\K^2$, by a generic electric ($I$) or magnetic ($J$) multipole
can be written as
\begin{equation}
\label{eq:emLO}
i \mathcal{A}_0(\omega,\K)=
\sum_r\frac{(-i)^{r+1}}{2\Lambda}\sigma_{ij}^{*}(\omega,\K) k_{R}\left[
{c_r}^{(I)} \omega^2 I^{ijR}(\omega)
+
c_r^{(J)}\omega\epsilon_{ikl} k_l J^{jkR}(\omega)
\right]\,,
\end{equation}
with its corresponding Feynman diagrams in Fig.~\ref{fig:emLO}.

By applying standard tools for Feynman diagram computations one can derive $O(G_N)$ and $O(G_N^2)$ corrections to the emission amplitude in Eq.~(\ref{eq:emLO}), which will be shown in the following sections. Explicit expression for propagators and interaction vertices can be read from \cite{RSSF4PNa} and will not be reported here, with the only modification that for emission processes retarded Green's functions have to be used, which can be represented as
\be
\label{eq:rgf}
G_R(\omega,\K)=\lim_{\ta\to 0^+}\frac 1{\pa{\omega+i\ta}^2-\K^2}\,,
\ee
and in all propagators in the rest of this chapter we will denote by $\ta$ an arbitrary small\footnote{Here, we have introduced this new notation to not be confused with the $\epsilon$ that parameterizes divergences, which, in their turn, play a particularly crutial role in the present chapter.}, positive quantity. The gravitational field can be obtained (at leading order) in Fourier space by multiplying the (leading order) amplitude (\ref{eq:emLO}) by the retarded Green's function (\ref{eq:rgf}), as it is causally determined by the source. Boundary conditions are specified by the pole displacement in the inverse space representation of the Green's function, hence their effect shows up only for the region of momenta having $|\K|=\omega$.\footnote{Note that in \cite{goldberger2010} Feynman Green's functions have been adopted instead. As pointed out in \cite{GalleyRoss2016}, such prescription does not generally allow to obtain the correct imaginary part of the amplitude (see also footnote before Eq.~(\ref{eq:tailE_bis})).}

\subsection{Tails}
\label{ssec:met_tail}
The computation of the tail amplitude involving the energy and the electric quadrupole was first derived in \cite{blanchet1988} and it has been re-derived in \cite{goldberger2010} with effective field theory methods; here we report the results involving generic electric and magnetic multipoles, as represented in Fig.~\ref{fig:tail}, as a warm up for subsequent calculations. Note that the gravitational mode attached to the conserved energy $E$ has vanishing time component.

\begin{figure}
\centering
\begin{equation}
i{\cal A}_{\rm tail}(\omega,\K) = \,\,
      \begin{tikzpicture}[baseline={([yshift=-.5ex]current bounding box.center)},scale=1.2]
      \draw [black, thick] (0,0) -- (3,0);
      \draw [black, thick] (0,-0.07) -- (3,-0.07);
      \draw[decorate, decoration={snake},segment length=6.5pt,segment amplitude=1.2pt, line width=1pt] (0.6,0) -- (1.5,1) node [pos=0.6, label=left:{$(\omega,\Q)$}] {};
      \draw[dashed, thick,line width=1pt] (2.37,0.03) -- (1.5,1) node [pos=0.56, label=right:{$\!\!(0,\K-\Q)$}] {};
      \draw[decorate, decoration={snake},segment length=6.5pt,segment amplitude=1.2pt, line width=1pt,shorten >= 0.8pt] (1.5,1.01) -- (1.5,1.8) node [pos=0.5, label=right:{$\!\!(\omega,\K)$}] {};
      \filldraw[black] (0.6,-0.035) circle (2pt) node[anchor=north] {$I\,,J$};
      \filldraw[black] (2.4,-0.035) circle (2pt) node[anchor=north] {$E$};
      \filldraw[black] (1.5,1) circle (2pt) node[anchor=north] {};
    \end{tikzpicture}\nn\,,
\end{equation}
\caption{Feynman diagram representing the tail emission amplitude.}
\label{fig:tail}
\end{figure}


Recalling the notation $\int_\Q\equiv \int \frac{{\rm d}^dq}{\pa{2\pi}^d}$,
in the electric case one has ($\omega^2=\K^2$)
\begin{align}
\label{eq:tailE}
i{\cal A}^{(e)}_{r-tail}(\omega,\K) = &(-i)^{r+1} \left(\frac{E c_r^{(I)}}{4\Lambda^3}\right)I^{ijR}(\omega)
\int_{\Q}\frac 1{\paq{\Q^2-\pa{\omega+i\ta}^2}}\frac 1{\pa{\K-\Q}^2}\times
q_{R}\nonumber\\
&\times\paq{\omega^4\delta_{ai}\delta_{bj}+2\omega^2q_i\pa{k-q}_a\delta_{bj}
+\frac {2}{c_d}q_iq_j\pa{k-q}_a\pa{k-q}_b} \sigma_{ab}^{*}(\omega,\K)\nn\\
  \simeq& i{\cal A}_{r0}^{(e)}(\omega,\K)\pa{i G_N E \omega}
  \paq{-\frac{\pa{\omega+i\ta}^2}{\tilde{\mu}^2}}^{\epsilon_{IR}/2}
        \paq{\frac{2}{\epsilon_{IR}}-2\kappa_{r+2}+O(\epsilon_{IR})}\,,
\end{align}
where ${\cal A}_{r0}^{(e)}$ is the electric part of the $2^{2+r}$-multipole in Eq.~(\ref{eq:emLO}), $\epsilon\equiv d-3$, $\tilde{\mu}^2\equiv \pi\mu^2e^{-\gamma_E}$, with $\gamma_E$ the Euler-Mascheroni constant,
\be
\kappa_{r+2}\equiv\frac{2r^2+13 r+22}{(r+2)(r+3)(r+4)}+H_r\,,
\ee
and $H_r$ is the $r$th \emph{Harmonic number} defined by $H_r\equiv \sum_{i=1}^r 1/i$. The second line in Eq.~(\ref{eq:tailE}) is determined by the bulk interactions of the tail diagram, which depends on the $\sigma^2\phi$, $\sigma A\phi$, and $\sigma\phi^2$ interactions contained in the Einstein-Hilbert action. Expanding also the factor $\paq{-\pa{\omega+i\ta}^2/\tilde\mu^2}^{\epsilon/2}$ in (\ref{eq:tailE}) for $\epsilon\to 0$, reminding the cut in the negative real semi-axis of the $\omega$ complex plane, one finally gets\footnote{Note the presence of the ${\rm sign}(\omega)$ term in Eq.~(\ref{eq:tailE_bis}), which is necessary to ensure that the tail corrections satisfy the reality property $\mathcal{A}^*(\omega)=\mathcal{A}(-\omega)$, to ensure a real waveform in direct space. Had one used Feynman Green's function, one would have had $(\omega^2+i\ta)$ replacing $\pa{\omega+i\ta}^2$ in Eq.~\eqref{eq:tailE}, then obtaining $-i\pi$ instead of $-i\pi{\rm sign}(\omega)$ in Eq.~(\ref{eq:tailE_bis}).}
\be
\label{eq:tailE_bis}
i{\cal A}^{(e)}_{r-tail}(\omega,\K) \simeq i {\cal A}_{r0}^{(e)}(\omega,\K)
\pa{i G_N E \omega}\paq{\frac{2}{\epsilon_{IR}}-2\kappa_{r+2}-i\pi\, {\rm sign}(\omega)+
\log\pa{\frac{\omega^2}{\tilde\mu^2}}}\,.
\ee

An analogous calculation for the magnetic multipole gives
\begin{align}
\label{eq:tailM}
i{\cal A}^{(m)}_{r-tail}(\omega,\K) =& (-i)^{r+1} \left(\frac{E c_r^{(J)}}{4\Lambda^3}\right)\omega\epsilon_{ikl} J^{jkR}(\omega)
\int_{\Q}\frac 1{[\Q^2-\pa{\omega+i\ta}^2]}\frac 1{\pa{\K-\Q}^2}\times
q_{R}\nn\\
&\times q_l\paq{\omega^2\delta_{aj}+q_j\pa{k-q}_a}\sigma_{ai}^{*}(\omega,\K)\nn\\
 \simeq& i{\cal A}_{r0}^{(m)}(\omega,\K)\pa{i G_N E\omega}
\paq{\frac 2{\epsilon_{IR}}-2\pi_{r+2}-i\pi\, {\rm sign}(\omega)+
\log\pa{\frac{\omega^2}{\tilde\mu^2}}}\,,
\end{align}
with
\be
\pi_{r+2}\equiv\frac{r+1}{(r+2)(r+3)}+H_{r+1}\,.
\ee
The integrals have been computed using the formulae reported in Appendix \ref{sec:app},
and the divergences encountered here are of the IR type, hence the index
``IR'' to $\epsilon$ in Eqs.~(\ref{eq:tailE},\ref{eq:tailE_bis},\ref{eq:tailM}). They are the leading order of an unobservable divergent phase term common to all multipoles; the finite terms proportional to $\kappa_{r+2},\pi_{r+2}$ (first computed in \cite{blanchet1995b}) are also exponentiated to a phase \cite{Porto2012}, which is however multipole-dependent, so in principle observable. Note that the contribution of the $-i\pi\, {\rm sign}(\omega)$ term in the square brackets is real relative to ${\cal A}_0$, hence it is the only contribution from the tail process to the emission flux at $G_NE\omega\sim v^3$ order.

The amplitudes (\ref{eq:tailE}) and (\ref{eq:tailM}) are proportional to waveforms, hence they can be inverse-Fourier transformed to give the waveforms in the time domain, with the result that the logarithmic terms in $\omega$ are responsible for non-local terms in direct space (i.e., in time) first individuated in \cite{blanchet1988}. Note that the IR divergence arises from the loop integral displayed in Eq.~(\ref{eq:tailE}), as it is clearly shown by changing the integration
variable to $\Q'\equiv\Q-\K$
\be
\label{eq:tail_qual}
\left.{\cal A}_{\rm tail}\right|_{\rm IR-div}(\omega)\propto \int_{\Q'}\frac 1{\pa{2\K\cdot \Q'+{\Q'}^2}\Q'^2}\,,
\ee
and it is present only for terms whose numerator, which is set to unity for clarity in Eq.~(\ref{eq:tail_qual}), is non-vanishing for $\Q'\to 0$. An analog process can be considered by replacing the energy $E$ insertion of the tail diagram with the angular momentum $L$, which however comes with one gradient, i.e., one power of $\Q'$, see Eq.~(\ref{eq:smult}), thus having no divergence and producing a local result both in Fourier and in direct space, as it can be explicitly checked in \cite{blanchet2008}; for this reason it has been dubbed ``failed'' angular momentum tail in \cite{RSSFhereditary}.

Another qualitatively different process, the (nonlinear) \emph{memory}, can be considered
at $O(G_N)$ order. It can be obtained by replacing the conserved quantity source insertion of
the tail diagram ($E$ or $L$) with a time-dependent multipole $I'$ or $J'$,
giving rise to an amplitude of the type
\be
\mathcal{A}_{\rm memory}(\omega)\propto\int\frac{d\omega'}{2\pi}\int_\Q
\frac{I(\omega-\omega')I'(\omega')}{\paq{\Q^2-\pa{\omega-\omega'+i\ta}^2}\paq{\pa{\K-\Q}^2-\pa{\omega'+i\ta}^2}}\,,
\ee
which is not divergent but gives rise to a product of (Fourier transformed)
dynamical multipoles, which in direct space involve a convolution in time
\cite{blanchet2008}.
In particular the contribution from
$I(\omega-\omega')I'(\omega')$ for $\omega\to 0$ gives rise to a
non-vanishing zero-frequency effect, the nonlinear memory effect \cite{christ1991}.

\section{Results for the tail-of-tail}
\label{sec:res}

We derive in this section the divergent and logarithmic parts of the more challenging tail-of-tail contributions, at second order in $G_N E \omega$ beyond leading order (equivalent to relative 3PN for binary systems), which is where UV divergences make their first appearance.

The tail-of-tail contribution to the radiative multipole has been derived in detail in \cite{blanchet2005b}, and in \cite{goldberger2010} for the electric quadrupole case only (terms $E^2\times I_{ij}$) within EFT methods, which we generalize in this section to the $E^2\times(I,J)$ case, for electric and magnetic multipoles of any order.

\begin{figure}
\begin{center}
      \begin{tikzpicture}[baseline={([yshift=-.5ex]current bounding box.center)}]
      \draw [black, thick] (0,0) -- (4.5,0);			
      \draw [black, thick] (0,-0.07) -- (4.5,-0.07);		
      \draw[decorate, decoration={snake},segment length=6.5pt,segment amplitude=1.2pt, line width=1pt] (0.7,0) -- (2.25,1.575) node [pos=0.6, label=left:${\bf q}$] {};
      \draw[dashed, thick,line width=1pt] (3.77,0.03) -- (2.20,1.6) node [pos=0.85, label=right:$\!\!{\bf k-q}$] {} node [pos=0.33, label=right:$\!\!{\bf k-p}$] {}; 
      \draw[dashed, thick,line width=1pt] (2.26,0.03) -- (2.965,0.78) node [pos=0.7, label=left:${\bf p-q}$] {}; 
      \draw[decorate, decoration={snake},segment length=6.5pt,segment amplitude=1.2pt, line width=1pt,shorten >= 0.8pt] (2.25,1.585) -- (2.25,2.5) node [pos=0.5, label=right:\!\!${\bf k}$] {};
      \filldraw[black] (0.68,-0.035) circle (2pt) node[anchor=north] {$I\,,J$};		
      \filldraw[black] (2.25,-0.035) circle (2pt) node[anchor=north] {$E$};			
      \filldraw[black] (3.8,-0.035) circle (2pt) node[anchor=north] {$E$};			
      \filldraw[black] (2.23,1.575) circle (2pt) node[anchor=north] {};				
      \filldraw[black] (2.995,0.8) circle (2pt) node[anchor=north] {};				
      \node[] at (2.25,-1.4) {(a)};
    \end{tikzpicture}\nn
\hspace{0.7cm}
      \begin{tikzpicture}[baseline={([yshift=-.5ex]current bounding box.center)}]
      \draw [black, thick] (0,0) -- (4.5,0);			
      \draw [black, thick] (0,-0.07) -- (4.5,-0.07);		
      \draw[decorate, decoration={snake},segment length=6.5pt,segment amplitude=1.2pt, line width=1pt] (0.7,0) -- (2.25,1.575) node [pos=0.6, label=left:${\bf q}$] {};
      \draw[dashed, thick,line width=1pt] (3.77,0.03) -- (2.20,1.6) node [pos=0.56, label=right:$\!\!{\bf k-p}$] {}; 
      \draw[dashed, thick,line width=1pt] (2.25,0.03) -- (2.25,1.575) node [pos=0.25, label=left:${\bf p-q}\!\!\!\!$] {}; 
      \draw[decorate, decoration={snake},segment length=6.5pt,segment amplitude=1.2pt, line width=1pt,shorten >= 0.8pt] (2.25,1.585) -- (2.25,2.5) node [pos=0.5, label=right:\!\!${\bf k}$] {};
      \filldraw[black] (0.68,-0.035) circle (2pt) node[anchor=north] {$I\,,J$};		
      \filldraw[black] (2.25,-0.035) circle (2pt) node[anchor=north] {$E$};			
      \filldraw[black] (3.8,-0.035) circle (2pt) node[anchor=north] {$E$};			
      \filldraw[black] (2.23,1.575) circle (2pt) node[anchor=north] {};				
      \node[] at (2.25,-1.4) {(b)};
    \end{tikzpicture}\nn
\hspace{0.7cm}
      \begin{tikzpicture}[baseline={([yshift=-.5ex]current bounding box.center)}]
      \draw [black, thick] (0,0) -- (4.5,0);			
      \draw [black, thick] (0,-0.07) -- (4.5,-0.07);		
      \draw[decorate, decoration={snake},segment length=6.5pt,segment amplitude=1.2pt, line width=1pt] (0.7,0) -- (2.25,1.575) node [pos=0.32, label=left:${\bf q}$] {} node [pos=0.85, label=left:${\bf p}$] {};
      \draw[dashed, thick,line width=1pt] (3.77,0.03) -- (2.20,1.6) node [pos=0.6, label=right:${\bf k-p}$] {}; 
      \draw[dashed, thick,line width=1pt] (2.22,0.03) -- (1.475,0.78) node [pos=0.7, label=right:${\bf p-q}$] {}; 
      \draw[decorate, decoration={snake},segment length=6.5pt,segment amplitude=1.2pt, line width=1pt,shorten >= 0.8pt] (2.25,1.585) -- (2.25,2.5) node [pos=0.5, label=right:\!\!${\bf k}$] {};
      \filldraw[black] (0.68,-0.035) circle (2pt) node[anchor=north] {$I\,,J$};		
      \filldraw[black] (2.25,-0.035) circle (2pt) node[anchor=north] {$E$}; 			
      \filldraw[black] (3.8,-0.035) circle (2pt) node[anchor=north] {$E$};			
      \filldraw[black] (2.23,1.575) circle (2pt) node[anchor=north] {};				
      \filldraw[black] (1.47,0.78) circle (2pt) node[anchor=north] {};				
      \node[] at (2.25,-1.4) {(c)};
    \end{tikzpicture}\nn
\end{center}
\caption{Feynman diagrams describing the tail-of-tail process. We label explicitly in the figure only the space components of the momenta, the
time component being $\omega$, with $\omega^2=\K^2$, for wavy lines, and
vanishing for dashed straight lines.}
\label{fig:tail2}
\end{figure}
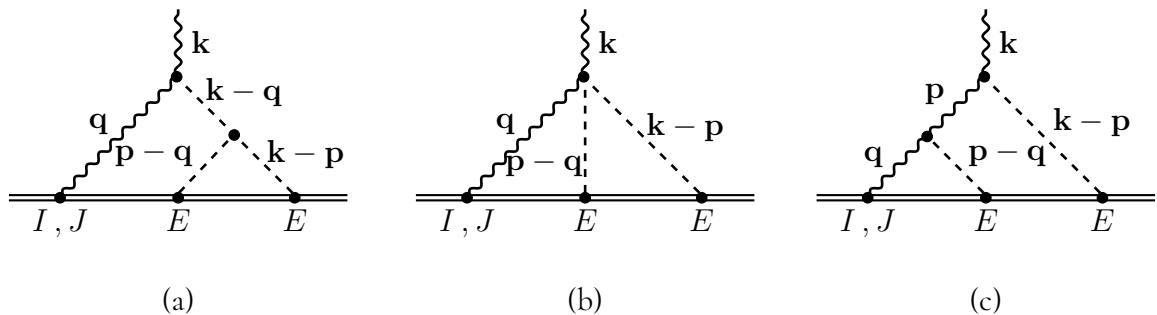

The tail-of-tail receives contributions from three different diagrams given in Fig.~\ref{fig:tail2}.
Diagrams (a) and (b) can be computed using standard integration techniques, bringing pure UV divergences for any multipole, as described in \cite{goldberger2010} for the quadrupole case, as it can be shown as follows. After the first loop integration over $\pp$, which can be performed via Eq.~\eqref{eq:app_std1}, and after dropping the tensor structure for clarity, one is left with an integral similar to the tail one of Eq.~(\ref{eq:tail_qual})
\be
\left.{\cal A}_{a,b-tail^2}\right|_{div}\sim\int_\Q\frac 1{\paq{\Q^2-\pa{\omega+i\ta}^2}\paq{\pa{\K-\Q}^2}^{m-d/2}}=
\int_{\Q'}\frac 1{\paq{2\K\cdot{\Q'}+{\Q'}^2}\paq{{\Q'}^2}^{m-d/2}}\,,
\ee
which however has the crucial difference from (\ref{eq:tail_qual}) of having $m$ a positive integer, giving a half-integer exponent for the ${\Q'}^2$ term, hence leading to a pure UV divergence when combined with the ${\Q'}^2$ part of the $(2\K\cdot\Q'+{\Q'}^2)$ propagator, and no IR divergence. As noted in \cite{goldberger2010}, such diagrams correspond to the scattering of the emitted radiation with the $1/r^2$ relativistic correction to the static potential.

The (a) and (b) diagrams in Fig.~\ref{fig:tail2} give for the electric and magnetic case (See Appendix \ref{apptailoftailabc} for detailed expressions)
\be
\label{eq:tail2_em_ab}
i {\mathcal A}^{(e,m)}_{a,b-tail^2}(\omega\,,\K) &\simeq&\ds i \mathcal{A}_{r0}^{(e,m)}(\omega\,,\K)\pa{G_N E\ \omega}^2\paq{-\frac{\pa{\omega+i\ta}^2}{\tilde{\mu}^2}}^{\epsilon_{UV}}\paq{\frac{\alpha^{(e,m)}_{a,b}(r)}{\epsilon_{UV}}+O(\epsilon^0)}\,,\\
\ds\alpha^{(e)}_a(r)&\equiv&\ds\frac{2r^3+3r^2-r+1}{(2 r-1) (2 r+1) (2 r+3) (2 r+5)}\,,\\
\ds\alpha^{(e)}_b(r)&\equiv&\ds -2\frac{\left(16 r^3+56 r^2+24 r -31\right)}{(2 r-1)
   (2 r+1) (2r +3) (2 r+5)}\,,\\
\alpha^{(m)}_a(r)&\equiv&\ds\frac{2r^3+11r^2+21r+17}{(2 r+1) (2 r+3) (2 r+5) (2 r+7)}\,,\\
\alpha^{(m)}_b(r)&\equiv&\ds -2\frac{\left(16 r^3+104 r^2+187 r +74\right)}{(2 r+1)
(2 r+3) (2r +5) (2 r+7) }\,.
\ee

For the more intricate diagram (c), which can be decomposed in terms of some
master integrals whose result can also be checked in Appendix \ref{apptailoftailabc}, we report its amplitude before integration,
split in terms of the gravitational polarization propagating in the internal wavy
lines of fig.~\ref{fig:tail2}c ($\omega^2=\K^2$).
For the electric case one has:
\be
\label{eq:c_e}
\ba{rllll}
\ds i{\cal A}^{(e)}_c(\omega\,,\K) &=&\ds (-i)^{r+1}&\ds
\pa{-\frac{E^2 c_r^{(I)}}{4\Lambda^5}}\omega^2 I^{ijR}(\omega)&\\
&&\ds\times\int_{\pp, \Q}&\ds\frac {q_{R}}{\paq{{\Q}^2-\pa{\omega+i\ta}^2}
\paq{{\pp}^2-\pa{\omega+i\ta}^2}(\pp-\Q)^2 (\pp-\K)^2}&\\
&&\ds\times \ds\sigma^*_{ab}(\omega,\K)&\ds
\Bigg\{-\frac{1}{2}\omega^4\delta_{ia}\delta_{jb}& \{\sigma^2\}\\
&&&\ds\quad+\omega^2\paq{q_aq_j-2p_a q_j +p_a p_j}\delta_{ib}&\{A\sigma\}\\
&&&\ds\quad -\frac{1}{c_d}q_i q_j p_a p_b &\{\phi^2\}\\
&&&\ds\quad+\frac{1}{c_d}q_i [q_j q_a -2 q_j p_a +p_j p_a]p_b&\{\phi A\}\\
&&&\ds\quad+\frac{1}{c_d}[(q-p)_i p_j p_a p_b-q_i q_j (q-p)_a q_b]&
\{\phi\sigma\}\\
&&&\ds\quad+q_i \paq{q_j p_b-q_bp_j + (\pp\cdot\Q)\delta_{bj}}p_a&\{A^2\}
\Bigg\}\\
&\simeq&\ds i \mathcal{A}_{r0}^{(e)}(\omega\,,\K)&\ds\pa{G_N E \omega}^2
\paq{-\frac{\pa{\omega+i\ta}^2}{\tilde{\mu}^2}}^{\epsilon}\paq{-\frac2{ \epsilon^2_{IR}}+\frac{\alpha^{(e)}_c(r)}{\epsilon}}&\,,\\
\ea
\ee
\begin{equation}
\label{eq:cc_e}
\alpha^{(e)}_c(r)\equiv 2\paq{(r+1)\frac{128r^6+1728r^5+8968 r^4+21490r^3+20607 r^2-1228 r-8628}{(r+2)(r+3)(r+4)(2r-1)(2r+1)(2r+3)(2r+5)(2r+7)}+2H_{r}}\,,
\end{equation}
and for the magnetic case:
\be
\label{eq:c_m}
\ba{rccll}
\ds i{\cal A}^{(m)}_c(\omega\,,\K) &=& (-i)^{r+1}&\ds
\pa{\frac{E^2 c_r^{(J)}}{8\Lambda^5}}\omega \epsilon_{ikl}J^{jkR}(\omega)&\\
&&\ds\int_{\pp,\Q}&\ds \frac {q_l q_{R}}{\paq{\Q^2-\pa{\omega+i\ta}^2}\paq{\pp^2-\pa{\omega+i\ta}^2}
\pa{\pp-\Q}^2 \pa{\pp-\K}^2}&\\
&&\ds\times \ds\sigma^*_{ab}(\omega,\K)&\ds
\Bigg\{+\omega^4 \delta_{ia}\delta_{jb}&\{\sigma^2\}\\
&&&\ds\quad+\omega^2 [ (q-p)_j p_a\delta_{ib}-p_i p_a\delta_{jb}-q_j (q-p)_a\delta_{ib} ]&\{A\sigma\}\\
&&&\ds\quad -\frac{1}{c_d} [p_i (q-p)_j + p_j (q-p)_i]p_a p_b&\{\phi \sigma\}\\
&&&\ds\quad -\frac{1}{c_d}q_j p_i p_a p_b&\{A\phi\}\\
&&&\ds\quad -q_j p_a [(\pp\cdot\Q)\delta_{bi} -p_i q_b]&\{A^2\}\Bigg\}\\
&\simeq&\ds i \mathcal{A}_{r0}^{(m)}(\omega\,,\K)&\ds\pa{G_N E \omega}^2
\paq{-\frac{\pa{\omega+i\ta}^2}{\tilde{\mu}^2}}^{\epsilon}\paq{-\frac2{ \epsilon^2_{IR}}+\frac{\alpha^{(m)}_c(r)}{\epsilon}}&\,,
\ea
\nonumber
\ee
\begin{equation}
\label{eq:cc_m}
\alpha^{(m)}_c(r)\equiv 4\paq{\frac{32r^6+448 r^5+2396 r^4+6268 r^3+8433 r^2+5430 r+1269}{(r+2)(r+3)(2r+1)(2r+3)(2r+5)(2r+7)(2r+9)}+H_{r+1}}\,.
\end{equation}

We highlight
that, contrarily to the single pole that contains both UV and IR divergences,
the double pole (due uniquely to the $\{\sigma^2\}$ contribution) is purely IR
and universal,\footnote{Here, by ``universal", we mean the fact that the coefficients of the IR divergences appearing in the tails series is the same for every multipole moment. This differs from the usual usage of this term in the context of group renormalization, the latter referring to the property that observables can be computed without the knowledge of the theory's UV physics.} as expected from the exponentiation of the simple tail IR
divergence.
Indeed, expanding the divergent phase at order $(G_NE\omega)^2$ one obtains schematically
\be
\label{eq:exp}
\ba{rcl}
\ds e^{iG_NE\omega\pa{\frac 2{\epsilon_{IR}}-2\rho^{(e,m)}}}&\simeq&\ds
1+iG_NE\omega\pa{\frac 2{\epsilon_{IR}}-2\rho^{(e,m)}}\\
&&\ds
-\pa{G_NE\omega}^2\pa{\frac 2{\epsilon^2_{IR}}-\frac{4\rho^{(e,m)}}{\epsilon_{IR}}+O(\epsilon^0_{IR})}
+O\pa{(G_NE\omega)^3}\,,
\ea
\ee
i.e., the knowledge of the $O(\epsilon^0_{IR})$ tail term, in Eq.~(\ref{eq:exp})
indicated generically with $\rho^{(e,m)}$ in the term linear in $G_NE\omega$, allows to isolate the simple pole IR
divergence of the tail-of-tail (quadratic piece in $G_NE\omega$), which in turn can be subtracted from
(\ref{eq:c_e}) and (\ref{eq:c_m}) to finally identify the UV one.

\section{Summary and Discussion}
\label{sec:sdisc}
The general structure of the emission amplitude, including post-Minkowskian multipolar corrections, is
\be
\label{eq:em_rad}
\ba{rcl}
\ds i \mathcal{A}(\omega,\K)&=&\ds e^{i\frac{\phi_{IR}(\omega)}{\epsilon_{IR}}}\sum_r\frac{(-i)^{r+1}}{2\Lambda}\sigma_{ij}^{*}(\omega,\K) k_{R}\\
&&\ds\qquad\times\left[{c_r}^{(I)} \omega^2 I^{ijR}_{rad}(\omega)+c_r^{(J)}\omega\epsilon_{ikl} k_l J^{jkR}_{rad}(\omega)\right]\,,
\ea
\ee
where $(I,J)^{jkR}_{rad}$ are the so-called \emph{radiative multipoles} and
\be
\phi_{IR}(\omega)\equiv 2G_N E \omega\pa{\frac{\omega^2}{\tilde{\mu}^2}}^{\epsilon_{IR}/2}
\ee
is the coefficient of the IR pole,
which is however unobservable because it represents a global phase shift common
to every multipolar contribution of the emission amplitude. Likewise unobservable is the logarithmic term
generated in $\phi_{IR}/\epsilon_{IR}$ at $\epsilon_{IR}^0$ order.

Differently from IR divergences, UV ones make their first appearence at second PM order, and have an important physical interpretation as they signal the breakdown of the point particle approximation for the composite object and must be regularized. Applying standard regularization and renormalization procedures one can obtain physical results from our UV divergent amplitude. Note that while such procedures have been first developed and are routinely used in \emph{quantum} field theory, they can be also applied here to our completely classical setting, as they depend on the \emph{field theory} nature of the problem.

The divergence can be absorbed in the definition of the
(divergent) \emph{bare} source multipoles $(I,J)^{ijR}_B$, related to
the renormalized, finite source multipoles $(I,J)^{ijR}_R$
by a divergent factor:
\be
\label{eq:bare}
I^{ijR}_B(\omega)=\paq{1-\frac{\beta^{(e)}(r)}{2\epsilon_{UV}}\pa{G_NE\omega}^2}I^{ijR}_R(\omega,\mu)\,,
\ee
and analogously for the magnetic multipoles.
From the calculation of the previous section we found
\begin{align}
\label{eq:beta_e}
\beta^{(e)}(r)&\equiv2\pa{\alpha^{(e)}_a+\alpha^{(e)}_b+\alpha^{(e)}_c-4\kappa_{r+2}}= -2\frac{15r^4+150 r^3+568r^2+965r+642}{(r +2)(r+3)(2r +3)(2r +5)(2r +7)}\,,\\
\beta^{(m)}(r)&\equiv2\pa{\alpha^{(m)}_a+\alpha^{(m)}_b+\alpha^{(m)}_c-4\pi_{r+2}}\nn\\
\label{eq:beta_m}
&=-2\frac{60 r^6+900 r^5+5535r^4+17306 r^3+28228 r^2+22101 r+5778}{(r+2)(r+3)(2r+1)(2r+3)(2r+5)(2r+7)(2r+9)}\,,
\end{align}
where the electric coefficients $\beta^{(e)}(r)$ have been first determined in \cite{blanchet1988}, and we have computed in this work for the first time the expression for the magnetic ones $\beta^{(m)}(r)$.

Substituting for $(I,J)$ in the amplitudes of the previous section the bare source multipoles $(I,J)_B$ in  Eq.~(\ref{eq:bare}), one finds finite expressions for the amplitudes in terms of the renormalized multipoles. Hence up to the second post-Minkowskian order, \emph{radiative multipoles} entering the physical amplitude (\ref{eq:em_rad}) can be related to renormalized source multipoles via
\be
\label{eq:Irr}
\ba{lcl}
\ds I^{ijR}_{rad}(\omega)&\simeq&\ds I_R^{ijR}(\omega,\mu)
e^{-2iG_NE\omega\kappa_{r+2}}\\
&&\ds\qquad\times\paq{1+\pi G_N |\omega|E+
\frac{\beta^{(e)}(r)}2\pa{G_N E\omega}^2\pa{\log\frac{\omega^2}{\tilde\mu^2}+{\cal O}(\epsilon^0)}}\,,
\ea
\ee
and analogously for the magnetic case. In this renormalization procedure, which relies on large scale physics and does not depend on the specific UV structure of the system, the finite ${\cal O}(\epsilon^0)$ contribution is left undetermined and must be fixed by comparison with experiments/observations or a fine-grained description of the source.

The leading order (real) tail correction $\pi G_N E|\omega|$ is multipole independent and is generated by the imaginary part of the $\epsilon_{IR}^{-1}\pa{-\pa{\omega+i\ta}^2}^{\epsilon_{IR}}$ term, which is finite for $\epsilon_{IR}\to 0$, as derived in subsection \ref{ssec:met_tail}. At the same post-Minkowskian order of the leading tail, there are further finite contributions, not displayed  in Eq.~(\ref{eq:Irr}), coming from the angular momentum failed tail and the memory effect, which for compact binaries are suppressed with respect to the leading order in the post-Newtonian expansion by a factor of $v^2$. The expression of such terms in time domain can be found in \cite{Faye2015} for the first multipoles ($r=0,1$). As to the finite phases proportional to $\kappa_{r+2}$ and $\pi_{r+2}$, they are in principle observable as discussed in the introduction because they are not universal, i.e., the renormalized radiative multipole have a phase characteristic of the particular multipole moment.

Note that as the physical emission amplitude is directly related to the radiative multipoles $(I,J)_{rad}$ and cannot depend on the arbitrary renormalization scale $\mu$, the renormalized multipoles must acquire at 2PM order a $\mu$ dependence to compensate the explicit dependence on $\mu$ of the expression (\ref{eq:Irr}), hence the argument $\mu$ added to $(I,J)_R$ already in Eq.~(\ref{eq:bare}).

This leads to the renormalization group equation
\be
\frac{{\rm d}I^{ijR}_R(\omega,\mu)}{{\rm d}\log\mu}=\beta^{(e)}(r) \pa{G_NE\omega}^2I^{ijR}_R(\omega,\mu)\,,
\ee
 which is solved by \cite{goldberger2010}
\be
I^{ijR}_R(\omega,\mu)=\pa{\frac{\mu}{\mu_0}}^{\beta^{(e)}(r)\pa{G_NE\omega}^2}I^{ijR}_R(\omega,\mu_0)\,,
\ee
and analogously for the magnetic multipoles $J^{ijR}_R(\omega,\mu)$.

The above equations make manifest the role of $\beta^{(e,m)}(r)$  as beta functions controlling the running of the radiative multipoles. The renormalization group equation of the electric quadrupole \cite{BlanchetRSSF2019} has been used to re-sum an infinite series of leading logarithmic terms in the gauge invariant expression for energy and angular momentum of compact binaries. While the phenomenological impact for gravitational waveforms is expected to be modest (we remind that the lowest order UV logarithms enter the waveform is 3PN), with the beta functions known at all multipole orders it is possible to compute the leading logarithmic terms in the energy, which are of the type $(M^2\log)^n\times \pa{d^{n+2}I^{ijR}/dt^{n+2}}^2$, at sub-leading PN orders. This allows the possibility of additional, highly non-trivial checks with the PN-expanded version of extreme mass ratio results, in analogy to what is done in \cite{BlanchetRSSF2019} at leading PN order, where terms given in \cite{kavanagh2015} for $n\leq 7$ (contributing to the energy of circular orbit up to 22PN order) could be explicitly checked.

Knowledge of all the beta functions allows for an extension of this approach: in particular, before the present work, only the first magnetic coefficient $\beta^{(m)}(0)$ coefficient was known and found to be equal to the electric one; according to our finding this equality is accidental and does not hold for other multipoles.

In the case of compact binaries, alternatively to the universal renormalization procedure, one can exploit the explicit knowledge of the system at small scales, as it has been done in the 4PN study of the conservative sector \cite{RSSF4PNb}, to cancel the UV divergence from the multipolar dynamics (also called far zone) with an IR divergence coming from the PN-expanded dynamics of individual binary components interacting via the exchange of longitudinal gravitational modes (near zone). In this case, the cancellation comes from the explicit determination of the \emph{source} multipoles in terms of the binary constituents' variables at 3PN order, and the previously undetermined ${\cal O}(\epsilon^0)$ term appearing in (\ref{eq:Irr}) is then unambiguously predicted in terms of the UV details of the system \cite{blanchet4PNrada,blanchet4PNradb}.

%% file: chapterx/perspectives.tex
\chapter{Conclusions and Perspectives}

As emphasized in the introduction of this thesis and also in the introduction of Chapter \ref{wardidentity}, one of the most important puzzles nowadays in the context of 
the two-body dynamics in general relativity
lies in understanding how to consistently take into account radiation-reaction effects to the compact binary dynamics at the 5PN order. This order is physically important since it is the first in which finite size effects start to contribute and that might give information about, for instance, the internal structure of neutron stars. We also mentioned that there are important physical effects that also start to contribute at this order, namely the memory effect and other higher-order tail contributions beyond the leading electric quadrupole case and, in particular, since these higher-order dissipative effects are highly non-trivial in nature, one has to be very careful on how to treat such time-asymmetric phenomena at the level of a path integral formulation.          

A thorough investigation of higher-order hereditary effects, both from the use of the in-in formalism in the context of NRGR and from the study of the scattering angle in the post-Minkowskian approximation (in which the in-in formalism continues to be important), considering, in particular, the memory interaction, is the main path we are currently following.
It is worthwhile mentioning at this point that the conservative dynamics at post-Minkowskian order $G_N^4$ was claimed to have been completed in Ref.~\cite{zvi2022}, which, in principle, include the memory effect, using scattering amplitude methods. Nevertheless, in this work, the authors used the unconventional prescription for the graviton propagators of taking its principal value, which, as argued in the paper, is equivalent to just accounting for the real part of the final classical amplitude computed with the Feynman boundary condition. This prescription, however, is usually not sufficient to include all the conservative terms stemming from dissipative phenomena \cite{porto2022}, and hence, the result obtained in Ref.~\cite{zvi2022} should not be fully trusted. Instead, a complete analysis should always rely on the in-in formalism.

As we have learned in Chapter \ref{wardidentity}, due to the simplicity found in the type of loop integrals involved in the tail effects, including the angular momentum failed-tail, all the information needed to fully understand radiation-reaction effects, in this case, is contained within the on-shell emission amplitudes. And, in particular, we saw that some of these emission diagrams were plagued with anomalies, but that could be easily fixed by the introduction of counter-terms, which is equivalent to including a homogeneous solution to Einstein's equation, in order to turn such \textit{bare} solutions into valid solutions of general relativity. 

It is important to notice that the appearance of anomalies in the quadrupole cases of the angular momentum failed-tail is by no means an exception. We expect that, as we move toward higher-order processes, the immediate compatibility of our bare solutions with the gauge condition equation will generally not hold, and hence the complete solution will be attained once we supplement our bare solution with a counter-term. In particular, the correction to gauge-violating amplitudes constitutes an essential step toward the complete account of radiation reaction for a given process, currently needed in the 5PN dynamics. 

As also discussed in the same chapter, for the memory effect, which possesses loop integrals with a  different propagator structure with respect to the tails, it is no longer enough to consider on-shell emission amplitudes. In this case, instead, a complete account of the off-shell version of the memory is needed. Thus, in order to tackle this problem and fully include the radiation-reaction contribution coming from it, with a special focus on the 5PN order, which involves only the quadrupolar memory, we must reformulate what we have learned in Chapter \ref{wardidentity} to account for the full range of four-momenta, to cover off-shell contributions.

In face of the 5PN program, the original framework presented in Chapter \ref{wardidentity} provides us with an indispensable tool for completing the 5PN order computation. After all, as previously discussed, past solutions for such processes are simply not fully compatible with general relativity. Besides this, it is worthwhile clarifying that, for any process, of any order, even for processes such as the memory, the formalism presented in this chapter gives us a way of computing the waveform in the distant field approximation, since they are directly computed from on-shell emission amplitudes, which are useful in the computation of gravitational-wave observables such as the energy and angular-momentum fluxes.

Finally, we are confident that the tools learned in Chapter \ref{wardidentity} constitute important advances in our current understanding of several aspects of our EFT for binary systems, including how to consistently compute self-energy and emission diagrams for higher-order processes, as well as the connection that these two types of diagrams have between themselves. We hope to finally solve the 5PN binary system dynamics in the near future by exploring the developments brought out in this thesis and specializing in the case of the memory effect.

%% file: appendices/svt.tex
\chapter{The scalar-vector-tensor (SVT) decomposition}\label{svtapp}

In this appendix we will study the scalar-vector-tensor decomposition of the metric perturbations in linearized gravity. We will follow \cite{flanagan} closely and will find that
out of the six independent components of the metric perturbation $h_{ab}$ satisfying the harmonic gauge condition, only two degrees of freedom are radiative (the transverse-traceless $h^{TT}_{ij}$) and four are longitudinal and tied to the matter sources. Out of these last four degrees of freedom, two of them are associated to a transverse and gauge-invariant vector and the remaining two are associated with two gauge-invariant scalar functions each. Only $h^{TT}_{ij}$ satisfies a wave equation in all gauges; the other aforementioned functions obey Poisson-type equations. 

The goal of the SVT decomposition in the case of linearized gravity is to focus our attention on gauge-invariant quantities.

\section{The SVT decomposition of the metric perturbation}

Let us decompose the metric perturbation $h_{\mu\nu}$ into a set of four scalars ($\phi, \gamma, H, \lambda$), two vectors $(\varepsilon_i, \beta_i)$ and one tensor ($h^{TT}_{ij}$): 
\begin{align}
    h_{00} &= -2\phi, \label{h00 SVT}\\
    h_{0i} &= \beta_i + \partial_{i} \gamma. \label{h0i SVT} \\
    h_{ij} &= h_{ij}^{TT} + \frac{1}{3} H \delta_{ij} + \partial_{(i}\varepsilon_{j)} + \left( \partial_{i} \partial_{j} - \frac{1}{3} \delta_{ij} \nabla^2  \right) \lambda \label{hij SVT}, 
\end{align}
together with the constraints
\begin{subequations}\label{constraints metric SVT}
\begin{align}
    \partial_i \beta_i = 0, \\
    \partial_i \varepsilon_i = 0, \\
    \partial_i h_{ij}^{TT} = 0, \\
    \delta^{ij} h_{ij}^{TT} = 0. 
\end{align}
\end{subequations}
As $r \to \infty$ we also impose that
\begin{equation}
    \gamma \to 0, \quad \varepsilon_i \to 0, \quad \lambda \to 0, \quad \nabla^2 \lambda \to 0.
\end{equation}
In our notation, $H$ is the trace of the spatial piece of the metric perturbation: $H = \delta^{ij} h_{ij}$.

\subsection{Uniqueness of the decomposition}

The uniqueness of this decomposition follows from a few observations. First, notice that if we take the $\partial_{i}$ derivative of equation \eqref{h0i SVT} and use \eqref{constraints metric SVT}, we get 
\begin{equation}
    \nabla^2 \gamma = \partial_{i} h_{0i}. 
\end{equation}
This equation can be uniquely solved for $\gamma$. Now take the $\partial_{i}\partial_{j}$ derivative of \eqref{hij SVT} and use the constraints to deduce that 
\begin{equation}
    \nabla^2 \nabla^2 \lambda = \frac{3}{2} \partial_{i} \partial_{j} h_{ij} - \frac{1}{2} \nabla^2 H. 
\end{equation}
With $\gamma$ determined, we can uniquely solve the equation above to obtain $\lambda$. 
Finally, we take the $\partial_j$ derivative of \eqref{hij SVT} to obtain
\begin{equation}
    \nabla^2 \varepsilon_i = 2 \partial_j h_{ij} - \frac{2}{3} \partial_i H - \frac{4}{3} \partial_i \nabla^2 \lambda. 
\end{equation}
This uniquely determines the vector $\varepsilon_i$. 

Let us count the number of independent variables. We have $4$ scalar components, $6$ vector components and $6$ tensor components from $h^{TT}_{ij}$, which gives a total of $16$ components. We also have $1 + 1 + 1 + 3 = 6$ constraints. Therefore, we have $16 - 6 = 10$ independent variables, as expected from a symmetric four-dimensional tensor.  

\subsection{Gauge transformations}

Consider a gauge transformation of the form
\begin{equation}\label{generic gauge transformation for x}
    x^a \to x^a + \xi^a,
\end{equation}
where $\xi^a \to 0$ as $r \to \infty$. Let us now decompose the spatial part of $\xi^a$ into a transverse and a longitudinal piece as follows:
\begin{equation}\label{SVT gauge transformation}
    \xi_a = (\xi_0, \xi_i) = (A, B_i + \partial_i C), 
\end{equation}
where $\partial_i B_i = 0$ and $C \to 0$ as $r \to \infty$. A simple computation shows that under the transformation \eqref{generic gauge transformation for x} the metric perturbation transforms as 
\begin{equation}
    h'_{ab} = h_{ab} - 2 \partial_{(a} \xi_{b)}. 
\end{equation}
Now we determine how each of the different pieces of the metric perturbation change under the gauge transformation determined by \eqref{SVT gauge transformation}. 

\paragraph{Investigating $h'_{00}$}

\ 

We find that 
\begin{equation}
    h'_{00} = -2 \phi' = -2\phi - 2 \Dot{A}. 
\end{equation}
Thus 
\begin{equation}
    \phi' = \phi + \Dot{A}.
\end{equation}

\paragraph{Investigating $h'_{0i}$}

\ 

Now we have
\begin{equation}
    \begin{split}
        h'_{0i} &= h_{0i} - \partial_0 \xi_i - \partial_i \xi_0 \\
                &= \beta_i + \partial_i \gamma - \partial_0 (B_i + \partial_i C) - \partial_i A \\
                &= \beta_i - \Dot{B}_i + \partial_i(\gamma - \Dot{C} - A).
    \end{split}
\end{equation}
Therefore
\begin{align}
    \beta´_i &= \beta_i - \Dot{B}_i, \\
    \gamma'  &= \gamma - A - \Dot{C}. 
\end{align}

\paragraph{Investigating $h'_{ij}$}

\

At last, we have
\begin{equation}
    \begin{split}
        h'_{ij} &= h_{ij}^{TT} + \frac{1}{3} H \delta_{ij} + \frac{1}{2}(\partial_i \varepsilon_j + \partial_j \varepsilon_i) + \left(\partial_i \partial_j - \frac{1}{3}\delta_{ij} \nabla^2 \right)\lambda - \partial_i(B_j + \partial_j C) - \partial_j (B_i + \partial_i C) \\[5pt]
                &= h_{ij}^{TT} + \frac{1}{3} H \delta_{ij} + \frac{1}{2}[\partial_i(\varepsilon_j-2B_j)+\partial_j(\varepsilon_i - 2B_i)] + \partial_i \partial_j(\lambda - 2C) - \frac{1}{3} \delta_{ij} \nabla^2 \lambda \\
                & \quad +\frac{2}{3}\delta_{ij}\nabla^2 C - \frac{2}{3}\delta_{ij}\nabla^2 C
                \\[5pt]
                &= h_{ij}^{TT} + \frac{1}{3} \delta_{ij}(H - 2 \nabla^2 C) + \left( \partial_i \partial_j - \frac{1}{3} \delta_{ij} \nabla^2 \right)(\lambda - 2C) + \frac{1}{2} [\partial_i(\varepsilon_j - 2B_j) + \partial_j(\varepsilon_i - 2B_i)]. 
    \end{split}
\end{equation}
Therefore 
\begin{align}
    h_{ij}^{'TT}   &= h_{ij}^{TT}, \\
    H'             &= H - 2 \nabla^2 C, \\
    \varepsilon'_i &= \varepsilon_i - 2B_i, \\
    \lambda'       &= \lambda - 2C. 
\end{align}

As a result, we can see that the following combinations of these functions are gauge-invariant:
\begin{align}
    \Phi   &\equiv  \phi + \Dot{\gamma} - \frac{1}{2} \Ddot{\lambda}, \\
    \Theta &\equiv \frac{1}{3}(H-\nabla^2 \lambda), \\
    \Xi_i  &\equiv \beta_i - \frac{1}{2} \Dot{\varepsilon}_i.
\end{align}

The $h^{TT}_{ij}$ component of the metric perturbation is gauge-invariant by construction. In the Newtonian limit, the function $\Phi$ reduces to the ordinary Newtonian potential $\Phi_N$, while $\Theta = -2 \Phi_N$. We will confirm this claim explicitly after we compute the Einstein equations in the SVT decomposition. 

It should be also noticed that, as a result of the constraints \eqref{constraints metric SVT}, we have
\begin{equation}
    \partial_i \Xi_i = 0. 
\end{equation}
Let's count the number of independent gauge-invariant functions. We have $2$ scalar components from $\Phi$ and $\Theta$ and $2$ vector components from $\Xi_i$. This gives us $4$ functions. The $h^{TT}_{ij}$ has $6$ components minus $4$ constraints (traceless and transverse conditions), giving us a total of $2$ functions. Therefore, we have $6$ independent gauge-invariant functions, as expected from a symmetric four-dimensional tensor $h_{ab}$ already satisfying the harmonic gauge. 

\section{Einstein's equations}

\subsection{The SVT decomposition of the energy-momentum tensor}

Similarly to the metric perturbation case, we split the different components of $T_{ab}$ introducing a set of four scalars $(\rho,S,P,\sigma)$, two vectors $(S_i,\sigma_i)$ and one tensor $\sigma_{ij}$. We write 

\begin{subequations}
\begin{align}
    T_{00} &= \rho, \label{T00 SVT}\\
    T_{0i} &= S_i + \partial_{i} S. \label{T0i SVT} \\
    T_{ij} &= P \delta_{ij} + \sigma_{ij} + \partial_{(i}\sigma_{j)} + \left(\partial_{i} \partial_{j} - \frac{1}{3} \delta_{ij} \nabla^2\right) \sigma \label{Tij SVT}, 
\end{align}
\end{subequations}
together with the constraints 
\begin{subequations}\label{constraints stress-energy SVT}
\begin{align}
    \partial_i S_i = 0, \\
    \partial_i \sigma_i = 0, \\
    \partial_i \sigma_{ij} = 0, \\
    \delta^{ij} \sigma_{ij} = 0. 
\end{align}
\end{subequations}
As $r \to \infty$ we demand that
\begin{equation}
    S \to 0, \quad \sigma_i \to 0, \quad \sigma \to 0, \quad \nabla^2 \sigma \to 0.
\end{equation}

The quantities introduced above are not all independent! The variables $\rho, P, S_i$ and $\sigma_{ij}$ can be specified arbitrarily; stress-energy conservation then determines the remaining  variables $S, \sigma$ and $\sigma_i$. Let's prove this claim. 

The stress-energy conservation condition is 
\begin{equation}
    0 = \partial^a T_{ab} = \partial^0 T_{0b} + \partial^{i} T_{ib} 
\end{equation}
On one hand, when $b=0$ we have
\begin{equation}
    0 = - \Dot{\rho} + \partial_i(S^i + \partial_i S) = - \Dot{\rho} + \nabla^2 S. 
\end{equation}
Therefore, 
\begin{equation}\label{Continuity T - scalar eq 1}
     \nabla^2 S = \Dot{\rho}.
\end{equation}
On the other hand, if $b = j$ then
\begin{equation}
    \begin{split}
        0 &= \partial^0(S_j + \partial_j S) + \partial^i [P \delta_{ij} + \sigma_{ij} + \frac{1}{2} \partial_i \sigma_j + \frac{1}{2} \partial_j \sigma_i + \left(\partial_i\partial_j - \frac{1}{3}\delta_{ij}\nabla^2 \right)\sigma] \\
          &= - \Dot{S}_j - \partial_j \Dot{S} + \partial_j P + \frac{1}{2} \nabla^2 \sigma_j + \partial_j(\nabla^2 \sigma) - \frac{1}{3} \partial_j(\nabla^2 \sigma) \\
          &= \frac{1}{2} \nabla^2 \sigma_j - \Dot{S}_j + \partial_j \left( \frac{2}{3} \nabla^2 \sigma - \Dot{S} + P \right). 
    \end{split}
\end{equation}
Therefore, 
\begin{align}
    \nabla^2 \sigma_j &= 2 \Dot{S}_j, \label{Continuity T - vector eq}\\
    \nabla^2 \sigma   &= \frac{3}{2}(\Dot{S}-P) \label{Continuity T - scalar eq 2}.
\end{align}

\subsection{Components of the Einstein tensor}

Now we can start computing all components of the Einstein tensor in the SVT decomposition. For reference, the linearized Einstein tensor is given by 
\begin{equation}
        G_{ab} = - \frac{1}{2} (\Box h_{ab} + \eta_{ab} \partial^{c} \partial^{d} h_{cd} - \partial_a \partial^{c} h_{cb} - \partial_b \partial^c h_{ca}) + \frac{1}{2}(\eta_{ab} \Box h - \partial_a \partial_b h).
\end{equation}

It is not difficult to compute $G_{00}$ and $G_{0i}$ directly, so we will simply quote the final result:
\begin{align}
    G_{00} &= - \nabla^2 \Theta, \\
    G_{0i} &= - \frac{1}{2} \nabla^2 \Xi_i - \partial_i \Dot{\Theta}.
\end{align}

Instead, we will offer a step-by-step derivation of $G_{ij}$, since is not so obvious to compute. First, notice that when $a=i$ and $b=j$ the Einstein tensor becomes
\begin{equation}
    G_{ij} =- \frac{1}{2} (\Box h_{ij} + \eta_{ij} \partial^{c} \partial^{d} h_{cd} - \partial_i \partial^{c} h_{cj} - \partial_j \partial^c h_{ci}) + \frac{1}{2}(\eta_{ij} \Box h - \partial_i \partial_j h).
\end{equation}

\paragraph{Finding the trace $h$}

\begin{equation}
        h = \delta^{ij} h_{ij} - h_{00} = H - 2\phi.
\end{equation}

\paragraph{Finding $\partial^c \partial^d h_{cd}$}

\begin{equation}
    \begin{split}
        \partial^c \partial^d h_{cd} &= \partial^0 \partial^0 h_{00} + 2 \partial^0 \partial^i h_{0i} + \partial^i \partial^j h_{ij} \\[3pt]
                                     &= 2 \Ddot{\phi} - 2 \partial^{i} \Dot{h}_{0i} + \partial^i \partial^j h_{ij} \\[3pt]
                                     &= 2 \Ddot{\phi} - 2 \partial^{i}(\Dot{\beta}_i + \partial_i \Dot{\gamma}) + \partial^i \partial^j h_{ij} \\[3pt]
                                     &= 2 \Ddot{\phi} - 2 \nabla^2 \Dot{\gamma} + \partial^i \partial^j h_{ij}. 
    \end{split}
\end{equation}

\paragraph{Finding $\partial_i \partial^c h_{cj}$}

\begin{equation}
    \begin{split}
        \partial_i \partial^c h_{cj} &= \partial_i \partial^0 h_{0j} + \partial_i \partial^k h_{kj} \\[3pt]
                                     &= - \partial^i \Dot{h}_{0j} + \partial_i \partial^k h_{kj} \\[3pt]
                                     &= - \partial^i \Dot{\beta}_j - \partial_i \partial_j \Dot{\gamma} + \partial_i \partial^k h_{kj}. 
    \end{split}
\end{equation}

All that is left now is to combine everything together and insert $h_{ij}$ explicitly in the terms above. We eventually find 
\begin{equation}
    \begin{split}
    G_{ij} =&- \frac{1}{2} \Box h^{TT}_{ij} - \partial_{(i} \Dot{\Xi}_{j)} - \frac{1}{2} \partial_i \partial_j (2\Phi + \Theta) \\[3pt] 
            &- \frac{1}{2} \delta_{ij} \left[ \nabla^2(-2\Dot{\gamma}) + \frac{2}{3} \Ddot{H} - \frac{1}{3} \nabla^2 H + 2 \nabla^2 \phi \, +\frac{1}{3} \nabla^2 \Ddot{\lambda} + \frac{1}{3} \nabla^2 \nabla^2 \lambda \right].
    \end{split}
\end{equation}
To simplify that horrible thing inside the brackets you gotta write that $\frac{1}{3}$ factor as $1-\frac{2}{3}$ in order to introduce two different terms proportional to $\nabla^2 \Ddot{\lambda}$. This will allow you to write
\begin{equation}
     G_{ij} =  - \frac{1}{2} \Box h^{TT}_{ij} - \partial_{(i} \Dot{\Xi}_{j)} - \frac{1}{2} \partial_i \partial_j (2\Phi + \Theta) - \frac{1}{2} \delta_{ij}[-\nabla^2(2\Phi + \Theta)+ 2 \Ddot{\Theta}].    
\end{equation}

\subsection{Investigating the different components of the Einstein equation}

We can now write all the different components of the Einstein equation and equate the scalar, vector and tensor pieces of each side. However, due to the stress-energy conservation equations \eqref{Continuity T - scalar eq 1}, \eqref{Continuity T - scalar eq 2} and \eqref{Continuity T - vector eq}, it happens that not all of these equations are independent. 

To make things more transparent, let us deduce the relevant components of the Einstein equations step by step: 

The $00$ Einstein equation is simply 
\begin{equation}
    G_{00} = - \nabla^2 \Theta = 8\pi G_N T_{00} = 8 \pi G_N \rho.
\end{equation}
Therefore, 
\begin{equation}\label{EFE E1}
\nabla^2 \Theta = -8\pi G_N \rho.
\end{equation}
Now consider the $0i$ Einstein equation:
\begin{equation}
    \begin{split}
        G_{0i} = - \frac{1}{2} \nabla^2 \Xi_i - \partial_i \Dot{\Theta} &= 8\pi G_N (S_i + \partial_i S) \\[3pt]
                                                                          &= 8\pi G_N S_i + 8 \pi G_N \partial_i S.
    \end{split}
\end{equation}
Equating the scalar piece of the equation above yields
\begin{equation}
    \partial_i \Dot{\Theta} = - 8 \pi G_N \partial_i S. 
\end{equation}
Using \eqref{Continuity T - scalar eq 1}, we find 
\begin{equation}
    \nabla^2 \Dot{\Theta} = - 8\pi G_N \nabla^2 S = -8\pi G_N \Dot{\rho}.
\end{equation}
The equation above is simply the time derivative of the $00$ Einstein equation.

From the vector piece of the $0i$ Einstein equation we also find 
\begin{equation}\label{EFE E2}
 \nabla^2 \Xi_i = -16\pi G_N S_i. 
\end{equation}

The $ij$ Einstein equation is
\begin{equation}
    \begin{split}
        &- \frac{1}{2} \Box h^{TT}_{ij} - \partial_{(i} \Dot{\Xi}_{j)} - \frac{1}{2} \partial_i \partial_j (2\Phi + \Theta) - \frac{1}{2} \delta_{ij}[-\nabla^2(2\Phi + \Theta)+ 2 \Ddot{\Theta}] \\
        &= 8\pi G_N \left[ P \delta_{ij} + \sigma_{ij} + \partial_{(i}\sigma_{j)} + \left(\partial_{i} \partial_{j} - \frac{1}{3} \delta_{ij} \nabla^2\right) \sigma  \right].         
    \end{split}
\end{equation}
If we equation the two scalar pieces we get
\begin{subequations}
\begin{align}
    - \frac{1}{2} \delta_{ij} [-\nabla^2(2\Phi + \Theta) + 2\Ddot{\Theta}] &= \delta_{ij} 8\pi G_N \left(P - \frac{1}{3} \nabla^2 \sigma \right) \label{scalar piece no.1 of the ij EFE}\\ 
    - \frac{1}{2} \partial_i \partial_j (2\Phi + \Theta)                     &= 8\pi G_N \partial_i \partial_j \sigma \label{scalar piece no.2 of the ij EFE}.
\end{align}
\end{subequations}
On one hand, the contraction of the spatial indices in \eqref{scalar piece no.2 of the ij EFE} gives us 
\begin{equation}\label{contraction of the scalar piece no.2 of the ij EFE}
    - \frac{1}{2} \nabla^2 (2\Phi + \Theta) = 8\pi G_N \nabla^2 \sigma. 
\end{equation}
On the other hand, \eqref{scalar piece no.1 of the ij EFE} is
\begin{equation}\label{scalar piece no.1 of the ij EFE (no delta)}
    \frac{1}{2} \nabla^2 (2\Phi + \Theta) - \Ddot{\Theta} = 8\pi G_N P - \frac{1}{3} 8\pi G_N \nabla^2 \sigma.
\end{equation}
To deal with the $\Ddot{\Theta}$ term, we notice that the time derivative of the $00$ Einstein equation says that
\begin{equation}
    \nabla^2 \dot{\Theta} = -8\pi G_N \Dot{\rho} = -8\pi G_N \nabla^2 S.
\end{equation}
This means that 
\begin{equation}
    \nabla^2(\dot{\Theta} + 8\pi G_N S) = 0.
\end{equation}
Therefore, 
\begin{equation}
    \dot{\Theta} = - 8\pi G_N S \implies \Ddot{\Theta} = - 8\pi G_N \Dot{S}. 
\end{equation}
Using the result above and \eqref{Continuity T - scalar eq 2}, we can rewrite \eqref{scalar piece no.1 of the ij EFE (no delta)} as 
\begin{equation}
    \begin{split}
        -\frac{1}{2} \nabla^2 (2\Phi + \Theta) &= 8\pi G_n \left(-P + \dot{S} + \frac{1}{3} \nabla^2 \sigma \right) \\
                                                 &= 8\pi G_n \left( \frac{2}{3} \nabla^2 \sigma + \frac{1}{3} \nabla^2 \sigma \right)\\
                                                 &= 8\pi G_n \nabla^2 \sigma. 
    \end{split}
\end{equation}
This is exactly \eqref{contraction of the scalar piece no.2 of the ij EFE}.

This means that the only remaining Einstein equations we need to investigate are the vector and tensor pieces of the $ij$ component and \eqref{contraction of the scalar piece no.2 of the ij EFE} itself. Let us deal with \eqref{contraction of the scalar piece no.2 of the ij EFE} first. If we use \eqref{Continuity T - scalar eq 2} once again, this equations says that
\begin{equation}
    - \nabla^2 \Phi - \frac{1}{2} \nabla^2 \Theta = 8 \pi G_N \left(\frac{3}{2} \dot{S} - \frac{3}{2}P\right).
\end{equation}
Using \eqref{EFE E1}, we obtain
\begin{equation}\label{EFE E3}
\nabla^2 \Phi = 4\pi G_N (\rho + 3P - 3\dot{S}).
\end{equation}

The vector piece of the $ij$ Einstein equation is 
\begin{equation}
    - \partial_{(i} \dot{\Xi}_{j)} = 8\pi G_N \partial_{(i} \sigma_{j)}. 
\end{equation}
If we apply the $\partial^{i}$ derivative on both sides of the equation above and use the constraints $\partial_i \Xi_i = \partial_i \sigma_i = 0$, we obtain
\begin{equation}
    - \nabla^2 \dot{\Xi}_j = 8\pi G_N \nabla^2 \sigma_j = 16 \pi G_N \dot{S}_j.
\end{equation}
The equation above is simply the time derivative of \eqref{EFE E2}! 

This finishes the investigation of the independent pieces of the Einstein field equations. Collecting all our results, we have
\begin{align}
    \nabla^2 \Theta &= - 8\pi G_N \rho, \\
    \nabla^2 \Phi   &= 4\pi G_N (\rho + 3P - 3\dot{S}), \\
    \nabla^2 \Xi_i  &= - 16\pi G_N S_i, \\
    \Box h^{TT}_{ij}  &= -16\pi G_N \sigma_{ij}. 
\end{align}

Thus only the $h^{TT}_{ij}$ metric components characterize the radiative degrees of freedom in the spacetime. 


%% file: appendices/jfailedtails.tex
\chapter{$\beta$ coefficients for simple and $J$-failed tail and Ward identity computations}\label{appward}

In this appendix, we report the expressions for the $\beta$ coefficients that define the emission amplitudes for the simple and angular momentum failed tail, for generic multipole moments of either parity. We also develop on the explicit checking of the corresponding Ward identities.

\section{Simple tails}

\subsection{Simple tail for the electric multipole moments}

Computation of the simple tail for arbitrary multipole moments of the electric type leads to
\begin{equation}\label{tailelecbetaXXX}
i\mathcal{A}^{(I)}_{E-\rm tail} = \frac{(-i)^{r+1} c_r^{(I)}E}{\Lambda^3} I^{ijR}(\omega) \int_\Q \frac{(k+q)_{R}}{[(\K+\Q)^2-\omega^2]\Q^2}\times \left( \beta_{\sigma}^{ab} \sigma^{*}_{ab} + \beta_{A}^a A^{*}_a + \beta_{\phi} \phi^{*}\right)\,,
\end{equation}
with
\begin{align}
\beta_\sigma^{ab} &= \frac14 \omega^4 \delta_{ia} \delta_{jb} -\frac14 \omega^2 (2\delta_{ib} q_a- \delta_{ab} q_i) (k+q)_j \nonumber\\
	 &\qquad\qquad\qquad\qquad+\frac{1}{4c_d}(k+q)_i (k+q)_j \left[ 2(k+q)_a q_b - (\K+\Q)\cdot\Q \delta_{ab}\right] \,, \label{betastail} \\
\beta_A^a	&= -\frac12\omega^3 \delta_{ja} q_i  +\frac12 \omega (k+q)_j \left[ k_a q_i-k_i q_a + (\K+\Q)\cdot\K\delta_{ia}\right] \label{betaatail} \nonumber\\
	&\qquad\qquad\qquad\qquad+\frac{1}{2c_d} \omega (k+q)_i(k+q)_j q_a \,,\\
\beta_\phi	&=  -\frac12\omega^2 \left[ (k+q)_i (k+q)_j + q_i q_j \right] \label{betaptail} \,. 
\end{align}

From this, and using Eqs.~\eqref{genericemissA}, \eqref{AbarwrtalphaS}, \eqref{AbarwrtalphaA}, \eqref{AbarwrtalphaP} we can derive the expressions for the scattering amplitudes for the tail processes of electric parity. It helps to define the following functional first:
\begin{equation}
\mathcal{F}^{(I)}[f] \equiv \frac{(-i)^{r+2} c_r^{(I)}E}{4\Lambda^4} I^{ijR}(\omega) \int_\Q \frac{(k+q)_{R}}{[(\K+\Q)^2-\omega^2]\Q^2}\times f(\omega,\K,\Q)\,,
\end{equation}
for arbitrary function $f =f(\omega,\K,\Q)$.
Then, with this definition, we have
\begin{align}
i\bar{\mathcal{A}}_{00} &= \mathcal{F}^{(I)}\left[ \omega^2 (k_i k_j+k_i q_j+q_i q_j) + \frac{1}{c_d} (k+q)_i (k+q)_j (\K+\Q)\cdot\Q \right]\,, \label{A00taileapp}
\\
i\bar{\mathcal{A}}_{0k} &= \mathcal{F}^{(I)}
\bigg[ \omega^3\delta_{jk} q_i + \omega (k+q)_i \left( k_j q_k - k_k q_j - (\K+\Q)\cdot\K\delta_{jk} -\frac{1}{c_d} (k+q)_j q_k \right)   \bigg]\,, \label{A0ktaileapp}
\\
i\bar{\mathcal{A}}_{kl} &= \mathcal{F}^{(I)}
\bigg[  \omega^4\delta_{ik}\delta_{jl} + \omega^2 (k+q)_i \left( q_j\delta_{kl} - q_k\delta_{jl}-q_l\delta_{jk} \right) \nonumber\\
&\,\,\quad\qquad\qquad\qquad+ \frac{1}{c_d} (k+q)_i (k+q)_j \left[ k_k q_l + k_l q_k + 2q_k q_l - (\K+\Q)\cdot\Q \delta_{kl}\right] \bigg]\,. \label{Akltaileapp}
\end{align}

\subsection{Ward Identity checking}\label{appwardtestingtaile}

Now, in hands of the above components, we can test whether the condition $k^\mu\bar{\mathcal{A}}_{\mu\nu} = 0$ is satisfied. In this case, we obtain:

$\bullet$ For $\nu=0$:
\begin{align}
k^\mu \bar{\mathcal{A}}_{\mu 0} &= \omega \bar{\mathcal{A}}_{00} + k_k \bar{\mathcal{A}}_{0k} = -i\omega \mathcal{F}^{(I)} \left[
	 (k+q)_i (k+q)_j \left( \omega^2-\K^2 + \frac{1}{c_d}\Q^2\right) \right]\,.
\end{align}

$\bullet$ For $\nu=l$:
\begin{align}
k^\mu \bar{\mathcal{A}}_{\mu l} &= \omega \bar{\mathcal{A}}_{0l} + k_k \bar{\mathcal{A}}_{kl} = i\mathcal{F}^{(I)}\bigg[
	(k+q)_i \left[ \frac{1}{c_d} (k+q)_j (k+q)_l - \omega^2 \delta_{jl}\right]\Q^2 \nonumber\\
	&\qquad\qquad\qquad\qquad\qquad\! -(k+q)_i \left[ \frac{1}{c_d} (k+q)_j q_l - \omega^2 \delta_{jl}\right] \left( (\K+\Q)^2-\omega^2\right) 
	\bigg]\,.
\end{align}
These results have been conveniently organized in terms of structures that appear in the denominator of the one-loop integral in $\Q$, present in $\tilde{F}^{(I)}$, obtained from the identity
\begin{equation}
2\K\cdot\Q = [(\K+\Q)^2-\omega^2] - \Q^2 + (\omega^2 - \K^2)\,.
\end{equation} 
This will help us simplify the expressions for $k^\mu \bar{\mathcal{A}}_{\mu l}$ and $k^\mu \bar{\mathcal{A}}_{\mu 0}$. 
In particular, neglecting terms $(\omega^2 - \K^2)$, which vanish on-shell, the remaining expression will have either of the structures $[(\K+\Q)^2-\omega^2]$ or $\Q^2$. In the case of terms proportional to $[(\K+\Q)^2-\omega^2]$, the remaining integral in $\Q$ is just a scaless integral, that vanishes in dimensional regularization. As for terms proportional to $\Q^2$, we will always have the structure
\begin{equation}
\int_\Q \frac{(k+q)_{R}}{[(\K+\Q)^2-\omega^2]\Q^2} (k+q)_{i_1} \dots (k+q)_{i_n} \times \Q^2 = \int_\Q \frac{q_{R} \, q_{i_1} \dots q_{i_n}}{(\Q^2-\omega^2)} \,,
\end{equation}
where, in the passage, we have made the change of variables $\Q \rightarrow \Q - \K$.
Now, the integral on the right-hand side of this equation is simple enough that has a closed form, given by
\begin{equation}
\int_\Q \frac{q_{i_1} \dots q_{i_{2n}}}{(\Q^2-\omega^2)} =  \frac{(d-2)!!}{(d+2n-2)!!} \omega^{2n} \delta_{i_1\dots i_{2n}} I_0(\omega)\,.
\end{equation}
In this expression, $\delta_{i_1\cdots i_{2n}}$ is the generalized Kronecker delta introduced in Chapter \ref{ChaptEFT}, and $I_0(\omega)$ is the same integral with no momenta $q_i$ in the numerator:
\begin{equation}\label{I0omega}
I_0(\omega) \equiv \int_\Q \frac{1}{(\Q^2 - \omega^2)} = \frac{\Gamma(1-d/2)}{(4\pi)^{d/2}} (-\omega^2)^{d/2-1}  \overset{d=3}{=} -\frac{\sqrt{-\omega^2}}{4\pi}\,.
\end{equation}
The result in $d=3$ on the RHS takes the value of $i\omega/(4\pi)$ in the retarded boundary condition. 


As a consequence of all this, we show below that, for both components $k^\mu \bar{\mathcal{A}}_{\mu 0}$ and $k^\mu \bar{\mathcal{A}}_{\mu l}$, the terms proportional to $\Q^2$ lead to contractions in the indices of the electric multipole moment $I^{ijR}$, which vanishes (for any $r$) due to the fact that $I^{ijR}$ is traceless in all its indices: 
\begin{multline}
I^{ijR}(\omega) \int_\Q \frac{(k+q)_{R}}{[(\K+\Q)^2-\omega^2]\Q^2}\times (k+q)_i (k+q)_j \Q^2 \\
=I^{ijR}(\omega) \int_\Q \frac{q_i q_jq_{R}}{(\Q^2-\omega^2)} \propto I^{ijR}(\omega) \delta_{ijR} = 0\,,
\end{multline}
and for the $\nu=l$ case, we have
\begin{multline}
I^{ijR}(\omega) \int_\Q \frac{(k+q)_{R}}{[(\K+\Q)^2-\omega^2]\Q^2}\times (k+q)_i \left[ \frac{1}{c_d} (k+q)_j (k+q)_l - \omega^2 \delta_{jl}\right]\Q^2 \\
= I^{ijR}(\omega) \int_\Q \frac{q_i q_{R}}{(\Q^2-\omega^2)}\times\left[ \frac{1}{c_d} q_j q_l - \omega^2 \delta_{jl}\right]
\propto I^{ijR}(\omega)\left(a_1\delta_{ijlR} +a_2 \delta_{iR}\delta_{jl} \right) = 0\,.
\end{multline}

Therefore, we have that the condition $k^\mu \bar{\mathcal{A}}_{\mu\nu}=0$ is indeed satisfied for any of the multipoles of the electric type, in the simple tail process, provided we neglect terms that are vanishing on-shell, $\K^2-\omega^2$, as it is justified in Sec.~\ref{onshelness}.

\subsection{Simple tail for the magnetic multipole moments}\label{appwardtestingtailm}

In the magnetic case for the simple tail, we have
\begin{equation}
i\mathcal{A}^{(J)}_{E-\rm tail} = \frac{(-i)^{r+1}c_r^{(J)}}{4\Lambda^3} E J^{b|iRa}  \int_\Q \frac{(k+q)_R (k+q)_b}{ \left[ (\K+\Q)^2 - \omega^2 \right]  \Q^2 } \times \left( \beta_\sigma^{cd} \sigma^{*}_{cd} + \beta_A^c A^{*}_{c} + \beta_\phi \phi^{*}  \right)\,,
\end{equation}
with
\begin{align}
\beta_\sigma^{cd} &= \omega^3 \delta_{ad} \delta_{ic} + \frac 12 \omega  (k+q)_i (q_a \delta_{cd}  - 2 q_d \delta_{ac})\,, \\
\beta_A^{d} &= -\omega^2 (q_i \delta_{ad} + q_a \delta_{id}) + (k+q)_i (k+q)_c (k_d \delta_{ac} + k_c \delta_{ad} - k_a \delta_{cd})\,,\\
\beta_\phi &= \omega (k_a q_i + k_i q_a) - \omega (k+q)_i q_a \,. 
\end{align}

From this, in complete analogy to the steps that we followed in the electric case, we obtain
\begin{align}
i \bar{\mathcal{A}}_{00} &= \mathcal{F}^{(J)}\left[\omega (k_a q_i + k_i q_a) \right]\,, \\
i \bar{\mathcal{A}}_{0k} &= \mathcal{F}^{(J)}  \left[  -\omega^2 (q_i \delta_{ak} + q_a \delta_{ik}) + (k+q)_i (k+q)_c (k_c \delta_{ak} - k_a \delta_{ck}) \right]\,, \\
i \bar{\mathcal{A}}_{kl} &= \mathcal{F}^{(J)} \left[  -\omega^3 (\delta_{ak} \delta_{il} + \delta_{ik} \delta_{al})
+ \omega (k+q)_i (q_l \delta_{ak} + q_k \delta_{al} - \delta_{kl} q_a)   \right]\,,
\end{align}
where we have defined
\begin{equation}
\mathcal{F}^{(J)}[f] \equiv \frac{(-i)^{r}c_r^{(J)}}{8\Lambda^4} E J^{b|iRa}  \int_\Q \frac{(k+q)_R (k+q)_b}{ \left[ (\K+\Q)^2 - \omega^2 \right]  \Q^2 }  \times  f(\omega,\K,\Q)\,.
\end{equation}

Regarding the Ward identity, we have immediately that $k^\mu \bar{\mathcal{A}}_{\mu 0} = 0$, for the $\nu = 0$ component, whereas we obtain the following expression for the spatial component $\nu = l$:
\begin{equation}
k^\mu \bar{\mathcal{A}}_{\mu l} = -\frac{(-i)^{r+1}c_r^{(J)}}{8\Lambda^4} E J^{b|iRa}  \int_\Q \frac{(k+q)_R (k+q)_b}{ \left[ (\K+\Q)^2 - \omega^2 \right]  \Q^2 }  \times 
\omega \delta_{al} (k+q)_i (\omega^2 - 2\K\cdot\Q - \K^2)\,.
\end{equation}
By writing $(\omega^2 - 2\K\cdot\Q - \K^2) = \Q^2-[(\K+\Q)^2-\omega^2]$, it is easy to show that 
\begin{equation}
k^\mu \bar{\mathcal{A}}_{\mu l} \propto J^{b|iRa} \int_\Q \frac{q_R q_b q_i}{(\Q^2-\omega^2)} \rightarrow 0\,.
\end{equation}
Therefore, like in the electric case, we also have  $k^\mu \bar{\mathcal{A}}_{\mu \nu} = 0$ for the simple tail with arbitrary magnetic multipole moments.

\section{The angular momentum failed tail}

\subsection{$J$-failed tail for the electric multipole moments}\label{jtaileapp}

Now, for the $J$-failed tail, assuming arbitrary multipole moments of the electric type, we have
\begin{equation}\label{ALarbitraryXXX}
i\mathcal{A}^{(I)}_{J-\rm tail} = (-i)^r\frac{c_r^{(I)}}{2\Lambda^3}J^{b|a} I^{ijR}(\omega) \int_\Q \frac{q_a(k+q)_R}{[(\K+\Q)^2-\omega^2]\Q^2}\times
\left( \beta_{\sigma}^{cd} \sigma^{*}_{cd} + \beta_{A}^c A^{*}_c + \beta_{\phi} \phi^{*}\right)\,,
\end{equation}
with coefficients $\beta_{\sigma}^{cd}, \beta_{A}^{c}$, and $\beta_{\phi}$ given by the following expressions:
\begin{align}
\beta_{\sigma}^{cd} &= \frac 14 \omega^3 \left[ \delta_{ib}\delta_{cd} q_j + \left( 2\delta_{bc} q_i -2\delta_{ib} q_c + 2\delta_{ic} k_b \right) \delta_{jd} \right]  
-\frac 12 \omega (k+q)_i (k+q)_j q_c \delta_{bd}  \nonumber\\
	&-\frac 14 \omega (k+q)_j \left[ 2(k_c q_i - k_i q_c)\delta_{bd} + 2k_b q_c \delta_{id} -2(k+q)_c q_d \delta_{ib} \right.\nonumber\\
 &\qquad \qquad \qquad \qquad \qquad \qquad \left.-2\delta_{bd}\delta_{ic} (\K+\Q)\cdot\Q + \left( (\K+\Q)\cdot\Q \delta_{ib}-k_b q_i\right) \delta_{cd}
\right] \,, \label{betasigma} \\
\beta_A^c &= \frac 12 \omega^2 \left[ k_i q_j \delta_{bc} -(k_i q_c - k_c q_i)\delta_{jb} - k_b q_i \delta_{jc} + \K\cdot\Q \delta_{ib}\delta_{jc} \right] \nonumber\\
	&+\frac 12 \omega^2 (k+q)_j \left( q_c \delta_{ib} - q_i \delta_{bc} +2k_b \delta_{ic} \right)  
	-\frac 12 (k+q)_i (k+q)_j \left( k_b q_c - \K\cdot\Q \delta_{bc} \right) \,, \label{betaA} \\
\beta_\phi &= -\frac12 c_d \omega^3 \delta_{jb} q_i 
		+\frac12 c_d \omega (k+q)_j \left[ k_b q_i - (\K+\Q)\cdot \Q \delta_{ib} \right] -\omega (k+q)_i (k+q)_j k_b \,. \label{betaphi}
\end{align}

From the above expressions for $\beta_{\sigma}^{cd}$, $\beta_A^c$, and $\beta_\phi$, we can immediately identify the corresponding $\alpha_{\sigma}^{cd}$, $\alpha_{A}^{c}$, and $\alpha_{\phi}$ as in Eq.~\eqref{genericemissA}, and, hence, use the expressions for the scattering amplitude components $i \bar{\mathcal{A}}_{\mu\nu}$ given in terms of the $\alpha$'s,  Eqs.~\eqref{AbarwrtalphaS}, \eqref{AbarwrtalphaA}, \eqref{AbarwrtalphaP}, to derive: 
\begin{align}
i \mathcal{\bar{A}}_{00}^{L-\rm tail} &= \mathcal{F}_J^{(I)}\Big\{ \delta_{jb} \omega^3 q_i  + \omega (k+q)_j \left[ k_b(2k_i-q_i) + 3(\K+\Q)\cdot\Q \delta_{ib} \right]  \Big\}\,,
\label{LtailAbar00}\\
i \mathcal{\bar{A}}_{0k}^{L-\rm tail} &= \mathcal{F}_J^{(I)}\Big\{ 
(k+q)_i (k+q)_j ( k_b q_k - \K\cdot\Q \delta_{bk})  \nonumber\\	
&\qquad\qquad\qquad  - \Big[\K\cdot\Q \delta_{ib}\delta_{jk} - \delta_{bk} q_i q_j + q_j(k+q)_k \delta_{ib} + (2k+q)_j k_b \delta_{ik} \Big] \omega^2
\Big\}\,,	\label{LtailAbar0k}\\
i \mathcal{\bar{A}}_{kl}^{L-\rm tail} &= \frac 12 (\delta_{kc}\delta_{ld}+\delta_{kd}\delta_{lc}) \times \mathcal{F}_J^{(I)}  
\Big\{ \omega (k+q)_j \Big[ k_b q_i \delta_{cd} -2(k+q)_c q_i  \delta_{bd} + 2(k+q)_c q_d \delta_{ib} \nonumber\\	
&\qquad\qquad\qquad - (\K+\Q)\cdot\Q \delta_{cd} \delta_{ib} - 2k_b q_d \delta_{ic} +2(\K+\Q)\cdot\Q \delta_{bd} \delta_{ic} \Big] \nonumber\\
&\qquad\qquad\qquad\qquad\qquad\qquad\quad\, + \omega^3 \Big[ q_j \delta_{ib}\delta_{cd} + 2(q_i \delta_{bc} - q_c \delta_{ib} + k_b \delta_{ic}) \delta_{jd} \Big] \Big\}\,, \label{LtailAbarkl}
\end{align}
from which the Ward identity can be tested. The functional $\tilde{F}_J^{(I)}$ used above is defined by
\begin{equation}
\mathcal{F}_J^{(I)}[f] \equiv (-i)^{r+1} \frac{c_r^{(I)}}{8\Lambda^4}J^{b|a} I^{ijR}(\omega) \int_\Q \frac{q_a(k+q)_R}{[(\K+\Q)^2-\omega^2]\Q^2} \times f(\omega,\K,\Q)\,.
\end{equation}

Similarly, we can plug the coefficients $\alpha$'s, obtained from the $\beta$'s above, into Eqs.~\eqref{nu0} and \eqref{num} to directly obtain expressions for $k^\mu \bar{\mathcal{A}}_{\mu 0}$ and $k^\mu \bar{\mathcal{A}}_{\mu l}$. Of course, since we have also constructed expressions for the components $i\bar{\mathcal{A}}_{\mu\nu}$, we can also use them directly. We obtain:

$\bullet$ For $\nu=0$:
\begin{align}
	k^\mu \mathcal{\bar{A}}_{\mu 0} &= \omega \mathcal{\bar{A}}_{00} + k_k \mathcal{\bar{A}}_{0k} \nonumber\\
		&= -i\omega^2  \delta_{ib} \mathcal{F}_J^{(I)}\Bigl[(k+q)_j (2\K\cdot\Q+3\Q^2) + q_{j}\left(\omega^{2}-\K^{2}\right)\Bigr]\,.
\end{align}
\indent$\bullet$ For $\nu=l$:
\begin{align}
	k^\mu \mathcal{\bar{A}}_{\mu l} &= \omega \mathcal{\bar{A}}_{0l} + k_k \mathcal{\bar{A}}_{kl} \nonumber\\
	&= i\omega \mathcal{F}_J^{(I)}  \bigg\{ 2\omega^2 \K\cdot\Q \delta_{ib} \delta_{jl}  +(k+q)_j (q_i \delta_{bl}-q_l\delta_{ib}) [(\K+\Q)^2-\omega^2] \nonumber\\
&\qquad\qquad\qquad\qquad+(k+q)_j  (k_l \delta_{ib} - k_b \delta_{il} - k_i \delta_{bl} - q_i \delta_{bl}+q_l\delta_{ib}) \Q^2  \bigg\}\,.
\end{align}

Now, we analyze this result in the same way we did in the cases above, for the mass tail. Namely, we rewrite the inner product $\K\cdot\Q$ in terms of denominators that appear in the one-loop integral for the tails and simplify the expressions. In particular, for the $\nu = 0$ component, we are left with just a term with the following behavior: 
\begin{align}
k^\mu \bar{\mathcal{A}}_{\mu\nu} \sim J^{i|a} I^{ijR}(\omega) \int_\Q \frac{(q-k)_a q_j q_R}{(\Q^2-\omega^2)}. 
\end{align}
The integral in $\Q$ on the right-hand side of this expression has two terms, proportional to $\delta_{ajR}$ and $\delta_{jR}$, that lead to zero when contracted to $J^{i|a} I^{ijR}$, irrespective of the value for $r$. For $\nu=l$, the complete remaining expression, after simplification, is given by
\begin{equation}
	k^\mu \mathcal{\bar{A}}_{\mu l} = (-i)^{r+1}\frac{c_r^{(I)}}{2\Lambda^4} \left( \frac{i\omega}{4} \right)  
\bigg[ k_a \omega^2 J^{i|a} I^{iRl}(\omega) \int_\Q \frac{q_R}{(\Q^2-\omega^2)}  \bigg]\,.
\end{equation}
Hence, we notice that, since the integral in $\Q$ is proportional to $\delta_R$, this result vanishes on account of the tracelessness of $I^{iRl}$, unless $r=0$, in which case we have:
\begin{align}
	k^\mu \mathcal{\bar{A}}_{\mu l}\big|_{r=0} &= \frac{1}{16\Lambda^4}   
k_a \omega^3 J^{i|a} I^{il}(\omega) \int_\Q \frac{1}{(\Q^2-\omega^2)} \nonumber\\
	&= 16\pi i G_N^2   k_a \omega^4 J^{i|a} I^{il}(\omega) \,.
\end{align}

In summary, we have shown that, while the Ward identity is satisfied for all electric multipole moments with $r \ge 1$, is is not for the quadrupole. For the latter, we have
\begin{align}
	k^\mu \mathcal{\bar{A}}_{\mu 0} \big|_{r=0} &= 0\,, \\
	k^\mu \mathcal{\bar{A}}_{\mu l} \big|_{r=0} &= 16\pi i G_N^2   k_a \omega^4 J^{i|a} I^{il}(\omega) \,.
\end{align}

\subsection{$J$-failed tail for the magnetic multipole moments}\label{jtailmapp}

For the magnetic case, we have
\begin{align}\label{offemJL}
i\mathcal{A}^{(J)}_{J-\rm tail} = (-i)^r \frac{c_r^{(J)}}{8\Lambda^3} J^{l|k} J^{b|iRa}(\omega) \int_\Q \frac{(k+q)_R (k+q)_b }{[(\K+\Q)^2 - \omega^2]\Q^2}  \times q_k 
\left( \beta_{\sigma}^{cd} \sigma^{*}_{cd}  + \beta_A^c A^{*}_c  + \beta_\phi \phi^{*} \right)\,,
\end{align}
with $\beta$ coefficients given by
\begin{align}
\beta_{\sigma}^{cd} &= 
\frac12 \omega^2 \left[ (q_i \delta_{al} - k_a \delta_{il}) \delta_{cd} + 2(q_i \delta_{ac} - k_a \delta_{ic}) \delta_{ld}
-2 q_c (\delta_{il} \delta_{ad} + \delta_{al}\delta_{id}) + 4 k_l \delta_{ic} \delta_{ad} \right] \nonumber\\
&\qquad -(k+q)_i \left[ k_l q_c \delta_{ad} - (k+q)_c q_d \delta_{al} - k_a (k+q)_c \delta_{ld} - (\K+\Q)\cdot \Q \delta_{ac} \delta_{ld} \right] \nonumber\\
&\qquad -\frac12 (k+q)_i \left[ k_a k_l + (\K+\Q)\cdot \Q \delta_{al} \right] \delta_{cd}\,, \label{jtailmsigmaapp}\\
\beta_{A}^{c} &= 
\omega \left[ (2k_i + q_i)k_l \delta_{ac} + (k+q)_c q_i \delta_{al} + (k_a k_l + \K\cdot\Q \delta_{al}) \delta_{ic} \right. \nonumber\\
& \qquad \left. +(\K\cdot\Q \delta_{ac} - k_a(k+q)_c) \delta_{il} + 2 k_a q_i \delta_{lc} \right]\,, \label{jtailmaapp} \\
\beta_\phi &= -c_d \left[ (q_i \delta_{al} - k_a \delta_{il}) \omega^2 + (k+q)_i \left( k_a(k+q)_l + (\K+\Q)\cdot \Q \delta_{al} \right)\right]\,. \label{jtailmphiapp}
\end{align}
From this, we can derive the expressions for the components $\bar{\mathcal{A}}_{\mu\nu}$. In this case, we have:
\begin{align}
i \bar{\mathcal{A}}_{00} &= \mathcal{F}_J^{(J)} 
\Big\{ \omega^2 (q_i \delta_{al} - k_a \delta_{il}) + (k+q)_i \left[ k_a (3k+2q)_l + 3 (\K+\Q)\cdot\Q \delta_{al}\right] \Big\}\,, \label{A00jtailmapp}\\
i \bar{\mathcal{A}}_{0c} &= - \mathcal{F}_J^{(J)} \Big\{ \omega \left[
(2k+q)_i k_l \delta_{ac} +(k+q)_c q_i \delta_{al} + (k_a k_l + \K\cdot\Q \delta_{al}) \delta_{ic} \right.\nonumber\\
&\qquad\qquad\quad \left. + (\K\cdot\Q \delta_{ac}-k_a(k+q)_c) \delta_{il} + 2k_a q_i \delta_{lc} \right] \Big\}\,, \label{A0kjtailmapp}\\
i \bar{\mathcal{A}}_{cd} &=  \mathcal{F}_J^{(J)} \Big\{
(k+q)_i [ (k_c q_d+k_d q_c) \delta_{al} + 2 q_c q_d \delta_{al} + k_a (k+q)_d \delta_{lc} + k_a (k+q)_c \delta_{ld} \nonumber\\
&\qquad\qquad\qquad\qquad+ (\K+\Q)\cdot\Q (\delta_{ad}\delta_{lc} + \delta_{ac}\delta_{ld}) - k_l (q_c \delta_{ad} + q_d \delta_{ac}) ]  \nonumber\\
&\qquad - \omega^2 [ (q_c \delta_{id} + q_d \delta_{ic}) \delta_{al} + (q_c \delta_{ad} + q_d \delta_{ac}) \delta_{il} -q_i (\delta_{ac}\delta_{ld}+\delta_{ad}\delta_{lc}) ] \nonumber\\
&\qquad\qquad\qquad\qquad+ k_a (\delta_{ic}\delta_{ld}+\delta_{id}\delta_{lc}) -2 k_l (\delta_{ad}\delta_{ic}+\delta_{ac}\delta_{id}) ] \nonumber\\
&\qquad +\delta_{cd}[
\omega^2 (q_i \delta_{al} - k_a \delta_{il}) - k_i \delta_{al} (\K+\Q)\cdot\Q - q_i (\K+\Q)\cdot\Q \delta_{al} - (k+q)_i k_a k_l ] \Big\} \,, \label{Akljtailmapp}
\end{align}
where we have defined
\begin{equation}
\mathcal{F}_J^{(J)}[f] = (-i)^{r+1} \frac{c_r^{(J)}}{16\Lambda^4} J^{l|k} J^{b|iRa}(\omega) \int_\Q \frac{(k+q)_R (k+q)_b q_k}{[(\K+\Q)^2 - \omega^2]\Q^2} \times f(\omega,\K,\Q)\,. 
\end{equation}
Now, for the Ward identity $k^\mu\bar{\mathcal{A}}_{\mu\nu} =0$, using the above expressions, we obtain

$\bullet$ For $\nu = 0$:
\begin{align}
k^\mu\bar{\mathcal{A}}_{\mu 0} &= -i \omega \mathcal{F}_J^{(J)}\Big\{  (\K^2-\omega^2) (k_a \delta_{il} - q_i \delta_{al}) + (k+q)_i (2\K+3\Q)\cdot\Q \delta_{al}  \Big\}\,.
\label{eq:m-l-tail-time}
\end{align}

$\bullet$ For $\nu = d$:
\begin{align}
k^\mu\bar{\mathcal{A}}_{\mu d} &=i \mathcal{F}_J^{(J)} \Big\{
(k+q)_i (q_d \delta_{al} + k_a \delta_{ld}) (\omega^2-\K^2) + 2 \omega^2 (\delta_{al} \delta_{id} + \delta_{ad} \delta_{il}) \K\cdot\Q \nonumber\\
&\qquad\quad -(k+q)_i [
(k_l \delta_{ad} - k_d \delta_{al}) \Q^2 + 2 q_d \delta_{al}\K\cdot\Q + k_a \delta_{ld} (2\K + \Q)\cdot\Q
]
\Big\}\,.
\end{align} 
Analyzing these results, we end up with the same qualitative behavior we found in the electric case, namely that no anomalies are present for $r\ge 1$, while the Ward identity is violated in the quadrupole case, in which case we have the following result:
\begin{align}
k^\mu\bar{\mathcal{A}}_{\mu 0} &=0\,, \\
k^\mu \bar{\mathcal{A}}_{\mu d} &= -\frac{128}{45}i G_N^2 \pi J^{l|k} J^{b|ia}(\omega) (\delta_{al}\delta_{id} + \delta_{ad} \delta_{il}) \delta_{bk} \omega^5\,.
\end{align}


%% file: appendices/eom.tex
\chapter{Equations of motion to third order in the fields}\label{eomapp}

In this appendix we present the expressions for the Einstein-Hilbert action expanded to third order in the fields and derive the equation of motion from this action. We also show the explicit expressions for the equations of motion evaluated on the angular momentum failed tail for the electric quadrupole moment, showing that, indeed, this provides a solution to the derived equations of motion. 

To this end, we use the Mathematica package \textit{xPert} \cite{xpert}, with the metric expanded as 
\begin{equation}
g_{\mu\nu} = \eta_{\mu\nu} + h_{\mu\nu} + \frac 12 h^{(2)}_{\mu\nu}
\end{equation}
to accommodate the Kaluza-Klein decomposition. In this case, the software yields 
\begin{align}
S_{\rm EH+GF} = \Lambda^2 \int d^4 x\, \bigg\{&
 -\frac 12 h_{ab,c} h^{ab,c} + \frac 14 h_{,b} h^{,b} \nonumber\\
&+ \frac 12 h^{ab} h^{ce}{}_{,a} h_{ce,b} - \frac14 h^{ab} h^c{}_{c,a} h^e{}_{e,b} -\frac12 h^{ab} h^e{}_{e,c} h_{ab}{}^{,c} + \frac18 h^a{}_a h^e{}_{e,c} h^b{}_b{}^{,c} \nonumber \\
&+ h^{ab} h_a{}^c{}_{,c} h_b{}^e{}_{,e} + 2h^{ab} h_a{}^c{}_{,b} h_c{}^e{}_{,e} - \frac 12 h^a{}_a h^{bc}{}_{,b} h_c{}^e{}_{,e} - 2 h^{ab} h_{ce,b} h_a{}^{c,e} \nonumber \\
&- h^{ab} h_{be,c} h_a{}^{c,e} + h^{ab} h_{bc,e} h_a{}^{c,e} + \frac 12 h^a{}_a h_{be,c} h^{bc,e} - \frac 14 h^a{}_a h_{bc,e} h^{bc,e} \nonumber \\
&+ \frac 14 h^a{}_{a,b} h^{(2)}{}^c{}_c{}^{,b} + h^{ab}{}_{,a} h^{(2)}_b{}^c{}_{,c} - h^{ab,c} h^{(2)}_{ac,b} - \frac 12 h^{ab,c} h^{(2)}_{ab,c} 
\bigg\} \,.\label{SEHGexpandedxxx}
\end{align}
Hence, in terms of the Kaluza-Klein variables $\sigma, A, \phi$, we have
\begin{equation}
h_{\mu\nu}=\frac{1}{\Lambda} \left(
  \begin{array}{ccc}
  -2\phi & & A_i\\
  A_j    & & \sigma_{ij} - (c_d - 2)\delta_{ij} \phi
  \end{array}
\right)\,,
\end{equation}
for the linear term order, and 
\begin{align}
h^{(2)}_{\mu\nu}&=\frac{1}{\Lambda^2} \left(
  \begin{array}{ccc}
  -4\phi^2 & & 4\phi A_i\\
  4\phi A_j    & & (c_d-2)\left[ (c_d-2)\delta_{ij}\phi^2 -2\phi \sigma_{ij} \right] - 2A_i A_j
  \end{array}
\right)\nonumber\\
&=\frac{1}{\Lambda^2} \left(
  \begin{array}{ccc}
  -h_{00}^2 & & -2h_{00}h_{0i}\\
  -2h_{00}h_{0i} & & \frac 14(c_d-2)h_{00}\left[ 4h_{ij} -(c_d-2)\delta_{ij}h_{00} \right] - 2 h_{0i} h_{0j}
  \end{array} 
\right)\,,
\end{align}
for the second-order terms in the Kaluza-Klein fields.

We then use the the xPert package again, this time to perform variations of the action, obtaining
\begin{align}\label{boxheqN}
\Box \bar{h}_{ab} & + \frac 12 \eta_{ab} h^{ce} h_{,ce} + \frac 12 h^{ce}{}_{,a} h_{ce,b} -  \frac 14 h_{,a} h_{,b} -\frac12 h_a{}^c{}_{,b} h_{,c} -\frac 12 h_b{}^c{}_{,a} h_{,c} + \frac 12 h_{,c} h_{ab}{}^{,c} -\frac 18  \eta_{ab} h_{,e} h^{,e} \nonumber\\
&+ h_a{}^{c,e} h_{be,c} - h_a{}^{c,e} h_{bc,e}- h_a{}^{c,e} h_{ce,b} - h_b{}^{c,e} h_{ce,a} - h^{ce} h_{ab,ce} - h_a{}^c{}_{,c} h_b{}^d{}_{,d}  + h_a{}^c{}_{,b} h_c{}^d{}_{,d}  \nonumber\\
&+ \frac 14 \eta_{ab} h_{ce,f} h^{ce,f} 
+ \frac 12 \eta_{ab} h{}^{,c} h_c{}^{d}{}_{,d} - \frac12 \eta_{ab} h^{cd}{}_{,c} h_d{}^{e}{}_{,e} +\frac12  h_{,a} h_b{}^c{}_{,c} +\frac12  h_{,b} h_a{}^c{}_{,c}   + \frac 12 \eta_{ab} h_{cf,e} h^{ce,f} \nonumber\\
&+ h_b{}^c{}_{,a} h_c{}^d{}_{,d} - h_{ab}{}^{,c} h_c{}^d{}_{,d}  +\frac 12 \left( h^{(2)}_{ab} - \frac 12 \eta_{ab} h^{(2)} \right)_{,c}{}^{,c} = 0\,.
\end{align}
Let us then rewrite this equation in the following more compact form, using a notation by Blanchet \cite{blanchetlivrev}:
\begin{equation}\label{heqNhhxxx}
\Box \bar{h}_{\mu\nu} = N_{\mu\nu}[h,h]\,.
\end{equation}

Thus, as mentioned before, in terms of perturbative expansions, this equation should be understood in the following sense: the $h_{\mu\nu}$ on the left-hand side of this equation represents processes of order $G_N^2$, while the two $h$ on the right-hand side are each of order $G_N$.

\subsubsection{Checking the $J$-failed tail solution}

For the electric $J$-failed tail, the $h_{\mu\nu}$ at order $G_N^2$ that enters the LHS of Eq.~\eqref{heqNhhxxx} is the one constructed out of components given in Eqs.~\eqref{LtailAbar00}, \eqref{LtailAbar0k}, and \eqref{LtailAbarkl}. As for the lower-order processes (i.e., of order $G_N$) that enter the right-hand side of Eq.~\eqref{heqNhhxxx}, we consider the $J$ coupling given in Eq.~\eqref{Lcoupling} and the electric quadrupole's in Eqs.~\eqref{hijIij}, with $r = 0$.

Below we present the results, which turn out to have matching expressions, either by computing $\Box \bar{h}_{\mu\nu}$ using the electric $J$-failed tail solution or $N[h,h]_{\mu\nu}$ from the lower-order solutions:\\

$\bullet$ For the $00$ component:
\begin{multline}
\Box \bar{h}_{00} = N_{00}[h,h] =  \frac{i}{16\Lambda^4} \frac{1}{4\pi r^3} J^{b|a} \int_\K\frac{d\omega}{2\pi}\frac{e^{-i\omega t+i \K\cdot \X}}{\K^2-\omega^2}  \times \omega I^{ij}(\omega)\\
\times\left[ -2i k_i k_j k_b r^a -3 k_j k_b (\delta_{ai} - 3 \hat{r}_a \hat{r}_i) +3 k_j k_c \delta_{ib} (\delta_{ac} - 3 \hat{r}_a \hat{r}_c) - 3 \delta_{jb} \omega^2  \hat{r}_i  \hat{r}_a
\right]\,.
\end{multline}

$\bullet$ For the $0k$ component:
\begin{multline}
\Box \bar{h}_{0k} = N_{0k}[h,h] = \frac{1}{16\Lambda^4} \frac{1}{4\pi r^3} J^{b|a} \int_\K\frac{d\omega}{2\pi}\frac{e^{-i\omega t+i \K\cdot \X}}{\K^2-\omega^2} I^{ij}(\omega) \times \Big\{ -2 \omega^2 \delta_{ik} k_j k_b r^a \\
+ \frac{3}{r} \left[ (\delta_{ai} - 3\hat{r}_a \hat{r}_i)\hat{r}_j + (\delta_{aj} - 3\hat{r}_a \hat{r}_j)\hat{r}_i + (\delta_{ij} - 3\hat{r}_i \hat{r}_j)\hat{r}_a 
+ 4 \hat{r}_a \hat{r}_i \hat{r}_j \right] \omega^2 \delta_{bk} \\
+ i \left[   (\delta_{aj} - 3\hat{r}_a \hat{r}_j) k_b \delta_{ik} -  (\delta_{al} - 3\hat{r}_a \hat{r}_l) k_l \delta_{ib} \delta_{jk} -  (\delta_{aj} - 3\hat{r}_a \hat{r}_j) k_k \delta_{ib} \right] \omega^2 \\
+ \GA{i} \left[   (\delta_{ak} - 3\hat{r}_a \hat{r}_k) k_i k_j k_b - (\delta_{al} - 3\hat{r}_a \hat{r}_l) k_i k_j k_l \delta_{bk} \right]
\Big\}\,.
\end{multline}

$\bullet$ For the $kl$ component:
\begin{multline}
\Box \bar{h}_{kl} = N_{kl}[h,h] = \frac{1}{16\Lambda^4} \frac{1}{4\pi r^3} J^{b|a} \int_\K\frac{d\omega}{2\pi}\frac{e^{-i\omega t+i \K\cdot \X}}{\K^2-\omega^2} I^{ij}(\omega) \times \\
\Big\{ 
2 \omega^3 \delta_{ik} \delta_{jl} k_b r^a  + i \omega \delta_{kl} ( \omega^2 \delta_{ib} \delta_{jn} + \delta_{in} k_j k_b - \delta_{ib} k_j k_n ) (\delta_{an} - 3 \hat{r}_a \hat{r}_n ) 
\\
+ i \omega \big[ \delta_{bk} \left( \omega^2 \delta_{il} \delta_{jm} + \delta_{il} k_j k_m - \delta_{im} k_j k_l \right) ( \delta_{am} - 3\hat{r}_a \hat{r}_m ) \\
- \left( \omega^2 \delta_{ik} \delta_{jb} + \delta_{ik} k_j k_b - \delta_{ib} k_j k_k \right) ( \delta_{al} - 3\hat{r}_a \hat{r}_l )  + (k \leftrightarrow l)
\big]
\Big\}\,.
\end{multline}
Thus, we have explicitly checked that, for our solution of the electric $J$-failed tail, we indeed have $\Box \bar{h}_{\mu\nu} = N_{\mu\nu}[h,h]$. The same also holds true for the magnetic-type solution.

%% file: appendices/integrals.tex
\chapter{Useful integrals}\label{sec:app}

In this appendix, we present the simple one-loop master integrals important for the computations of the tails and tails of tails in Chapters \ref{wardidentity} and \ref{renormachapter}. 
In particular, all the integrals involved in these processes (apart from the ones appearing in the tail-of-tail diagram (c); See below), can be derived (eventually after iteration) from the following standard one-loop scalar master integrals:
\begin{align}
&\ds J_{ab}(\Q) \equiv \ds\int_\Pp\frac 1{\Pp^{2a}\pa{\Pp-\Q}^{2b}}=\ds(\Q^2)^{d/2-a-b}\frac{\Gamma(a+b-d/2)\Gamma(d/2-a)\Gamma(d/2-b)}{(4\pi)^{d/2}\Gamma(a)\Gamma(b)\Gamma(d-a-b)}\,, \label{eq:app_std1}\\
&\ds I_a(\omega)\equiv \int_\Q\frac 1{\paq{\pa{\K-\Q}^2}^a[\Q^2-\pa{\omega+i\ta}^2]}
=\paq{-\pa{\omega+i\ta}^2}^{d/2-a-1}\dfrac{\Gamma(a+1-d/2)\Gamma(d-2a-1)}
{\pa{4\pi}^{d/2}\Gamma(d-a-1)}\,, \label{eq:app_std2}
\end{align}
where in the $I_a$ equation it is understood that $\K^2=\omega^2$.
The eventual presence of tensorial structures at the numerator is accounted by
the usual scalarization procedure plus some combinatorics.
For instance, borrowing notation from Ref. \cite{smirnov2004},
\begin{align}
&\ds\int_\Q\frac{q_{i_1}\dots q_{i_n}}{\paq{\pa{\K-\Q}^2}^a\paq{\Q^2-(\omega+i\ta)^2}^b}=
  \sum_{m=0}^{[n/2]}S_{a,b}(n,m)\,, \label{eq:Strick}\\
&\ds S_{a,b}(n,m)\equiv
  \frac{\paq{-\pa{\omega+i\ta}^2}^{d/2-a-b+m}}{2^m\pa{4\pi}^{d/2}}
  \frac{\Gamma(a+b-d/2-m)\Gamma(a+n-2m)\Gamma(d+2m-2a-b)}{\Gamma(a)\Gamma(b)\Gamma(d+n-a-b)} \nonumber\\
&\qquad\qquad\qquad\times \{[\delta]^m[k]^{n-2m}\}_{i_1\cdots i_n}\,,  
\end{align}
where $\{[\delta]^m[k]^{n-2m}\}_{i_1\cdots i_{n}}$ is symmetric in its $n$ indices,
and involves $m$ Kronecker deltas and $n-2m$ occurrences of $k$ vectors, and $[n/2]$ is the integer part of $n/2$.

%% file: appendices/tailoftail.tex
\chapter{Details for computation of the tail-of-tail diagrams (a), (b), and (c)}\label{apptailoftailabc}

In this appendix, we present the extensive expressions for diagrams (a), (b), and (c) of the tail-of-tail process. 
To write the amplitude of the diagrams in Figs. 3(a) and 3(b)
we preliminarily define
\begin{align}
\tilde{\delta}_{abcd}&\equiv \delta_{ac}\delta_{bd}+\delta_{ad}\delta_{bc}-\frac{2}{d-2}\delta_{ab}\delta_{cd}\,,\\
D^{(1)}_{abcd} &= \delta_{ab}\delta_{cd}-\frac 12 \delta_{ac}\delta_{bd}\,,\\
D^{(2)}_{abcdef} &= \frac 14 \delta_{ab}\delta_{ce}\delta_{df}+\frac 12 \delta_{cd}\delta_{ae}\delta_{bf}-\delta_{ac}\delta_{be}\delta_{df}\,,
\\
D^{(3)}_{abcdefmn} &= -\frac 14 \delta_{ab}\delta_{mn}\delta_{ce}\delta_{df}-\frac{1}{2}\delta_{mn}\delta_{cd}\delta_{ae}\delta_{bf} + \delta_{mn}\delta_{ac}\delta_{be}\delta_{df} + \frac 12 \delta_{am}\delta_{bn}\delta_{ce}\delta_{df}
\nonumber\\
&+D^{(1)}_{arbs}\times
(\delta_{rc}\delta_{se}\delta_{mf}\delta_{nd}-\delta_{rc}\delta_{se}\delta_{md}\delta_{nf}
+\delta_{rm}\delta_{se}\delta_{cd}\delta_{nf}-\delta_{rc}\delta_{sm}\delta_{ef}\delta_{nd})\,,\\
D^{(4)}_{abcdef} & = 2 \delta_{ae}\delta_{bc}\delta_{df} -\delta_{ad}\delta_{be}\delta_{cf} -2\delta_{af}\delta_{be}\delta_{cd} +\delta_{af}\delta_{bc}\delta_{de}\,.
\end{align}
In diagram (a) the propagator labeled by $q$ can carry a $\sigma$ or an $A$
polarization (the others are fixed, as only $\phi$ couples to the conserved energy $E$), the two separate contributions being
\begin{align}
i \mathcal{A}^{(e)}_{a\,,\sigma}(\omega\,,\K) &= \frac{(-i)^{r+1}E^2 c_{r}^{(I)}}{16 c_d\Lambda^5} \omega^2 I^{iji_1\cdots i_r}(\omega)
\int_{\Q}\frac {q_{i_1}\cdots q_{i_r}}{\pa{\K-\Q}^2[\Q^2-\pa{\omega+i\ta}^2]}
\int_{\pp}\frac 1{\pa{\pp-\Q}^2\pa{\K-\pp}^2} \nonumber \\
&\quad \times \pa{p-q}_{\beta}(p-k)_{\delta}D^{(1)}_{\alpha\beta\gamma\delta}\nonumber\\
&\quad\times
\Bigg\{ D^{(2)}_{abcdef}\omega^2 \tilde{\delta}_{ab\alpha\gamma}
[\delta_{ic}\delta_{jd}\sigma^{*}_{ef}(\omega\,,\K)+\delta_{ie}\delta_{jf}\sigma^{*}_{cd}(\omega\,,\K)] \\
&\qquad +D^{(3)}_{abcdefmn}\bigg[-\delta_{ia}\delta_{jb}(\pa{q-k}_m k_n \tilde{\delta}_{cd\alpha\gamma}\sigma^{*}_{ef}(\omega\,,\K)+k_m\pa{q-k}_n  \tilde{\delta}_{ef\alpha\gamma}\sigma^{*}_{cd}(\omega\,,\K))    \nonumber\\
&\qquad\qquad\qquad\qquad\qquad+\delta_{ic}\delta_{jd}q_m (\tilde{\delta}_{ab\alpha\gamma}k_n\sigma^{*}_{ef}(\omega\,,\K)+\tilde{\delta}_{ef\alpha\gamma}\pa{q-k}_n\sigma^{*}_{ab}(\omega\,,\K))      \nonumber\\
&\qquad\qquad\qquad\qquad\qquad+\delta_{ie}\delta_{jf}q_n (\tilde{\delta}_{ab\alpha\gamma}k_m\sigma^{*}_{cd}(\omega\,,\K)+\tilde{\delta}_{cd\alpha\gamma}\pa{q-k}_m\sigma^{*}_{ab}(\omega\,,\K))
\bigg]
\Bigg\}\nn
\end{align} 
and
\begin{align}
  i \mathcal{A}^{(e)}_{a\,,A}(\omega\,,\K) &= \frac{-(-i)^{r+1}E^2 c_{r}^{(I)}}{32 c_d\Lambda^5} \omega^2\! I^{iji_1\cdots i_r}(\omega)
\int_{\Q}\frac {q_{i_1}\cdots q_{i_r}}{\pa{\K-\Q}^2[\Q^2-\pa{\omega+i\ta}^2]}\int_{\pp}\frac 1{(\pp-\Q)^2\pa{\K-\pp}^2} \nonumber \\
&\quad \times
D^{(1)}_{\alpha\beta\gamma\delta} D^{(4)}_{abcdei}
q_j [\pa{p-q}_{\beta}(p-k)_{\delta}+(p-q)_{\delta}(p-k)_{\beta}] (q-k)_c \tilde{\delta}_{\alpha\gamma de}
\sigma^{*}_{ab}(\omega\,,\K)\,.
\end{align}
Similarly, for the magnetic case
\begin{align}
  i \mathcal{A}^{(m)}_{a\,,\sigma}(\omega\,,\K) &= \frac{(-i)^{r+1}E^2 c_{r}^{(J)}}{64 c_d\Lambda^5} \omega\epsilon_{ikl} J^{jki_1\cdots i_r}(\omega)
\int_{\Q}\frac {q_{i_1}\cdots q_{i_r}}{(\K-\Q)^2[\Q^2-\pa{\omega+i\ta}^2]}\int_{\pp}\frac 1{\pa{\pp-\Q}^2(\K-\pp)^2} \nonumber \\
&\quad \times   D^{(1)}_{\alpha\beta\gamma\delta}q_l [(p-q)_{\beta}(p-k)_{\delta}+(p-q)_{\delta}(p-k)_{\beta}]\nonumber\\
&\qquad\times\Bigg\{ D^{(2)}_{abcdef}\omega^2 \tilde{\delta}_{ab\alpha\gamma}
[\tilde{\delta}_{ijcd}\sigma^{*}_{ef}(\omega\,,\K)+\tilde{\delta}_{ijef}\sigma^{*}_{cd}(\omega\,,\K)] \nonumber \\
&\qquad \qquad+D^{(3)}_{abcdefmn}\bigg[-\tilde{\delta}_{ijab}((q-k)_m k_n \tilde{\delta}_{cd\alpha\gamma}\sigma^{*}_{ef}(\omega\,,\K)+k_m(q-k)_n  \tilde{\delta}_{ef\alpha\gamma}\sigma^{*}_{cd}(\omega\,,\K))    \nonumber\\
  &\qquad\qquad\qquad\qquad\qquad+\tilde{\delta}_{ijcd}q_m (\tilde{\delta}_{ab\alpha\gamma}k_n\sigma^{*}_{ef}(\omega\,,\K)+\tilde{\delta}_{ef\alpha\gamma}(q-k)_n\sigma^{*}_{ab}(\omega\,,\K))\nn\\
  &\qquad\qquad\qquad\qquad\qquad+\tilde{\delta}_{ijef}q_n (\tilde{\delta}_{ab\alpha\gamma}k_m\sigma^{*}_{cd}(\omega\,,\K)+\tilde{\delta}_{cd\alpha\gamma}(q-k)_m\sigma^{*}_{ab}(\omega\,,\K))
\bigg]\Bigg\}
\end{align}
and 
\begin{align}
  i \mathcal{A}^{(m)}_{a\,,A}(\omega\,,\K)\! &=\! \frac{-(-i)^{r+1}E^2 c_{r}^{(J)}}{64 c_d\Lambda^5} \omega\epsilon_{ikl} J^{jki_1\cdots i_r}(\omega)
\int_{\Q}\!\frac {q_{i_1}\cdots q_{i_r}}{\pa{\K-\Q}^2\![\Q^2-\pa{\omega+i\ta}^2]}\!\!\int_{\pp}\!\frac 1{\pa{\pp-\Q}^2\!(\K-\pp)^2} \nonumber \\
&\quad \times D^{(1)}_{\alpha\beta\gamma\delta}\! D^{(4)}_{abcdei}
q_lq_j [\pa{p-q}_{\beta}(p-k)_{\delta}+\pa{p-q}_{\delta}(p-k)_{\beta}](q-k)_c \tilde{\delta}_{\alpha\gamma de}
\sigma^{*}_{ab}(\omega\,,\K)\,.
\end{align}
The calculation of diagram (b) is similar and gives
\begin{align}
i{\cal A}^{(e)}_b(\omega\,,\K) &= (-i)^{r+1} \left(\frac{E^2 c_r^{(I)}}{16\Lambda^5}\right)I^{iji_1\cdots i_r}(\omega)
\int_{\Q}\frac {q_{i_1}\cdots q_{i_r}}{[\Q^2-\pa{\omega+i\ta}^2]}
\int_{\pp}\frac 1{\pa{\pp-\Q}^2\pa{\K-\pp}^2}
\nonumber\\
&\Bigg\{\delta_{ib}\omega^4+\delta_{ib}\frac {\omega^2}{c_d}\pa{\pp-\Q}\cdot(\pp-\K) 
-\frac {4}{c_d}\omega^2 (p-q)_i(p-k)_b\Bigg\}\sigma_{bj}^{*}(\omega\,,\K)\,,
\end{align} 
\begin{align}
&i{\cal A}^{(m)}_b(\omega\,,\K) = (-i)^{r+1} \left(\frac{E^2 c_r^{(J)}}{16\Lambda^5}\right)\omega\epsilon_{ikl}J^{jki_1\cdots i_r}(\omega)
\int_{\Q}\frac {q_{i_1}\cdots q_{i_r}}{[\Q^2-\pa{\omega+i\ta}^2]}
\int_{\bf p}\frac 1{\pa{\pp-\Q}^2(\K-\pp)^2}
\nonumber\\
&\qquad \times q_l\Bigg\{\omega^2\delta_{ai}\delta_{bj}+\frac1{c_d}\pa{\pp-\Q}\cdot(\pp-\K)\delta_{ai}\delta_{bj}
-\frac2{c_d}(p-q)_a \left[(p-k)_i \delta_{bj}+(p-k)_j \delta_{bi}\right] \Bigg\}\sigma_{ab}^{*}(\omega\,,\K)\,.
\end{align}

Tail-of-tail amplitude (c) is more complicated, as it involves the following
family of two-loop integrals (always $\omega^2=\K^2$):
\be
I_{in}\paq{a_1,a_2,a_3,a_4,a_5}\equiv\int_{\pp,\Q}
\frac 1{\paq{\Q^2-\pa{\omega+i\ta}^2}^{a_1}\paq{\pp^2-\pa{\omega+i\ta}^2}^{a_2}\pa{\pp-\Q}^{2a_3} \pa{\pp-\K}^{2a_4}\pa{\Q-\K}^{2a_5}}\,.\nn\\
\ee
The general expression is long and complicated; here, we focus only on the  part which is singular in the $d\rightarrow 3$ limit, which is the relevant one in the renormalization procedure.
Using the standard technique of integration by parts implemented by the software Reduze \cite{vonManteuffel:2012np}, one can express the main scalar integral as
\be
\ba{rcl}
\ds I_{in}\paq{1,1,1,1,0}&=&\ds
\frac 1{\omega^2}\frac{3d-8}{4\pa{d-3}}\int_{\pp,\Q}
\frac 1{\pa{\pp^2-\pa{\omega+i\ta}^2}\pa{\pp-\Q}^2\pa{\pp-\K}^2}+\\
&&\ds\frac 1{4\omega^4}\frac{d^2+4d-4}{\pa{d-3}^2}
\int_{\pp,\Q}\frac 1{\pa{\Q^2-(\omega+i\ta)^2}\pa{\pp^2-(\omega+i\ta)^2}}\,.
\ea
\ee
When reducing tensor integral to scalar ones, the following results are needed, for $m\,,n\in\mathbb{N}$:
\be
\ba{rcl}
\ds I_{in}\paq{1,1,1,1,0}&\simeq& \ds-\paq{128 \pi^2 \pa{\omega+i\ta}^2\epsilon^2}^{-1}+O(\epsilon^0)\\
\ds I_{in}\paq{1,1,1,1,-n}&\simeq&\ds -\frac{\paq{4\pa{\omega+i\ta}^2}^n}{n}
  \frac{1}{128 \pi^2 \pa{\omega+i\ta}^2\epsilon}+O(\epsilon^0)\,\quad {\rm for}\ n\geq 1\\
I_{in}\paq{1, -m, 1, 1, -n}&\simeq&\ds
-\frac{(-1)^m \paq{\pa{\omega+i\ta}^2}^{m+n} \Gamma (m+2 n+1)}{64 \pi ^{3/2} \epsilon  \Gamma (n+1) \Gamma \left(m+n+\frac{3}{2}\right)}+O(\epsilon^0)\\
\ds I_{in}\paq{1,1,-m,1,-n}&\simeq&\ds -\frac{\paq{4\pa{\omega+i\ta}^2}^{m+n}}{32\pi^2\pa{m+n+1}\epsilon}+O(\epsilon^0)\\
\ds I_{in}\paq{1,1,1,-m,-n}&\simeq &\ds I_{in}\paq{1,1,-m,1,-n}+O(\epsilon^0)\\
\ds I_{in}\paq{1,1,0,0,-n}&\simeq &\ds O(\epsilon^0)\,.
\ea
\ee
From there, one can compute the only unknown parameter involved in the following equation:
\be
\label{eq:app}
\int_{\pp,\Q}\frac {q_{(i_1\cdots i_r)}}{{\cal D}_{(tail)^2}}\equiv\int_{\pp,\Q}\frac {q_{(i_1\cdots i_r)}}{\paq{\Q^2-\pa{\omega+i\ta}^2}\paq{\pp^2-\pa{\omega+i\ta}^2}\pa{\pp-\Q}^2 \pa{\pp-\K}^2}\simeq\frac{{\cal A}_r}{\omega^2} k_{(i_1\cdots i_r)}\,,
\ee
$k_{(i_1\cdots i_r)}$ being (still following the notation of Ref. \cite{smirnov2004}) the symmetric traceless (STF) combination  of $k^i$'s.
In detail:
\be
\ds\frac{{\cal A}_r}{\omega^2} k_{(i_1\cdots i_r)}\times k_{i_1}\cdots k_{i_r}&=&\ds {\cal A}_r C_r\pa{\omega^2}^{r-1}=\int_{\pp, \Q}\frac {q_{(i_1\cdots i_r)}k_{i_1}\cdots k_{i_r}}{{\cal D}_{(tail)^2}}\nn\\
&=&\ds \sum_{j=0}^{\paq{\frac r2}}b_{r,j}\pa{\omega^2}^j\int_{\pp,\Q}\frac{\pa{\Q^2}^j\pa{{\Q}\cdot\K}^{r-2j}}{{\cal D}_{(tail)^2}}\nn\\
&=&\ds\pa{\omega^2}^r \sum_{j=0}^{\paq{\frac r2}}\sum_{a_1=0}^{r-2j}b_{r,j}\pa{-2\omega^2}^{-a_1}\binom{r-2j}{a_1}I_{in}\paq{1,1,1,1,-a_1}\nn\\
&\simeq&\ds -\frac{\pa{\omega^2}^{r-1}C_r}{128 \pi^2}\paq{\frac1{\epsilon^2}-\frac{2 H_r}{\epsilon}}\,,
\ee
with $H_r$ the harmonic number and 
\be
b_{r,j} \equiv \frac{r!}{4^j j! (r-2j)! (2-r-d/2)_j}\,,\quad C_r\equiv \sum_{i=0}^{\paq{\frac r2}}b_{r,i}=\frac{\Gamma\pa{d+r-2}}{\pa{d-3}!!\pa{d+2r-4}!!}\,,\nn
\ee
$(a)_b$ being the Pochhammer symbol.
Moreover, one needs to compute
  integrals like the one above, with the addition of up to four $p_i$'s and up to two extra $q_j$'s (not involved in the STF combination with the other $q^{i_r}$'s), and this can be achieved
via a tedious but straightforward scalarization procedure.

For instance, for one extra $p_i$, one can write
\be
&&\int_{\pp, \Q}\frac {q_{(i_1\cdots i_r)}p_i}{{\cal D}_{(tail)^2}}=\frac{{\cal A}^{(p)}_r}{\omega^2}k_{(i_1\cdots i_r)k_i}+{\cal B}^{(p)}_r \delta_{i(i_1}k_{i_2\cdots i_r)}\,,
\ee
and two independent contractions are needed to solve the linear system. One is the same as above, while another can be obtained by contracting the index $i$ with one of the STF indices. The integrals are just slightly more complicated with
respect to the one needed in Eq. (\ref{eq:app}). Adding extra factors to the
integrand does not introduce insurmountable complications.

For one extra, non-STF, $q$ factor, one can proceed in the same way:
\be
&&\int_{\pp, \Q}\frac {q_{(i_1\cdots i_r)}q_j}{{\cal D}_{(tail)^2}}=\frac{{\cal A}^{(q)}_r}{\omega^2}k_{(i_1\cdots i_r)k_j}+{\cal B}^{(q)}_r \delta_{j(i_1}k_{i_2\cdots i_r)}\,,
\ee
and solve the associated linear system. Actually, by noticing that
\be
q_{(i_1\cdots i_r)}q_j=q_{(i_1\cdots i_rj)}+\frac r{d+2r -2}\Q^2\delta_{j(i_1}q_{i_2\cdots i_r)}
\ee
one can straightforwardly derive
\be
\int_{\pp, \Q}\frac {q_{(i_1\cdots i_r)}q_j}{{\cal D}_{(tail)^2}}&=&\int_{\pp, \Q}\frac {q_{(i_1\cdots i_rj)}}{{\cal D}_{(tail)^2}}+\frac r{d+2r -2}\int_{\pp, \Q}\Q^2\frac{\delta_{j(i_1}q_{i_2\cdots i_r)}}{{\cal D}_{(tail)^2}}\nn\\
&=&\frac{{\cal A}_{r+1}}{\omega^2}k_{(i_1\cdots i_r j)} +\frac r{d+2r -2}{\cal A}_{r-1} \delta_{j(i_1}k_{i_2\cdots i_r)}\\
&=&\frac{{\cal A}_{r+1}}{\omega^2}k_{(i_1\cdots i_r) j} +\frac r{d+2r -2}\paq{{\cal A}_{r-1}-{\cal A}_{r+1}} \delta_{j(i_1}k_{i_2\cdots i_r)}\nn\,.
\ee